\documentclass[apj,twocolumn,revtex4,twocolappendix]{emulateapj-rtx4}
\usepackage{natbib}
\bibpunct[:]{(}{)}{;}{a}{}{,}
\usepackage{graphicx}
\usepackage{bm}
\usepackage[dvips]{color}
\usepackage{ulem}
\usepackage{longtable}
\usepackage{lscape}
\usepackage{xspace}
\usepackage{subfigure}
\usepackage{MnSymbol}
\renewcommand\thefootnote{\alpha{footnote}}
\renewcommand{\arraystretch}{1.3}

\newcommand{\lirlx}{$L_{\rm K\alpha}/L_{\rm 10-50\hspace{1mm} keV}$}

\newcommand{\lirlxB}{$L_{\rm K\alpha}/L^{\rm BAT}_{\rm 10-50\hspace{1mm} keV}$}
\newcommand{\oiv}{[O\hspace{1mm}{\footnotesize IV}] }
\newcommand{\oiii}{[O\hspace{1mm}{\footnotesize III}] }
\shorttitle{{\it SUZAKU} OBSERVATIONS OF SWIFT/BAT SELECTED COMPTON-THIN AGNs} 
\shortauthors{Kawamuro et al.}

\begin{document}

\bibliographystyle{apj}

\title{
Suzaku Observations of Moderately Obscured (Compton-thin) Active Galactic Nuclei Selected 
by Swift/BAT Hard X-ray Survey 
}

\author{
 Taiki Kawamuro\altaffilmark{1},
 Yoshihiro Ueda\altaffilmark{1}, 
 Fumie Tazaki\altaffilmark{2}, 
 Claudio Ricci\altaffilmark{3,}\altaffilmark{4},
 Yuichi Terashima\altaffilmark{5} 
}

\altaffiltext{1}{Department of Astronomy, Kyoto University, Kyoto 606-8502, Japan}
\altaffiltext{2}{Mizusawa VLBI Observatory, National Astronomical Observatory of Japan,
Osawa, Mitaka, Tokyo 181-8588, Japan} 
\altaffiltext{3}{Instituto de Astrof\'{\i}sica, Facultad de F\'{i}sica, Pontificia Universidad Cat\'{o}lica de Chile, Casilla 306, Santiago 22, Chile} 
\altaffiltext{4}{EMBIGGEN anillo, Concepcion, Chile}
\altaffiltext{5}{Department of Physics, Ehime University, Matsuyama 790-8577, Japan}

\begin{abstract} 

We report the results obtained by a systematic, broadband (0.5--150 keV) X-ray
spectral analysis of moderately obscured (Compton-thin; $22 \leq \log
N_{\rm H} < 24$) active galactic nuclei (AGNs) observed with {\it Suzaku} and 
{\it Swift}/Burst Alert Telescope (BAT). Our sample consists 
of 45 local AGNs at $z<0.1$ with $\log L_{\rm 14-195\hspace{1mm}keV} > 42$ 
detected in the {\it Swift}/BAT 70-month survey, whose {\it Suzaku} archival data are 
available as of 2015 December. All spectra are uniformly fit with a baseline model composed of 
an absorbed cutoff power-law component, reflected emission accompanied by a
narrow fluorescent iron-K$\alpha$ line from cold matter (torus), and
scattered emission. Main results based on the above analysis are as follows. 
(1) The photon index is correlated with Eddington ratio, but not with luminosity
or black hole mass.
(2) The ratio of the iron-K$\alpha$ line to X-ray luminosity, a torus covering fraction 
indicator, shows significant anti-correlation with luminosity. 
(3) The averaged reflection strength derived from stacked spectra above 14 keV 
is larger in less luminous ($\log L_{\rm 10-50\hspace{1mm}keV} \leq 43.3$; 
$R= 1.04^{+0.17}_{-0.19}$) or highly obscured AGNs ($\log N_{\rm H} > 23$; 
$R = 1.03^{+0.15}_{-0.17}$) than in more luminous ($\log L_{\rm 10-50\hspace{1mm}keV} 
> 43.3$; $R= 0.46^{+0.08}_{-0.09}$) or lightly obscured
objects ($\log N_{\rm H} \leq 23$; $R = 0.59^{+0.09}_{-0.10}$), respectively.
(4) The \oiv 25.89 $\mu$m line to X-ray luminosity ratio 
is significantly smaller in AGNs with lower soft X-ray scattering fractions, 
suggesting that the \oiv 25.89 $\mu$m luminosity 
underestimates the intrinsic power of an AGN buried in a small opening-angle torus.

\end{abstract}


\section{Introduction}
High-quality broadband X-ray spectral observations are essential to
unveil the structure of active galactic nuclei (AGNs). The main X-ray 
continuum, which can be well approximated by a power law with an
exponential cutoff, is thought to be Comptonized photons by a hot corona
in the vicinity of the supermassive black hole (SMBH). This emission
interacts with the surrounding cold matter, the putative ``dusty torus''
invoked by the AGN unified model \citep{Ant93}. When one observes the
central engine through the torus (so-called type-2 or Compton-thin obscured AGNs; 
$22 \leq \log N_{\rm H} <24$), the spectrum shows a low energy cutoff due to 
photoelectric absorption. The torus also produces a reflected component, which is 
seen as a hump at $\sim 30$ keV, accompanied by a narrow iron-K$\alpha$ fluorescent 
line at $\simeq$ 6.4 keV \citep[e.g.,][]{Geo91,Mat91}. 
The reflection component from the inner accretion disk with 
a relativistically broadened iron-K$\alpha$ line is often reported
in type-1 AGNs \citep[e.g.,][]{Tan95,Nan07,Pat12},
although it is more difficult to robustly confirm its existence in type-2
AGNs. It is because when we see an AGN through the torus with an edge-on
view, the features are smeared out by the absorption and more significant broadening.
A scattered component by gas 
surrounding the torus is present, which is observed as a weak unabsorbed continuum 
in the soft X-ray band in obscured AGNs. 
The column density, the reflection strengths from the torus and disk,
the equivalent width (EW) of an iron-K$\alpha$ line, and the scattering fraction
all carry information on the distribution of surrounding matter. 

Moderately obscured (Compton-thin) AGNs, defined as those with
line-of-sight column densities of $22 \leq \log N_{\rm H} < 24$, are
the most abundant AGN population in the universe \citep{Ued14,Air15,Ric15}.
Also, they are ideal targets to study gas distribution around the nucleus.
Unlike the case of type-1 or unobscured AGNs ($\log N_{\rm H} < 22$),
the photoelectric absorption feature enables us to accurately measure
$N_{\rm H}$ and to observe the scattered X-ray light thanks to the
suppression of the direct component. 
Because effects by Compton scattering can be neglected, it is possible to accurately estimate the 
intrinsic X-ray luminosity in Compton-thin AGNs, whose measurement would
become unavoidably somewhat model-dependent in Compton-thick AGNs ($\log
N_{\rm H} > 24$).
Moreover, AGNs sometimes show time variation of the intrinsic luminosity 
and/or absorption \citep[changing look AGNs; e.g.,][]{Ris02,Gua05}. 
In that case, they provide us with valuable information such as the 
locus and structure of the dusty torus.

Hard X-ray all-sky surveys performed with {\it INTEGRAL} IBIS/ISGRI and {\it Swift}/Burst 
Alert Telescope (BAT) give least biased AGN samples against the obscuration 
\citep[e.g.,][]{Bec09,Bau13}, thanks to the high penetrating power of hard X-ray 
photons ($> 10$ keV). The {\it Suzaku} observatory \citep[2005--2015;][]{Mit07} 
was capable of simultaneously observing broadband X-ray spectra of AGNs, covering 
typically the 0.5--40 keV band. It achieved the best sensitivity at energies 
above 10 keV before {\it NuSTAR} as a pointing observatory \citep{Har13}. The 
combination of {\it Suzaku} data and time averaged {\it Swift}/BAT 
spectra covering the 14--195 keV band is very powerful for studying the
broadband X-ray spectra of local AGNs selected by {\it Swift}/BAT, 
allowing to improve the understanding of the absorbing and reprocessing material 
for individual obscured AGNs 
\citep[e.g.,][]{Ued07,Win09b,Egu09,Egu11,Taz11,Gan13,Gan15,Tan16}

This article is a summary paper reporting the data of essentially all 
local moderately-obscured AGNs observed with both {\it Suzaku} and {\it Swift}/BAT,
except for a few objects whose data have been already intensively
analyzed and published. The number of the targets is 45, all originally
selected from the {\it Swift}/BAT 70-month catalog 
\citep{Bau13}. Among them, the {\it Suzaku} broadband spectra 
of 19 objects are reported here for the first time. 
Our main goal is to investigate the properties of matter around 
the nuclei through a uniform analysis of the broadband X-ray spectra. {\it Suzaku} 
summary papers for low luminosity AGNs and 
Compton-thick AGNs are presented by \cite{Kaw16}
and Tanimoto et al.\ (in prep.), respectively.

This paper is organized as follows. Section~\ref{sec:obs} describes the
details of our sample and the overview of the data. We explain our
procedure of the spectral analysis in Section~\ref{sec:broad_spec}. The
results and discussion are presented in Section~\ref{sec:res_dis}. 
Section~\ref{sec:con} summarizes our findings. We adopt the cosmological 
parameters of ($H_0$, $\Omega_{\rm m}$, $\Omega_{\rm lambda}$) = 
(70 km s$^{-1}$ Mpc$^{-1}$, 0.3, 0.7) when calculating a distance from a 
redshift. Unless otherwise noted, all errors are quoted at the 1$\sigma$ 
confidence level for a single parameter of interest.

\section{Observation and Data Reduction}\label{sec:obs}

\subsection{Sample} 

Our sample of moderately-obscured (Compton-thin) AGNs consists of 
45 {\it Swift}/BAT selected AGNs \citep{Bau13} at $z<0.1$ whose 
{\it Suzaku} archival data are available as of 2015 December. The 
advantage of this sample is its high-quality broadband X-ray spectra 
(0.5--150 keV) that allow us to robustly constrain the X-ray spectral 
features. As for the sample selection by $N_{\rm H}$, we firstly 
check previous {\it Suzaku} papers that are complied by \cite{Ich12} 
and \cite{Fuk11}. For the rest of objects, we refer to the results
with other satellites listed in \cite{Ich12} and \cite{Mal12}. We 
exclude those that turned
out to be not moderately-obscured ($22 \leq \log N_{\rm H} < 24$) AGNs
from our spectral analysis. As described in Section~\ref{sec:res_dis},
time variation of $N_{\rm H}$ does not affect our sample selection. 
The hard X-ray luminosity averaged for 70 months is limited to $\log
L_{\rm 14-195\hspace{1mm}keV} > 42 $, since objects with $\log L_{\rm
14-195\hspace{1mm}keV} < 42 $ are reported in \cite{Kaw16}.
Radio-loud (e.g., PKS, 3C, or 4C sources) or blazar type objects, which
possibly possess jets, are not included because of possible
contamination of the X-ray emission due to the presence of a jet. Also,
we exclude 3 bright objects with complex spectra, NGC 3227, NGC 3516,
and NGC 4151, which have been intensively analyzed with different models
such as relativistic reflection components from the
inner disk, complex absorbers, and multiple power-law components
\citep[e.g.,][]{Pat12,Kec15,Nod14,Beu15} .

\begin{figure}[t]
\begin{center}
\includegraphics[scale=0.53,angle=-90]{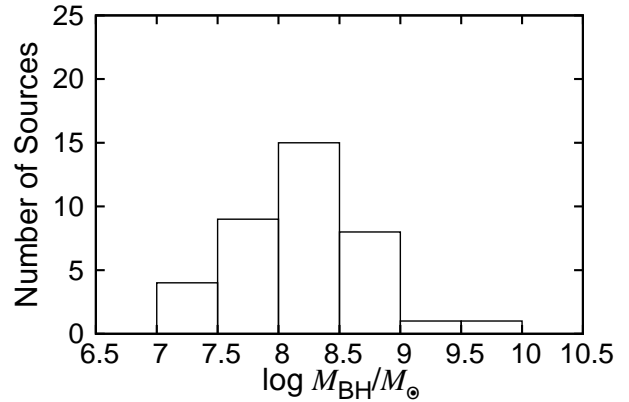}
\caption{
Distribution of black hole mass. 
}
\label{fig:mass}
\end{center}
\end{figure}

The basic information of the sample (i.e., galaxy name, position,
redshift, distance, black hole mass $M_{\rm BH}$) is listed in
Table~\ref{tab:info_srcs}. 
The distances are the mean value available in
the NASA/IPAC Extragalactic Database (NED), or calculated from the
redshift when the distances are not available in the NED. 
We compile the black hole masses estimated by the gas dynamics around 
the SMBH, reverberation mapping, empirical 
formula using the broad-line width and luminosity in the optical band
(e.g., H$\beta$ and $\lambda$5100), or relations of $M_{\rm BH}$ with bulge properties 
(e.g., velocity dispersion and $K$-band luminosity).
When multiple SMBH masses are available for a single object, the mean value is taken.
The $M_{\rm BH}$ distribution of our sample is represented in Figure~\ref{fig:mass}.
The mean and standard deviation of $\log (M_{\rm BH}/M_{\odot})$, 
where $M_{\odot}$ is the solar mass, is $8.1\pm0.1$ and $0.6\pm0.1$, respectively. 
Throughout this paper, we adopt the 2--10 keV 
to bolometric correction factor of $20$ \citep{Vas09}, 
which is applicable to AGNs with Eddington ratios of $\lambda_{\rm Edd}
< 0.1$.
As described in Section~\ref{sec:x-ray_pro}, 
almost all of our objects show $\lambda_{\rm Edd} < 0.1$. 
Adoption of the luminosity-dependent bolometric correction factor 
of \cite{Mar04} does not affect our main conclusions, except for the 
hard X-ray luminosity and Eddington ratio 
correlation (Section~\ref{sec:x-ray_pro}).

% references, 
% Awaki+94, Brightman+11, Claudio+10, Comastri+10, Comastri+92, Fukazawa+11, Gandhi+13, 
% Gozalez-martin+09 Ichikawa+12, Levenson+05, Malizia+07, Malizia+12, Matt+09, Parisi+12, 
% Pariya+14, Patrick+12, Raynolds+09 
% Reeves, Ricci+10, Svoboda+12, Teng+09, Vasudevan+13, Winter+12, winter+09

\subsection{Data reduction}

{\it Suzaku} carries the X-ray Imaging Spectrometers
\citep[XIS;][]{Koy07} and the Hard X-ray Detector \citep[HXD;][]{Tak07},
sensitive to soft ($< 10 $ keV) and hard X-ray photons ($> 10$ keV),
respectively. Three of the four XISs are frontside-illuminated camera
(FI-XISs; XIS-0, XIS-2, and XIS-3), and the other is a
backside-illuminated one (BI-XIS; XIS-1). The HXD consists of the PIN
diodes and GSO scintillators. Table~\ref{tab:info_srcs} lists the IDs of
the {\it Suzaku} observation data we analyze. If {\it Suzaku} observed
an object on several occasions, we adopt the data with the longest
exposure. For 2MASX J0350-5018, we utilize all observations because
the exposure of a single observation is found to be too short for
meaningful spectral analysis.

FTOOLS v6.15.1 and the {\it Suzaku} calibration database released on
2015 Jan 5 are used for the data reduction. We reprocess the unfiltered
XIS event data in the standard manner, as described in the ABC 
guide\footnotemark[1]. The XIS source events are extracted from a
circular region with radii of 1'--4' depending on the flux, whereas the
background is taken from an off-source region within the XIS
field-of-view, where no other source is present. All the FI-XISs spectra available 
in each observation are combined into one to increase the signal-to-noise ratio. 
We generate the
XIS response matrix and ancillary response files with {\tt xisrmfgen}
and {\tt xissimarfgen} \citep{Ish07}, respectively. 
We bin the XIS spectra with minimum counts of 100 per bin. 
For HXD/PIN, we start with the ``cleaned'' event files provided by the {\it Suzaku}/HXD
team.  We make the background spectrum including the ``tuned'' non X-ray
background model \citep{Fuk09} and the simulated Cosmic 
X-ray background spectrum based on \citet{Gru99}.  
We basically use the HXD/PIN data in the 16--40 keV band for the spectral analysis.
We further limit them to an energy band where the source signals 
are sufficiently higher than the uncertainty in the non X-ray background 
model: $\sim$ 3\% for exposures of less than 40 ksec and $\sim $ 1\% for
longer exposures \citep{Fuk09}.

\footnotetext[1]{https://heasarc.gsfc.nasa.gov/docs/suzaku/analysis/abc/}

\tabletypesize{\footnotesize}
\setlength{\tabcolsep}{0.03in}
\begin{deluxetable*}{lllllrccccc}
\tablecaption{Information of Targets\label{tab:info_srcs}}
\tablewidth{0pt}
\tablehead{\colhead{Galaxy Name} 
& \colhead{Swift ID} 
& \colhead{RA.} 
& \colhead{Dec.} 
& \colhead{Redshift} 
& \colhead{$D$} 
& \colhead{$\log M_{\rm BH}/M_{\odot}$} 
& \colhead{$M_{\rm BH}$ Ref.} 
& \colhead{{\it Suzaku} ID} 
& \colhead{{\it Suzaku} Ref.} \\ 
\colhead{(1)} & \colhead{(2)} & \colhead{(3)} & \colhead{(4)} &
\colhead{(5)} & \colhead{(6)} & \colhead{(7)} & \colhead{(8)} & \colhead{(9)} & \colhead{(10)}
} 
\startdata
\thispagestyle{empty}
2MASX J0216+5126 & J0216.3+5128 & 34.124333  & 51.440194  & 0.0288   & 126.1 & $...$& $...$& 705006010   & $\largestar$ \\ 
2MASX J0248+2630 & J0249.1+2627 & 42.247199  & 26.510890  & 0.057997 & 259.3 & $...$& $...$& 704013010   & $\largestar$ \\ 
2MASX J0318+6829 & J0318.7+6828 & 49.579079  & 68.492062  & 0.090100 & 412.0 & $...$& $...$& 702075010   & 1  \\ 
2MASX J0350-5018 & J0350.1-5019 & 57.599042  & -50.309917 & 0.036492 & 160.6 & 8.8  & 1    & 701017010$^\dagger$   & 2  \\ 
2MASX J0444+2813 & J0444.1+2813 & 71.037542  & 28.216861  & 0.011268 & 38.6  & 7.4  & 2    & 703021010   & $\largestar$ \\ 
2MASX J0505-2351 & J0505.8-2351 & 76.440542  & -23.853889 & 0.035041 & 154.1 & 7.5  & 1    & 701014010   & 2 \\ 
2MASX J0911+4528 & J0911.2+4533 & 137.874863 & 45.468331  & 0.026782 & 117.1 & 7.5  & 1    & 703008010   & $\largestar$ \\ 
2MASX J1200+0648 & J1200.8+0650 & 180.241393 & 6.806423   & 0.036045 & 158.6 & 8.5  & 1    & 703009010   & 1  \\ 
Ark 347          & J1204.5+2019 & 181.1236551& 20.3162130 & 0.022445 & 97.8  & 8.1  & 1    & 705002010   & $\largestar$ \\ 
ESO 103-035      & J1838.4-6524 & 279.584750 & -65.427556 & 0.013286 & 57.5  & 7.5  & 1,3  & 703031010   & 3  \\ 
ESO 263-G013     & J1009.3-4250 & 152.450875 & -42.811222 & 0.033537 & 147.3 & 8.0  & 4    & 702120010   & 1,4 \\ 
ESO 297-G018     & J0138.6-4001 & 24.654833  & -40.011417 & 0.025227 & 110.1 & 9.7  & 1    & 701015010   & 2  \\ 
ESO 506-G027     & J1238.9-2720 & 189.727458 & -27.307833 & 0.025024 & 109.2 & 8.6  & 1    & 702080010   & 1,5  \\ 
Fairall 49       & J1836.9-5924 & 279.242875 & -59.402389 & 0.020021 & 87.1  & $...$& $...$& 702118010   & 1,6,7 \\ 
Fairall 51       & J1844.5-6221 & 281.224917 & -62.364833 & 0.014178 & 45.9  & 8.0  & 5    & 708046010   & 8 \\ 
IC 4518A         & J1457.8-4308 & 224.421583 & -43.132111 & 0.016261 & 70.5  & 7.5  & 3    & 706012010   & $\largestar$ \\ 
LEDA 170194      & J1239.3-1611 & 189.72     & -16.23     & 0.040000 & 161.5 & 8.9  & 1,3  & 703007010   & $\largestar$ \\ 
MCG +04-48-002   & J2028.5+2543 & 307.146083 & 25.733333  & 0.013900 & 60.2  & 7.1  & 4    & 702081010   & 1,5 \\ 
MCG -01-05-047   & J0152.8-0329 & 28.204167  & -3.446833  & 0.017197 & 68.5  & 7.6  & 4    & 704043010   & $\largestar$ \\ 
MCG -02-08-014   & J0252.7-0822 & 43.097481  & -8.510413  & 0.016752 & 72.7  & $...$& $...$& 704045010   & $\largestar$ \\ 
MCG -05-23-016   & J0947.6-3057 & 146.917319 & -30.948734 & 0.008486 & 36.6  & 7.4  & 1,3  & 700002010   & 1,9,10,11 \\ 
Mrk 1210         & J0804.2+0507 & 121.0244092& 5.1138450  & 0.013496 & 58.4  & 7.9  & 6    & 702111010   & 1,12 \\ 
Mrk 1498         & J1628.1+5145 & 247.016937 & 51.775390  & 0.054700 & 244.0 & 8.6  & 1    & 701016010   & 1,2 \\ 
Mrk 18           & J0902.0+6007 & 135.493323 & 60.151709  & 0.011088 & 47.9  & 7.5  & 1    & 705001010   & $\largestar$ \\ 
Mrk 348          & J0048.8+3155 & 12.1964225 & 31.9569681 & 0.015034 & 65.1  & 8.0  & 1    & 703029010   & 13 \\ 
Mrk 417          & J1049.4+2258 & 162.378861 & 22.964555  & 0.032756 & 143.8 & 8.0  & 1    & 702078010   & 1,5 \\ 
Mrk 520          & J2200.9+1032 & 330.17458  & 10.54972   & 0.026612 & 108.0 & 8.3  & 7    & 407014010   & $\largestar$ \\ 
Mrk 915          & J2236.7-1233 & 339.193768 & -12.545162 & 0.024109 & 105.2 & 8.1  & 5    & 708029010   & $\largestar$ \\ 
NGC 1052         & J0241.3-0816 & 40.2699937 & -8.2557642 & 0.005037 & 19.7  & 8.7  & 8    & 702058010   & 1,10,14 \\ 
NGC 1142         & J0255.2-0011 & 43.8008169 & -0.1835573 & 0.028847 & 126.3 & 9.2  & 1,3,4& 701013010   & 1,2 \\ 
NGC 2110         & J0552.2-0727 & 88.047420  & -7.456212  & 0.007789 & 35.6  & 8.3  & 1    & 100024010   & 1,9,10,15 \\ 
NGC 235A         & J0042.9-2332 & 10.720042  & -23.541028 & 0.022229 & 96.8  & 8.8  & 1    & 708026010   & $\largestar$ \\ 
NGC 3081         & J0959.5-2248 & 149.873080 & -22.826277 & 0.007976 & 26.5  & 7.7  & 1,4,9& 703013010   & 1,16 \\ 
NGC 3431         & J1051.2-1704A& 162.812667 & -17.008028 & 0.017522 & 76.1  & $...$& $...$& 707012010   & $\largestar$ \\ 
NGC 4388         & J1225.8+1240 & 186.444780 & 12.662086  & 0.008419 & 20.5  & 8.0  & 1,3,4,9 & 800017010& 1,10,17 \\ 
NGC 4507         & J1235.6-3954 & 188.9026308& -39.9092628& 0.011801 & 51.0  & 8.0  & 1,3,4,9 & 702048010& 1,18 \\ 
NGC 4992         & J1309.2+1139 & 197.2733500& 11.6341550 & 0.025137 & 109.7 & 8.4  & 1,3,4   & 701080010& 1,4 \\ 
NGC 5252         & J1338.2+0433 & 204.5665139& 4.5425817  & 0.022975 & 83.6  & 8.9  & 1,3  & 707028010   & $\largestar$ \\ 
NGC 526A         & J0123.8-3504 & 20.9766408 & -35.0655289& 0.019097 & 83.0  & 8.0  & 1,5  & 705044010   & $\largestar$ \\ 
NGC 5506         & J1413.2-0312 & 213.3120500& -3.2075769 & 0.006181 & 23.8  & 7.5  & 1,3,4,9 & 701030020& 1,3,10,11\\ 
NGC 6300         & J1717.1-6249 & 259.247792 & -62.820556 & 0.003699 & 13.9  & 7.3  & 3,4  & 702049010   & 1 \\ 
NGC 7172         & J2201.9-3152 & 330.5078800& -31.8696658& 0.008683 & 33.9  & 8.0  & 1,3,4,9 & 703030010& 1 \\ 
NGC 788          & J0201.0-0648 & 30.2768639 & -6.8155172 & 0.013603 & 58.9  & 8.2  & 1,3,4   & 703032010& $\largestar$ \\ 
UGC 03142        & J0443.9+2856 & 70.944958  & 28.971917  & 0.021655 & 94.3  & 8.3  & 10   & 707032010   & $\largestar$ \\ 
UGC 12741        & J2341.8+3033 & 355.481083 & 30.581750  & 0.017445 & 76.4  & $...$& $...$& 704014010   & $\largestar$ 
\enddata
\tablecomments{(1) Galaxy name. 
(2) {\it Swift}/BAT name in the 70-month catalog \citep{Bau13}. 
(3)--(5) Position in units of degree and redshift taken from the NED. 
(6) Distance in units of Mpc. 
(7)--(8) Black hole mass and the reference. 
(9) Observation ID of the {\it Suzaku} data we analyze. 
(10) Paper already reporting the {\it Suzaku} spectral analysis. \\  
References for black hole masses.  \\ 
(1) \citet{Win09}
(2) \citet{Vas10}
(3) \citet{Pan15}
(4) \citet{Kho12}
(5) \citet{Ben06}
(6) \citet{Zha08}
(7) \citet{Win10}
(8) \citet{Dop15}
(9) \citet{Dia12}
(10) \citet{Wan07}
 \\
References for papers. \\
(1) \citet{Fuk09} 
(2) \citet{Egu09}
% (3) \citet{Ued07} 
(3) \citet{Gof13} 
(4) \citet{Com10} 
(5) \citet{Win09b} 
(6) \citet{Tri13} 
(7) \citet{Lob14} 
(8) \citet{Svo15} 
(9) \citet{Ree07} 
(10) \citet{Miy09}
(11) \citet{Pat12}
(12) \citet{Mat09}
(13) \citet{Mar14}
(14) \citet{Bre09}
(15) \citet{Riv14}
(16) \citet{Egu11}
(17) \citet{Shi08}
(18) \citet{Bra13}
($\largestar$) The {\it Suzaku} spectra are reported for the first time in this paper. \\
\dagger We also analyze the data, whose observation IDs are 701017020 and 701017030. 
}
\end{deluxetable*}
\newpage

\renewcommand{\arraystretch}{1.3}
\begin{deluxetable*}{llcrrrrr}
\scriptsize
\tablecaption{Correlations \label{tab:cor}}
\tablehead{\colhead{Y} & \colhead{X} 
& \colhead{Sample}
& \colhead{$N$}
& \colhead{$\rho($X$,$Y$)$}
& \colhead{$P($X$,$Y$)$}
& \colhead{$a$}
& \colhead{$b$} \\
\colhead{(1)} & \colhead{(2)} & \colhead{(3)} & \colhead{(4)} & 
\colhead{(5)} & \colhead{(6)} & \colhead{(7)} & \colhead{(8)}  
} 
\startdata 
$\log \lambda^{\rm BAT}_{\rm Edd}$ & $\log L^{\rm BAT}_{\rm 10-50\hspace{1mm}keV}$ 
& All & 38 & $0.17$ & $3.1\times10^{-1}$ & $ ... $ & $ ... $  \\  % OK 

$\Gamma$ & $\log L^{\rm BAT}_{\rm 10-50\hspace{1mm}keV}$ 
& All & 45 & $-0.26$ & $8.6\times10^{-2}$ & $...$ & $...$  \\  % OK 

$\log N_{\rm H}$ & $\log L^{\rm BAT}_{\rm 10-50\hspace{1mm}keV}$ 
& All & 45 & $0.08$ & $5.9\times10^{-1}$ & $ ... $ & $ ... $  \\ % OK

$\Gamma$ & $\log \lambda^{\rm BAT}_{\rm Edd}$ 
& All & 38 & $0.41$ & $9.7\times10^{-3}$ & $2.11\pm0.01$ & $0.20\pm0.01$  \\  % OK

$\log N_{\rm H}$ & $\log \lambda^{\rm BAT}_{\rm Edd}$ 
& All & 38 & $-0.04$ & $7.9\times10^{-1}$ & $ ... $ & $ ... $   \\ % OK 

$\Gamma$ & $\log N_{\rm H}$  
& All & 45 & $-0.03$ & $8.6\times10^{-1}$ & $ ... $ & $ ... $   \\  
\hline % 

$\log L_{\rm K\alpha}$ & $\log L^{\rm BAT}_{\rm 10-50\hspace{1mm}keV} $ 
& All & 45 & $0.89$ & $4.1\times10^{-16}$ & $0.2\pm1.8$ & $0.94\pm0.04$ \\  % 
\hline % OK 

$R$ & $\log L^{\rm BAT}_{\rm 10-50\hspace{1mm}keV} $ 
& All & 45 & $-0.29$ & $5.6\times10^{-2}$ & $ ... $ & $ ... $ \\ 

$R$ & $\log (L_{\rm K\alpha}/L^{\rm BAT}_{\rm 10-50\hspace{1mm}keV}) $ 
& All & 45 & $0.38$ & $1.0\times10^{-2}$ & $ ... $ & $ ... $ \\ 

$R$ & $\log N_{\rm H}$ 
& All & 45 & $0.04$ & $7.9\times10^{-1}$ & $ ... $ & $ ... $ \\ 
\hline

$\log \lambda L_{\lambda\hspace{1mm}12 \mu {\rm m}}$ & 
$\log  L^{\rm BAT}_{\rm 10-50\hspace{1mm}keV} $ 
& All & 43  & $0.67$ & $1.0\times10^{-6}$ & $3.7\pm3.0$ & $0.92\pm0.07$ \\

$\log \lambda L_{\lambda\hspace{1mm}12 \mu {\rm m}}$ & 
$\log  L^{\rm BAT}_{\rm 10-50\hspace{1mm}keV} $ 
& Nuc. & 28 & $0.53$ & $3.8\times10^{-3}$ & $3.8\pm4.1$ & $0.91\pm0.10$ \\ 
\hline 

$\log \lambda F_{\lambda\hspace{1mm}12 \mu {\rm m}}$ & 
$\log  F^{\rm BAT}_{\rm 10-50\hspace{1mm}keV} $ 
& All & 43 & $0.63$ & $5.8\times10^{-6}$ & $0.9\pm1.0$ & $1.06\pm0.09$ \\

$\log \lambda F_{\lambda\hspace{1mm}12 \mu {\rm m}}$ & 
$\log  F^{\rm BAT}_{\rm 10-50\hspace{1mm}keV} $ 
& Nuc. & 28 & $0.60$ & $6.9\times10^{-4}$  & $1.8\pm1.3$ & $1.15\pm0.13$ \\ 
\hline 

$\log L_{\rm \oiv}$ & $\log  L^{\rm BAT}_{\rm 10-50\hspace{1mm}keV} $ 
& All & 33 & $0.10$ & $5.8\times10^{-1}$ & $-7.8\pm4.9$ & $1.13\pm0.12$ \\ 

$\log (L_{\rm \oiv}/L^{\rm BAT}_{\rm 10-50\hspace{1mm}keV}) $ & 
$\log  f_{\rm scat} $ 
& All & 32 & $0.35$ & $4.9\times10^{-2}$ & $-2.25\pm0.11$ & $0.98\pm0.07$ 

\enddata
\tablecomments{
(1) Y varibale.
(2) X variable. 
(3) Sample used for the fitting. Nucleus (Nuc.) corresponds to the sample whose 12 $\mu$m luminosities 
were measured at high spatial resolution. 
(4) Number of objects of the sample. 
(5) Spearman's Rank coefficient for the correlation. 
(6) Null hypothesis probability of obtaining no correlation.  
(7)-(8) Fitting parameters of Y $= a + b$X, which are derived for the correlations with the 
null hypothesis probabilities smaller than 5\% except for the correlations of $R$, and that of the \oiv versus 
X-ray luminosity. 
}
\end{deluxetable*}

\renewcommand{\arraystretch}{1.0}

\section{Broadband Spectral Analysis}\label{sec:broad_spec} 

In addition to the {\it Suzaku} spectra, we also utilize the {\it Swift}/BAT 
spectra averaged for 70 months \citep{Bau13}. XSPEC (version 12.8.1.g) is 
used for the spectral analysis. We perform a simultaneous fit to the FI-XISs, 
BI-XIS, HXD/PIN, and {\it Swift}/BAT spectra, which cover the 1--10 keV, 
0.5--8 keV, 16--40 keV, and 14--150 keV bands, respectively. The 1.7--1.9 
keV XIS spectra are excluded to avoid the response uncertainties at the Si-K 
edge energy. Also, we do not use energy ranges where source photons are not 
significantly detected at 1$\sigma$. The cross-normalizations of the 
{\it Swift}/BAT and HXD/PIN spectra with respect to the FI-XISs spectrum are 
set to 1.0 and 1.16 (1.18) for the XIS (HXD) nominal position observation, 
respectively, whereas that of the BI-XIS one is left as a free parameter. 
By using the {\tt phabs} model, we always consider the Galactic absorption 
($N^{\rm Gal}_{\rm H}$), whose value is estimated from the 
H\hspace{1mm}{\footnotesize I} map of \citet{Kal05}. Solar abundances given by 
\citet{And89} are assumed.

\subsection{Baseline Model}\label{sec:base_model}

To reproduce the broadband X-ray spectra covering the 0.5--150 keV band, 
we start with the following baseline model:
\begin{small}
 \begin{itemize}
 \item {\tt constant*zphabs*zpowerlw*zhighect \\ 
+constant*zhighect*zpowerlw+pexrav+zgauss}. 
 \end{itemize}
\end{small}  
This model consists of an absorbed cutoff power law (transmitted
component), a scattered component, and a reflection component from
distant, cold matter accompanied by a narrow fluorescence iron-K$\alpha$
line. We fix the cutoff energy at 300 keV, a canonical value for nearby
AGNs \citep{Dad08}. Through the first {\tt constant} model ($N_{\rm
XIS}$), we take into account possible time variation of the cutoff
power-law component between the {\it Suzaku} and {\it Swift}/BAT spectra. 
The {\tt zphabs} model is used to represent photoelectric absorption.
The second term represents scattered emission of the primary X-ray 
component by gas located outside the torus. This unabsorbed component 
is assumed to have the same shape as the transmitted 
component with a fractional normalization of $f_{\rm scat}$. 
The {\tt pexrav} code \citep{Mag95} reproduces reflected 
continuum emission, whose relative strength to the transmitted component
is defined by $R = \Omega/2\pi$ ($\Omega$ is the solid angle of the
reflector). The inclination angle to the reflector is fixed at
60$^\circ$. To avoid unphysical fitting results, we impose an upper
limit of $R = 2$, corresponding to the extreme case where the nucleus is
covered by the reflector in all directions.  
The {\tt zgauss} component
represents an iron-K$\alpha$ fluorescent line, where the line width and
energy is fixed at 20 eV and 6.4 keV, respectively. The width corresponds 
to a typical velocity dispersion of $\sim$ 2000 km s$^{-1}$ measured in 
local Seyfert galaxies with {\it Chandra}/HETGS by \cite{Shu10}. 
If the line energy is allowed to vary, the resultant value is 
consistent with 6.4 keV within the 99\% confidence interval except for Fairall 51, 
Mrk 1498, and NGC 5506. For the three objects, we leave the line energy as a
free parameter. 
We assume that the
reflection components did not vary in accordance with the primary
emission between the {\it Suzaku} and {\it Swift}/BAT observations,
considering the large size of the reflector ($\sim$ a pc scale).

After fitting the spectra with the above baseline model, we
systematically test if inclusion of other model components improves the
fit. We adopt a new model if the improvement is found to be significant
at a 99\% confidence level (i.e., $\Delta \chi^2 < -6.64$ and $< -9.21$, which  
correspond to the 99\% limits of the $\chi^2$ distribution with degrees of 
freedom of 1 and 2, respectively.). The additional 
model components we consider are as follows: (1) optically-thin thermal
emission from the host galaxy ({\tt apec} in XSPEC), (2) partial
absorption of the cutoff power-law component ({\tt zpcfabs}), (3)
absorption of the reflection components ({\tt zphabs}), 
by considering that emission from a large-scale reflector like a 
dusty torus \citep[e.g., see Figure 2 of][]{Ike09} may be subject to 
absorption different from that in the line of sight,  
and (4) emission/absorption lines ({\tt zgauss}) of He-like iron ions at $6.70$
keV, H-like iron ions at $6.97$ keV, iron-K$\beta$ at 7.06 keV, and
nickel-K$\alpha$ at 7.48 keV. 
The Compton shoulder of an iron-K$\alpha$
line is also considered, which is modelled by a gaussian ({\tt zgauss})
at 6.31 keV \citep[e.g.,][]{Mat02,Shi08}. The line width ($1 \sigma$) of 
these lines is set to 20 eV.
Moreover, we systematically survey other emission or absorption lines 
(e.g., those of ultra fast outflow) at energies above 6.4 keV by adding a line 
component ({\tt zgauss}) with two additional free parameters (line 
energy and normalization). 
We include lines if the improvement of $\chi^2$ is larger than $\Delta \chi^2$ = 9.21. 
We also check the absorption lines 
reported by \cite{Tom11}, who analyzed {\it XMM-Newton} data, but we do not
detect any of them in our {\it Suzaku} spectra at $>$99\% confidence level.
It would be because they are too weak or variable.

In the Appendix, Table~\ref{tab:info_para} summarizes the results of the spectral 
analysis. We obtain good fits for all targets, which fulfill either 
$\chi^2/d.o.f < 1.2$ or null hypothesis probability larger than 
$1\%$. Figure~\ref{fig:spec} and Figure~\ref{fig:narrow_spec} show the unfolded 
spectra and best-fit models in the 0.5--150 keV and 4--9 keV bands, 
respectively. Table~\ref{tab:info_fxlx} lists the flux, absorption-corrected
luminosity, Eddington ratio, and EW of the iron-K$\alpha$ line with respect to the total contnuum. 
Here, we define the Eddington luminosity as $L_{\rm Edd} = 1.26 \times 10^{38}
(M_{\rm BH}/M_{\odot})$ erg s$^{-1}$. 
The information of the detected emission/absorption lines is
listed in Table~\ref{tab:info_lines}.

\subsection{Relativistic Reflection Component from the Accretion Disk} \label{sec:disk}

We further examine whether the spectra statistically require
relativistically blurred reflection component from the inner
optically-thick accretion disk. For this purpose, we use the model {\tt
constant*kdblur*reflionx}, where {\tt reflionx} calculates a reflected
continuum and emission lines from an ionized disk \citep{Ros05} and {\tt
kdblur} reproduces relativistic effects in the vicinity of a SMBH.
Compared with type-1 AGNs, this component, if any, would be more
difficult to detect and characterize in type-2 AGNs because of the
absorption and lower flux contribution in edge-on geometry. Thus, we
minimize the number of the free parameters.  Among the parameters of 
the {\tt reflionx} model, the photon index ($\Gamma$) and normalization
are liked to those of the primary cutoff power-law component in the
baseline model. We assume two ionization parameters of the disk ($\xi =
10$ and $100$). The parameters of the {\tt kdblur} model are a radial
emissivity index $q$ (emissivity $\propto r^{-q}$), inner and outer
radii ($r_{\rm in}$ and $r_{\rm out}$), and inclination angle
($\theta_{\rm inc}$). We allow $r_{\rm in}$ to vary within 1--100$r_{\rm g}$ 
($r_{\rm g}$ is the Gravitational radius); this upper limit is 
imposed to avoid strong coupling with the narrow iron line from distant
matter. We fix $q$, $r_{\rm out}$, and $\theta_{\rm inc}$ at $3$,
$400r_{\rm g}$, and $60^\circ$, respectively.  
Because the {\tt reflionx} model does not 
have an explicit parameter of the reflection strength ($\Omega/2\pi$), 
we quantify it by multiplying the {\tt constant} model. 
Its upper limit is set to $3.2\times10^{-3}$ and $3.2\times10^{-4}$ for $\xi = 10$ and $100$,
respectively, which reproduces the same 10--100 keV flux as the {\tt
pexrav} model with $\Omega/2\pi$ = 1 for a photon index of 1.7. The disk
reflection component is subject to the same (partial) absorption models
as for the primary component. In summary, only the inner radius ($r_{\rm
in}$) and normalization ({\tt constant}) are left as free parameters.

Adding the disk-reflection component to the best-fit models obtained in 
Section~\ref{sec:base_model}, we find that fits are significantly 
improved at $>$99\% confidence level in  four sources 
(2MASX J1200+0648, Fairall 49, NGC 526A, and NGC 6300).
Their unfolded spectra that give a smaller $\chi^2$ value 
between the assumptions of $\xi=10$ or $100$ are shown in Figure~\ref{fig:disk}. 
Except for NGC 6300, the disk-reflection component better improves the fit at 
energies below 10 keV than above 10 keV. That is, a broad 
iron-K$\alpha$ line feature is more essential to 
reproduce the spectra than the reflection hump at $\sim$ 30 keV in the
first three objects. 
Possible presence of an ionized-disk reflection component in 2MASX J1200+648 
and NGC 6300 is suggested here for the first time, while
it was reported for Fairall 49 by \cite{Iwa04}, consistent with 
our result. \cite{Nan07} analyzed the {\it XMM-Newton} spectra of NGC 526A and 
found that the relativistic reflection can well reproduce the spectra but is 
indistinguishable from absorption models in terms of statistics.

Table~\ref{tab:info_disk_para} summarizes the 
resultant parameters of the disk-reflection component. For convenience,
the normalization factor is converted into equivalent reflection
strength in units of $\Omega/2\pi$. The fraction of these AGNs (4 out of
45) should be regarded as a conservative lower limit, because of the
small ranges of parameters we have investigated and of the
limitation in the spectral quality. We leave detailed discussion on the
relativistic reflection components in obscured AGNs for future studies.

%\if0 %%%%IRU file to be incorporated %%%%
\begin{figure*}[t]
\begin{center}
\includegraphics[scale=0.55,angle=-90]{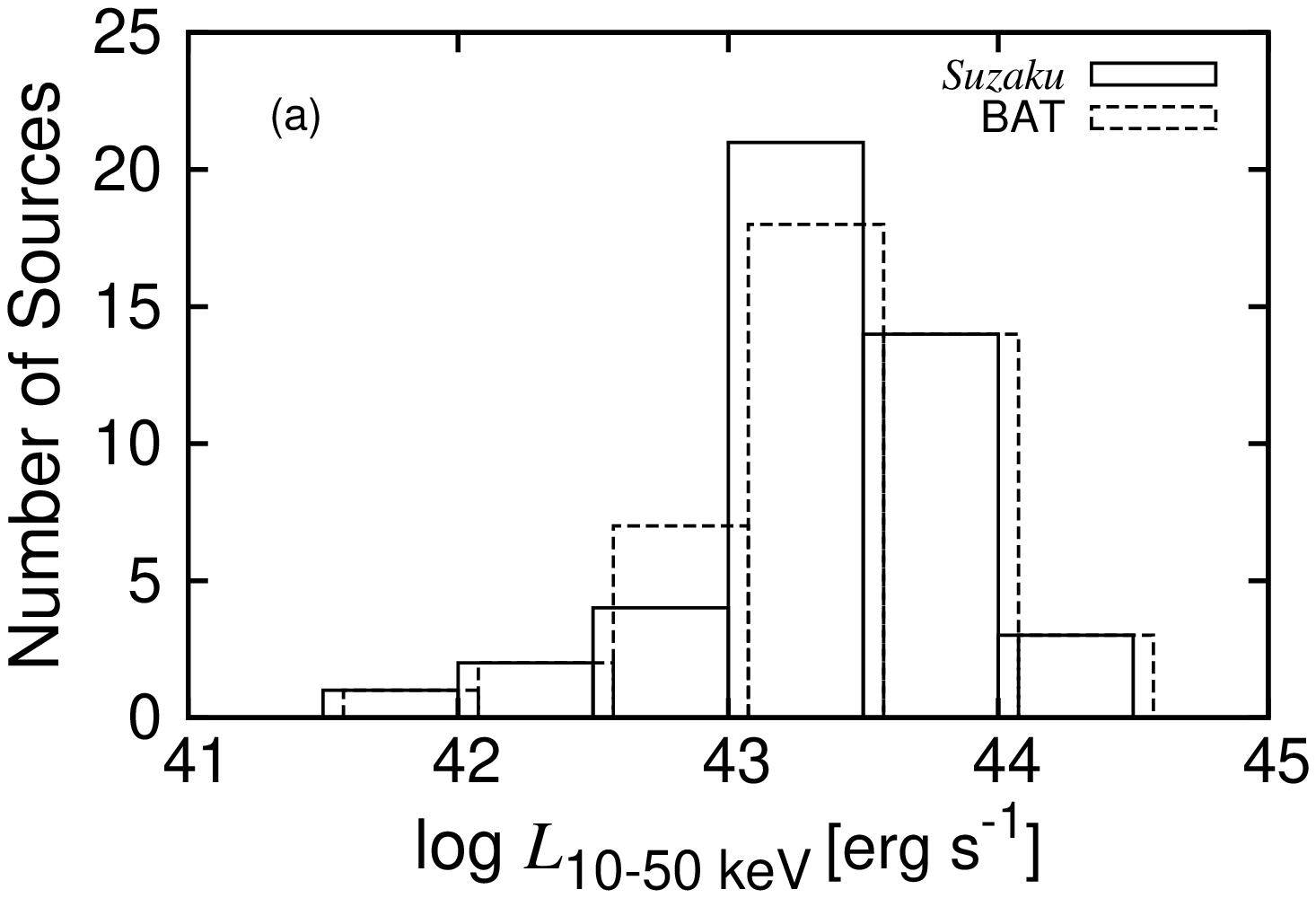}\hspace{5mm}
\includegraphics[scale=0.55,angle=-90]{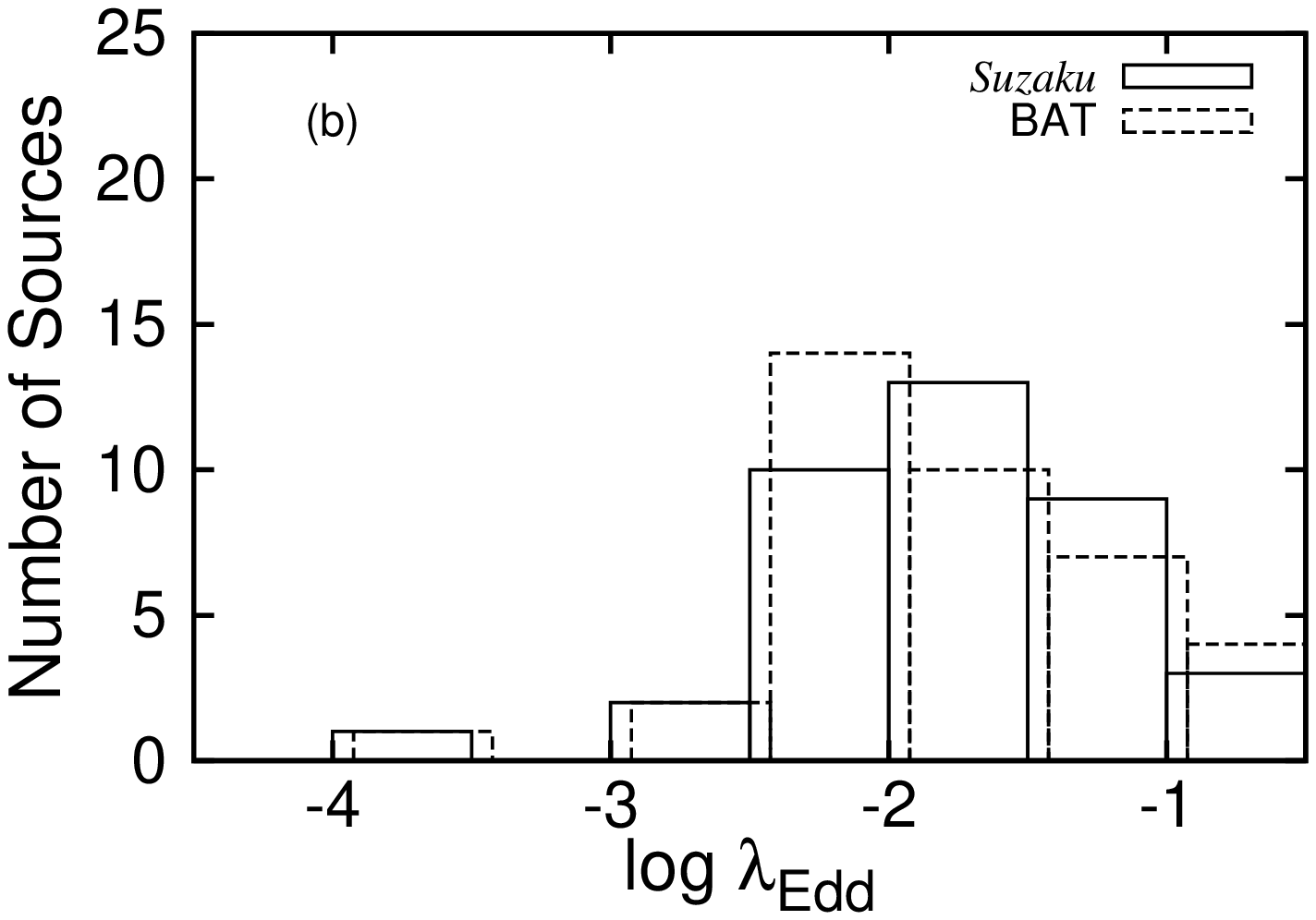} \\
\hspace{2mm}
\includegraphics[scale=0.54,angle=-90]{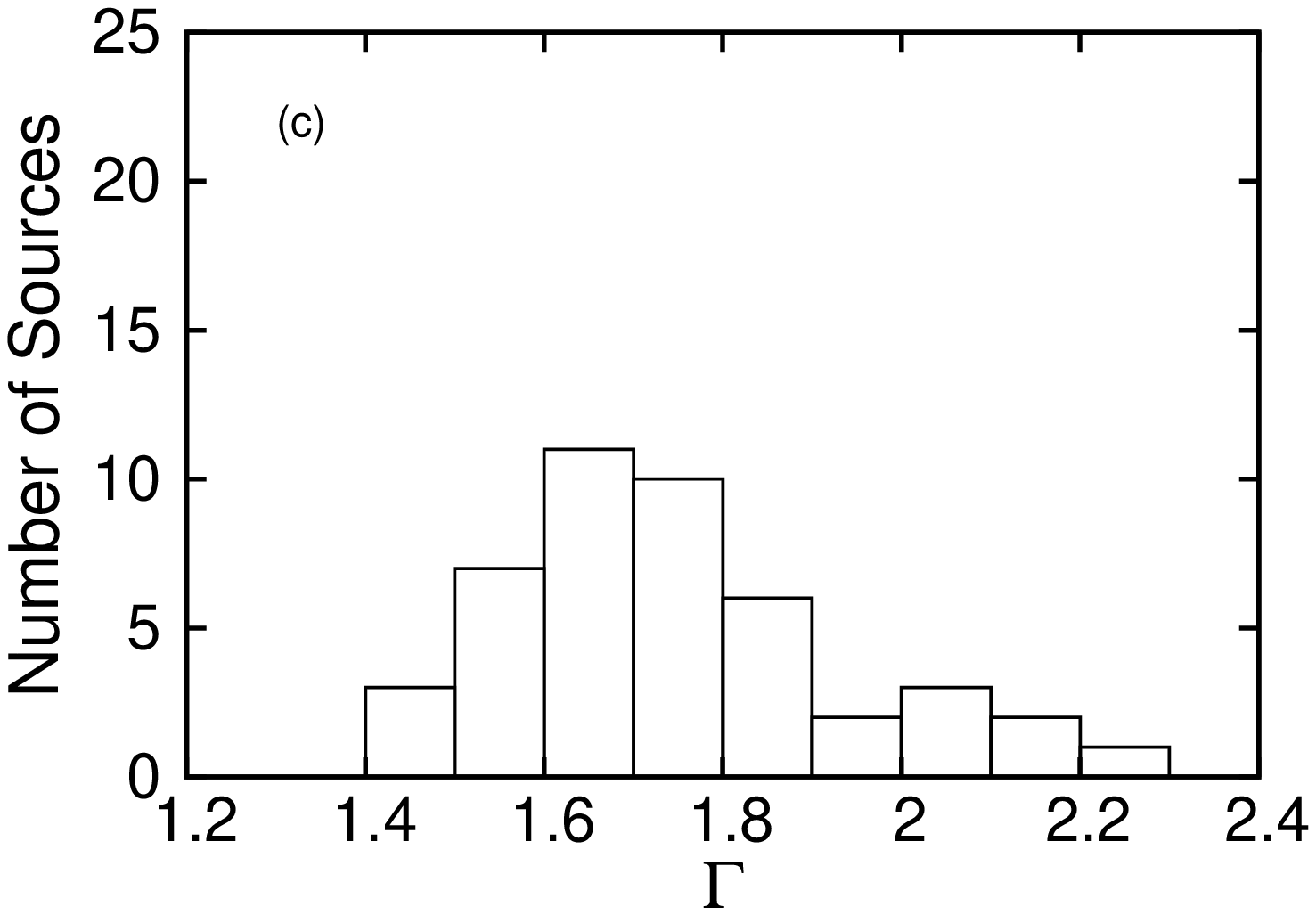} \hspace{5mm}
\includegraphics[scale=0.56,angle=-90]{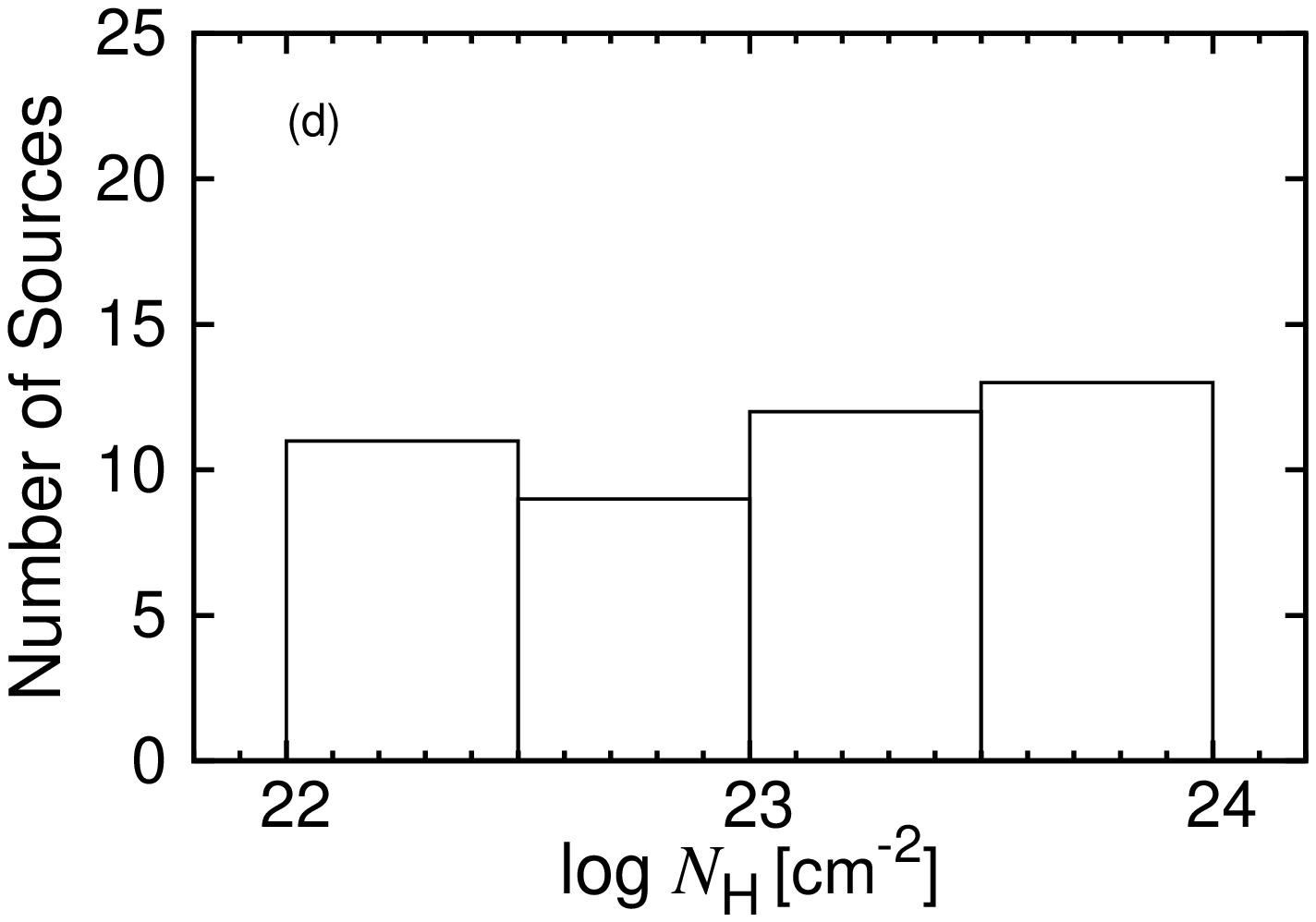}
\caption{
(a) distribution of absorption-corrected 10--50 keV luminosity.
(b) distribution of Eddington ratio. 
(c) distribution of photon index. 
(d) distribution of hydrogen column density. 
The solid and dashed histograms in the upper figures refer to the luminosities measured 
with {\it Suzaku} and {\it Swift}/BAT, respectively. For clarity, the dashed histograms are
slightly shifted to the right.
}
\label{fig:mean_std}
\end{center}
\end{figure*}
%\fi 

%\if0 %%%%IRU file to be incorporated %%%%
\begin{figure}[t]
\begin{center}
\includegraphics[scale=0.53,angle=-90]{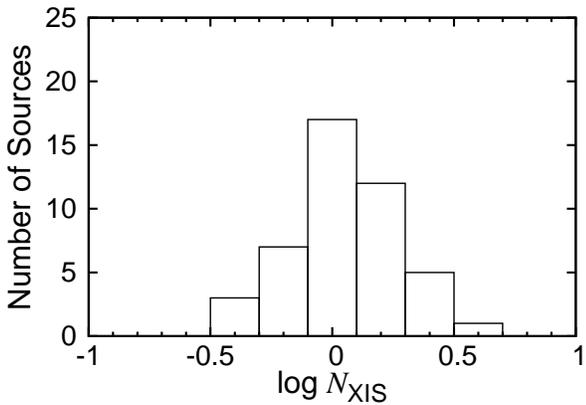}
\caption{
Distribution of the time variation constant of the cutoff power-law component 
between the {\it Suzaku} and {\it Swift}/BAT observations. 
}
\label{fig:hist_n_xis}
\end{center}
\end{figure}
%\fi 

%\if0 %%%%IRU file to be incorporated %%%%
\begin{figure*}[t]
\begin{center}\hspace{-3mm}
\includegraphics[scale=0.38,angle=-90]{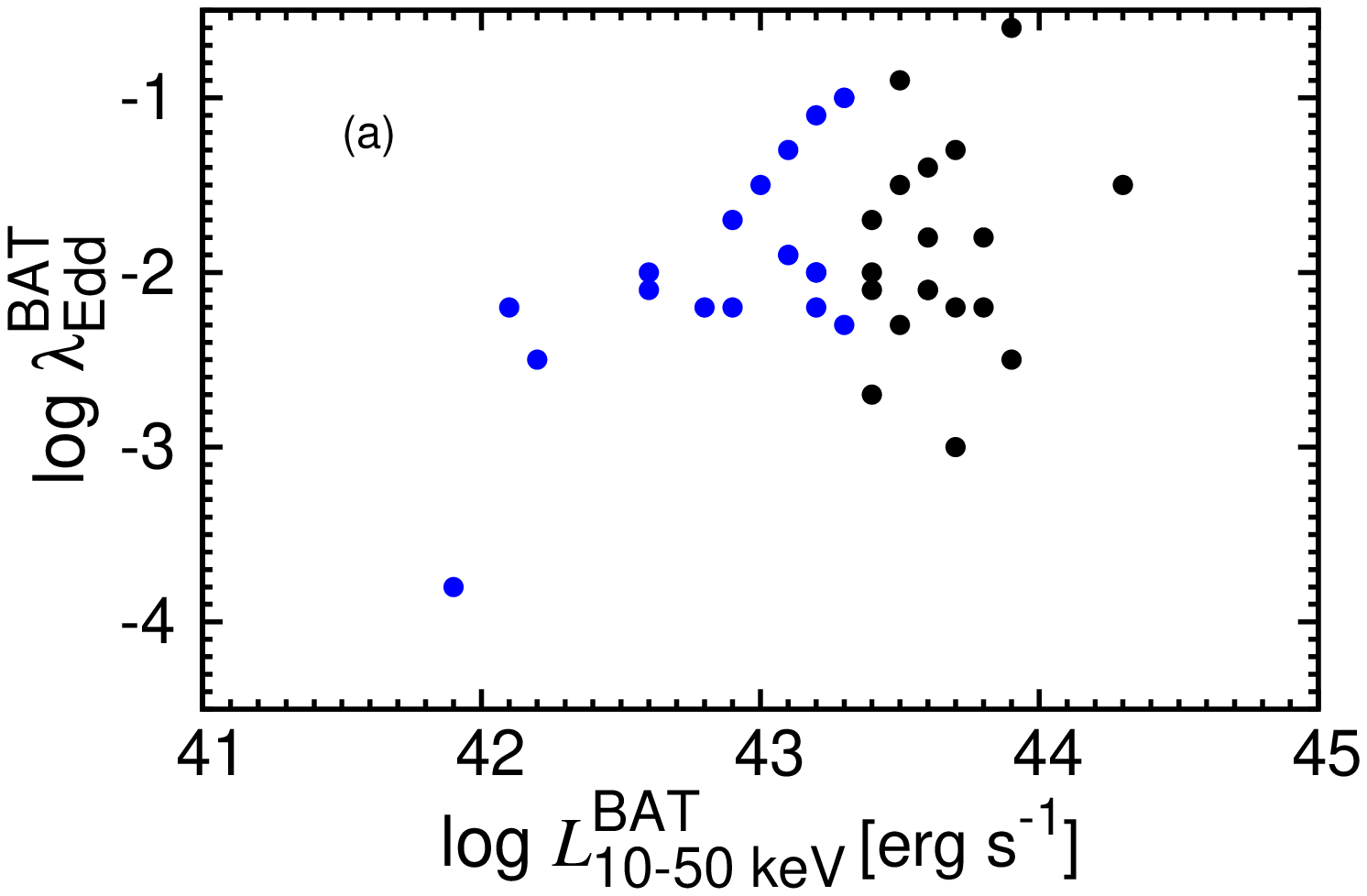} 
\includegraphics[scale=0.38,angle=-90]{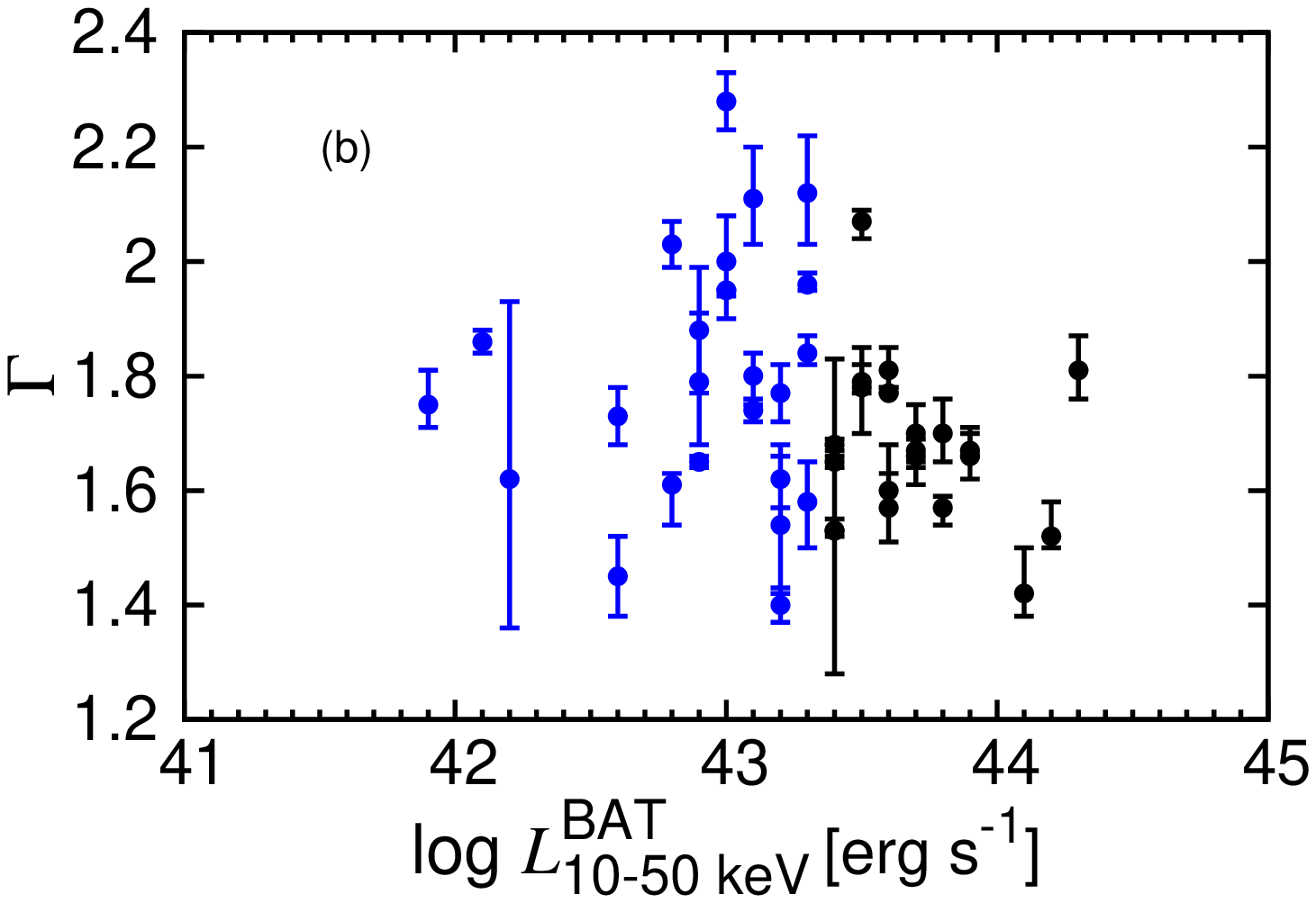} 
\includegraphics[scale=0.38,angle=-90]{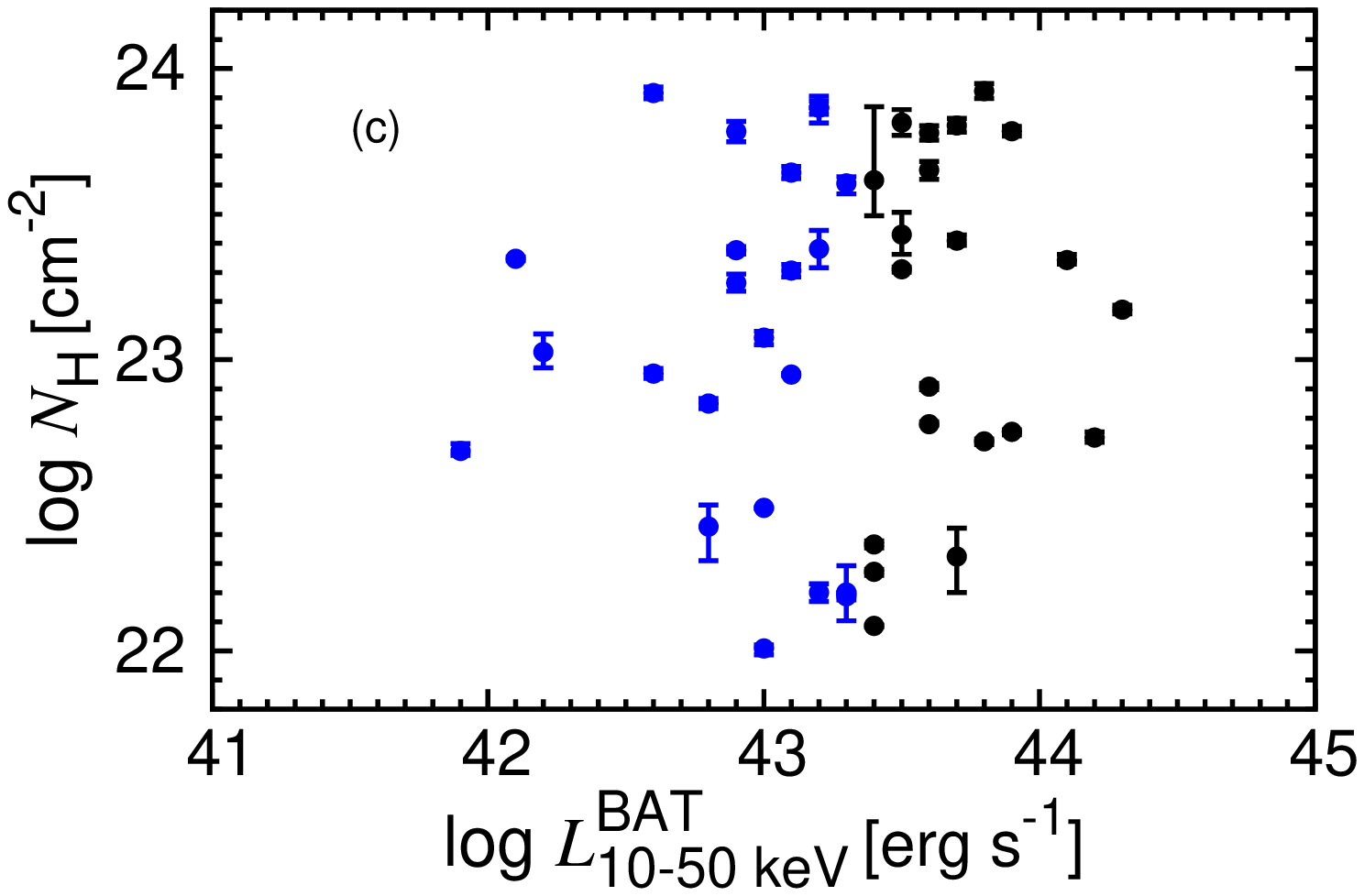} 
\includegraphics[scale=0.38,angle=-90]{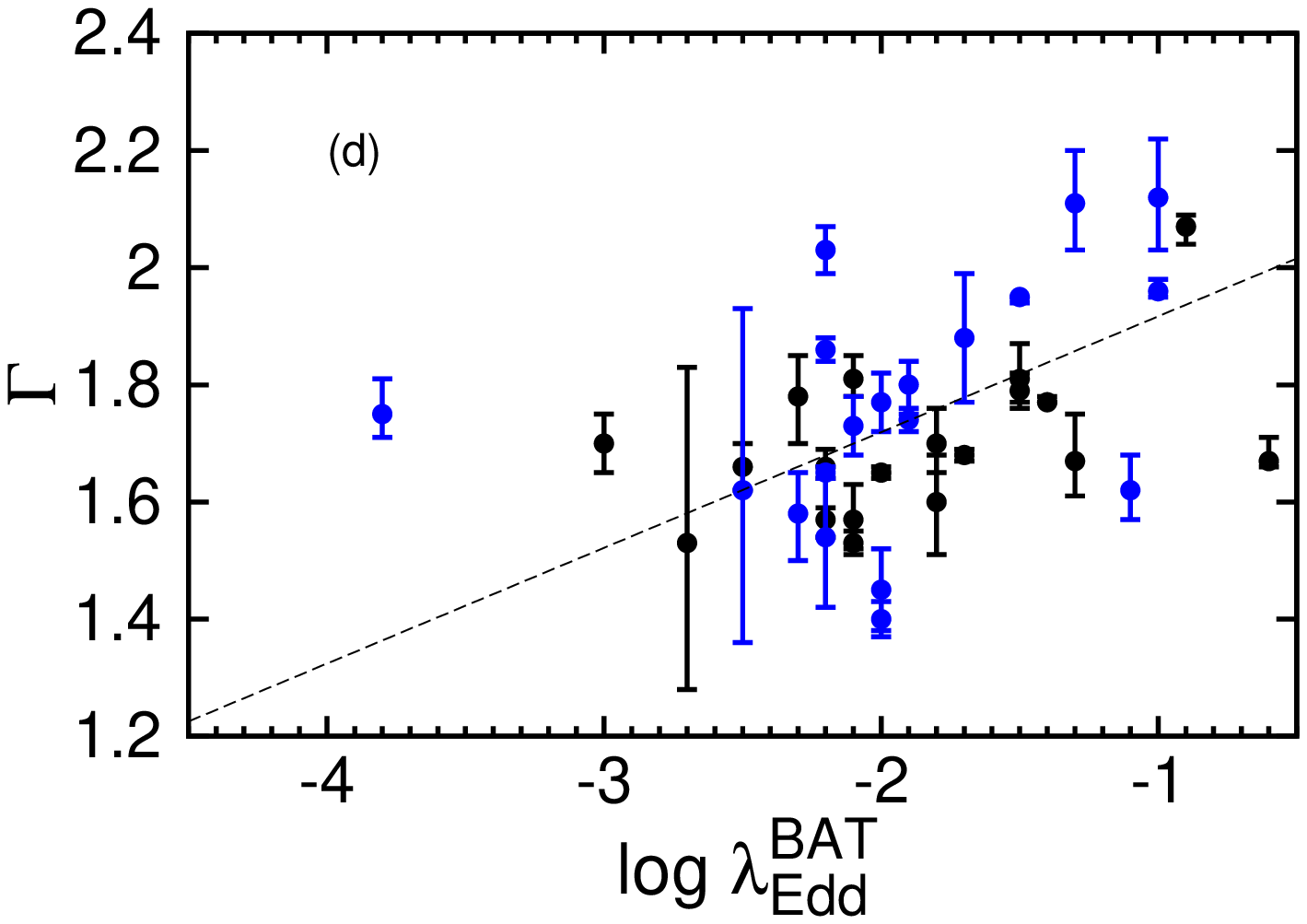} 
\includegraphics[scale=0.38,angle=-90]{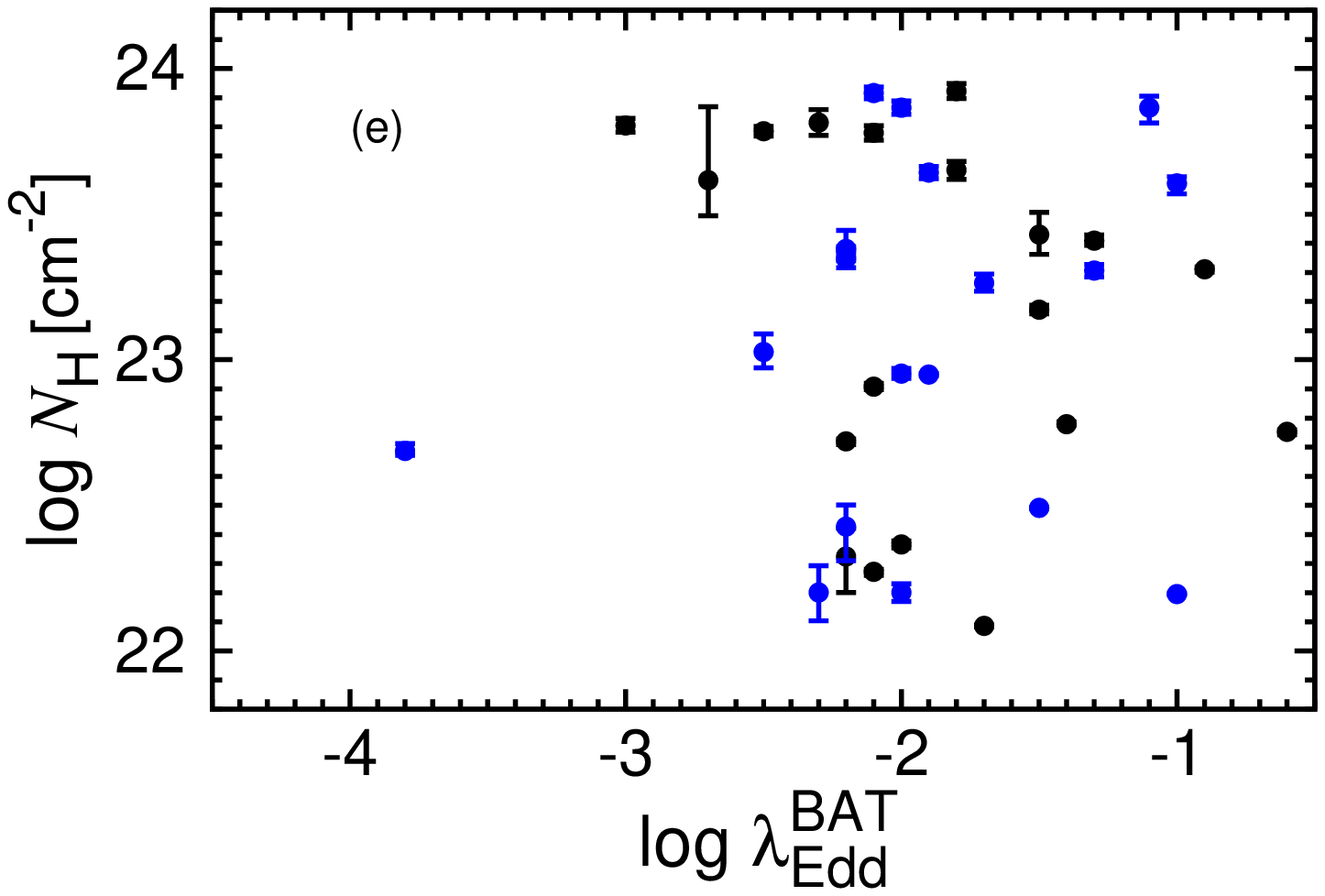} 
\includegraphics[scale=0.38,angle=-90]{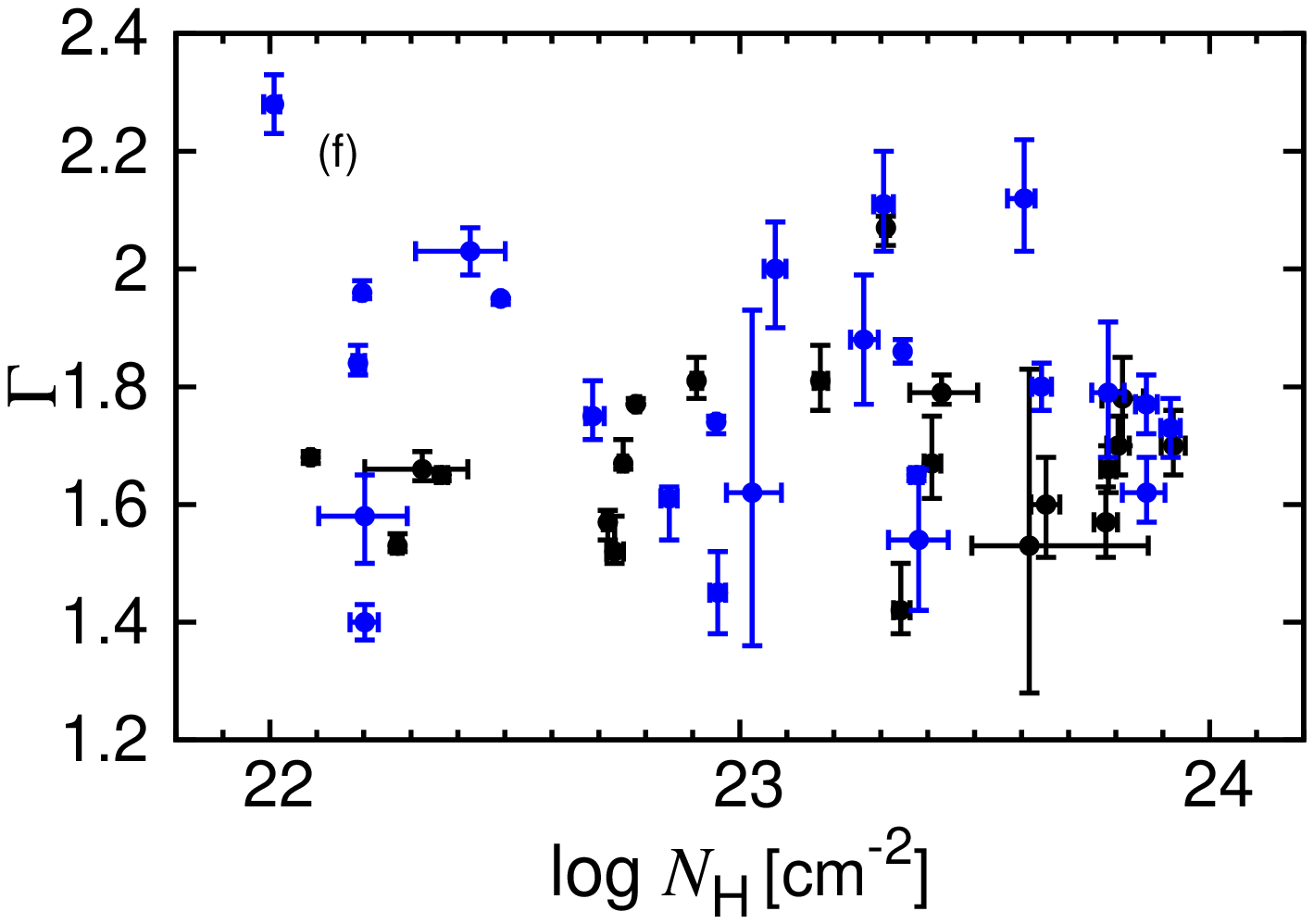} 
\caption{
(a) correlation of Eddington ratio with 10--50 keV luminosity 
and regression function (dashed line). 
(b) correlation of photon index with 10--50 keV luminosity 
of 
$\Gamma = 8.6 - 0.16 \log L^{\rm BAT}_{\rm 10-50\hspace{1mm}keV}$.
(c) correlation of absorption column density with 10--50 keV luminosity. 
(d) correlation of photon index with Eddington ratio and regression function (dashed line) 
of 
$\Gamma = 2.11 + 0.20 \log \lambda^{\rm BAT}_{\rm Edd}$.
(e) correlation of absorption column density with Eddington ratio. 
(f) correlation of photon index with absorption column density. 
The blue and black circles represent the MLAGNs 
($L^{\rm BAT}_{\rm 10-50\hspace{1mm}keV} \leq 43.3$) and HLAGNs 
($L^{\rm BAT}_{\rm 10-50\hspace{1mm}keV} > 43.3$), respectively. 
}
\label{fig:basic_cor}
\end{center}
\end{figure*}
%\fi 

\section{Results and Discussion}\label{sec:res_dis}

In this section, we summarize X-ray properties of our sample obtained
from the spectral analysis described in Section~\ref{sec:base_model},
and investigate correlations among them. We always refer to the results 
with the best-fit model without the relativistic disk-reflection component 
(Section~\ref{sec:base_model}) for all targets. Inclusion of the 
disk-reflection components little affects the other spectral parameters 
for the four objects reported in Section~\ref{sec:disk}. Then we also 
compare them with the MIR properties (12 $\mu$m and \oiv 25.89 $\mu$m, 
hereafter \oiv\hspace{-1mm}, luminosity). To statistically quantify the 
correlation strength between two variables $($X$,$Y$)$, we calculate the 
Spearman's rank coefficients $\rho ($X$,$Y$)$ and standard Student's t-null 
significance levels $P($X$,$Y$)$.  To derive a linear regression line with 
a form of Y$ = a + b$X, we adopt the ordinary least-square (OLS) bisector 
method \citep{Iso90} for luminosity-luminosity correlations, or the least
chi-square method for the others, unless otherwise noted. Table~\ref{tab:cor} 
gives the results for different combinations of parameters.  If the 
correlation is found to be significant at $>$95\% confidence level, then 
its best-fit regression line is plotted in the corresponding figure. 

We examine possible time variability of $N_{\rm H}$ 
by compiling previous results in the literature obtained 
with {\it XMM-Newton}, {\it Chandra}, or {\it NuSTAR}.
Also, if an object was observed with {\it Suzaku} on multiple 
occasions, we analyze all available data with the same spectral 
model and derive $N_{\rm H}$ for each epoch. 
These results are summarized in Table~\ref{tab:info_nh}. 
Although we have to bear in mind that the best-fit $N_{\rm H}$ depends on the
continuum model adopted, 4 objects (Mrk 1210, Mrk 348, NGC 1052, and NGC 4507) 
seem to show significant time variability of $N_{\rm H}$ by a factor of $>2$ within 15 years. 
Nevertheless, 
we confirm that even if we adopt the averaged value of $N_{\rm H}$
instead of the {\it Suzaku} only result, it does not affect the sample
selection $(22 \leq \log N_{\rm H} < 24)$ and our results on 
the correlations of $N_{\rm H}$ with other X-ray properties 
(Sections~\ref{sec:x-ray_pro} and \ref{sec:ref}) and on 
the stacked X-ray spectral analysis (Section~\ref{sec:ave_ref}).

\subsection{Basic X-ray Properties}\label{sec:x-ray_pro}

Figure~\ref{fig:mean_std} shows distributions of the absorption-corrected
10--50 keV luminosity ($L_{\rm 10-50\hspace{1mm}keV}$), Eddington ratio 
($\lambda_{\rm Edd}$), photon index ($\Gamma$), and hydrogen column density 
($N_{\rm H}$) of the whole sample. The mean and standard deviation are
summarized in Table~\ref{tab:mean_std}. In Figure~\ref{fig:hist_n_xis}, we 
plot the distribution of the time-variation constant, $N_{\rm XIS}$, which 
represents the luminosity change of the primary cutoff power-law component 
between the {\it Suzaku} and {\it Swift}/BAT observations. The mean and 
standard deviation are $0.05\pm0.03$ and $0.21\pm0.04$. This suggests 
that a typical level of variability of the primary X-ray emission 
on timescales of $\sim$ a day to several years is $\sim$ 0.2 dex.

\renewcommand{\arraystretch}{1.3}
\begin{deluxetable}{ccc}
\tabletypesize{\small}
\tablecaption{Mean and Standard Deviation of Spectral Parameters\label{tab:mean_std}}
\tablewidth{0pt}
\tablehead{ X & $r($X$)$ & $\sigma($X$)$ \\
(1) & (2) & (3)
}
\startdata
$\log L^{\it Suzaku}_{\rm 10-50\hspace{1mm}keV}$ & $43.3\pm0.08$ & $0.53\pm0.11$\\ 
$\log L^{\rm BAT}_{\rm 10-50\hspace{1mm}keV}$   & $43.3\pm0.08$ & $0.51\pm0.11$\\ 
$\log \lambda^{\it Suzaku}_{\rm Edd}$ & $-1.9\pm0.1$ & $0.6\pm0.1$\\ 
$\log \lambda^{\rm BAT}_{\rm Edd}$ & $-1.9\pm0.1$ & $0.6\pm0.1$ \\ 
$\Gamma$ & $1.74\pm0.03$ & $0.19\pm0.04$ \\ 
$\log N_{\rm H}$ & $23.1\pm0.09$ & $0.59\pm0.12$ \\
$\log N_{\rm XIS}$ & $0.05\pm0.03$ & $0.21\pm0.04$ 
\enddata
\tablecomments{
Columns: (1) Parameter. (2) Mean of X. 
(3) Standard deviation of X. 
}
\end{deluxetable}
\renewcommand{\arraystretch}{1.0}

Figure~\ref{fig:basic_cor} shows correlations among 
$\log L^{\rm BAT}_{\rm 10-50\hspace{1mm}keV}$, 
$\log \lambda^{\rm BAT}_{\rm Edd}$, $\Gamma$, and $\log N_{\rm H}$. 
For easy check of any luminosity dependence, we divide our sample into
two groups, moderate luminosity AGNs (MLAGNs) with 
$\log L^{\rm BAT}_{\rm 10-50\hspace{1mm}keV} \leq 43.3$ and high luminosity 
ones (HLAGNs) with $\log L^{\rm BAT}_{\rm 10-50\hspace{1mm}keV}  
> 43.3$, which consists of 24 and 21 objects, respectively. 
The criterion of $\log L^{\rm BAT}_{\rm 10-50\hspace{1mm}keV}$ is determined so that 
the source numbers of the two subsamples become the same when we make 
spectral stacking analysis (Section~\ref{sec:ave_ref}).

As inferred from Figure~\ref{fig:basic_cor}(a), there is no 
 significant $\log \lambda^{\rm BAT}_{\rm Edd}$--$\log L^{\rm BAT}_{\rm
10-50\hspace{1mm}keV}$ correlation. However, when adopting the 
luminosity-dependent correction factor of \cite{Mar04}, 
we find a significant correlation with 
$P (\log L^{\rm BAT}_{\rm 10-50\hspace{1mm}keV}, \log \lambda_{\rm Edd}) 
= 2.9\times10^{-2}$.

We find that photon index increases with Eddington ratio ($P (\log
\lambda^{\rm BAT}_{\rm Edd}, \Gamma) = 9.7 \times10^{-3}$), but does not
significantly correlates with luminosity ($P (\log L^{\rm BAT}_{\rm
10-50\hspace{1mm}keV}, \Gamma) = 8.6\times10^{-2}$). The dependence 
of photon index on black hole mass is found to be rather weak, with 
$P (\log M_{\rm BH}, \Gamma) = 5.0 \times10^{-2}$ and 
$\rho(\log M_{\rm BH}, \Gamma) = -0.32$. The positive 
$\Gamma$--$\log \lambda_{\rm Edd}$ correlation for luminous AGNs was 
also reported previously \citep[e.g.,][]{She08,Bri13,Yan15b}. The 
slope we obtain ($b = 0.20\pm0.01$) is flatter than those obtained in 
previous studies \citep[$\approx 0.3$; e.g.,][]{She08}, however. It may
be because our sample lacks high-Eddington ratio, high-luminosity AGNs
($\log L_{\rm X} > 44$), which likely show much softer spectra. Most of them
are identified as type-1 AGNs, as expected from the luminosity
dependence of the type-1 AGN fraction
\citep[e.g.,][]{Ued03,Ued14,Has08,Bec09,Bri11a,Bur11,Ric13a}. 
We also confirm that even if we adopt the bolometric correction factor of \cite{Mar04},  
the positive correlation remains tight with $P (\log \lambda^{\rm BAT}_{\rm Edd}, 
\Gamma) = 2.1 \times10^{-2}$. Hence, the Eddington ratio may be an important 
parameter that determines the nature of the X-ray emitting corona 
(optical depth end electron temperature). Theoretically, this correlation 
can be explained if Compton cooling of 
the corona by seed photons becomes more efficient with increasing accretion rate, 
leading to a smaller Compton y-parameter, and hence a softer spectrum.

The hydrogen column density correlates with neither $\log L^{\rm
BAT}_{\rm 10-50\hspace{1mm}keV}$ nor $\log \lambda^{\rm BAT}_{\rm Edd}$.
Also, there is no significant correlation between $\log N_{\rm H}$ and
$\Gamma$, supporting that the underlying continuum shape is properly
determined without strong coupling with the absorption unlike the case
of narrow band spectral analysis with limited photon statistics.

%\if0 %%%%IRU file to be incorporated %%%%
\begin{figure}[t]
\begin{center}
\includegraphics[scale=0.53,angle=-90]{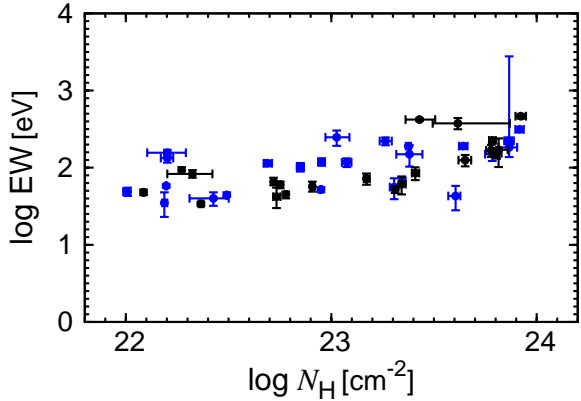}
\caption{
Correlation between the iron-K$\alpha$ EW and absorption column density.
The blue and black circles represent the MLAGNs 
($L^{\rm BAT}_{\rm 10-50\hspace{1mm}keV} \leq 43.3$) and HLAGNs 
($L^{\rm BAT}_{\rm 10-50\hspace{1mm}keV} > 43.3$), respectively. 
}
\label{fig:x-ray_baldwin}
\end{center}
\end{figure}
%\fi 

\subsection{Reflected Emission from Distant, Cold Matter} \label{sec:ref}

The narrow iron-K$\alpha$ fluorescence line at $\approx$ 6.4 keV is
originated from distant, cold matter around the nucleus. We
conventionally call it ``torus'', although its shape and size are still 
largely unknown. We significantly detect the iron-K$\alpha$ line for all the 
objects of our sample. 
Except for 5 objects, the EW with respect to the reflection continuum
is found to be in a range of 0.5--3 keV, which is consistent with a 
theoretical prediction from reflection by optically thick, cold matter
within variations of inclination (EW $\sim$ 1--2 keV; \citealt{Mat91}) and
of iron abundance by a factor of 2 \citep{Mat97}.
In the rest of objects (NGC 4388, NGC 5252, LEDA 170194, Mrk 520, and Mrk
915), the EW is very large ($>3$ keV) because the reflection continuum 
(i.e., $R$) is apparently very weak. We interpret that 
the tori in these objects are Compton-thin, thus producing a 
much weaker hump at $\sim$ 30 keV and absorption iron-K edge features 
than what the {\tt pexrav} model predicts. In fact, we confirm that a 
Monte-Carlo based torus model 
by \cite{Ike09} can well reproduce the spectra of these objects
including the continuum and iron-K$\alpha$ emission line, as done in
\cite{Taz13}, \cite{Kaw13}, and \cite{Kaw16}. Systematic 
application of numerical torus models 
to all spectra of our sample is a subject of future works.

The relative intensity of the iron-K$\alpha$ line to the
underlying continuum contains critical information on the covering
fraction and column density of the torus. Figure~\ref{fig:x-ray_baldwin}
plots the observed EW with respect to the total continuum
against $N_{\rm H}$. A systematic increase in EW
with $N_{\rm H}$ is seen in log $N_{\rm H} > 23$, confirming previous
results \citep[e.g.,][]{Fuk11,Bri11a}. This can be explained by the attenuation
of the transmitted component by absorption at the same energy, which
makes the EW of the iron K$\alpha$ line larger. Thus, EW is not a good
indicator to discuss the torus structure.

\begin{figure*}[t]
\begin{center}
\includegraphics[scale=0.55,angle=-90]{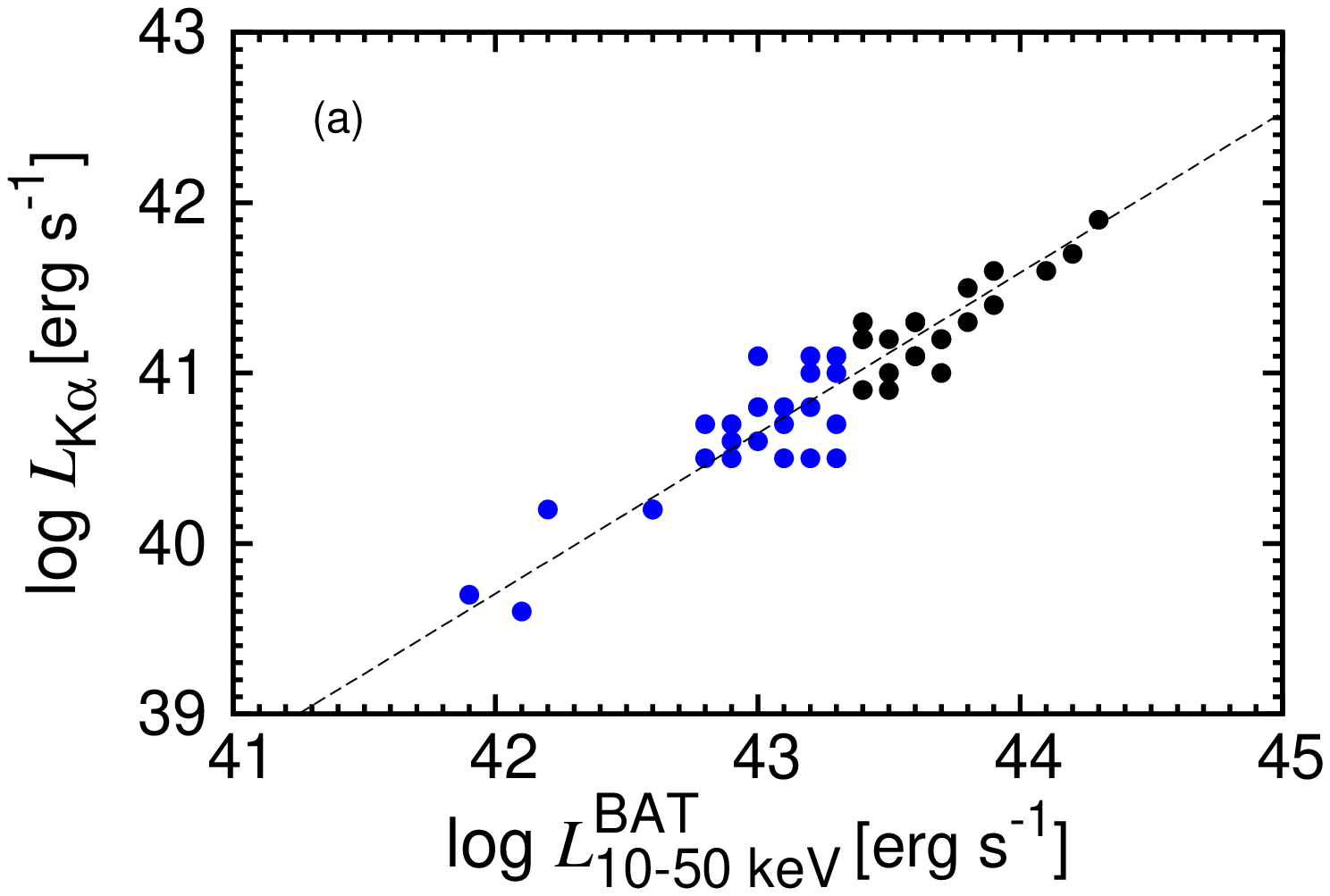} 
\hspace{2mm}
\includegraphics[scale=0.55,angle=-90]{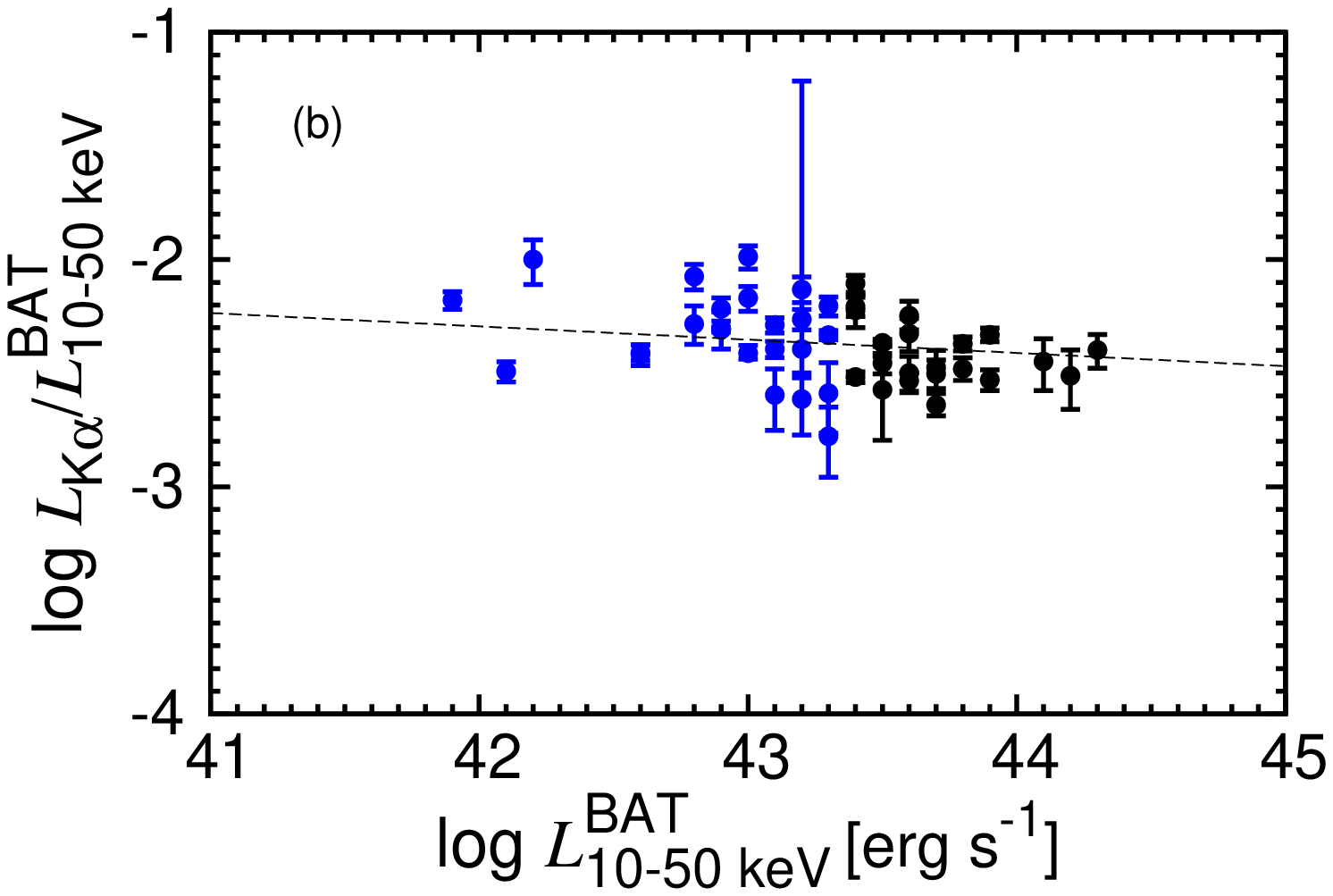}
\caption{
(a) correlation of iron-K$\alpha$ line luminosity with 10--50 keV luminosity 
and regression function (dashed line) of 
$\log L_{\rm K\alpha} = 0.2 + 0.94 \log L^{\rm BAT}_{\rm 10-50\hspace{1mm}keV}$.
(b) correlation of the iron-K$\alpha$ to 
10--50 keV luminosity ratio with 10--50 keV luminosity 
and regression function (dashed line) of 
$\log (L_{\rm K\alpha}/L^{\rm BAT}_{\rm 10-50\hspace{1mm}keV}) = 
0.2 - 0.06 \log L^{\rm BAT}_{\rm 10-50\hspace{1mm}keV}$.
The blue and black circles represent the MLAGNs 
($L^{\rm BAT}_{\rm 10-50\hspace{1mm}keV} \leq 43.3$) and HLAGNs 
($L^{\rm BAT}_{\rm 10-50\hspace{1mm}keV} > 43.3$), respectively. 
The dashed lines represent the regression line. 
}
\label{fig:cor_x-ray_baldwin}
\end{center}
\end{figure*}

To correct for the absorption effect in the continuum flux, we adopt the
\lirlx\hspace{1mm} ratio instead of the EW as proposed by \citet{Ric14b}.
Figure~\ref{fig:cor_x-ray_baldwin} shows the correlations of 
$\log L^{\rm BAT}_{\rm 10-50\hspace{1mm}keV}$ with 
$\log L_{\rm K\alpha}$ and $\log($\lirlxB$)$. We first
calculate the OLS bisector regression line of the $\log L_{\rm
K\alpha}$--$\log L^{\rm BAT}_{\rm 10-50\hspace{1mm}keV}$ correlation. 
The correlation is significant with $P(\log L^{\rm BAT}_{\rm 10-50\hspace{1mm}keV}, 
\log L_{\rm K\alpha}) = 4.1\times10^{-16} $. The regression line we obtain gives 
a negative correlation between $\log($\lirlxB$)$ and $\log L^{\rm
BAT}_{\rm 10-50\hspace{1mm}keV}$ as shown in Figure~\ref{fig:cor_x-ray_baldwin}(b). 
The slope ($-0.06\pm0.04$) is consistent with $-0.11\pm0.01$ derived by 
\cite{Ric14b}. 

%\if0 %%%%IRU file to be incorporated %%%%
%\fi 

\begin{figure*}[t]
\begin{center}
\includegraphics[scale=0.39,angle=-90]{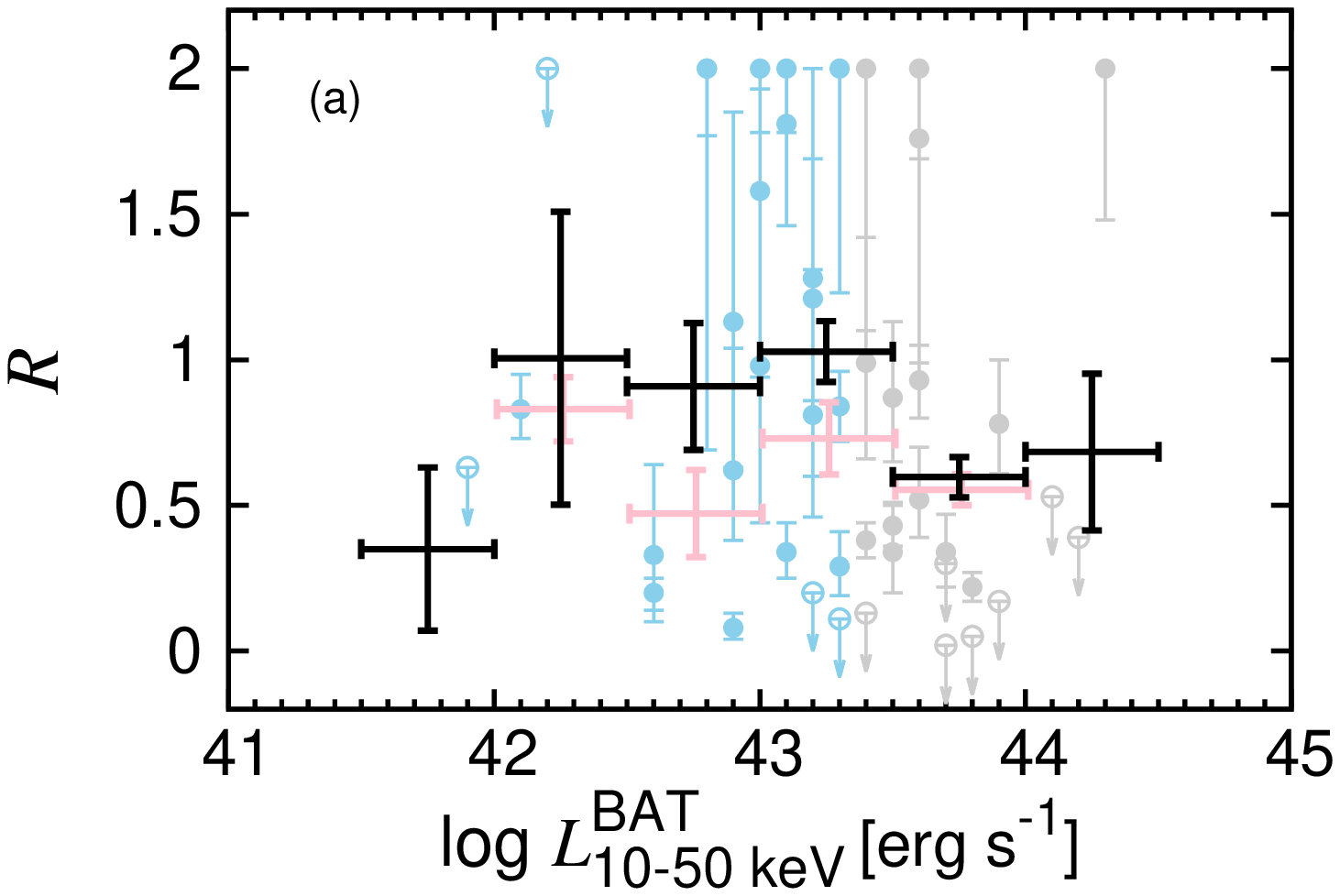} 
\hspace{-4mm}
\includegraphics[scale=0.39,angle=-90]{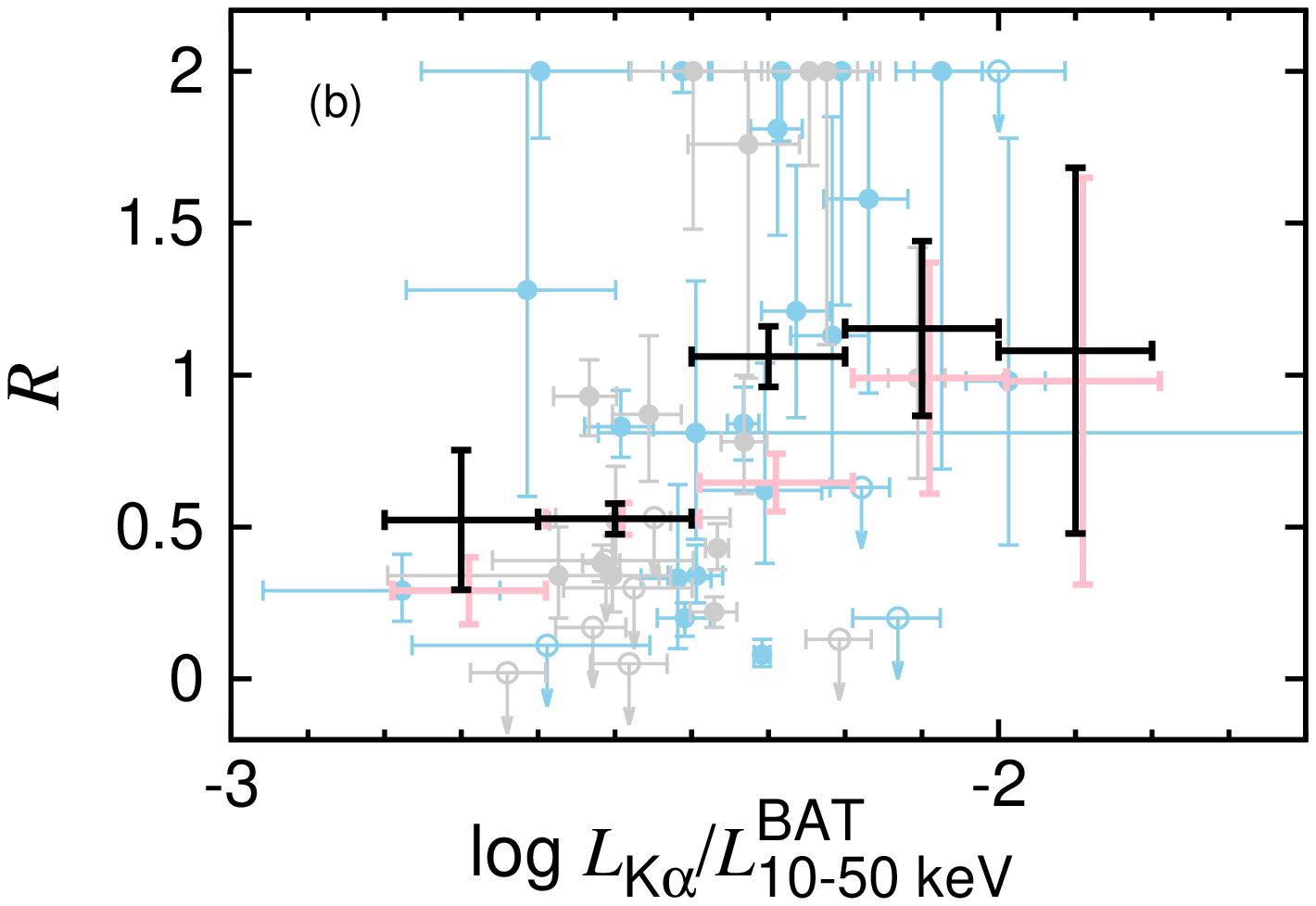} 
\hspace{-4mm}
\includegraphics[scale=0.39,angle=-90]{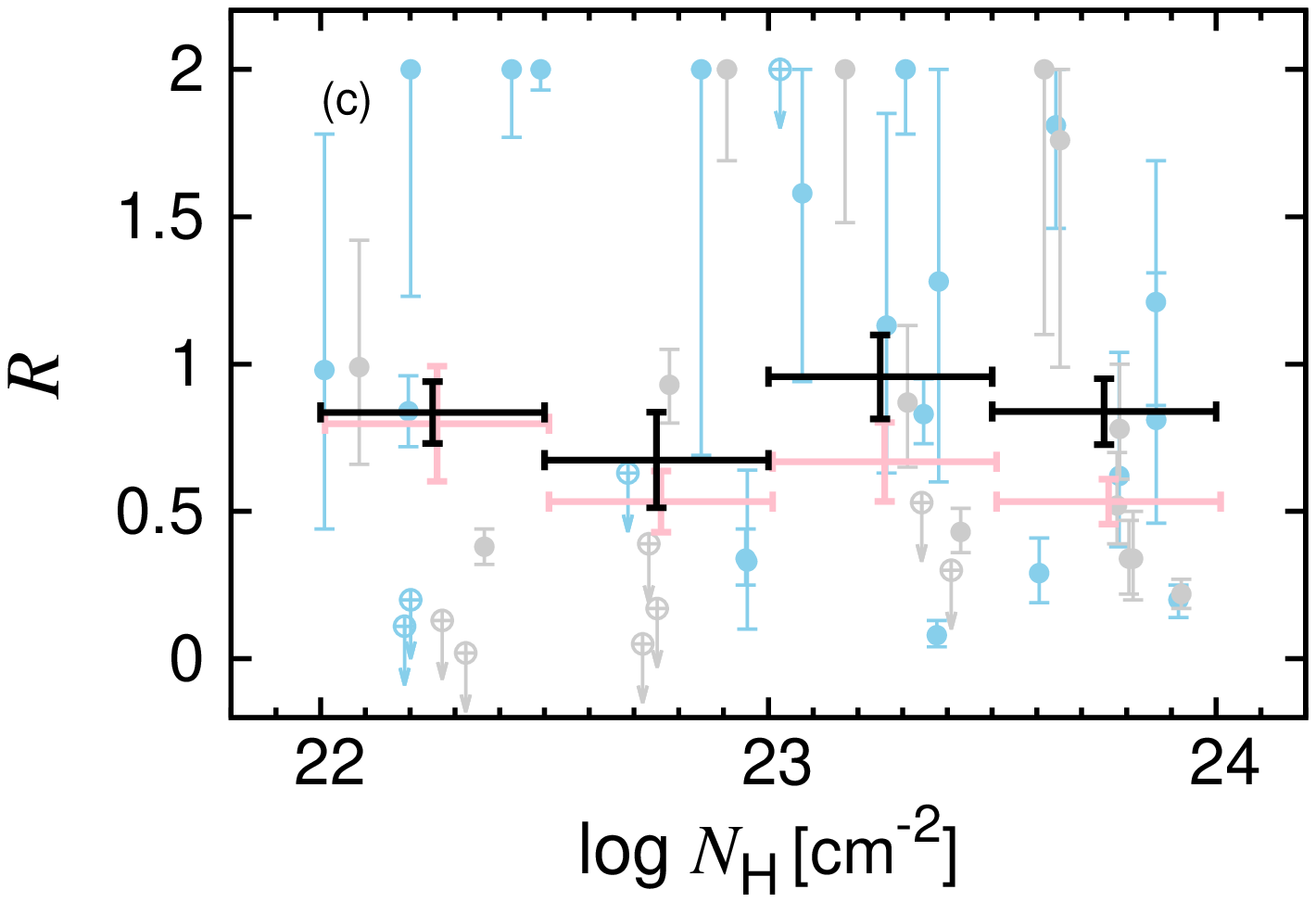} 
\caption{
(a) correlation of reflection strength with 10--50 keV luminosity.
(b) correlation of reflection strength with the iron-K$\alpha$ to 10--50 keV luminosity ratio.
(c) correlation of reflection strength with absorption column density. 
The cyan and gray circles represent the MLAGNs
 ($L^{\rm BAT}_{\rm 10-50\hspace{1mm}keV} \leq 43.3$) and HLAGNs 
($L^{\rm BAT}_{\rm 10-50\hspace{1mm}keV} > 43.3$), respectively.
The open circles represent data with upper limits. 
The black lines represent the mean and $1\sigma$ error of $R$ in each region 
for all objects, while the pink lines 
(slightly shifted to the right for clarity) for only objects whose 
$R$ values are significantly measured.
}
\label{fig:r_cor}
\end{center}
\end{figure*}

Strength of the reflection component from the torus, $R$ $(=
\Omega/2\pi)$, can be also used as an indicator of the torus covering
fraction. Figure~\ref{fig:r_cor} shows
correlations of $R$ with $\log L^{\rm BAT}_{\rm 10-50\hspace{1mm}keV}$,
$\log($\lirlxB$)$, and $\log N_{\rm H}$. 
We overplot the mean values of $R$ with 1$\sigma$ errors calculated 
in each region of the X-axis parameter from all objects (black) and 
from only objects whose $R$ values are significantly measured (pink) 
(i.e., the lower and upper limits of $R$ are larger than 0 and smaller
than 2, respectively), by adopting the best-fit value listed in
Table~\ref{tab:info_para}. 
Most of the results are consistent within the errors between the
two calculations and show the same trends against the X-axis parameter.
Although the negative correlation of $R$ with luminosity is not strong 
($P(\log L^{\rm BAT}_{\rm 10-50\hspace{1mm}keV}, R) = 5.6\times10^{-2}$), 
this trend is confirmed in Section~\ref{sec:ave_ref} by analyzing the stacked
hard X-ray spectra of MLAGNs  ($L^{\rm BAT}_{\rm 10-50\hspace{1mm}keV}
\leq 43.3$) and HLAGNs ($L^{\rm BAT}_{\rm 10-50\hspace{1mm}keV} > 43.3$).
On the other hand, the correlation between $R$ and $\log($\lirlxB$)$ is found to be
significant with $P(\log ($\lirlxB$), R) =
1.0\times10^{-2}$ and $\rho(\log ($\lirlxB$), R) = 0.38$. 
This result supports that the reflection continuum and narrow iron-K$\alpha$
emission line originate from the same material (i.e., torus). 
We also find the trend that $R$ is larger in more obscured AGNs, although
the significance is not high in this plot (but see 
Section~\ref{sec:ave_ref} for an analysis of averaged hard X-ray
spectra of the moderately obscured AGNs with $\log N_{\rm H} \leq 23$ and 
 highly obscured ones with $\log N_{\rm H} > 23$; hereafter MOAGNs and HOAGNs, 
respectively). This is expected if the direction-averaged column density and/or 
covering fraction of the torus is larger in AGNs with larger
line-of-sight absorptions.

%\if0 %%%%IRU file to be incorporated %%%%
%\fi 

\subsection{Average Hard X-ray Spectra}\label{sec:ave_ref}

To investigate the average reflection strength in an alternative way, we
analyze the {\it Swift}/BAT and HXD/PIN stacked hard X-ray spectra for the 
subsamples of MLAGNs and HLAGNs or those of MOAGNs and HOAGNs. Because
only continuum emission is present above 14 keV, we do not make any
K-correction in the summation, for simplicity, by excluding 3 distant
AGNs at $z>0.05$. A cutoff power-law plus its reflected component ({\tt
zpowerlw*zhighect+pexrav} in the XSPEC terminology) is used to reproduce
the continuum. The redshift is fixed to the mean value of each
subsample. Even at energies above 14 keV, large absorption with $\log
N_{\rm H} \gtrsim 23.5$ is not negligible. To take into account this
effect, we multiply the partial covering model, {\tt zpcfabs}, to the
above continuum. The covering fraction is set to the fraction of the
integrated fluxes of AGNs with $\log N_{\rm H} > 23.5$ in each subsample
and the column density is fixed at their average value. Thus, the
reflection strength, photon index, and normalization are left as free
parameters.

\begin{figure*}[t]
\begin{center}
\includegraphics[scale=0.22]{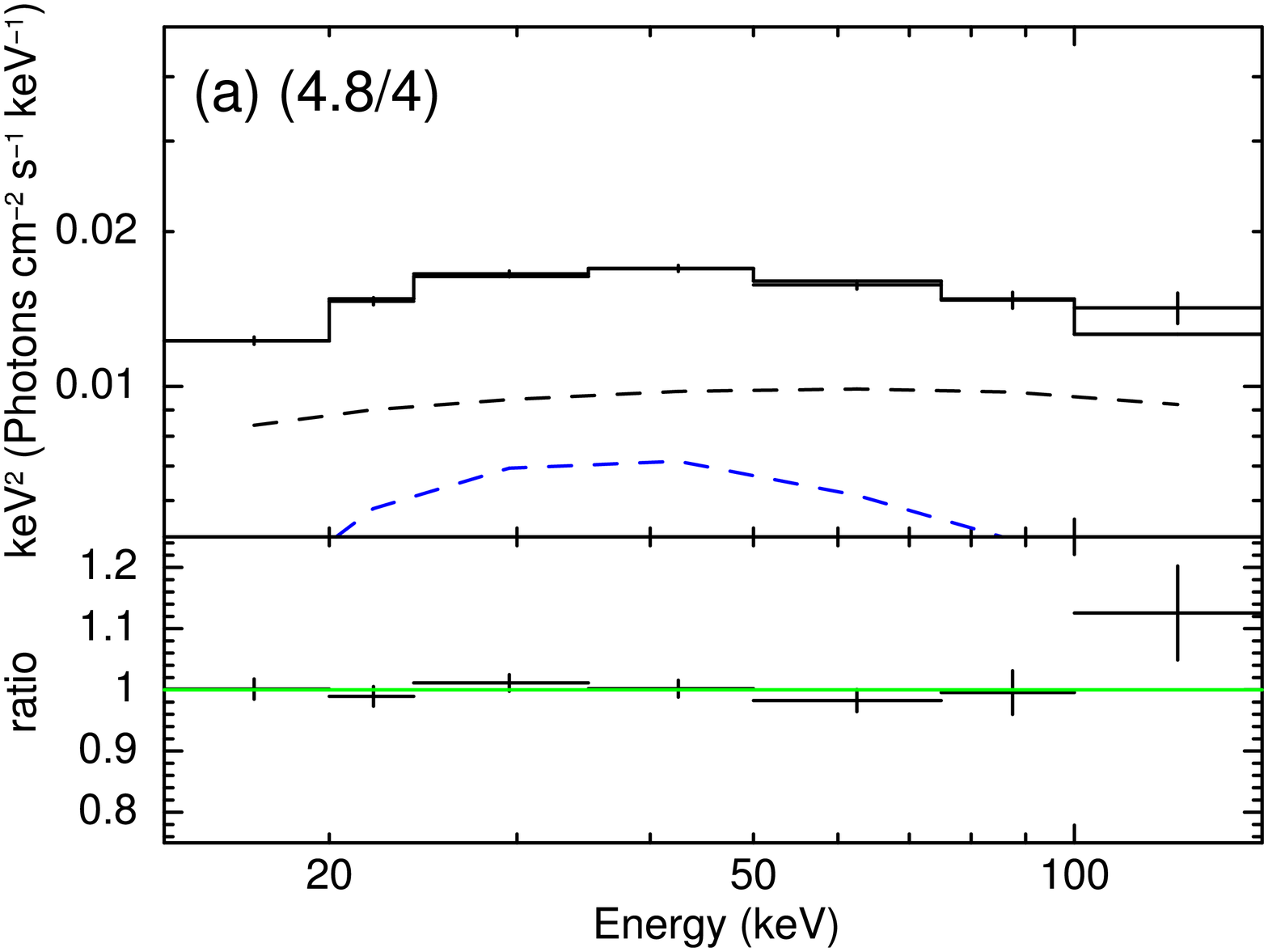} \hspace{2mm}
\includegraphics[scale=0.22]{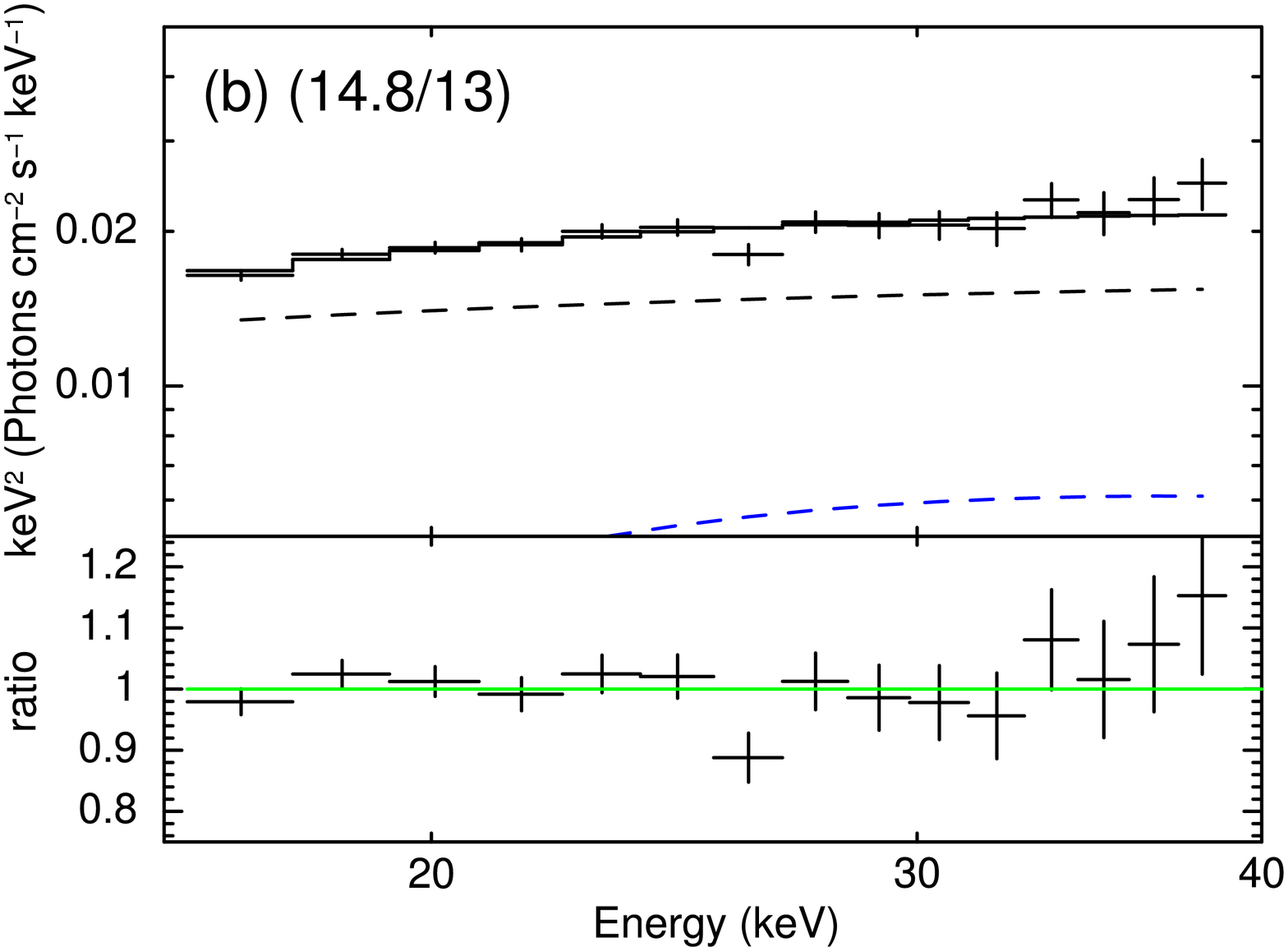} \hspace{3mm}
\includegraphics[scale=0.22]{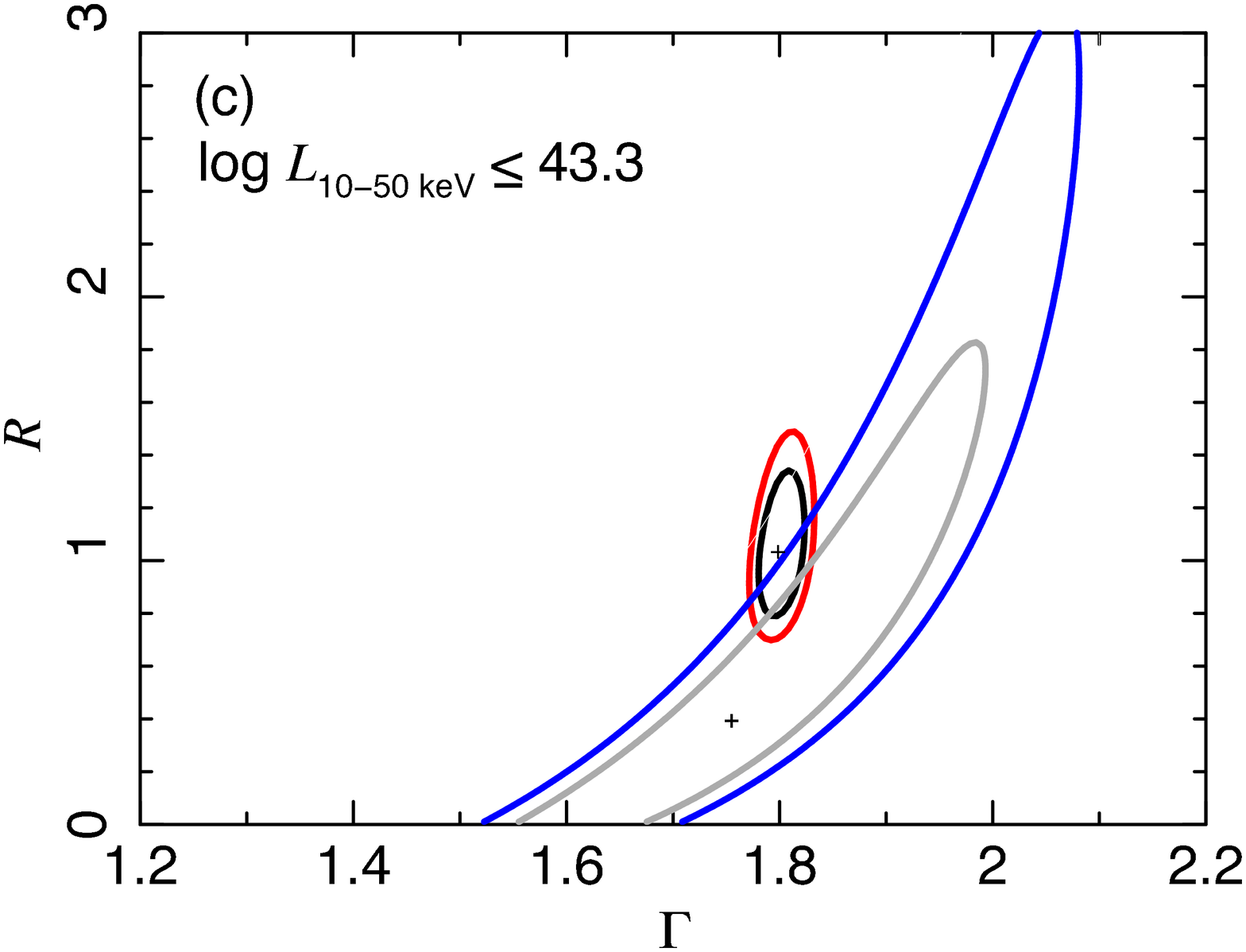}
\includegraphics[scale=0.22]{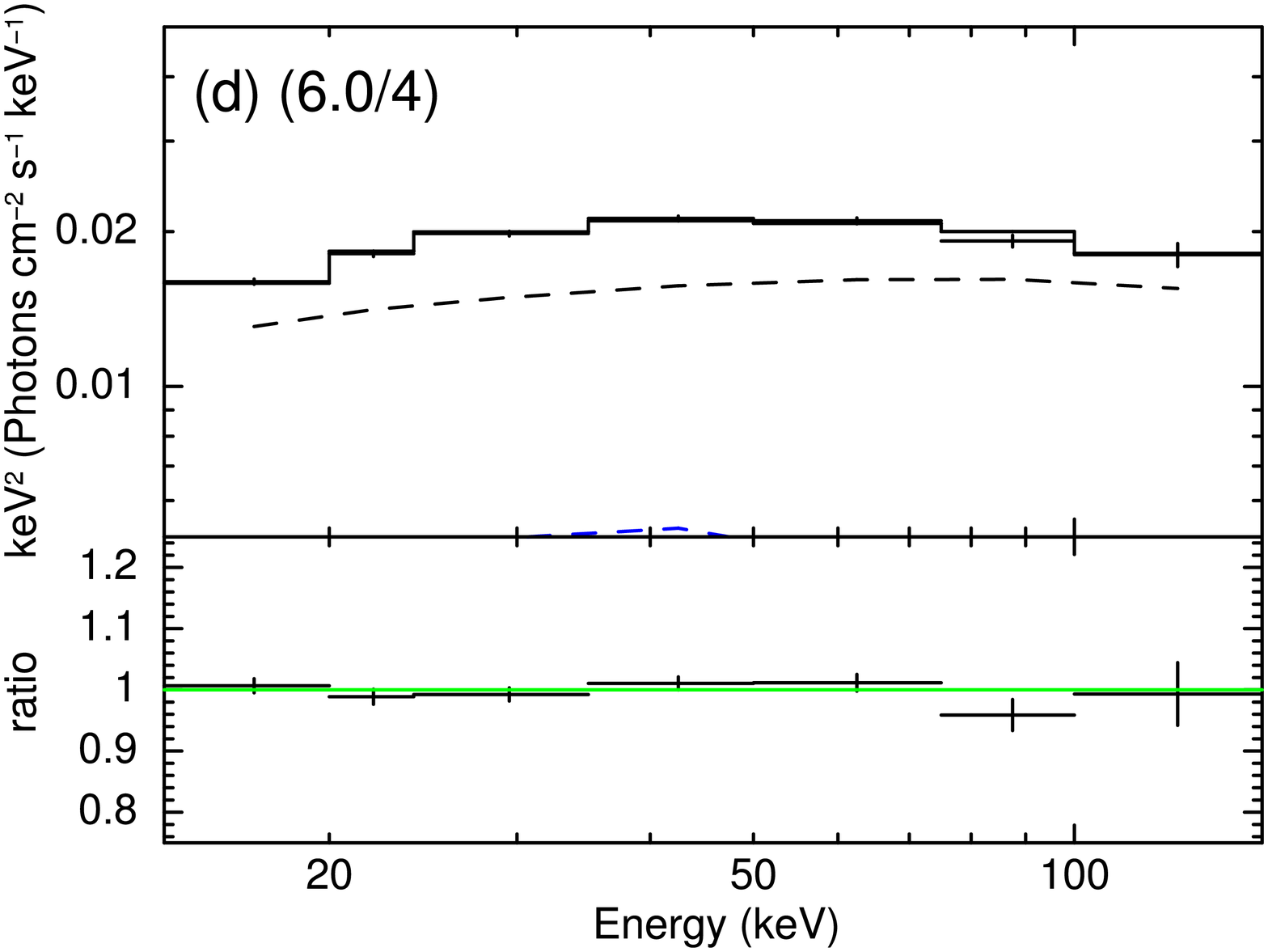} \hspace{2mm}
\includegraphics[scale=0.22]{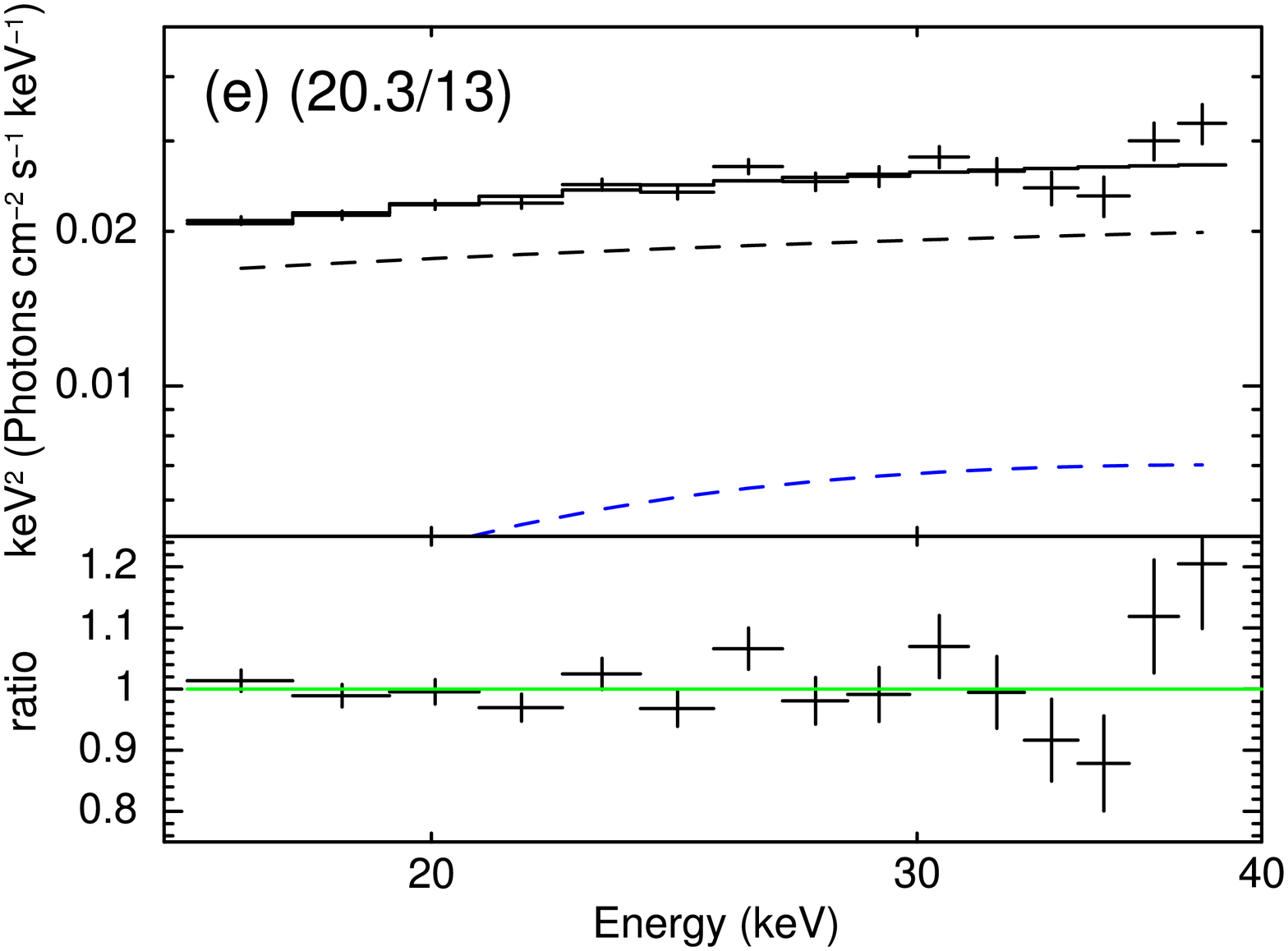} \hspace{3mm}
\includegraphics[scale=0.22]{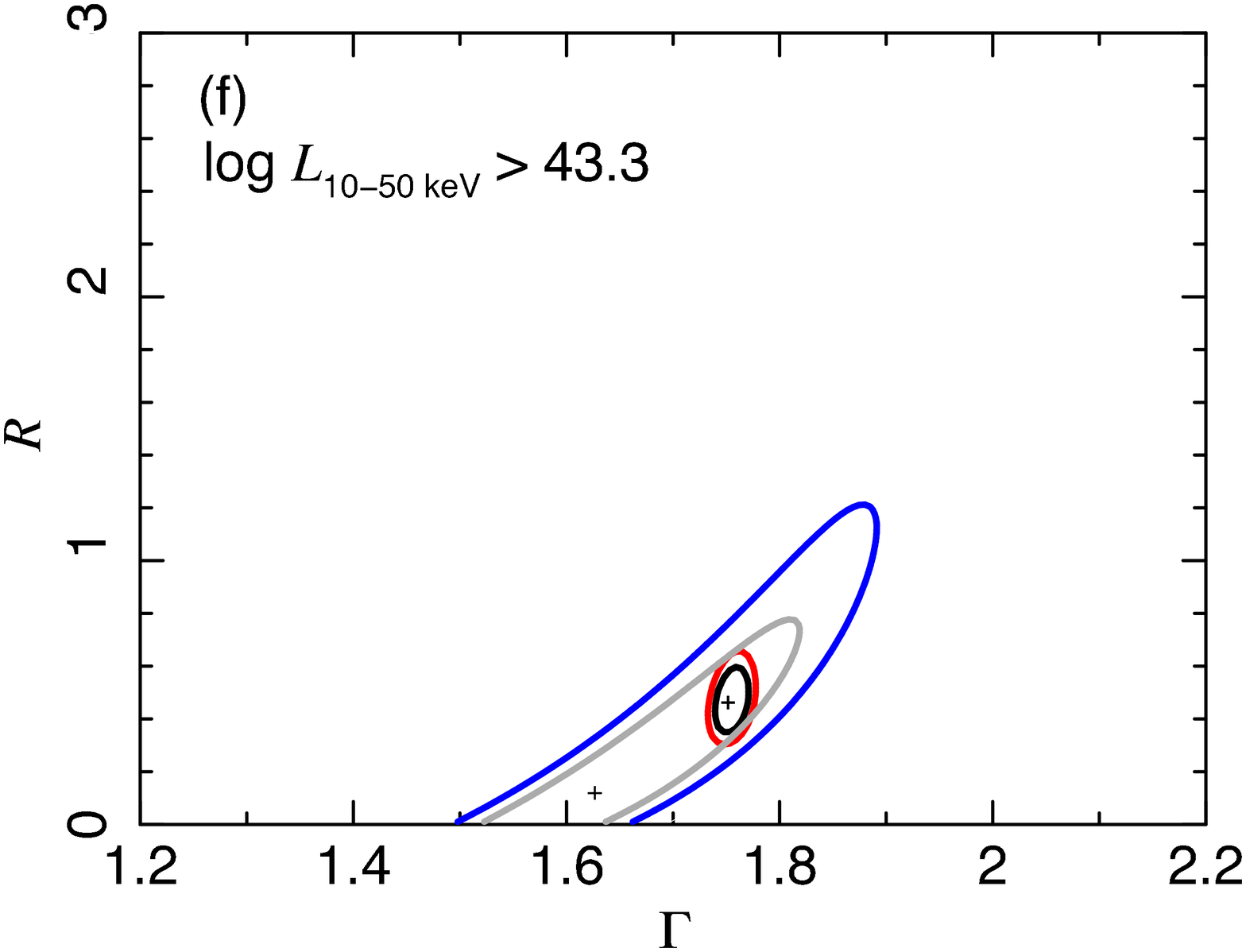}
\caption{
Upper: for the MLAGNs ($\log L^{\rm BAT}_{\rm 10-50\hspace{1mm}keV} \leq 43.3$). 
Lower: for HLAGNs ($\log L^{\rm BAT}_{\rm 10-50\hspace{1mm}keV} > 43.3$). 
(a), (d) the stacked {\it Swift}/BAT spectra.
(b), (e) the stacked HXD/PIN spectra.
(c), (f) confidence contours between photon index and reflection strength. 
In the left and central figures, the upper 
panel shows the spectrum with the best-fit model consisting of a
cutoff power-law (black dashed line) and its reflection component (blue
dashed line), whereas the lower panel plots the ratio
of the data to the model. The reduced chi-squared statistic 
($\chi^2/dof$) for each 
fitting result is represented in the parenthesis.
In the right figures, the constraint obtained 
from the {\it Swift}/BAT (HXD/PIN) spectrum is represented with the
black (gray) and red (blue) lines, corresponding to $\Delta \chi^2=2.3$
and $4.6$, or the 68\% and 90\% confidence levels, respectively. 
}
\label{fig:ave_spec_lhx}
\end{center}
\vspace{0.5cm}
\end{figure*}

Figure~\ref{fig:ave_spec_lhx} shows the unfolded spectra of {\it Swift}/BAT
and HXD/PIN for the MLAGN ($L^{\rm BAT}_{\rm 10-50\hspace{1mm}keV} \leq 43.3$) and HLAGN 
($L^{\rm BAT}_{\rm 10-50\hspace{1mm}keV} > 43.3$) subsamples, which contain
the same number of sources (21), together with their best-fit models. 
Here, when fitting the HXD/PIN spectra, we fix the photon index at 
the best-fitting value obtained from each {\it Swift}/BAT spectrum. 
The MLAGN spectrum shows significantly stronger
reflection strength ($R = 1.04^{+0.17}_{-0.19}$) than that of the HLAGNs
($R = 0.46^{+0.08}_{-0.09}$), consistent with the results suggected in 
Section~\ref{sec:ref}. The confidence contours between photon index 
and $R$ obtained from the {\it Swift}/BAT and HXD/PIN spectra are plotted 
in Figure~\ref{fig:ave_spec_lhx}(c) and (f). As noticed, the two results 
are compatible with each other, although the constraints obtained with 
the HXD/PIN data are much weaker owing to the limited energy band. Our 
findings support the luminosity-dependent unified AGN scheme.

\begin{figure*}[t]
\begin{center}
\includegraphics[scale=0.22]{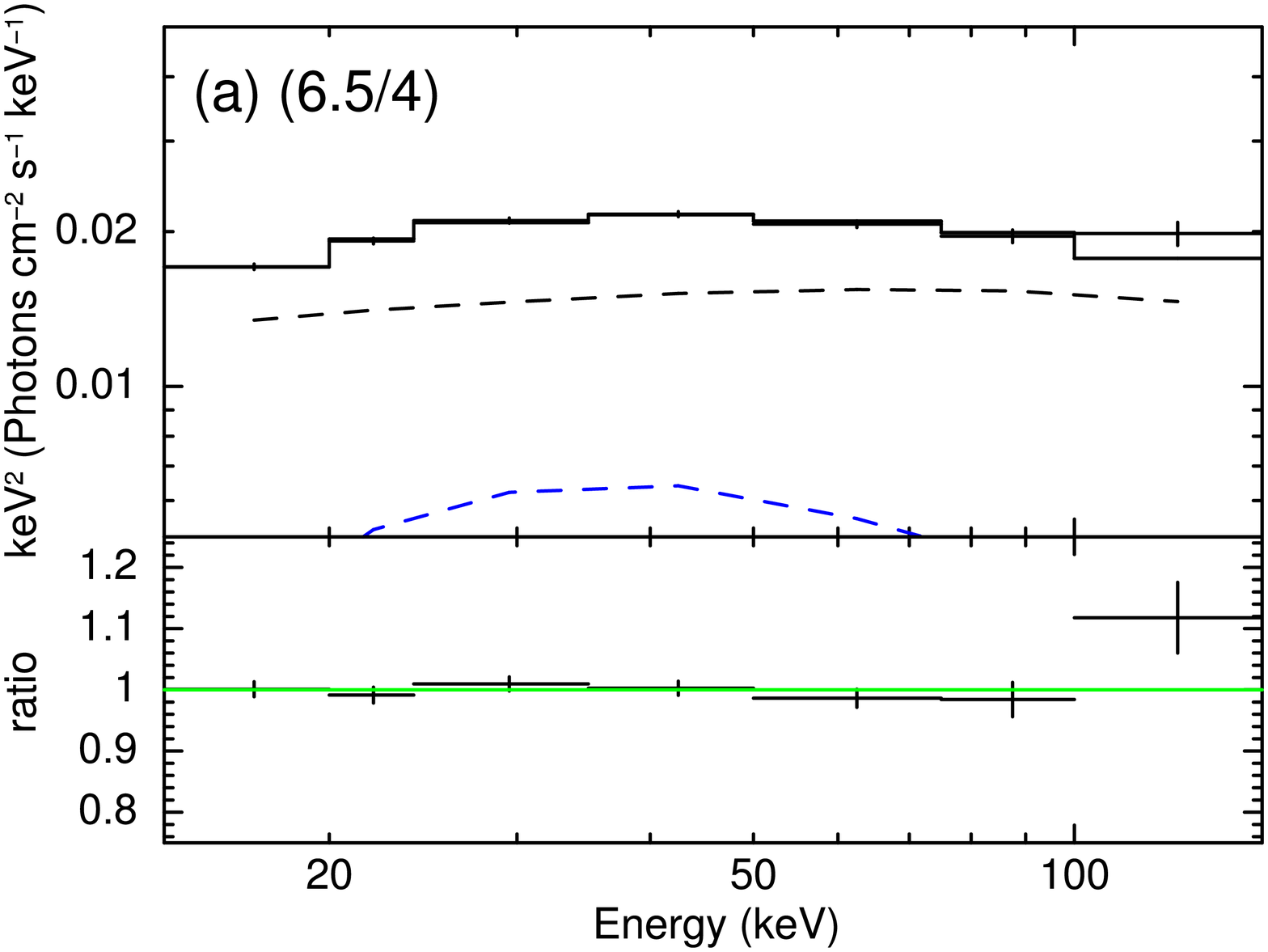} \hspace{2mm}
\includegraphics[scale=0.22]{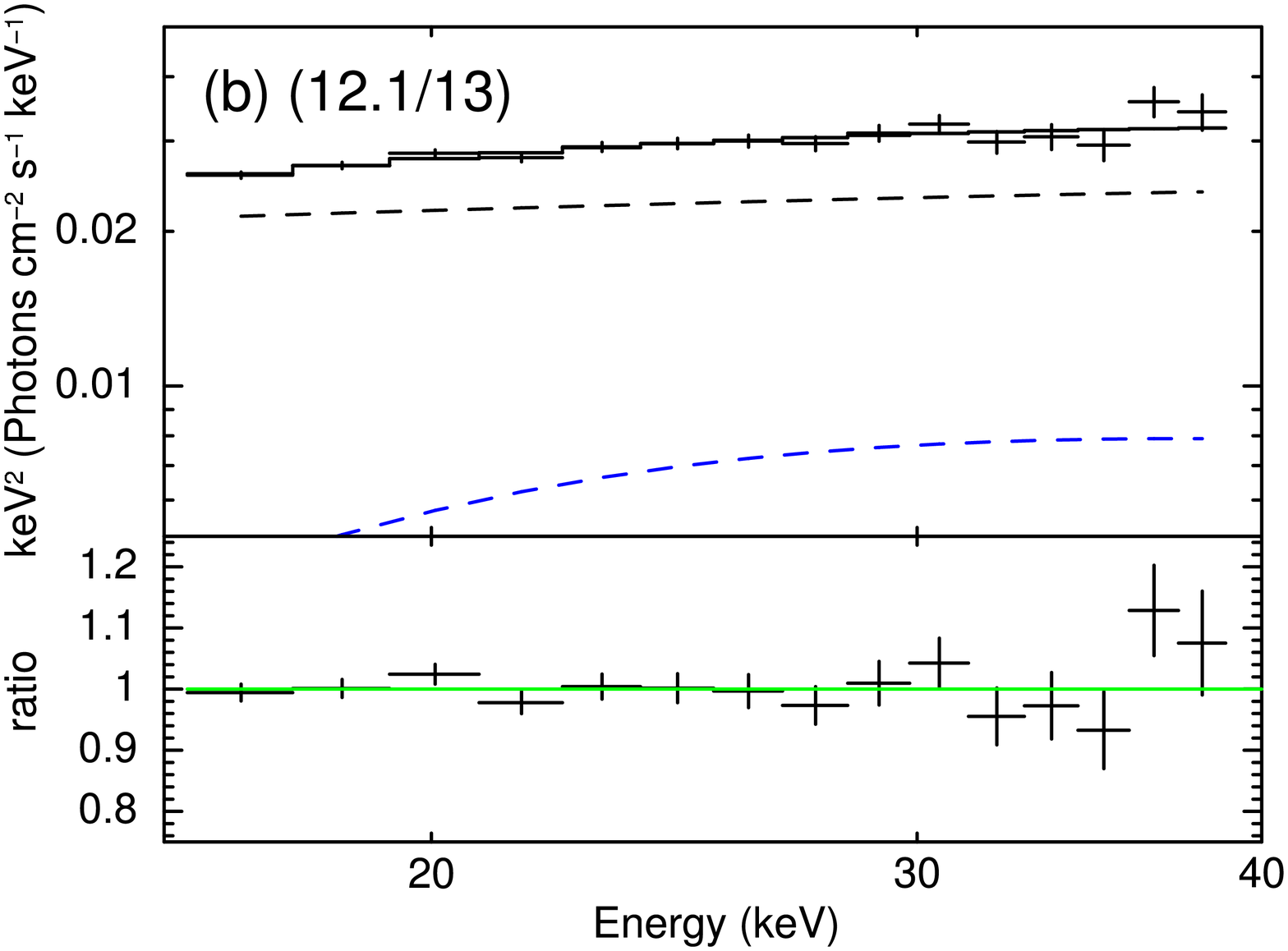} \hspace{3mm}
\includegraphics[scale=0.22]{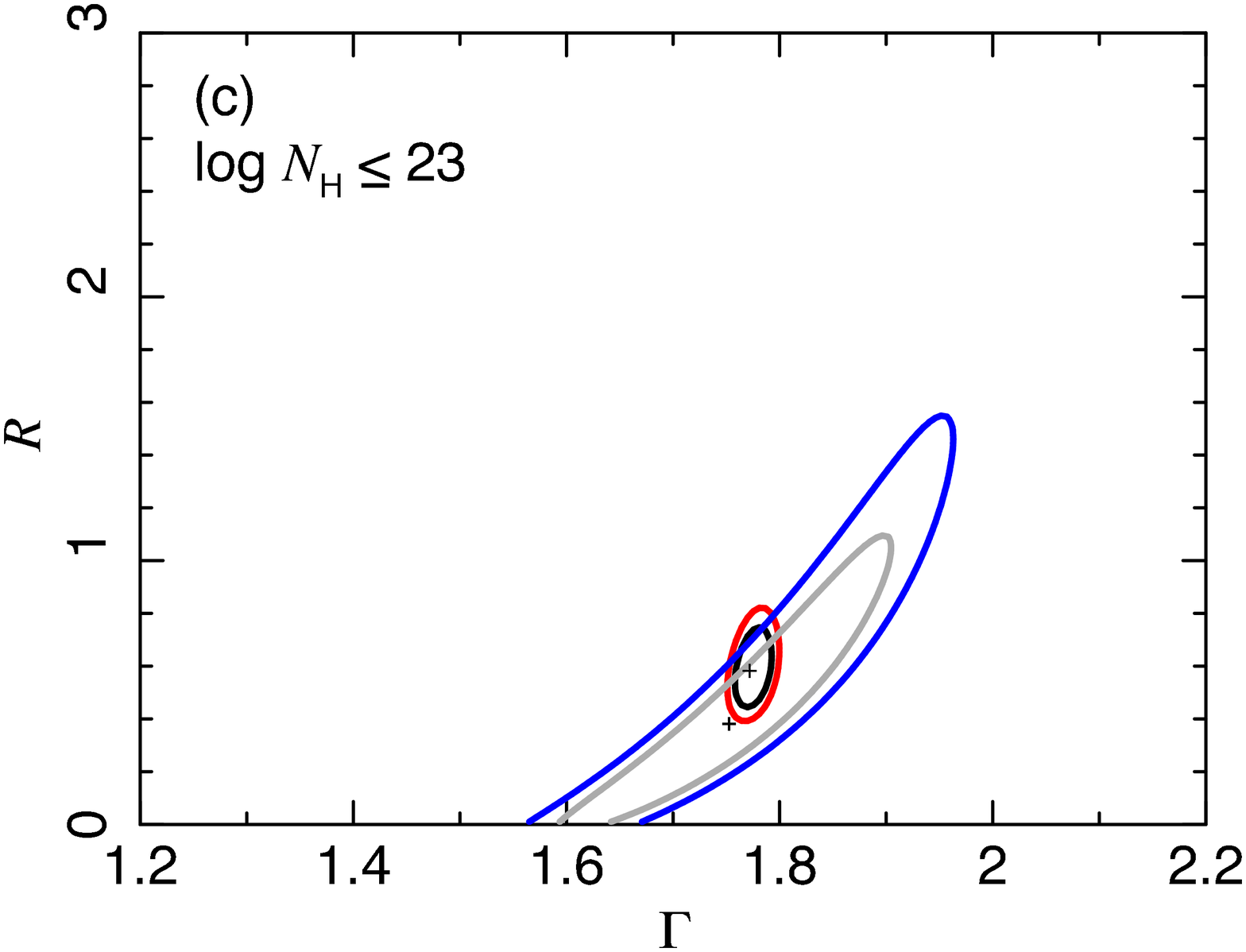}
\includegraphics[scale=0.22]{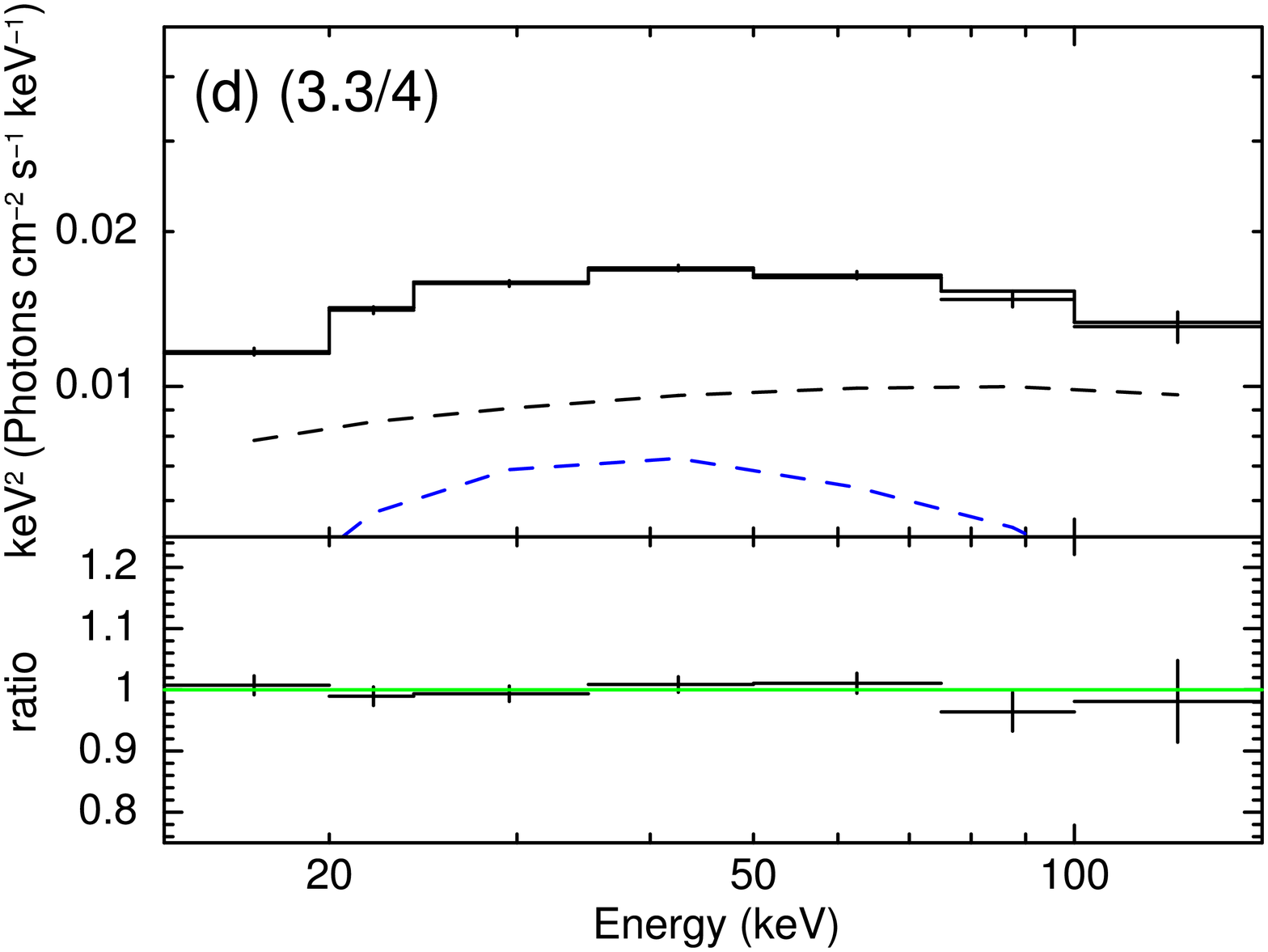} \hspace{2mm}
\includegraphics[scale=0.22]{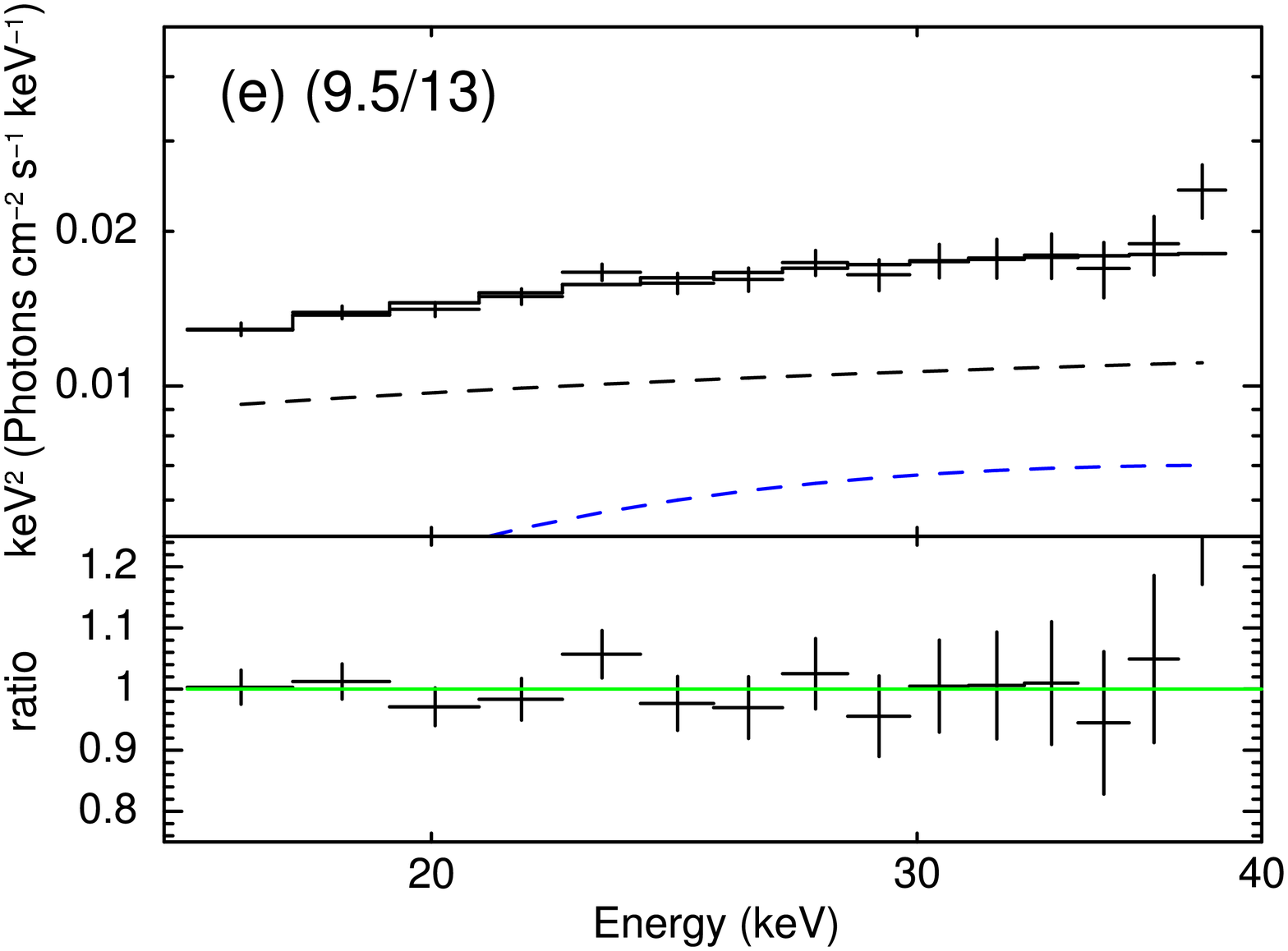} \hspace{3mm}
\includegraphics[scale=0.22]{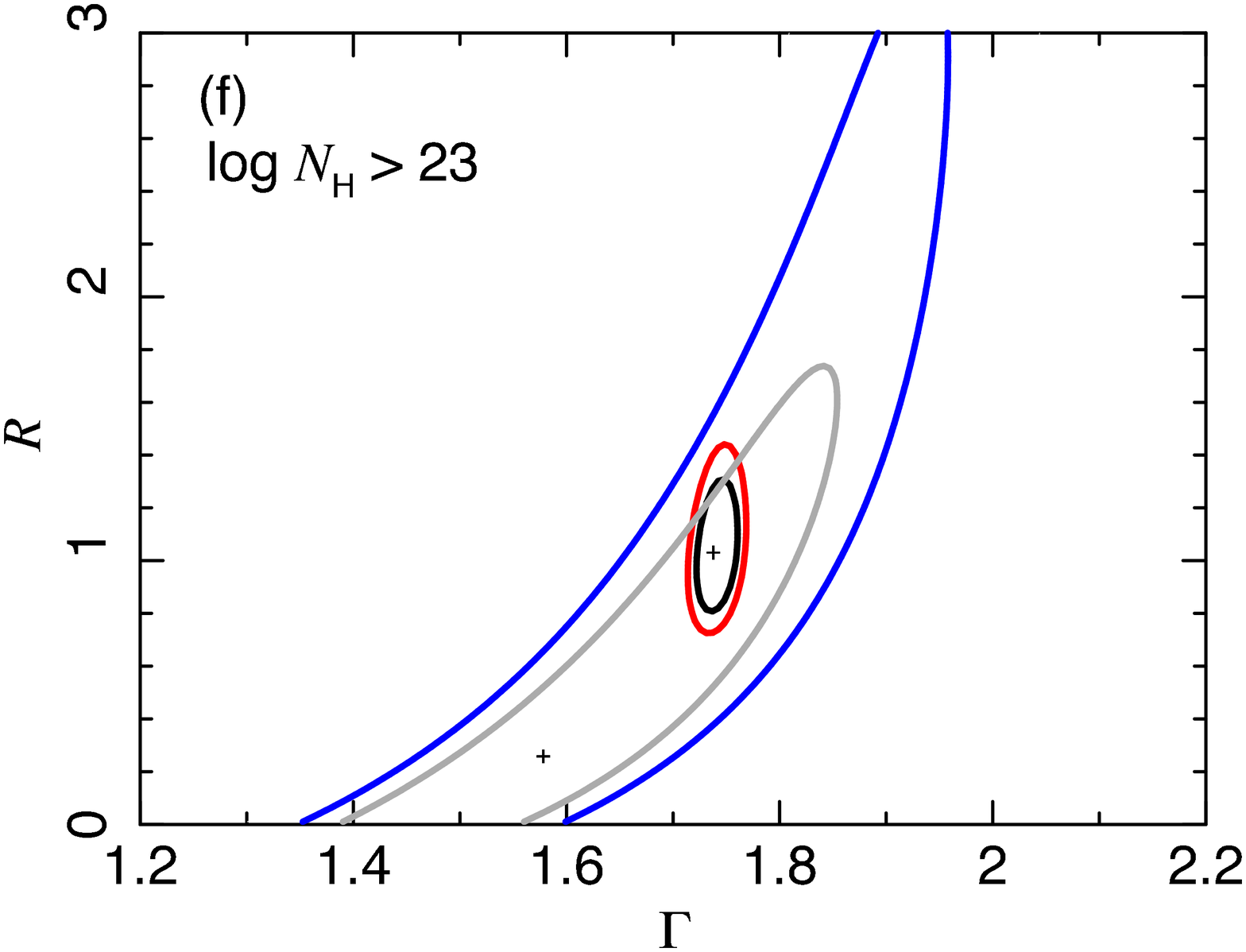}
\caption{
Same as Figure~\ref{fig:ave_spec_lhx} but for the MOAGNs ($\log N_{\rm H} \leq 23$) in
the upper figures and for the HOAGNs ($\log N_{\rm H} > 23$) in the
lower figures.
}
\label{fig:ave_spec_nh}
\end{center}
\end{figure*}

We also make the same analysis to the subsamples of MOAGNs ($\log N_{\rm
H} \leq 23$) and HOAGNs ($\log N_{\rm H} > 23$).  The unfolded spectra
with best-fit models and confidence contours between photon index and
$R$ are plotted in Figure~\ref{fig:ave_spec_nh}. We find a significant difference in $R$
between the MOAGNs ($R = 0.59^{+0.09}_{-0.10}$) and HOAGNs ($R =
1.03^{+0.15}_{-0.17}$) from the {\it Swift}/BAT spectra, confirming the trend
already reported in Section~\ref{sec:ref} more robustly. This is consistent with 
the previous work by \citet{Ric11}, who carried out stacking {\it INTEGRAL} data. 
As we already mentioned, this implies that on average the covering fraction and/or average
column density of a torus is larger in AGNs with larger line-of-sight
absorptions. 

In summary, the average reflection strength of the
MLAGNs and HOAGNs is larger than that of the HLAGNs and MOAGNs,
respectively. 

% These results are consistent with that a large fraction of
% AGNs with $\log L^{\rm BAT}_{\rm 10-50\hspace{1mm}keV} \lesssim 43$ have
% high hydrogen column densities ($\log N_{\rm H} \gtrsim 23$) as shown in
% Figure~\ref{fig:basic_cor}(c).  

%\if0 % IRU 
%\fi 

%\if0 % IRU
%\fi 

\subsection{Correlations among the X-ray and MIR properties}

\begin{figure*}[t]
\begin{center}
\includegraphics[scale=0.55,angle=-90]{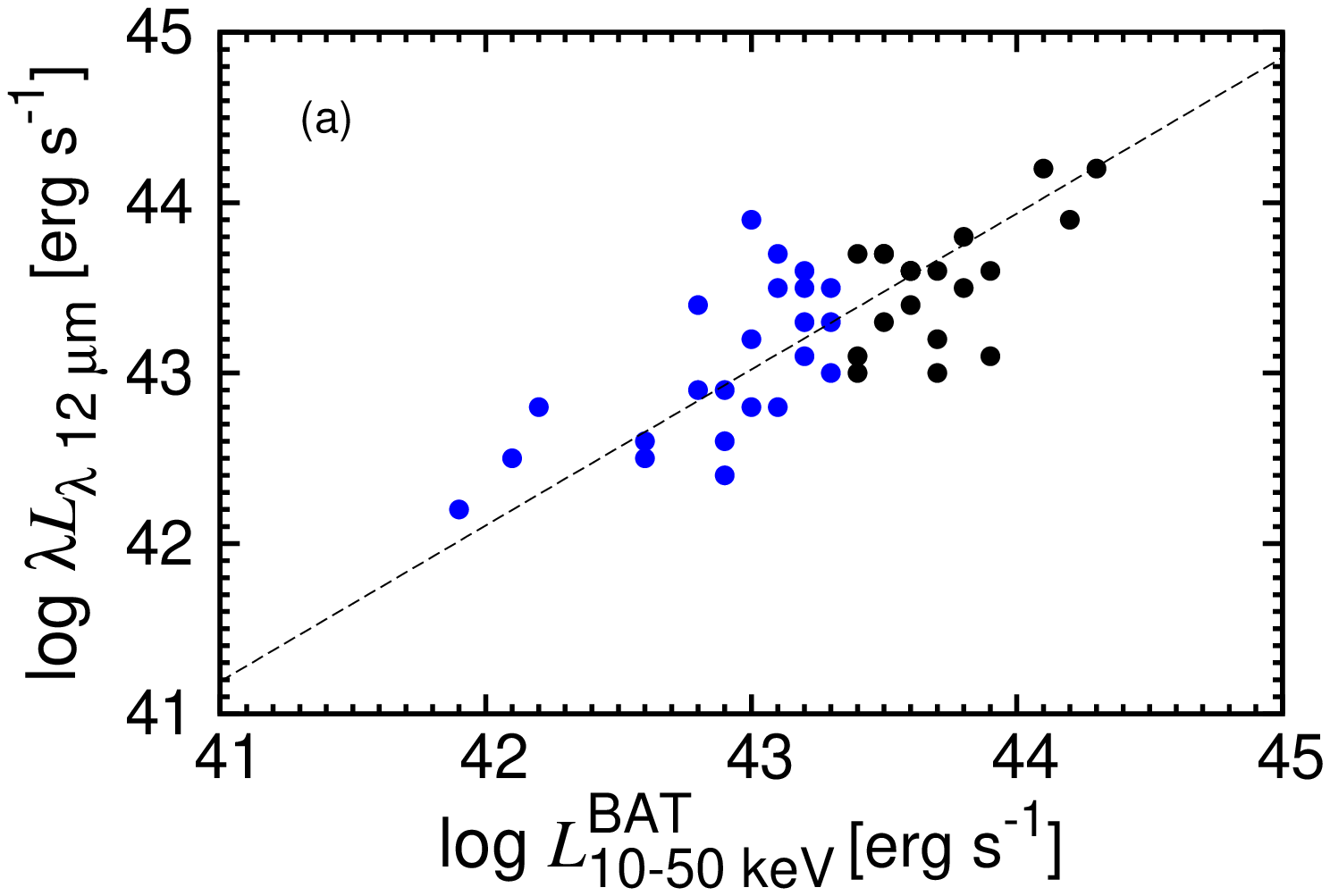}\hspace{2mm}
\includegraphics[scale=0.55,angle=-90]{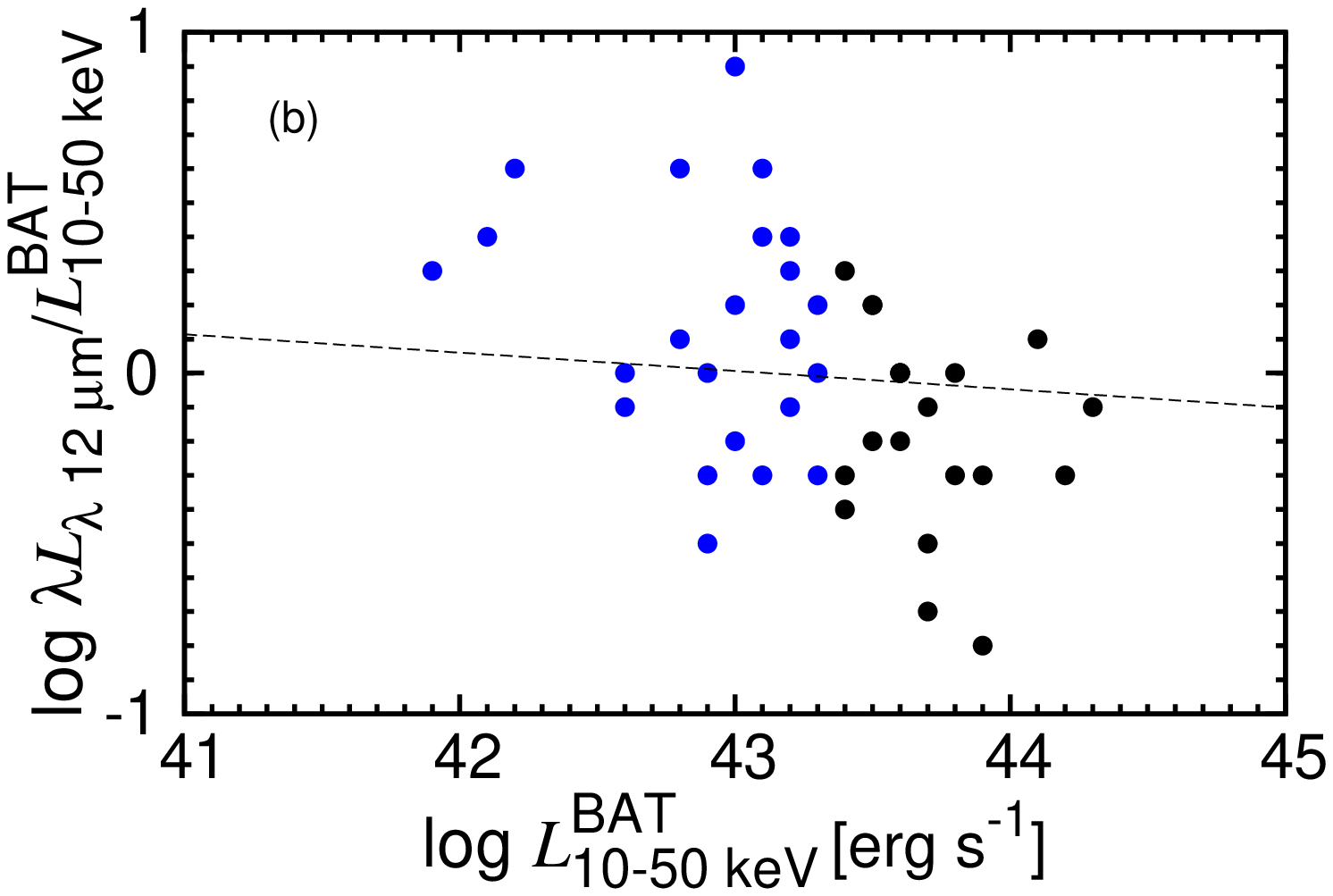} 
\caption{
(a) correlation between $12$ $\mu$m and 10--50 keV luminosities
and regression function (dashed line) of 
$\log \lambda L_{\rm 12\hspace{1mm}\mu m} = 3.7 + 0.92 \log L^{\rm BAT}_{\rm 10-50\hspace{1mm}keV}$. 
(b) correlation between the $12$ $\mu$m to 10--50 keV luminosity ratio and 10--50 keV luminosity 
and regression function (dashed line) of 
$\log (\lambda L_{\rm 12\hspace{1mm}\mu m}/L^{\rm BAT}_{\rm 10-50\hspace{1mm}keV}) = 3.7 - 0.08 
\log L^{\rm BAT}_{\rm 10-50\hspace{1mm}keV}$. 
The blue and black circles represent the MLAGNs 
($L^{\rm BAT}_{\rm 10-50\hspace{1mm}keV} \leq 43.3$) and HLAGNs 
($L^{\rm BAT}_{\rm 10-50\hspace{1mm}keV} > 43.3$), respectively. 
The dashed lines represent the regression line. 
}
\label{fig:mir_x-ray}
\end{center}
\end{figure*}

AGNs are also bright in the MIR band owing to emission of hot dust in
the torus heated by the primary radiation from the central engine. 
In fact, a good correlation between the MIR and X-ray luminosities in 
AGNs has been reported by several works \citep{Gan09,Ich12,Mat12,Asm15}. Using our sample, 
we examine correlations of $\log L^{\rm BAT}_{\rm 10-50\hspace{1mm}keV}$ 
with 12 $\mu$m MIR luminosity $\log \lambda L_{\rm 12\hspace{1mm}\mu m}$ 
and $\log (\lambda L_{\rm 12\hspace{1mm}\mu m}/L^{\rm 
BAT}_{\rm 10-50\hspace{1mm}keV})$ as plotted in Figure~\ref{fig:mir_x-ray}. 
Here we refer to the nucleus 12 $\mu$m
luminosities compiled by \cite{Asm15} based on subarcsecond resolution
imaging.  For AGNs whose luminosities are not available in \cite{Asm15},
we calculate those using the photometric data of Wide-field Infrared Survey
Explorer \citep[{\it WISE};][]{Wri10}. Because the spatial resolution of
{\it WISE} is limited ($\sim 6''.5$ in the 12 $\mu$m band),
contamination from the host galaxy may not be ignorable in these data.
Table~\ref{tab:info_mir_oiv} summarizes the compiled 
$\lambda L_{\rm 12\hspace{1mm}\mu m}$ luminosities. 

We obtain a tight $\log \lambda L_{\rm 12\hspace{1mm}\mu m}$--$\log
L^{\rm BAT}_{\rm 10-50\hspace{1mm}keV}$ correlation with a slope $b
= 0.92\pm0.07$ $ (0.91\pm0.10) $ from the whole sample (that only with the high 
resolution MIR data). The slope is consistent with previous results \citep{Ich12,Asm15}.
We confirm that the flux-flux correlation is also significant,
indicating that the luminosity-luminosity correlation is robust against
the Malmquist bias. The regression line yields a negative slope 
($b = -0.08\pm0.07$) in the correlation between the MIR to 10--50 keV 
luminosity ratio $(\lambda L_{\rm 12\hspace{1mm}\mu m}/L^{\rm BAT}_{\rm 10-50\hspace{1mm}keV})$ and
the 10--50 keV luminosity. As investigated in detail by \cite{Sta16}, 
$\lambda L_{\rm 12\hspace{1mm}\mu m}/L^{\rm BAT}_{\rm 10-50\hspace{1mm}keV}$ 
is predicted to monotonically increase with increasing covering factor, or decreasing 
half opening-angle, of the torus in the edge-on view (i.e., type-2 AGN) case. 
Here, the half opening-angle is defined as that between the polar axis and
the upper edge of the torus.  
It is because the solid angle of the torus illuminated by the 
accretion disk increases with decreasing half opening-angle and 
consequently the reprocessed emission in the infrared band becomes stronger. 
Hence, the negative correlation of the MIR to X-ray luminosity ratio 
with X-ray luminosity is consistent with the luminosity-dependent AGN unified model 
\citep[e.g.,][]{Ued03,Mai07,Lus13,Ric13a}.

%\if0 %%%%IRU file to be incorporated %%%%
%\fi 

%\if0 %%%%IRU file to be incorporated %%%%
\begin{figure*}[!t]
\begin{center}
\includegraphics[scale=0.55,angle=-90]{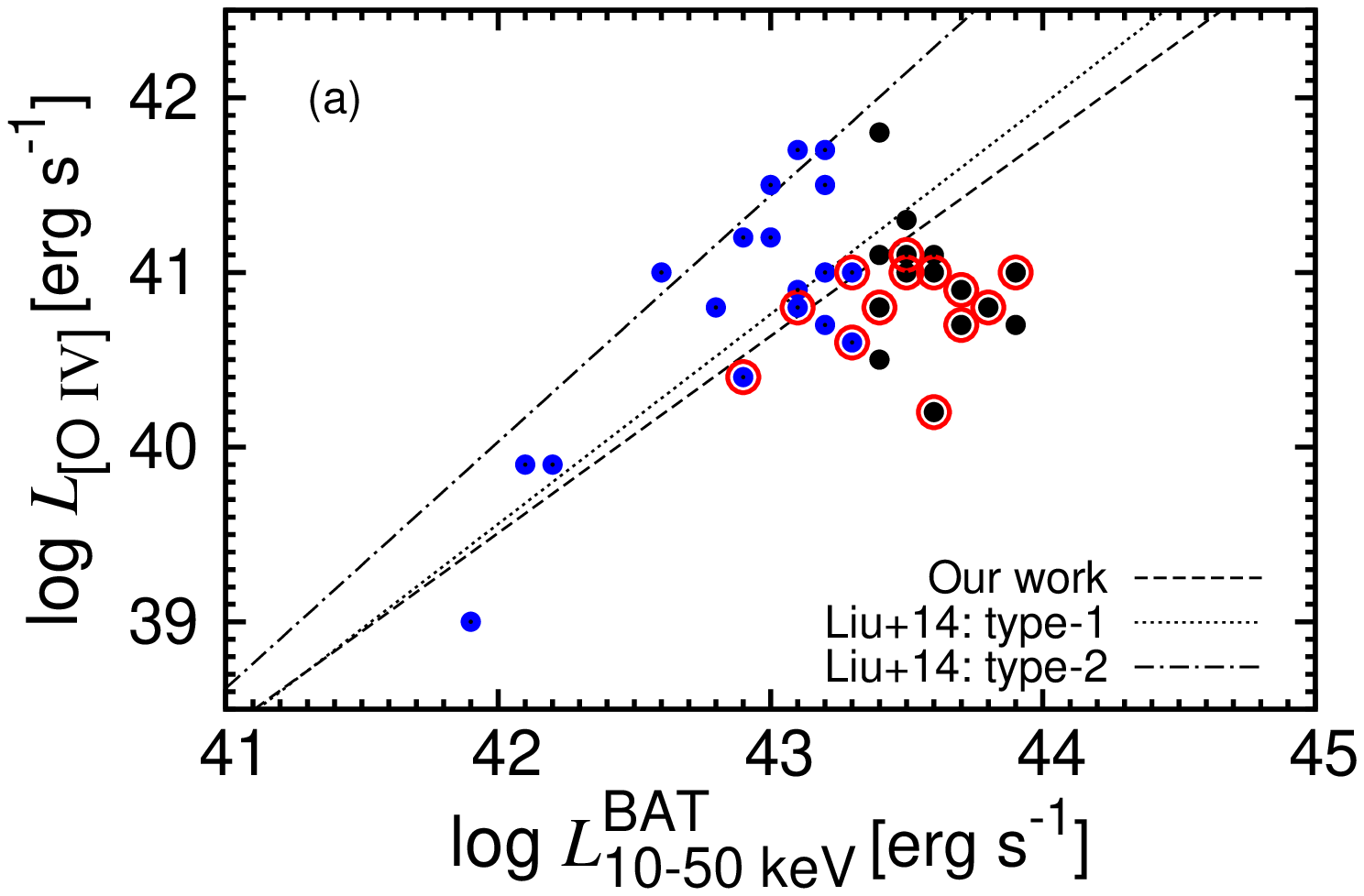} \hspace{2mm}
\includegraphics[scale=0.55,angle=-90]{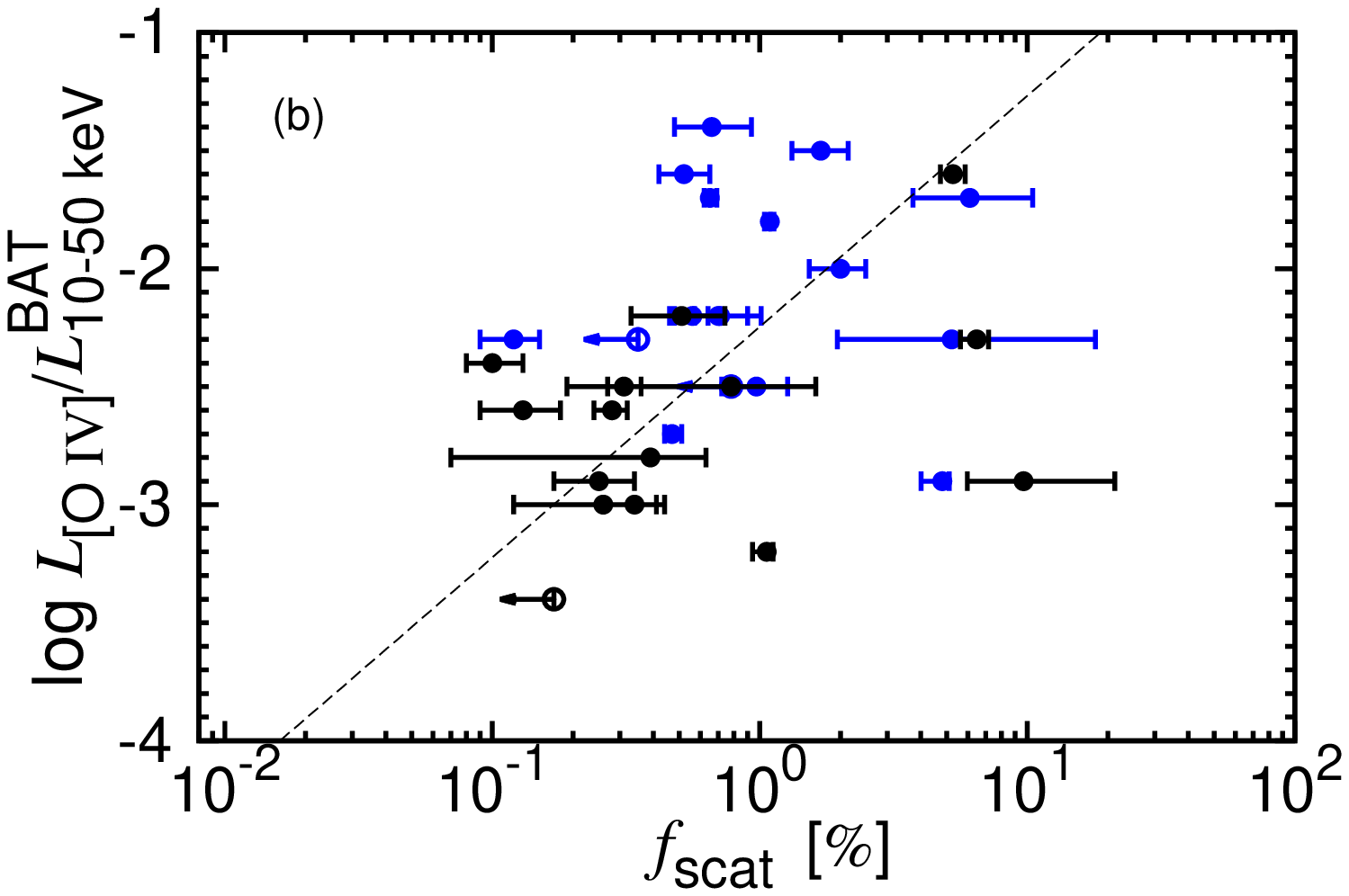}
\caption{
(a) correlation between [O IV] and 10--50 keV luminosities 
and regression function (dashed line) of 
$\log L_{\rm \oiv} = -7.8 + 1.13 \log L^{\rm BAT}_{\rm 10-50\hspace{1mm}keV}$. 
The dotted and dot-dashed lines represent the regression lines
obtained from the type-1 and type-2 AGN samples of \citet{Liu14}, respectively.
(b) correlation between the [O IV] to 10--50 keV luminosity ratio and scattered fraction 
and regression function (dashed line) of 
$\log (L_{\rm \oiv}/L^{\rm BAT}_{\rm 10-50\hspace{1mm}keV}) = -2.25 + 0.98 \log f_{\rm scat}$. 
The blue, black, and red circles (only in the left figure)
represent the MLAGNs ($L^{\rm BAT}_{\rm 10-50\hspace{1mm}keV} \leq 43.3$), HLAGNs 
($L^{\rm BAT}_{\rm 10-50\hspace{1mm}keV} > 43.3$), and low scattering-fraction AGNs 
($f_{\rm scat} < 0.5\%$), respectively. 
}
\label{fig:oiv}
\end{center}
\end{figure*}
%\fi 

\subsection{X-ray and [O IV] Luminosity as AGN Power Indicator} 

Gas around the torus is excited by irradiation from the central engine and
scatters a part of incident photons. Hence, it provides us with 
the information of the intrinsic luminosity even in obscured AGNs. 
To estimate the intrinsic AGN power, some authors proposed
the usage of the \oiv line \citep[e.g.,][]{Mel08,Rig09}. This is 
because, compared with optical emission lines, \oiv is much less affected 
by dust extinction in the interstellar matter and is less contaminated by 
starlight from the host galaxy due to the high ionization potential ($54.9$ eV). 
To investigate the correlation of the \oiv line to
10--50 keV luminosity, another proxy of AGN luminosity, we compile the
\oiv luminosities ($L_{\rm \oiv}$) from the literature
\citep{Wea10,Wee12,Ina13,Liu14}, as listed in Table~\ref{tab:info_mir_oiv}. 
Figure~\ref{fig:oiv}(a) plots $\log L_{\rm \oiv}$ against 
$\log L^{\rm BAT}_{\rm 10-50\hspace{1mm}keV}$. 
The correlation is insignificant with $P (\log L^{\rm
BAT}_{\rm 10-50\hspace{1mm}keV}, \log L_{\rm \oiv}) = 5.8\times10^{-1}$.

Previous studies reported that the $L_{\rm \oiv}/L_{\rm X}$ ratio of 
type-2 AGNs may be higher than that of type-1 AGNs
\citep[e.g.,][]{Mel08,Rig09,Liu14}. They suggested that the difference
is ascribed to an underestimation of the X-ray luminosity due to
obscuration in type-2 AGNs or to anisotropy of the intrinsic X-ray
emission. For comparison, we overplot the relations of \cite{Liu14} by
converting the 14--195 keV luminosity into the 10--50 keV one with
a power-law photon index of 1.7. As shown in Figure~\ref{fig:oiv}(a),
the $L_{\rm \oiv}/L_{\rm X}$ ratio of our sample is more similar to that
of type-1 AGNs than that of the type-2 AGNs in \cite{Liu14}. As a
result, we do not see significant difference in the $L_{\rm \oiv}/L_{\rm
X}$ ratio between type-1 and Compton-thin type-2 AGNs. This suggests
that the anisotropy of X-ray emission is unlikely. 

A notable finding is that most of low scattering-fraction AGNs (with
best-fit $f_{\rm scat} < 0.5\%$) show systematically low values of
$L_{\rm \oiv}/L^{\rm BAT}_{\rm 10-50\hspace{1mm}keV}$ ratio than the
average. Indeed, we find a significant correlation between the $L_{\rm
\oiv}/L^{\rm BAT}_{\rm 10-50\hspace{1mm}keV}$ ratio and the scattered
fraction (see Figure~\ref{fig:oiv}(b)).  The regression line is
calculated with the OLS bisector method by ignoring objects without
significant detection of the scattered component.  We exclude Mrk 915
because of its apparently very high scattering fraction ($\sim 40\%$),
which is much higher than typical values in obscured AGNs \citep[$\sim
3\%$;][]{Bia07} and should be attributed to leaky or ionized absorbers.
These results well agree with that by \citet{Ued15} that low
scattering-fraction AGNs show low \oiii\hspace{-1mm}\ to hard X-ray
luminosity ratios on average.  This supports the picture that a
significant fraction of this population of AGNs are deeply ``buried'' in
small opening-angle tori. This also implies that the \oiv\ luminosity
may not be an ideal indicator of the intrinsic AGN power for the whole
AGN population.

\section{Conclusion} \label{sec:con}

We have analyzed the broadband (0.5--150 keV) X-ray spectra of 45 local,
moderately obscured (Compton-thin) AGNs observed with {\it Suzaku} and
{\it Swift}/BAT in a uniform manner. The broadband X-ray spectra are
basically well reproduced with the baseline model composed of an
absorbed cutoff power-law component, a scattered component, and a
(unabsorbed/absorbed) reflection component with a fluorescent
iron-K$\alpha$ line. Additional components such as emission/absorption
lines and optically-thin thermal emission in the host galaxy are also
taken into account if required. The main conclusions of our work 
are summarized as follows.

\begin{enumerate}

\item We evaluate time variation of 
      the luminosity of the primary power-law component
      between the {\it Suzaku} and 70-month averaged {\it Swift}/BAT observations. 
      The standard deviation is $\sim$ 0.2 dex, which can be regarded as typical 
      variability on timescales of $\sim$ a day to 
      several years. 

\item We find a significant correlation of
      photon index with Eddington ratio, but not with luminosity
      or black hole mass. 
      This is consistent with previous results \citep{She08,Bri13,Yan15b}. 

\item A narrow iron-K$\alpha$ line is significantly detected in all
            objects. The \lirlx\hspace{1mm} ratio decreases with luminosity, 
            supporting the luminosity-dependent AGN unified 
            model where the covering fraction of tori decreases with
            luminosity. 

\item The average reflection strength derived from stacked spectra above
      14 keV is found to be larger in less luminous ($\log L_{\rm 10-50\hspace{1mm}keV} 
      \leq 43.3$) or heavily obscured ($\log N_{\rm H} > 23$) AGNs than in more luminous 
      ($\log L_{\rm 10-50\hspace{1mm}keV} > 43.3$) or lightly obscured 
      AGNs ($\log N_{\rm H} \leq 23$), respectively.

\item We confirm strong correlation between the X-ray and MIR
      luminosities ($\log \lambda L_{\rm \lambda 12\hspace{1mm}\mu
      m}$--$\log L_{\rm 10-50\hspace{1mm}keV}$), which results in a negative
      $\log (\lambda L_{\rm \lambda 12\hspace{1mm}\mu m}/L_{\rm
      10-50\hspace{1mm}keV})$--$\log L_{\rm 10-50\hspace{1mm}keV}$
      correlation. This is again consistent with the
      luminosity-dependent unified model.

\item The average \oiv line to hard X-ray luminosity ratio obtained from
      our sample is lower than previous estimates using other samples of
      type-2 AGNs. In particular, this ratio is found to be
      significantly lower in low scattering-fraction AGNs. This suggests
      that the \oiv luminosity may significantly underestimate the
      intrinsic luminosity of AGNs deeply buried in small opening-angle
      tori.

\end{enumerate}

%\begin{landscape}
%\LongTables
%\setlength{\topmargin}{2cm}
\tabletypesize{\normalsize}
\setlength{\tabcolsep}{0.05in}
\begin{deluxetable*}{crcccc}
\tablecaption{MIR luminosity\label{tab:info_mir_oiv}}
\tablewidth{0pt}
\tablehead{\colhead{Target Name} 
& \colhead{$\log L_{12\hspace{1pt}\mu{\rm m}}$}  
& \colhead{Ref. $\log L_{12\hspace{1pt}\mu{\rm m}}$}  
& \colhead{$\log L_{\rm \oiv}$}  
& \colhead{Ref. $\log L_{\rm \oiv}$}  
\\
\colhead{(1)} & \colhead{(2)} & \colhead{(3)} & \colhead{(4)} & \colhead{(5)}
} 
\startdata
2MASX J0216+5126 &  $ ... $  & $ ... $ & $ ... $ & $   ... $ \\
 2MASX J0248+2630 &  $ 44.17\pm0.01 $ & W & $ ...  $ & $  ... $ \\
 2MASX J0318+6829 &  $ 43.90\pm0.01 $ & W & $ ...  $ & $  ... $ \\
 2MASX J0350-5018 &  $ 43.04\pm0.01 $ & W & $ 40.5  $ & $ 1 $ \\
 2MASX J0444+2813 &  $ 42.58\pm0.01 $ & W & $ ...  $ & $  ... $ \\
 2MASX J0505-2351 &  $ 43.57\pm0.13 $ & A & $ 40.7  $ & $ 1 $ \\
 2MASX J0911+4528 &  $ 43.03\pm0.01 $ & W & $ 41.0  $ & $ 1 $ \\
 2MASX J1200+0648 &  $ 43.56\pm0.01 $ & W & $ ...  $ & $  ... $ \\
 Ark 347 &  $ 43.27\pm0.11 $ & A & $ 41.5  $ & $ 1 $ \\
 ESO 103-035 &  $ 43.68\pm0.19 $ & A & $ 41.1  $ & $ 2 $ \\
 ESO 263-G013 &  $ 43.56\pm0.03 $ & A & $ ...  $ & $  ... $ \\
 ESO 297-G018 &  $ 43.03\pm0.07 $ & A & $ 40.7  $ & $ 1 $ \\
 ESO 506-G027 &  $ 43.80\pm0.04 $ & A & $ 40.8  $ & $ 1 $ \\
 Fairall 49 &  $ 43.93\pm0.20 $ & A & $ 41.5  $ & $ 2 $ \\
 Fairall 51 &  $ 43.39\pm0.04 $ & A & $ 40.8  $ & $ 2 $ \\
 IC 4518A &  $ 43.47\pm0.06 $ & A & $ 41.7  $ & $ 3 $ \\
 LEDA 170194 &  $ 43.52\pm0.08 $ & A & $ ...  $ & $  ... $ \\
 MCG +04-48-002 &  $ 43.59\pm0.01 $ & W & $ 40.7  $ & $ 4 $ \\
 MCG -01-05-047 &  $ 42.88\pm0.15 $ & A & $ ...  $ & $  ... $ \\
 MCG -02-08-014 &  $ 42.84\pm0.07 $ & A & $ ...  $ & $  ... $ \\
 MCG -05-23-016 &  $ 43.45\pm0.04 $ & A & $ 40.6  $ & $ 2 $ \\
 Mrk 1210 &  $ 43.67\pm0.01 $ & W & $ 40.9  $ & $ 1 $ \\
 Mrk 1498 &  $ 44.24\pm0.01 $ & W & $ ...  $ & $  ... $ \\
 Mrk 18 &  $ 42.84\pm0.01 $ & W & $ 39.9  $ & $ 2 $ \\
 Mrk 348 &  $ 43.59\pm0.01 $ & W & $ 41.0  $ & $ 2 $ \\
 Mrk 417 &  $ 43.58\pm0.01 $ & W & $ 41.1  $ & $ 1 $ \\
 Mrk 520 &  $ ... $  & $ ... $ & $ 41.8 $ & $  1 $ \\
 Mrk 915 &  $ 43.45\pm0.04 $ & A & $ 41.7  $ & $ 1 $ \\
 NGC 1052 &  $ 42.20\pm0.06 $ & A & $ 39.0  $ & $ 2 $ \\
 NGC 1142 &  $ 43.08\pm0.20 $ & A & $ 41.0  $ & $ 3 $ \\
 NGC 2110 &  $ 43.08\pm0.06 $ & A & $ 40.8  $ & $ 2 $ \\
 NGC 235A &  $ 43.30\pm0.16 $ & A & $ 41.3  $ & $ 4 $ \\
 NGC 3081 &  $ 42.49\pm0.07 $ & A & $ 41.0  $ & $ 2 $ \\
 NGC 3431 &  $ 42.91\pm0.01 $ & W & $ ...  $ & $  ... $ \\
 NGC 4388 &  $ 42.38\pm0.07 $ & A & $ 41.2  $ & $ 2 $ \\
 NGC 4507 &  $ 43.68\pm0.04 $ & A & $ 41.0  $ & $ 2 $ \\
 NGC 4992 &  $ 43.44\pm0.09 $ & A & $ 40.2  $ & $ 1 $ \\
 NGC 5252 &  $ 43.16\pm0.04 $ & A & $ 40.9  $ & $ 1 $ \\
 NGC 526A &  $ 43.68\pm0.05 $ & A & $ 41.1  $ & $ 2 $ \\
 NGC 5506 &  $ 43.16\pm0.03 $ & A & $ 41.2  $ & $ 2 $ \\
 NGC 6300 &  $ 42.51\pm0.11 $ & A & $ 39.9  $ & $ 3 $ \\
 NGC 7172 &  $ 42.81\pm0.04 $ & A & $ 40.8  $ & $ 2 $ \\
 NGC 788 &  $ 43.15\pm0.05 $ & A & $ 41.0  $ & $ 2 $ \\
 UGC 03142 &  $ 43.25\pm0.01 $ & W & $ ...  $ & $  ... $ \\
 UGC 12741 &  $ 42.63\pm0.01 $ & W & $ 40.4  $ & $ 2 $  
\enddata
\tablecomments{
(1) Galaxy name. 
(2) 12 $\mu$m luminoisty. 
(3) References for the 12 $\mu$m luminosity: 
(A) \cite{Asm15}, 
(W) the data taken from the ALLWISE Source Catalog \citep{Wri10}.
(4) \oiv luminosity. 
(5) References for the \oiv luminosity:
(1) \cite{Wee12} (2) \cite{Wea10} (3) \cite{Liu14} (4) \cite{Ina13}. 
 }
\end{deluxetable*}
\clearpage
%\end{landscape}

\acknowledgments

Part of this work was financially supported by the Grant-in-Aid for JSPS
Fellows for young researchers (T.K.) and for Scientific Research
26400228 (Y.U.). We acknowledge financial support from the CONICYT-Chile grants 
``EMBIGGEN" Anillo ACT1101 (CR), FONDECYT 1141218 (CR), and Basal-CATA PFB--06/2007 (CR). 
This research has made use of the NASA/ IPAC Infrared
Science Archive, which is operated by the Jet Propulsion Laboratory,
California Institute of Technology, under contract with the National
Aeronautics and Space Administration.

% \bibliography{myref.bib}

\clearpage 
\newpage 
\newpage

%%%%%%%%%%%%%%%%%%%
%    INFOMATION   %
%%%%%%%%%%%%%%%%%%%

\clearpage 
\newpage 

%\if0 %%%%%%%%%%%%%%%%%%%%%%%%%%%%%%%%%%%%%% IRUIRU 
\setcounter{table}{0}
\setcounter{figure}{0}
\appendix
\renewcommand{\thetable}{\Alph{table}} % .\arabic{table}} 
\renewcommand{\thefigure}{\Alph{figure}} % .\arabic{figure}}

\section{Broadband X-ray spectra and best-fitting models}

% TABLE.1 
%%%%%%%%%
\begin{landscape}
\LongTables
\setlength{\topmargin}{1.8cm}
\begin{deluxetable}{cccccccccccccccccccc}[h] 
\tabletypesize{\scriptsize}
\tabletypesize{\footnotesize}
\setlength{\tabcolsep}{0.01in}
\tablecaption{Best-fit parameters\label{tab:info_para}}
\tablewidth{-0.5pt}
\tablehead{\colhead{Target Name} 
& \colhead{$N^{\rm Gal}_{\rm H}$}  
& \colhead{$A_{\rm pl}$} 
& \colhead{$N_{\rm XIS}$} 
& \colhead{$N_{\rm H}$}  
& \colhead{$N^{\rm ref}_{\rm H}$}  
& \colhead{$N^{\rm pc}_{\rm H}$}  
& \colhead{$f_{\rm pc}$}  
& \colhead{$\Gamma$} 
& \colhead{$f_{\rm scat}$}  
& \colhead{$R$} 
& \colhead{$kT_1$} 
& \colhead{$kT_2$} 
& \colhead{$\chi^2/dof$}  \\ 
\colhead{(1)} & \colhead{(2)} & \colhead{(3)} & \colhead{(4)} &
\colhead{(5)} & \colhead{(6)} & \colhead{(7)} & \colhead{(8)}  & \colhead{(9)} &
\colhead{(10)} & \colhead{(11)}  & \colhead{(12)} & \colhead{(13)} & \colhead{(14)}  
} 
\startdata
\thispagestyle{empty}
  2MASX J0216+5126 &  $ 14.2 $ & $ 2.96^{+0.40}_{-0.35} $ & $ 0.90^{+0.11}_{-0.07} $ & $ 1.54^{+0.05}_{-0.04} $ &  $ ... $ & $ ... $ & $ ... $ &$ 1.84^{+0.03}_{-0.02} $ & $ 0.64^{+0.37}_{-0.38} $ & $ 0.00^{+0.11}_{} $ &  $ ... $ & $ ... $ &$ 430.3/403  $ \\
 2MASX J0248+2630 &  $ 10.3 $ & $ 1.02^{+0.27}_{-0.22} $ & $ 1.84^{+0.42}_{-0.16} $ & $ 22.01^{+1.01}_{-0.74} $ &  $ ... $ & $ ... $ & $ ... $ &$ 1.42^{+0.08}_{-0.04} $ & $ 2.41^{+0.67}_{-0.53} $ & $ 0.05^{+0.48}_{-0.05} $ &  $ ... $ & $ ... $ &$ 96.0/115  $ \\
 2MASX J0318+6829 &  $ 30.8 $ & $ 0.88^{+0.10}_{-0.16} $ & $ 1.45^{+0.33}_{-0.12} $ & $ 5.41^{+0.25}_{-0.20} $ &  $ ... $ & $ ... $ & $ ... $ &$ 1.52^{+0.06}_{-0.02} $ & $ 3.63^{+0.92}_{-0.56} $ & $ 0.00^{+0.39}_{} $ &  $ ... $ & $ ... $ &$ 278.6/234  $ \\
 2MASX J0350-5018 &  $ 1.16 $ & $ 0.40^{+0.85}_{-0.23} $ & $ 0.73^{+0.51}_{-0.37} $ & $ 41^{+33}_{-10} $ &  $ ... $ & $ ... $ & $ ... $ &$ 1.53^{+0.30}_{-0.25} $ & $ 9.7^{+11.5}_{-3.7} $ & $ 2.00^{}_{-0.90} $ &  $ ... $ & $ ... $ &$ 31.4/41  $ \\
 2MASX J0444+2813 &  $ 17.8 $ & $ 1.47^{+0.31}_{-0.24} $ & $ 0.86^{+0.09}_{-0.08} $ & $ 8.97^{+0.35}_{-0.34} $ &  $ ... $ & $ ... $ & $ ... $ &$ 1.45\pm0.07 $ & $ 0.83^{+0.25}_{-0.22} $ & $ 0.33^{+0.31}_{-0.23} $ &  $ ... $ & $ ... $ &$ 239.1/230  $ \\
 2MASX J0505-2351 &  $ 2.12 $ & $ 4.47^{+0.29}_{-0.20} $ & $ 0.89^{+0.04}_{-0.05} $ & $ 5.65^{+0.10}_{-0.12} $ &  $ ... $ &$ 12.0^{+5.0}_{-3.9} $ & $ 0.29\pm0.01 $ & $ 1.67^{+0.04}_{-0.01} $ & $ 1.06^{+0.06}_{-0.12} $ & $ 0.12^{+0.05}_{-0.12} $ &  $ ... $ & $ ... $ &$ 722.3/711  $ \\
 2MASX J0911+4528 &  $ 1.23 $ & $ 6.7^{+2.8}_{-2.0} $ & $ 0.69^{+0.14}_{-0.10} $ & $ 40.3^{+2.2}_{-3.2} $ &  $ ... $ & $ ... $ & $ ... $ &$ 2.12^{+0.10}_{-0.09} $ & $ 0.00^{+0.35}_{} $ & $ 0.29^{+0.12}_{-0.10} $ &  $ ... $ & $ ... $ &$ 110.2/96  $ \\
 2MASX J1200+0648 &  $ 1.18 $ & $ 1.44^{+0.32}_{-0.21} $ & $ 2.78^{+0.34}_{-0.38} $ & $ 8.09^{+0.21}_{-0.18} $ &  $ ... $ & $ ... $ & $ ... $ &$ 1.81^{+0.04}_{-0.03} $ & $ 0.19^{+0.33}_{-0.19} $ & $ 2.00^{}_{-0.31} $ &  $ ... $ & $ ... $ &$ 406.9/435  $ \\
 Ark 347 &  $ 2.30 $ & $ 0.73^{+0.50}_{-0.32} $ & $ 0.50^{+0.17}_{-0.14} $ & $ 24.0^{+3.8}_{-3.3} $ &  $ ... $ & $ ... $ & $ ... $ &$ 1.54\pm0.12 $ & $ 6.1^{+4.4}_{-2.4} $ & $ 1.28^{+0.72}_{-0.68} $ & $ 1.26^{+0.35}_{-0.20} $ &  $ ... $ &$ 56.7/46  $ \\
 ESO 103-035 &  $ 5.71 $ & $ 28.6^{+3.4}_{-3.6} $ & $ 1.27^{+0.07}_{-0.05} $ & $ 20.45^{+0.37}_{-0.44} $ & $ 4.6^{+1.3}_{-1.1} $ & $ 55.9^{+8.8}_{-9.5} $ & $ 0.33^{+0.03}_{-0.04} $ & $ 2.07^{+0.02}_{-0.03} $ & $ 0.10^{+0.03}_{-0.02} $ & $ 0.87^{+0.26}_{-0.22} $ & $ 1.07^{+0.27}_{-0.15} $ &  $ ... $ &$ 1126.1/1096  $ \\
 ESO 263-G013 &  $ 10.2 $ & $ 2.80^{+0.79}_{-0.61} $ & $ 0.94^{+0.13}_{-0.11} $ & $ 25.64^{+1.17}_{-0.96} $ &  $ ... $ & $ ... $ & $ ... $ &$ 1.67^{+0.08}_{-0.06} $ & $ 0.71^{+0.24}_{-0.25} $ & $ 0.05^{+0.25}_{-0.05} $ & $ 0.93^{+0.13}_{-0.17} $ &  $ ... $ &$ 96.7/108  $ \\
 ESO 297-G018 &  $ 1.63 $ & $ 5.23^{+1.13}_{-0.95} $ & $ 0.96^{+0.10}_{-0.08} $ & $ 63.8^{+3.6}_{-3.4} $ &  $ ... $ & $ ... $ & $ ... $ &$ 1.70\pm0.05 $ & $ 0.26^{+0.15}_{-0.14} $ & $ 0.34^{+0.13}_{-0.12} $ &  $ ... $ & $ ... $ &$ 42.9/48  $ \\
 ESO 506-G027 &  $ 5.45 $ & $ 7.4^{+1.7}_{-1.4} $ & $ 0.54^{+0.05}_{-0.04} $ & $ 83.7^{+5.1}_{-4.8} $ &  $ ... $ & $ ... $ & $ ... $ &$ 1.70^{+0.06}_{-0.05} $ & $ 0.34^{+0.10}_{-0.08} $ & $ 0.22\pm0.05 $ &  $ ... $ & $ ... $ &$ 70.0/60  $ \\
 Fairall 49 &  $ 6.47 $ & $ 8.5^{+2.1}_{-2.4} $ & $ 2.06^{+0.88}_{-0.43} $ & $ 1.02^{+0.03}_{-0.05} $ &  $ ... $ &$ 3.20^{+0.85}_{-0.95} $ & $ 0.22^{+0.03}_{-0.04} $ & $ 2.28\pm0.05 $ & $ 1.69^{+0.45}_{-0.37} $ & $ 0.98^{+0.80}_{-0.54} $ &  $ ... $ & $ ... $ &$ 1949.8/1753  $ \\
 Fairall 51 &  $ 6.97 $ & $ 5.98^{+1.12}_{-0.89} $ & $ 2.56^{+0.26}_{-0.24} $ & $ 2.67^{+0.50}_{-0.63} $ &  $ ... $ &$ 5.30^{+0.99}_{-0.47} $ & $ 0.77^{+0.10}_{-0.12} $ & $ 2.03\pm0.04 $ & $ 2.00^{+0.49}_{-0.47} $ & $ 2.00^{}_{-0.23} $ & $ 0.19^{+0.02}_{-0.01} $ &  $ ... $ &$ 554.8/534  $ \\
 IC 4518A &  $ 8.78 $ & $ 6.7^{+2.5}_{-1.9} $ & $ 0.71^{+0.18}_{-0.14} $ & $ 20.25^{+0.97}_{-0.98} $ & $ 3.53^{+1.09}_{-0.93} $ &  $ ... $ & $ ... $ &$ 2.11^{+0.09}_{-0.08} $ & $ 0.66^{+0.27}_{-0.18} $ & $ 2.00^{}_{-0.22} $ & $ 0.24^{+0.04}_{-0.05} $ & $ 1.06^{+0.21}_{-0.12} $ & $ 234.1/212  $ \\
 LEDA 170194 &  $ 3.00 $ & $ 2.27^{+0.23}_{-0.24} $ & $ 0.92^{+0.08}_{-0.06} $ & $ 5.24^{+0.10}_{-0.12} $ &  $ ... $ & $ ... $ & $ ... $ &$ 1.57^{+0.02}_{-0.03} $ & $ 1.84^{+0.22}_{-0.18} $ & $ 0.00^{+0.05}_{} $ &  $ ... $ & $ ... $ &$ 738.4/643  $ \\
 MCG +04-48-002 &  $ 20.7 $ & $ 3.38^{+0.95}_{-0.74} $ & $ 0.95^{+0.12}_{-0.10} $ & $ 73.5^{+6.9}_{-8.4} $ & $ 26^{+152}_{-13} $ &  $ ... $ & $ ... $ &$ 1.62^{+0.06}_{-0.05} $ & $ 0.97^{+0.30}_{-0.25} $ & $ 0.81^{+0.50}_{-0.35} $ & $ 1.01^{+0.18}_{-0.28} $ &  $ ... $ &$ 69.9/57  $ \\
 MCG -01-05-047 &  $ 2.72 $ & $ 2.41^{+1.26}_{-0.83} $ & $ 0.50^{+0.16}_{-0.11} $ & $ 18.4^{+1.4}_{-1.2} $ &  $ ... $ & $ ... $ & $ ... $ &$ 1.88\pm0.11 $ & $ 2.65^{+1.37}_{-0.92} $ & $ 1.13^{+0.72}_{-0.50} $ &  $ ... $ & $ ... $ &$ 110.0/87  $ \\
 MCG -02-08-014 &  $ 4.46 $ & $ 3.7^{+1.2}_{-1.0} $ & $ 1.04^{+0.22}_{-0.18} $ & $ 11.90^{+0.63}_{-0.64} $ &  $ ... $ & $ ... $ & $ ... $ &$ 2.00^{+0.08}_{-0.10} $ & $ 0.00^{+0.63}_{} $ & $ 1.58^{+0.42}_{-0.64} $ &  $ ... $ & $ ... $ &$ 178.9/160  $ \\
 MCG -05-23-016 &  $ 8.70 $ & $ 36.1^{+1.4}_{-1.5} $ & $ 1.29^{+0.04}_{-0.03} $ & $ 1.57\pm0.01 $ &  $ ... $ &$ 50.6\pm3.0 $ & $ 0.27\pm0.02 $ & $ 1.96^{+0.02}_{-0.01} $ & $ 0.47^{+0.04}_{-0.03} $ & $ 0.84\pm0.12 $ &  $ ... $ & $ ... $ &$ 3845.9/3479  $ \\
 Mrk 1210 &  $ 3.45 $ & $ 3.75^{+1.01}_{-0.72} $ & $ 1.04^{+0.11}_{-0.14} $ & $ 43.9^{+2.1}_{-2.0} $ &  $ ... $ & $ ... $ & $ ... $ &$ 1.80\pm0.04 $ & $ 0.70^{+0.31}_{-0.24} $ & $ 1.81^{+0.19}_{-0.35} $ & $ 0.27\pm0.03 $ & $ 1.11^{+0.08}_{-0.07} $ & $ 193.8/158  $ \\
 Mrk 1498 &  $ 1.83 $ & $ 2.97^{+0.68}_{-0.30} $ & $ 1.64^{+0.18}_{-0.23} $ & $ 14.85^{+0.54}_{-0.45} $ &  $ ... $ & $ ... $ & $ ... $ &$ 1.81^{+0.06}_{-0.05} $ & $ 1.59^{+0.39}_{-0.42} $ & $ 2.00^{}_{-0.52} $ & $ 0.14^{+0.08}_{-0.06} $ &  $ ... $ &$ 183.4/168  $ \\
 Mrk 18 &  $ 4.37 $ & $ 0.40^{+0.57}_{-0.29} $ & $ 1.12^{+1.18}_{-0.46} $ & $ 10.6^{+1.6}_{-1.3} $ &  $ ... $ & $ ... $ & $ ... $ &$ 1.62^{+0.31}_{-0.26} $ & $ 5.2^{+12.7}_{-3.3} $ & $ 1.18^{+0.82}_{-1.18} $ &  $ ... $ & $ ... $ &$ 40.4/34  $ \\
 Mrk 348 &  $ 5.79 $ & $ 11.94^{+0.26}_{-0.14} $ & $ 1.40^{+0.03}_{-0.04} $ & $ 6.01^{+0.11}_{-0.05} $ &  $ ... $ &$ 7.64^{+0.83}_{-0.66} $ & $ 0.60^{0.00}_{-0.02} $ & $ 1.77^{+0.01}_{} $ & $ 0.13^{+0.05}_{-0.04} $ & $ 0.93^{+0.12}_{-0.13} $ & $ 0.85^{+0.10}_{-0.08} $ &  $ ... $ &$ 1699.4/1767  $ \\
 Mrk 417 &  $ 1.88 $ & $ 0.94^{+0.40}_{-0.30} $ & $ 2.39^{+0.30}_{-0.51} $ & $ 44.9\pm3.2 $ &  $ ... $ & $ ... $ & $ ... $ &$ 1.60^{+0.08}_{-0.09} $ & $ 0.78^{+0.84}_{-0.59} $ & $ 1.76^{+0.24}_{-0.77} $ & $ 1.04^{+0.20}_{-0.23} $ &  $ ... $ &$ 74.3/70  $ \\
 Mrk 520 &  $ 4.30 $ & $ 1.63^{+0.16}_{-0.14} $ & $ 1.25^{+0.11}_{-0.07} $ & $ 1.87^{+0.04}_{-0.05} $ &  $ ... $ & $ ... $ & $ ... $ &$ 1.53^{+0.02}_{-0.01} $ & $ 5.27^{+0.59}_{-0.54} $ & $ 0.00^{+0.13}_{} $ &  $ ... $ & $ ... $ &$ 699.6/688  $ \\
 Mrk 915 &  $ 5.35 $ & $ 0.60^{+0.13}_{-0.11} $ & $ 1.28^{+0.22}_{-0.18} $ & $ 1.59\pm0.11 $ &  $ ... $ & $ ... $ & $ ... $ &$ 1.40\pm0.03 $ & $ 44.1^{+10.1}_{-7.8} $ & $ 0.00^{+0.20}_{} $ &  $ ... $ & $ ... $ &$ 304.5/277  $ \\
 NGC 1052 &  $ 2.83 $ & $ 2.41^{+0.32}_{-0.37} $ & $ 1.21\pm0.14 $ & $ 4.86^{+0.29}_{-0.16} $ &  $ ... $ &$ 19.8^{+2.5}_{-2.6} $ & $ 0.71^{+0.02}_{-0.01} $ & $ 1.75^{+0.06}_{-0.04} $ & $ 4.81^{+0.29}_{-0.80} $ & $ 0.35^{+0.28}_{-0.35} $ & $ 0.82\pm0.05 $ &  $ ... $ &$ 416.1/425  $ \\
 NGC 1142 &  $ 5.81 $ & $ 4.48^{+0.74}_{-0.64} $ & $ 0.96\pm0.07 $ & $ 60.9^{+2.4}_{-2.2} $ & $ 7.0^{+2.5}_{-2.3} $ &  $ ... $ & $ ... $ &$ 1.66\pm0.04 $ & $ 0.25^{+0.09}_{-0.08} $ & $ 0.78^{+0.22}_{-0.17} $ & $ 0.19\pm0.07 $ & $ 0.97\pm0.05 $ & $ 200.6/204  $ \\
 NGC 2110 &  $ 2.18 $ & $ 18.70\pm0.42 $ & $ 1.74\pm0.04 $ & $ 2.32^{+0.07}_{-0.06} $ &  $ ... $ &$ 2.80\pm0.09 $ & $ 0.64\pm0.03 $ & $ 1.65\pm0.01 $ & $ 0.28\pm0.04 $ & $ 0.38\pm0.06 $ & $ 0.98\pm0.02 $ &  $ ... $ &$ 4276.3/3823  $ \\
 NGC 235A &  $ 1.41 $ & $ 5.0^{+1.8}_{-1.3} $ & $ 0.61^{+0.10}_{-0.09} $ & $ 65.2^{+7.0}_{-6.3} $ &  $ ... $ & $ ... $ & $ ... $ &$ 1.78^{+0.07}_{-0.08} $ & $ 0.51^{+0.23}_{-0.18} $ & $ 0.34^{+0.16}_{-0.14} $ & $ 0.59^{+0.07}_{-0.08} $ &  $ ... $ &$ 37.8/28  $ \\
 NGC 3081 &  $ 3.88 $ & $ 7.6^{+1.6}_{-1.3} $ & $ 0.72\pm0.05 $ & $ 82.5^{+4.0}_{-3.8} $ &  $ ... $ & $ ... $ & $ ... $ &$ 1.73\pm0.05 $ & $ 0.52^{+0.13}_{-0.10} $ & $ 0.20^{+0.05}_{-0.06} $ & $ 0.19^{+0.01}_{-0.02} $ & $ 0.98\pm0.08 $ & $ 80.1/69  $ \\
 NGC 3431 &  $ 4.17 $ & $ 0.52^{+0.26}_{-0.12} $ & $ 3.01^{+0.68}_{-1.05} $ & $ 7.08^{+0.28}_{-0.29} $ &  $ ... $ & $ ... $ & $ ... $ &$ 1.61^{+0.02}_{-0.07} $ & $ 1.42^{+0.78}_{-0.75} $ & $ 2.0^{}_{-1.3} $ &  $ ... $ & $ ... $ &$ 245.9/234  $ \\
 NGC 4388 &  $ 2.58 $ & $ 19.2\pm1.1 $ & $ 0.88^{+0.01}_{-0.02} $ & $ 23.78^{+0.69}_{-0.70} $ &  $ ... $ &$ 43.1\pm3.6 $ & $ 0.56\pm0.03 $ & $ 1.65\pm0.01 $ & $ 0.65^{+0.04}_{-0.03} $ & $ 0.08^{+0.05}_{-0.04} $ & $ 0.24^{+0.01}_{-0.02} $ & $ 0.97\pm0.02 $ & $ 1669.1/1703  $ \\
 NGC 4507 &  $ 7.04 $ & $ 19.7^{+2.5}_{-2.3} $ & $ 0.57^{+0.03}_{-0.02} $ & $ 26.9^{+5.2}_{-3.9} $ &  $ ... $ &$ 79.9^{+5.1}_{-5.7} $ & $ 0.91\pm0.02 $ & $ 1.79^{+0.03}_{-0.02} $ & $ 0.31^{+0.05}_{-0.04} $ & $ 0.43^{+0.08}_{-0.07} $ & $ 0.15\pm0.01 $ & $ 0.81\pm0.02 $ & $ 387.0/360  $ \\
 NGC 4992 &  $ 1.93 $ & $ 2.17^{+0.65}_{-0.51} $ & $ 1.05^{+0.15}_{-0.13} $ & $ 60.1^{+3.6}_{-3.4} $ &  $ ... $ & $ ... $ & $ ... $ &$ 1.57\pm0.06 $ & $ 0.00^{+0.17}_{} $ & $ 0.52^{+0.18}_{-0.13} $ &  $ ... $ & $ ... $ &$ 74.0/61  $ \\
 NGC 5252 &  $ 2.14 $ & $ 8.88^{+0.87}_{-0.77} $ & $ 0.47\pm0.02 $ & $ 2.11^{+0.53}_{-0.52} $ &  $ ... $ &$ 5.86^{+0.36}_{-0.33} $ & $ 0.81^{+0.05}_{-0.06} $ & $ 1.66^{+0.03}_{-0.02} $ & $ 0.39^{+0.24}_{-0.32} $ & $ 0.00^{+0.02}_{} $ & $ 0.16\pm0.02 $ & $ 0.84^{+0.06}_{-0.05} $ & $ 472.6/443  $ \\
 NGC 526A &  $ 2.31 $ & $ 3.12\pm0.39 $ & $ 3.78^{+0.58}_{-0.43} $ & $ 1.22\pm0.01 $ &  $ ... $ & $ ... $ & $ ... $ &$ 1.68\pm0.01 $ & $ 6.46^{+0.71}_{-0.84} $ & $ 0.99^{+0.43}_{-0.33} $ &  $ ... $ & $ ... $ &$ 2467.0/2335  $ \\
 NGC 5506 &  $ 4.08 $ & $ 26.29^{+0.83}_{-0.59} $ & $ 1.73\pm0.03 $ & $ 3.10^{+0.01}_{-0.02} $ &  $ ... $ & $ ... $ & $ ... $ &$ 1.95^{0.00}_{-0.01} $ & $ 1.09^{+0.04}_{-0.05} $ & $ 2.00^{}_{-0.07} $ &  $ ... $ & $ ... $ &$ 3398.7/3185  $ \\
 NGC 6300 &  $ 7.79 $ & $ 11.43^{+1.00}_{-0.93} $ & $ 1.04^{+0.06}_{-0.05} $ & $ 22.21^{+0.35}_{-0.34} $ &  $ ... $ & $ ... $ & $ ... $ &$ 1.86\pm0.02 $ & $ 0.56\pm0.08 $ & $ 0.83^{+0.12}_{-0.10} $ & $ 0.85^{+0.06}_{-0.07} $ &  $ ... $ &$ 765.9/723  $ \\
 NGC 7172 &  $ 1.95 $ & $ 14.18^{+0.53}_{-0.51} $ & $ 1.44^{+0.04}_{-0.05} $ & $ 8.90\pm0.07 $ &  $ ... $ & $ ... $ & $ ... $ &$ 1.74^{+0.01}_{-0.02} $ & $ 0.12\pm0.03 $ & $ 0.34^{+0.10}_{-0.09} $ & $ 0.33^{+0.14}_{-0.04} $ &  $ ... $ &$ 2056.7/2033  $ \\
 NGC 788 &  $ 2.12 $ & $ 5.35^{+1.16}_{-0.99} $ & $ 1.19^{+0.13}_{-0.11} $ & $ 73.4^{+4.0}_{-3.8} $ & $ 11.9^{+3.4}_{-3.1} $ &  $ ... $ & $ ... $ &$ 1.77\pm0.05 $ & $ 0.71^{+0.19}_{-0.14} $ & $ 1.21^{+0.48}_{-0.35} $ & $ 0.75^{+0.08}_{-0.11} $ &  $ ... $ &$ 88.8/87  $ \\
 UGC 03142 &  $ 17.6 $ & $ 1.11^{+0.34}_{-0.17} $ & $ 1.45^{+0.17}_{-0.26} $ & $ 1.59^{+0.37}_{-0.32} $ &  $ ... $ &$ 9.67^{+0.94}_{-0.65} $ & $ 0.79^{+0.01}_{-0.03} $ & $ 1.58^{+0.07}_{-0.08} $ & $ 3.92^{+0.81}_{-1.45} $ & $ 2.00^{}_{-0.77} $ &  $ ... $ & $ ... $ &$ 205.8/231  $ \\
 UGC 12741 &  $ 5.79 $ & $ 1.88^{+1.28}_{-0.80} $ & $ 1.47^{+0.55}_{-0.33} $ & $ 60.8^{+5.0}_{-4.7} $ &  $ ... $ & $ ... $ & $ ... $ &$ 1.79^{+0.12}_{-0.11} $ & $ 0.00^{+0.78}_{} $ & $ 0.62^{+0.42}_{-0.24} $ &  $ ... $ & $ ... $ &$ 52.5/41  $ 
\enddata 
\tablecomments{
(1) Galaxy name. 
(2) Galactic absorption in units of $10^{20}$ cm$^{-2}$. 
(3) Normalization of the cutoff power-law component at 1 keV in units of $10^{-3}$ photons 
keV$^{-1}$ cm$^{-2}$ s$^{-1}$.  
(4) Time variability of the cutoff power-law component between the {\it Suzaku} and {\it Swift}/BAT spectra.
(5) Intrinsic absorption in units of $10^{22}$ cm$^{-2}$. 
(6) Absorption of the reflection components in units of $10^{22}$ cm$^{-2}$. 
(7) Partial absorption of the cutoff power-law component in units of $10^{22}$ cm$^{-2}$. 
(8) Covering fraction of the partial absorption of the cutoff power-law component. 
(9) Photon index of the cutoff power-law component. 
(10) Scattered fraction in units of \%. 
(11) Relative reflection strength ($R = \Omega/2\pi$) of the pexrav model. 
(12)--(13) Temperatures of the {\tt apec} models in units of keV. 
(14) Chi squared and degrees of freedom. 
 }
\end{deluxetable}
\clearpage
\end{landscape}

 % IRU 
\begin{landscape}
\LongTables
\setlength{\topmargin}{2cm}
\tabletypesize{\normalsize}
\setlength{\tabcolsep}{0.03in}
\begin{deluxetable}{cccccccccccccccccccc}
\tablecaption{Flux and luminosity\label{tab:info_fxlx}}
\tablewidth{0pt}
\tablehead{\colhead{Target Name} 
& \colhead{$\log F^{\rm BI-XIS}_{\rm 0.5-2}$}  
& \colhead{$\log F^{\rm FI-XIS}_{\rm 2-10}$}  
& \colhead{$\log F^{\rm PIN\ast}_{\rm 10-50}$}  
& \colhead{$\log F^{\rm BAT}_{\rm 10-50}$}  
& \colhead{$\log L^{\rm BI-XIS}_{\rm 0.5-2}$}  
& \colhead{$\log L^{\rm FI-XIS}_{\rm 2-10}$}  
& \colhead{$\log L^{\rm PIN\ast}_{\rm 10-50}$}  
& \colhead{$\log L^{\rm BAT}_{\rm 10-50}$}  
& \colhead{EW}  
& \colhead{$L_{{\rm K}\alpha}/L^{\rm BAT}_{\rm 10-50}$}  
& \colhead{$\lambda^{\it Suzaku}_{\rm Edd}/\lambda^{\rm BAT}_{\rm Edd}$}  \\
\colhead{(1)} & \colhead{(2)} & \colhead{(3)} & \colhead{(4)} & 
\colhead{(5)} & \colhead{(6)} & \colhead{(7)} & \colhead{(8)}  & \colhead{(9)} & 
\colhead{(10)} & \colhead{(11)}  & \colhead{(12)}
} 
\startdata
2MASX J0216+5126 & $ -12.2 $ & $ -11.1 $ & $ -11.0 $ & $ -11.0 $ & $ 43.0 $ & $ 43.2 $ & $ 43.3 $ & $ 43.3 $ & $ 35^{+13}_{-12} $ & $ 2.58^{+0.93}_{-0.86} $ & $ .../ ... $ \\ 
 2MASX J0248+2630 & $ -13.4 $ & $ -11.3 $ & $ -10.6 $ & $ -10.9 $ & $ 43.5 $ & $ 43.9 $ & $ 44.3 $ & $ 44.1 $ & $ 61\pm16  $ & $ 3.56\pm0.91 $ & $ .../ ... $ \\ 
 2MASX J0318+6829 & $ -13.3 $ & $ -11.4 $ & $ -10.9 $ & $ -11.1 $ & $ 43.7 $ & $ 44.1 $ & $ 44.4 $ & $ 44.2 $ & $ 42^{+13}_{-12} $ & $ 3.08^{+0.92}_{-0.89} $ & $ .../ ... $ \\ 
 2MASX J0350-5018 & $ -13.1 $ & $ -12.1 $ & $ -11.1 $ & $ -11.1 $ & $ 42.3 $ & $ 42.7 $ & $ 43.4 $ & $ 43.4 $ & $ 376^{+65}_{-61} $ & $ 5.97^{+1.03}_{-0.96} $ & $ -2.8/-2.7 $ \\ 
 2MASX J0444+2813 & $ -13.6 $ & $ -11.3 $ & $ -10.7 $ & $ -10.6 $ & $ 41.7 $ & $ 42.1 $ & $ 42.6 $ & $ 42.6 $ & $ 119\pm13  $ & $ 3.82\pm0.40 $ & $ -2.0/-2.0 $ \\ 
 2MASX J0505-2351 & $ -12.8 $ & $ -11.0 $ & $ -10.6 $ & $ -10.5 $ & $ 43.4 $ & $ 43.7 $ & $ 43.9 $ & $ 43.9 $ & $ 60\pm6  $ & $ 2.96\pm0.31 $ & $ -0.7/-0.6 $ \\ 
 2MASX J0911+4528 & $ -14.2 $ & $ -11.7 $ & $ -11.1 $ & $ -11.0 $ & $ 43.2 $ & $ 43.2 $ & $ 43.2 $ & $ 43.3 $ & $ 43\pm15  $ & $ 1.67\pm0.57 $ & $ -1.2/-1.0 $ \\ 
 2MASX J1200+0648 & $ -13.5 $ & $ -11.1 $ & $ -10.7 $ & $ -10.9 $ & $ 43.4 $ & $ 43.6 $ & $ 43.8 $ & $ 43.6 $ & $ 57^{+9}_{-8} $ & $ 5.67^{+0.88}_{-0.76} $ & $ -1.7/-2.1 $ \\ 
 Ark 347 & $ -12.9 $ & $ -12.0 $ & $ -11.0 $ & $ -10.9 $ & $ 42.0 $ & $ 42.4 $ & $ 43.0 $ & $ 43.2 $ & $ 149\pm46  $ & $ 2.43\pm0.74 $ & $ -2.5/-2.2 $ \\ 
 ESO 103-035 & $ -13.1 $ & $ -10.6 $ & $ -10.1 $ & $ -10.1 $ & $ 43.5 $ & $ 43.5 $ & $ 43.6 $ & $ 43.5 $ & $ 52\pm5  $ & $ 3.50\pm0.36 $ & $ -0.8/-0.9 $ \\ 
 ESO 263-G013 & $ -13.2 $ & $ -11.4 $ & $ -10.8 $ & $ -10.8 $ & $ 43.2 $ & $ 43.4 $ & $ 43.7 $ & $ 43.7 $ & $ 85\pm16  $ & $ 3.35\pm0.64 $ & $ -1.3/-1.3 $ \\ 
 ESO 297-G018 & $ -13.5 $ & $ -11.5 $ & $ -10.5 $ & $ -10.5 $ & $ 43.2 $ & $ 43.5 $ & $ 43.7 $ & $ 43.7 $ & $ 154\pm23  $ & $ 3.14\pm0.47 $ & $ -3.0/-3.0 $ \\ 
 ESO 506-G027 & $ -13.3 $ & $ -11.7 $ & $ -10.6 $ & $ -10.4 $ & $ 43.1 $ & $ 43.3 $ & $ 43.6 $ & $ 43.8 $ & $ 465^{+33}_{-32} $ & $ 4.26\pm0.30 $ & $ -2.0/-1.8 $ \\ 
 Fairall 49 & $ -11.3 $ & $ -10.6 $ & $ -10.7 $ & $ -10.9 $ & $ 43.5 $ & $ 43.4 $ & $ 43.3 $ & $ 43.0 $ & $ 49\pm6  $ & $ 10.3\pm1.2 $ & $ .../ ... $ \\ 
 Fairall 51 & $ -12.1 $ & $ -10.7 $ & $ -10.3 $ & $ -10.6 $ & $ 42.9 $ & $ 42.9 $ & $ 43.1 $ & $ 42.8 $ & $ 40\pm8 $ & $ 5.21^{+1.03}_{-0.98} $ & $ -1.8/-2.2 $ \\ 
 IC 4518A & $ -12.8 $ & $ -11.3 $ & $ -10.7 $ & $ -10.7 $ & $ 42.8 $ & $ 42.8 $ & $ 43.0 $ & $ 43.1 $ & $ 56\pm17 $ & $ 2.53^{+0.77}_{-0.76} $ & $ -1.5/-1.3 $ \\ 
 LEDA 170194 & $ -12.9 $ & $ -11.2 $ & $ -10.7 $ & $ -10.7 $ & $ 43.1 $ & $ 43.5 $ & $ 43.8 $ & $ 43.8 $ & $ 66^{+8}_{-7} $ & $ 3.30^{+0.40}_{-0.36} $ & $ -2.2/-2.2 $ \\ 
 MCG +04-48-002 & $ -13.1 $ & $ -11.7 $ & $ -10.5 $ & $ -10.5 $ & $ 42.5 $ & $ 42.8 $ & $ 43.2 $ & $ 43.2 $ & $ 183^{+2590}_{-46} $ & $ 4.0^{+57.1}_{-1.0} $ & $ -1.2/-1.1 $ \\ 
 MCG -01-05-047 & $ -12.9 $ & $ -11.7 $ & $ -11.0 $ & $ -10.9 $ & $ 42.2 $ & $ 42.4 $ & $ 42.7 $ & $ 42.9 $ & $ 221\pm26  $ & $ 6.07\pm0.71 $ & $ -2.0/-1.7 $ \\ 
 MCG -02-08-014 & $ -13.7 $ & $ -11.3 $ & $ -10.8 $ & $ -10.8 $ & $ 42.7 $ & $ 42.8 $ & $ 43.0 $ & $ 43.0 $ & $ 117\pm15  $ & $ 6.77\pm0.85 $ & $ .../ ... $ \\ 
 MCG -05-23-016 & $ -11.1 $ & $ -10.1 $ & $ -9.8 $ & $ -9.9 $ & $ 43.2 $ & $ 43.3 $ & $ 43.4 $ & $ 43.3 $ & $ 58\pm3  $ & $ 4.65\pm0.22 $ & $ -0.9/-1.0 $ \\ 
 Mrk 1210 & $ -12.6 $ & $ -11.3 $ & $ -10.5 $ & $ -10.5 $ & $ 42.6 $ & $ 42.8 $ & $ 43.1 $ & $ 43.1 $ & $ 190\pm15  $ & $ 5.15\pm0.40 $ & $ -1.9/-1.9 $ \\ 
 Mrk 1498 & $ -12.3 $ & $ -11.1 $ & $ -10.5 $ & $ -10.6 $ & $ 43.9 $ & $ 44.1 $ & $ 44.4 $ & $ 44.3 $ & $ 72\pm12  $ & $ 4.00\pm0.68 $ & $ -1.3/-1.5 $ \\ 
 Mrk 18 & $ -13.4 $ & $ -11.8 $ & $ -11.3 $ & $ -11.3 $ & $ 41.4 $ & $ 41.8 $ & $ 42.2 $ & $ 42.2 $ & $ 248^{+55}_{-56} $ & $ 10.0\pm2.2 $ & $ -2.4/-2.5 $ \\ 
 Mrk 348 & $ -12.7 $ & $ -10.5 $ & $ -10.0 $ & $ -10.1 $ & $ 43.3 $ & $ 43.5 $ & $ 43.7 $ & $ 43.6 $ & $ 45^{+4}_{-5} $ & $ 2.93^{+0.25}_{-0.30} $ & $ -1.3/-1.4 $ \\ 
 Mrk 417 & $ -13.3 $ & $ -11.5 $ & $ -10.6 $ & $ -10.8 $ & $ 43.1 $ & $ 43.4 $ & $ 43.8 $ & $ 43.6 $ & $ 125\pm21  $ & $ 4.72\pm0.78 $ & $ -1.4/-1.8 $ \\ 
 Mrk 520 & $ -12.2 $ & $ -11.1 $ & $ -10.7 $ & $ -10.8 $ & $ 42.8 $ & $ 43.1 $ & $ 43.5 $ & $ 43.4 $ & $ 92\pm9 $ & $ 6.20^{+0.62}_{-0.59} $ & $ -2.0/-2.1 $ \\ 
 Mrk 915 & $ -12.1 $ & $ -11.2 $ & $ -10.8 $ & $ -10.9 $ & $ 42.5 $ & $ 42.9 $ & $ 43.3 $ & $ 43.2 $ & $ 133^{+18}_{-17} $ & $ 7.39^{+1.01}_{-0.93} $ & $ -2.0/-2.0 $ \\ 
 NGC 1052 & $ -12.5 $ & $ -11.3 $ & $ -10.8 $ & $ -10.8 $ & $ 41.5 $ & $ 41.7 $ & $ 41.9 $ & $ 41.9 $ & $ 114\pm10 $ & $ 6.63^{+0.58}_{-0.60} $ & $ -3.7/-3.8 $ \\ 
 NGC 1142 & $ -13.0 $ & $ -11.4 $ & $ -10.4 $ & $ -10.4 $ & $ 43.2 $ & $ 43.6 $ & $ 43.9 $ & $ 43.9 $ & $ 226^{+16}_{-15} $ & $ 4.66^{+0.33}_{-0.31} $ & $ -2.5/-2.5 $ \\ 
 NGC 2110 & $ -11.6 $ & $ -10.0 $ & $ -9.6 $ & $ -9.8 $ & $ 43.0 $ & $ 43.3 $ & $ 43.6 $ & $ 43.4 $ & $ 34\pm2  $ & $ 3.04\pm0.17 $ & $ -1.7/-2.0 $ \\ 
 NGC 235A & $ -12.8 $ & $ -11.7 $ & $ -10.8 $ & $ -10.6 $ & $ 42.9 $ & $ 43.1 $ & $ 43.3 $ & $ 43.5 $ & $ 171\pm69  $ & $ 2.7\pm1.1 $ & $ -2.5/-2.3 $ \\ 
 NGC 3081 & $ -12.6 $ & $ -11.5 $ & $ -10.5 $ & $ -10.4 $ & $ 42.0 $ & $ 42.3 $ & $ 42.5 $ & $ 42.6 $ & $ 313\pm25  $ & $ 3.90\pm0.31 $ & $ -2.2/-2.1 $ \\ 
 NGC 3431 & $ -13.5 $ & $ -11.3 $ & $ -10.8 $ & $ -11.1 $ & $ 42.4 $ & $ 42.7 $ & $ 43.1 $ & $ 42.8 $ & $ 102\pm13  $ & $ 8.4\pm1.1 $ & $ .../ ... $ \\ 
 NGC 4388 & $ -12.3 $ & $ -10.7 $ & $ -9.9 $ & $ -9.9 $ & $ 42.3 $ & $ 42.6 $ & $ 42.8 $ & $ 42.9 $ & $ 190\pm5  $ & $ 4.92\pm0.12 $ & $ -2.2/-2.2 $ \\ 
 NGC 4507 & $ -12.3 $ & $ -11.2 $ & $ -10.2 $ & $ -10.0 $ & $ 42.9 $ & $ 43.1 $ & $ 43.3 $ & $ 43.5 $ & $ 421\pm14  $ & $ 4.30\pm0.15 $ & $ -1.7/-1.5 $ \\ 
 NGC 4992 & $ -14.3 $ & $ -11.6 $ & $ -10.6 $ & $ -10.6 $ & $ 42.8 $ & $ 43.3 $ & $ 43.6 $ & $ 43.6 $ & $ 168\pm30  $ & $ 3.17\pm0.57 $ & $ -2.0/-2.1 $ \\ 
 NGC 5252 & $ -12.4 $ & $ -11.0 $ & $ -10.6 $ & $ -10.2 $ & $ 42.9 $ & $ 43.1 $ & $ 43.4 $ & $ 43.7 $ & $ 83^{+10}_{-9} $ & $ 2.29^{+0.28}_{-0.24} $ & $ -2.5/-2.2 $ \\ 
 NGC 526A & $ -11.3 $ & $ -10.4 $ & $ -10.1 $ & $ -10.5 $ & $ 43.3 $ & $ 43.6 $ & $ 43.9 $ & $ 43.4 $ & $ 48\pm4  $ & $ 7.85\pm0.67 $ & $ -1.2/-1.7 $ \\ 
 NGC 5506 & $ -11.4 $ & $ -10.0 $ & $ -9.7 $ & $ -9.8 $ & $ 42.8 $ & $ 43.0 $ & $ 43.1 $ & $ 43.0 $ & $ 44^{+4}_{-2} $ & $ 3.87^{+0.32}_{-0.22} $ & $ -1.3/-1.5 $ \\ 
 NGC 6300 & $ -12.7 $ & $ -10.8 $ & $ -10.2 $ & $ -10.2 $ & $ 41.8 $ & $ 42.0 $ & $ 42.2 $ & $ 42.1 $ & $ 67\pm7  $ & $ 3.22\pm0.33 $ & $ -2.1/-2.2 $ \\ 
 NGC 7172 & $ -12.8 $ & $ -10.3 $ & $ -9.9 $ & $ -10.1 $ & $ 42.8 $ & $ 43.0 $ & $ 43.2 $ & $ 43.1 $ & $ 52\pm4  $ & $ 4.04\pm0.33 $ & $ -1.7/-1.9 $ \\ 
 NGC 788 & $ -12.9 $ & $ -11.4 $ & $ -10.4 $ & $ -10.4 $ & $ 42.8 $ & $ 43.0 $ & $ 43.3 $ & $ 43.2 $ & $ 223^{+24}_{-22} $ & $ 5.45^{+0.58}_{-0.54} $ & $ -2.0/-2.0 $ \\ 
 UGC 03142 & $ -12.8 $ & $ -11.3 $ & $ -10.6 $ & $ -10.7 $ & $ 42.6 $ & $ 42.9 $ & $ 43.4 $ & $ 43.3 $ & $ 156\pm15 $ & $ 6.24^{+0.60}_{-0.59} $ & $ -2.1/-2.3 $ \\ 
 UGC 12741 & $ -14.4 $ & $ -11.8 $ & $ -10.9 $ & $ -11.0 $ & $ 42.6 $ & $ 42.8 $ & $ 43.0 $ & $ 42.9 $ & $ 150\pm28 $ & $ 4.96^{+0.92}_{-0.93} $ & $ .../ ... $ 
\enddata
\tablecomments{
(1) Galaxy name. 
(2)--(5) Logarithmic observed flux in the 0.5--2 keV (BI-XIS), 2--10 
keV (FI-XISs), 10--50 keV (PIN), and 10--50 keV (BAT) bands in units of erg cm$^{-2}$ s$^{-1}$. 
(6)--(9) Logarithmic absorption-corrected luminosity in the same energy bands as (2)--(5) 
in units of erg s$^{-1}$, respectively. 
(10) Equivalent width of the iron-K$\alpha$ line in units of eV. 
(11) Ratio of the iron-K$\alpha$ line to 10--50 keV continuum luminosity in units of $\times 10^{-3}$. 
(12) Logarithmic Eddington ratio based on the 2--10 keV luminosity measured with {\it Suzaku} and {\it Swift}/BAT. 
* According to the XIS or HXD nominal position observation, the flux and luminosity are divided 
by 1.16 or 1.18 to take into account the instrumental cross-calibration factor between the 
FI-XISs and HXD/PIN spectra.
 }
\end{deluxetable}
\clearpage
\end{landscape}
 % IRU 
% TABLE.1 
%%%%%%%%%
\begin{landscape}
\LongTables
\setlength{\topmargin}{1.8cm}
\begin{deluxetable}{cccccccccccccccccccc}
\tabletypesize{\footnotesize}
\setlength{\rightmargin}{5cm}
\setlength{\tabcolsep}{0.04in}
\tablecaption{Information of detected emission/absorption lines\label{tab:info_lines}}
\tablewidth{-2pt}
\tablehead{\colhead{Target Name} 
& \colhead{$N_{\rm 6.4\hspace{1mm}keV}$}  
& \colhead{$N_{\rm 6.31\hspace{1mm}keV}$}  
& \colhead{$N_{\rm 6.70\hspace{1mm}keV}$}  
& \colhead{$N_{\rm 6.97\hspace{1mm}keV}$}  
& \colhead{$N_{\rm 7.06\hspace{1mm}keV}$}  
& \colhead{$N_{\rm 7.48\hspace{1mm}keV}$}  
& \colhead{$E_{1}$}  & \colhead{$N_{1}$}  
& \colhead{$E_{2}$}  & \colhead{$N_{2}$}  
& \colhead{$E_{3}$}  & \colhead{$N_{3}$}  
& \colhead{$E_{4}$}  & \colhead{$N_{4}$}  \\
\colhead{(1)} & \colhead{(2)} & \colhead{(3)} & \colhead{(4)} &
\colhead{(5)} & \colhead{(6)} & \colhead{(7)} & \colhead{(8)}  & \colhead{(9)} &
\colhead{(10)} & \colhead{(11)}  & \colhead{(12)} & \colhead{(13)} & \colhead{(14)} & \colhead{(15)} 
} 
\startdata
\thispagestyle{empty}
2MASX J0216+5126 &  $ 2.9^{+1.1}_{-1.0}  $ &  $ ... $ &  $ ... $ &  $ ... $ &  $ ... $ &  $ ... $ &  $ ... $ &  $ ... $ &  $ ... $ &  $ ... $ &  $ ... $ &  $ ... $ &  $ ... $ &  $ ... $ \\ 
 2MASX J0248+2630 &  $ 5.7\pm1.4  $ &  $ ... $ &  $ ... $ &  $ ... $ &  $ ... $ &  $ ... $ &  $ ... $ &  $ ... $ &  $ ... $ &  $ ... $ &  $ ... $ &  $ ... $ &  $ ... $ &  $ ... $ \\ 
 2MASX J0318+6829 &  $ 3.0\pm0.9  $ &  $ ... $ &  $ ... $ &  $ ... $ &  $ ... $ &  $ ... $ &  $ ... $ &  $ ... $ &  $ ... $ &  $ ... $ &  $ ... $ &  $ ... $ &  $ ... $ &  $ ... $ \\ 
 2MASX J0350-5018 &  $ 5.7^{+1.0}_{-0.9}  $ &  $ ... $ &  $ ... $ &  $ ... $ &  $ ... $ &  $ ... $ &  $ ... $ &  $ ... $ &  $ ... $ &  $ ... $ &  $ ... $ &  $ ... $ &  $ ... $ &  $ ... $ \\ 
 2MASX J0444+2813 &  $ 9.1\pm1.0  $ &  $ ... $ &  $ ... $ &  $ ... $ &  $ ... $ &  $ ... $ &  $ ... $ &  $ ... $ &  $ ... $ &  $ ... $ &  $ ... $ &  $ ... $ &  $ ... $ &  $ ... $ \\ 
 2MASX J0505-2351 &  $ 9.3\pm1.0  $ &  $ ... $ &  $ ... $ &  $ ... $ &  $ 3.3\pm0.9  $ &  $ ... $ &  $ ... $ &  $ ... $ &  $ ... $ &  $ ... $ &  $ ... $ &  $ ... $ &  $ ... $ &  $ ... $ \\ 
 2MASX J0911+4528 &  $ 2.0\pm0.7  $ &  $ ... $ &  $ ... $ &  $ ... $ &  $ ... $ &  $ ... $ &  $ ... $ &  $ ... $ &  $ ... $ &  $ ... $ &  $ ... $ &  $ ... $ &  $ ... $ &  $ ... $ \\ 
 2MASX J1200+0648 &  $ 7.4^{+1.1}_{-1.0}  $ &  $ ... $ &  $ ... $ &  $ ... $ &  $ ... $ &  $ ... $ &  $ ... $ &  $ ... $ &  $ ... $ &  $ ... $ &  $ ... $ &  $ ... $ &  $ ... $ &  $ ... $ \\ 
 Ark 347 &  $ 3.2\pm1.0  $ &  $ ... $ &  $ ... $ &  $ ... $ &  $ ... $ &  $ ... $ &  $ ... $ &  $ ... $ &  $ ... $ &  $ ... $ &  $ ... $ &  $ ... $ &  $ ... $ &  $ ... $ \\ 
 ESO 103-035 &  $ 27.1\pm2.8  $ &  $ 6.4\pm2.4  $ &  $ ... $ &  $ ... $ &  $ ... $ &  $ ... $ &  $ ... $ &  $ ... $ &  $ ... $ &  $ ... $ &  $ ... $ &  $ ... $ &  $ ... $ &  $ ... $ \\ 
 ESO 263-G013 &  $ 6.4\pm1.2  $ &  $ ... $ &  $ ... $ &  $ ... $ &  $ ... $ &  $ ... $ &  $ ... $ &  $ ... $ &  $ ... $ &  $ ... $ &  $ ... $ &  $ ... $ &  $ ... $ &  $ ... $ \\ 
 ESO 297-G018 &  $ 11.7\pm1.7  $ &  $ ... $ &  $ -4.5^{+1.5}_{-1.6}  $ &  $ ... $ &  $ ... $ &  $ ... $ &  $ ... $ &  $ ... $ &  $ ... $ &  $ ... $ &  $ ... $ &  $ ... $ &  $ ... $ &  $ ... $ \\ 
 ESO 506-G027 &  $ 21.4\pm1.5  $ &  $ ... $ &  $ ... $ &  $ ... $ &  $ ... $ &  $ ... $ &  $ ... $ &  $ ... $ &  $ ... $ &  $ ... $ &  $ ... $ &  $ ... $ &  $ ... $ &  $ ... $ \\ 
 Fairall 49 &  $ 12.7\pm1.5  $ &  $ ... $ &  $ 8.7\pm1.5  $ &  $ 7.2\pm1.5  $ &  $ ... $ &  $ ... $ &  $ ... $ &  $ ... $ &  $ ... $ &  $ ... $ &  $ ... $ &  $ ... $ &  $ ... $ &  $ ... $ \\ 
 Fairall 51 &  $ ... $ &  $ ... $ &  $ -24.7\pm2.3  $ &  $ -14.1^{+2.5}_{-2.4}  $ &  $ ... $ &  $ ... $ & $ 6.31\pm0.03  $ & $ 14.1^{+2.8}_{-2.7}  $ & $ 8.00\pm0.04  $ & $ -13.2^{+2.9}_{-3.2}  $ & $ 8.29^{+0.06}_{-0.05}  $ & $ -11.1^{+3.6}_{-3.2}  $ & $ 8.67^{+0.04}_{-0.05}  $ & $ -13.4^{+3.5}_{-3.3}  $  \\ 
 IC 4518A &  $ 5.5^{+1.7}_{-1.6}  $ &  $ 4.7\pm1.5  $ &  $ -5.9^{+0.8}_{-0.9}  $ &  $ -2.6\pm0.9  $ &  $ ... $ &  $ ... $ &  $ ... $ &  $ ... $ &  $ ... $ &  $ ... $ &  $ ... $ &  $ ... $ &  $ ... $ &  $ ... $ \\ 
 LEDA 170194 &  $ 6.9\pm0.8  $ &  $ ... $ &  $ ... $ &  $ ... $ &  $ ... $ &  $ ... $ & $ 0.88\pm0.02  $ & $ 5.5^{+2.1}_{-1.9}  $ &  $ ... $ &  $ ... $ &  $ ... $ &  $ ... $ &  $ ... $ &  $ ... $ \\ 
 MCG +04-48-002 &  $ 15.3^{+217.2}_{-3.9}  $ &  $ ... $ &  $ ... $ &  $ ... $ &  $ ... $ &  $ ... $ &  $ ... $ &  $ ... $ &  $ ... $ &  $ ... $ &  $ ... $ &  $ ... $ &  $ ... $ &  $ ... $ \\ 
 MCG -01-05-047 &  $ 8.2\pm1.0  $ &  $ ... $ &  $ ... $ &  $ ... $ &  $ ... $ &  $ ... $ &  $ ... $ &  $ ... $ &  $ ... $ &  $ ... $ &  $ ... $ &  $ ... $ &  $ ... $ &  $ ... $ \\ 
 MCG -02-08-014 &  $ 10.6\pm1.3  $ &  $ ... $ &  $ ... $ &  $ ... $ &  $ ... $ &  $ ... $ & $ 7.46^{+0.09}_{-0.05}  $ & $ -4.8\pm1.2  $ &  $ ... $ &  $ ... $ &  $ ... $ &  $ ... $ &  $ ... $ &  $ ... $ \\ 
 MCG -05-23-016 &  $ 62.6\pm3.0  $ &  $ 14.5\pm2.9  $ &  $ ... $ &  $ ... $ &  $ 7.0\pm2.2  $ &  $ 8.4\pm2.2  $ &  $ ... $ &  $ ... $ &  $ ... $ &  $ ... $ &  $ ... $ &  $ ... $ &  $ ... $ &  $ ... $ \\ 
 Mrk 1210 &  $ 16.8\pm1.3  $ &  $ ... $ &  $ ... $ &  $ ... $ &  $ ... $ &  $ ... $ & $ 1.18\pm0.01  $ & $ 4.2\pm1.0  $ &  $ ... $ &  $ ... $ &  $ ... $ &  $ ... $ &  $ ... $ &  $ ... $ \\ 
 Mrk 1498 &  $ ... $ &  $ ... $ &  $ ... $ &  $ ... $ &  $ 6.6\pm1.9  $ &  $ ... $ & $ 6.29\pm0.02  $ & $ 11.3\pm1.9  $ &  $ ... $ &  $ ... $ &  $ ... $ &  $ ... $ &  $ ... $ &  $ ... $ \\ 
 Mrk 18 &  $ 5.3\pm1.2  $ &  $ ... $ &  $ ... $ &  $ ... $ &  $ ... $ &  $ ... $ &  $ ... $ &  $ ... $ &  $ ... $ &  $ ... $ &  $ ... $ &  $ ... $ &  $ ... $ &  $ ... $ \\ 
 Mrk 348 &  $ 25.3^{+2.2}_{-2.6}  $ &  $ 10.3^{+1.9}_{-2.6}  $ &  $ -5.9^{+1.7}_{-1.8}  $ &  $ ... $ &  $ ... $ &  $ ... $ &  $ ... $ &  $ ... $ &  $ ... $ &  $ ... $ &  $ ... $ &  $ ... $ &  $ ... $ &  $ ... $ \\ 
 Mrk 417 &  $ 7.6\pm1.3  $ &  $ ... $ &  $ ... $ &  $ ... $ &  $ ... $ &  $ ... $ &  $ ... $ &  $ ... $ &  $ ... $ &  $ ... $ &  $ ... $ &  $ ... $ &  $ ... $ &  $ ... $ \\ 
 Mrk 520 &  $ 10.9^{+1.1}_{-1.0}  $ &  $ ... $ &  $ ... $ &  $ ... $ &  $ ... $ &  $ ... $ &  $ ... $ &  $ ... $ &  $ ... $ &  $ ... $ &  $ ... $ &  $ ... $ &  $ ... $ &  $ ... $ \\ 
 Mrk 915 &  $ 10.0^{+1.4}_{-1.3}  $ &  $ ... $ &  $ ... $ &  $ ... $ &  $ ... $ &  $ ... $ &  $ ... $ &  $ ... $ &  $ ... $ &  $ ... $ &  $ ... $ &  $ ... $ &  $ ... $ &  $ ... $ \\ 
 NGC 1052 &  $ 10.1\pm0.9  $ &  $ ... $ &  $ -1.3\pm0.8  $ &  $ ... $ &  $ ... $ &  $ ... $ &  $ ... $ &  $ ... $ &  $ ... $ &  $ ... $ &  $ ... $ &  $ ... $ &  $ ... $ &  $ ... $ \\ 
 NGC 1142 &  $ 20.8^{+1.5}_{-1.4}  $ &  $ ... $ &  $ ... $ &  $ ... $ &  $ ... $ &  $ ... $ &  $ ... $ &  $ ... $ &  $ ... $ &  $ ... $ &  $ ... $ &  $ ... $ &  $ ... $ &  $ ... $ \\ 
 NGC 2110 &  $ 48.6\pm2.8  $ &  $ 18.1\pm2.7  $ &  $ ... $ &  $ ... $ &  $ ... $ &  $ ... $ &  $ ... $ &  $ ... $ &  $ ... $ &  $ ... $ &  $ ... $ &  $ ... $ &  $ ... $ &  $ ... $ \\ 
 NGC 235A &  $ 7.5\pm3.0  $ &  $ 8.1\pm3.1  $ &  $ ... $ &  $ ... $ &  $ ... $ &  $ ... $ &  $ ... $ &  $ ... $ &  $ ... $ &  $ ... $ &  $ ... $ &  $ ... $ &  $ ... $ &  $ ... $ \\ 
 NGC 3081 &  $ 17.8\pm1.4  $ &  $ ... $ &  $ ... $ &  $ ... $ &  $ ... $ &  $ ... $ &  $ ... $ &  $ ... $ &  $ ... $ &  $ ... $ &  $ ... $ &  $ ... $ &  $ ... $ &  $ ... $ \\ 
 NGC 3431 &  $ 7.6\pm1.0  $ &  $ ... $ &  $ ... $ &  $ ... $ &  $ ... $ &  $ ... $ &  $ ... $ &  $ ... $ &  $ ... $ &  $ ... $ &  $ ... $ &  $ ... $ &  $ ... $ &  $ ... $ \\ 
 NGC 4388 &  $ 70.3\pm1.7  $ &  $ 6.9\pm1.5  $ &  $ -8.1\pm1.1  $ &  $ -6.1\pm1.5  $ &  $ 10.5\pm1.4  $ &  $ 8.7\pm1.1  $ &  $ ... $ &  $ ... $ &  $ ... $ &  $ ... $ &  $ ... $ &  $ ... $ &  $ ... $ &  $ ... $ \\ 
 NGC 4507 &  $ 47.0\pm1.6  $ &  $ 5.4\pm1.3  $ &  $ ... $ &  $ ... $ &  $ 6.6\pm0.9  $ &  $ 5.1\pm0.8  $ & $ 1.21\pm0.01  $ & $ 5.6\pm0.8  $ & $ 1.35\pm0.01  $ & $ 5.8\pm0.6  $ & $ 2.45^{+0.03}_{-0.04}  $ & $ 1.6\pm0.5  $ & $ 3.72^{+0.01}_{-0.02}  $ & $ 1.9\pm0.5  $ \\ 
 NGC 4992 &  $ 8.3\pm1.5  $ &  $ 4.5\pm1.4  $ &  $ ... $ &  $ ... $ &  $ ... $ &  $ ... $ &  $ ... $ &  $ ... $ &  $ ... $ &  $ ... $ &  $ ... $ &  $ ... $ &  $ ... $ &  $ ... $ \\ 
 NGC 5252 &  $ 13.9^{+1.7}_{-1.5}  $ &  $ ... $ &  $ ... $ &  $ ... $ &  $ ... $ &  $ ... $ &  $ ... $ &  $ ... $ &  $ ... $ &  $ ... $ &  $ ... $ &  $ ... $ &  $ ... $ &  $ ... $ \\ 
 NGC 526A &  $ 25.2\pm2.2  $ &  $ ... $ &  $ ... $ &  $ ... $ &  $ ... $ &  $ ... $ &  $ ... $ &  $ ... $ &  $ ... $ &  $ ... $ &  $ ... $ &  $ ... $ &  $ ... $ &  $ ... $ \\ 
 NGC 5506 &  $ ... $ &  $ 42.8^{+3.3}_{-6.1}  $ &  $ 24.1^{+3.5}_{-3.0}  $ &  $ 11.8^{+4.2}_{-4.7}  $ &  $ 14.7^{+4.8}_{-4.2}  $ &  $ ... $ & $ 6.45\pm0.01  $ & $ 56.6^{+4.7}_{-3.2}  $ &  $ ... $ &  $ ... $ &  $ ... $ &  $ ... $ &  $ ... $ &  $ ... $ \\ 
 NGC 6300 &  $ 19.2\pm2.0  $ &  $ 6.9\pm1.9  $ &  $ -4.0\pm1.3  $ &  $ ... $ &  $ ... $ &  $ ... $ &  $ ... $ &  $ ... $ &  $ ... $ &  $ ... $ &  $ ... $ &  $ ... $ &  $ ... $ &  $ ... $ \\ 
 NGC 7172 &  $ 36.5\pm3.0  $ &  $ 8.0\pm2.9  $ &  $ ... $ &  $ ... $ &  $ ... $ &  $ ... $ &  $ ... $ &  $ ... $ &  $ ... $ &  $ ... $ &  $ ... $ &  $ ... $ &  $ ... $ &  $ ... $ \\ 
 NGC 788 &  $ 23.6^{+2.5}_{-2.3}  $ &  $ ... $ &  $ ... $ &  $ ... $ &  $ ... $ &  $ ... $ &  $ ... $ &  $ ... $ &  $ ... $ &  $ ... $ &  $ ... $ &  $ ... $ &  $ ... $ &  $ ... $ \\ 
 UGC 03142 &  $ 13.4\pm1.3  $ &  $ ... $ &  $ ... $ &  $ ... $ &  $ ... $ &  $ ... $ &  $ ... $ &  $ ... $ &  $ ... $ &  $ ... $ &  $ ... $ &  $ ... $ &  $ ... $ &  $ ... $ \\ 
 UGC 12741 &  $ 5.6\pm1.1  $ &  $ ... $ &  $ ... $ &  $ ... $ &  $ ... $ &  $ ... $ &  $ ... $ &  $ ... $ &  $ ... $ &  $ ... $ &  $ ... $ &  $ ... $ &  $ ... $ &  $ ... $
\enddata 
\tablecomments{
(1) Galaxy name. 
(2)-(7) Normalization of the emission lines at 6.40 keV, 6.31 keV, 6.68 keV, 6.93 keV, 7.06 keV, and 7.48 keV 
in units of $\times10^{-6}$ photons cm$^{-2}$ s$^{-1}$. 
(8)-(13) Line Energy and normalization of emission/absorption lines in units of keV and $\times10^{-6}$ photons cm$^{-2}$ s$^{-1}$, respectively. 
}
\end{deluxetable}
\clearpage
\end{landscape}

\begin{figure*}
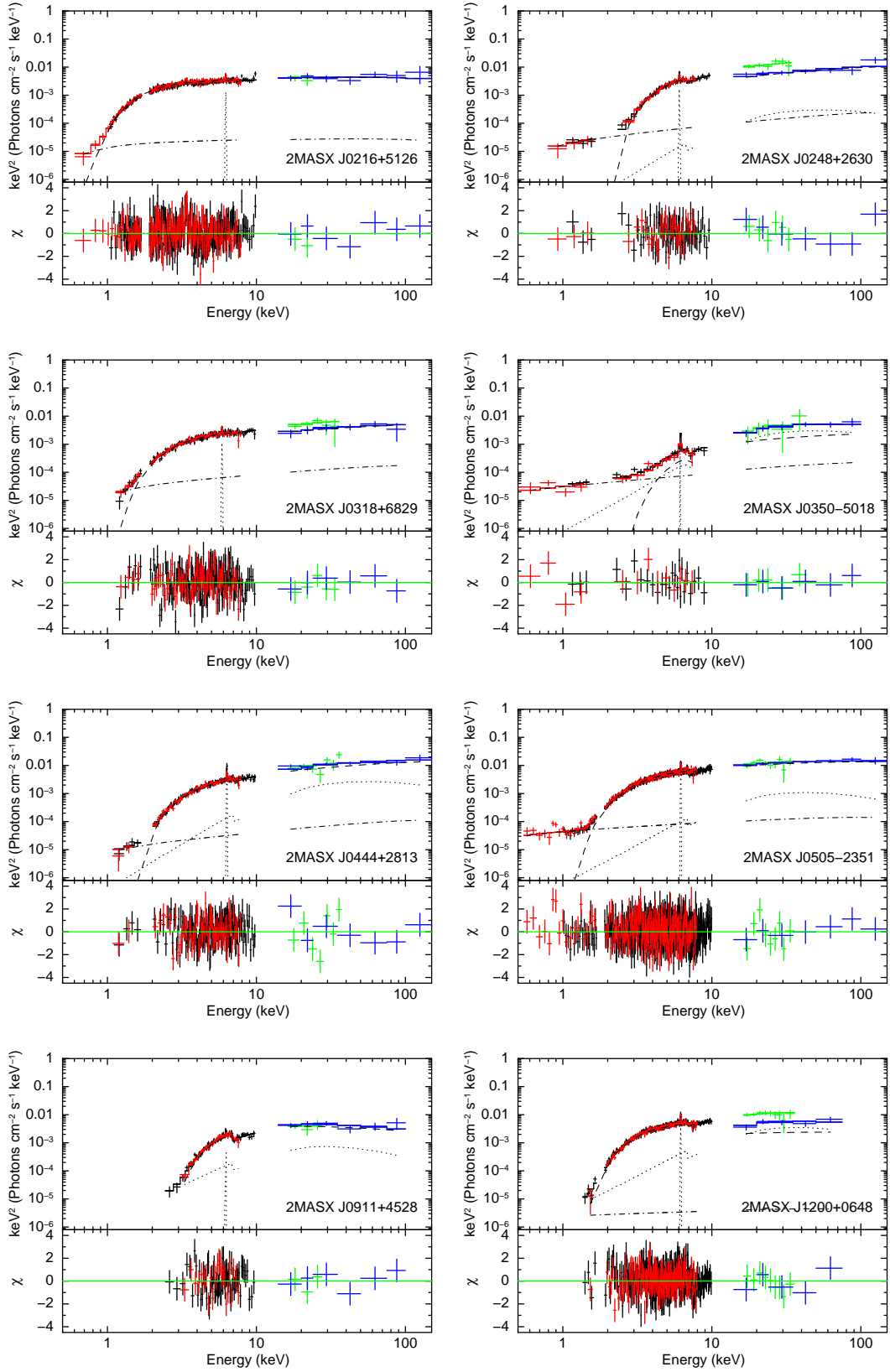

\centering
\subfigure{
\resizebox{7.1cm}{!}{\includegraphics[scale=0.8,angle=-90]{fig_a01.ps}}
\resizebox{7.1cm}{!}{\includegraphics[scale=0.8,angle=-90]{fig_a02.ps}}
}\vspace{0.0cm}
\subfigure{
\resizebox{7.1cm}{!}{\includegraphics[scale=0.8,angle=-90]{fig_a03.ps}}
\resizebox{7.1cm}{!}{\includegraphics[scale=0.8,angle=-90]{fig_a04.ps}}
}\vspace{0.0cm}
\subfigure{
\resizebox{7.1cm}{!}{\includegraphics[scale=0.8,angle=-90]{fig_a05.ps}}
\resizebox{7.1cm}{!}{\includegraphics[scale=0.8,angle=-90]{fig_a06.ps}}
}\vspace{0.0cm}
\subfigure{
\resizebox{7.1cm}{!}{\includegraphics[scale=0.8,angle=-90]{fig_a07.ps}}
\resizebox{7.1cm}{!}{\includegraphics[scale=0.8,angle=-90]{fig_a08.ps}}
}\vspace{0.0cm}
\caption{
Unfolded spectra and best-fitting models obtained in Section~\ref{sec:base_model}. 
The FI-XISs, BI-XIS, HXD/PIN and {\it Swift}/BAT spectra are represented with the black, red, green, and blue 
crosses in the upper panel, respectively, whereas the fit residuals in the lower panel. 
The solid, dashed, dotted, dot-dashed, and dot-dot-dashed lines correspond to the total, cutoff power-law 
component, reflection components (e.g., iron-K$\alpha$ emission line and reflection continuum), scattered 
component, and optically-thin thermal emission, respectively. 
}
\label{fig:spec}
\end{figure*}
\addtocounter{figure}{-1}
\begin{figure*}
\centering
\subfigure{
\resizebox{7.1cm}{!}{\includegraphics[scale=0.8,angle=-90]{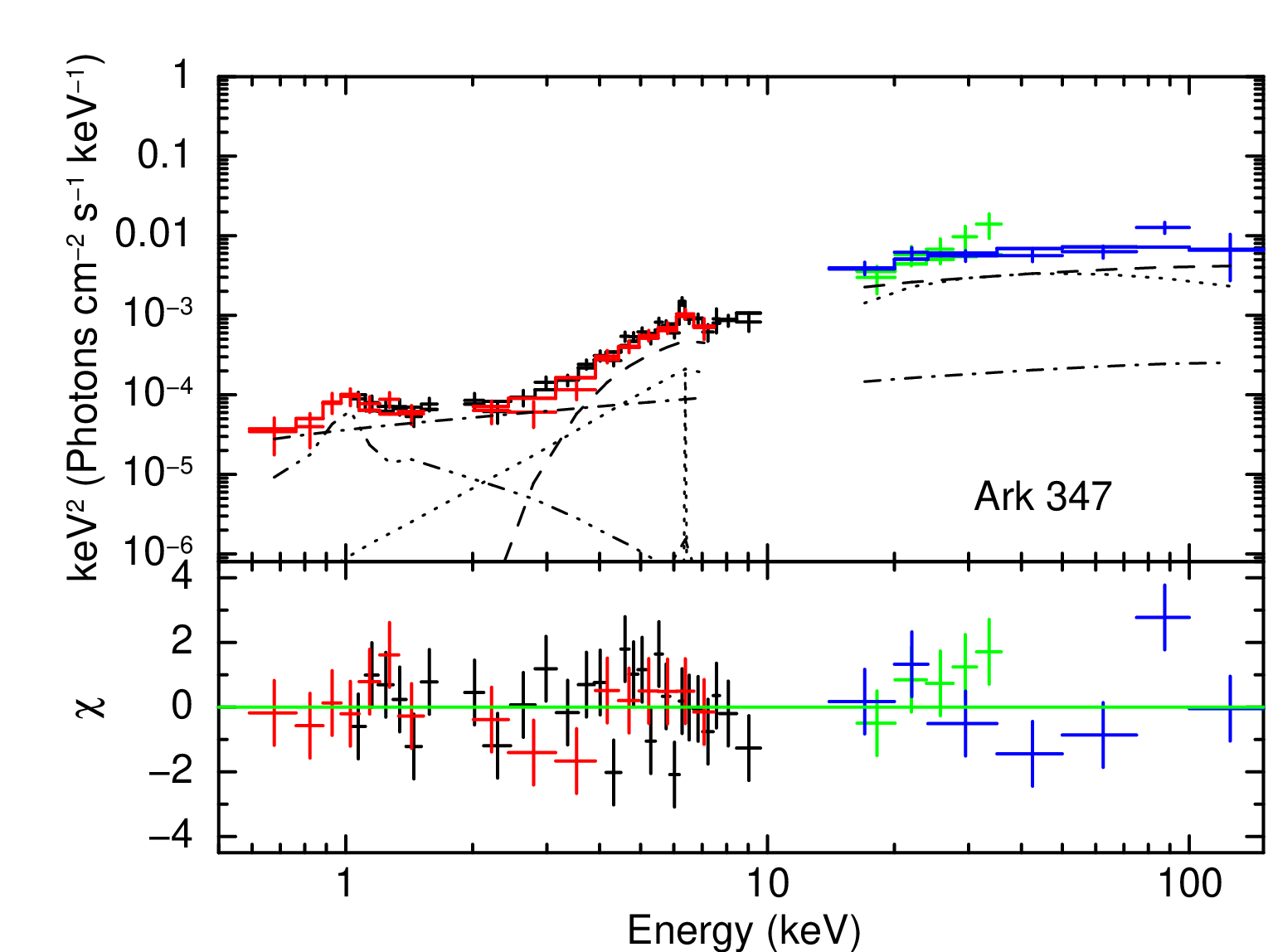}}
\resizebox{7.1cm}{!}{\includegraphics[scale=0.8,angle=-90]{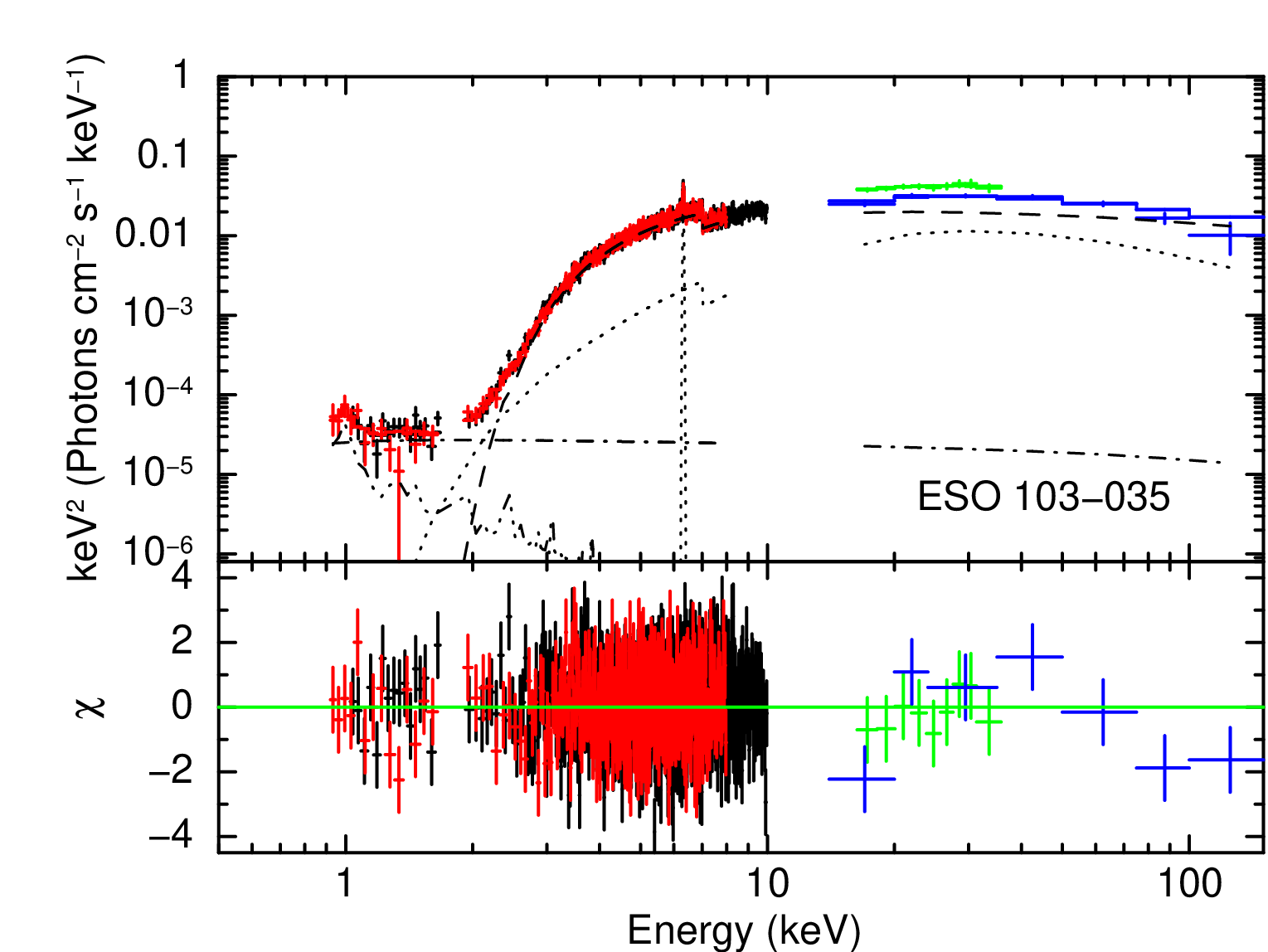}}
}\vspace{0.0cm}
\subfigure{
\resizebox{7.1cm}{!}{\includegraphics[scale=0.8,angle=-90]{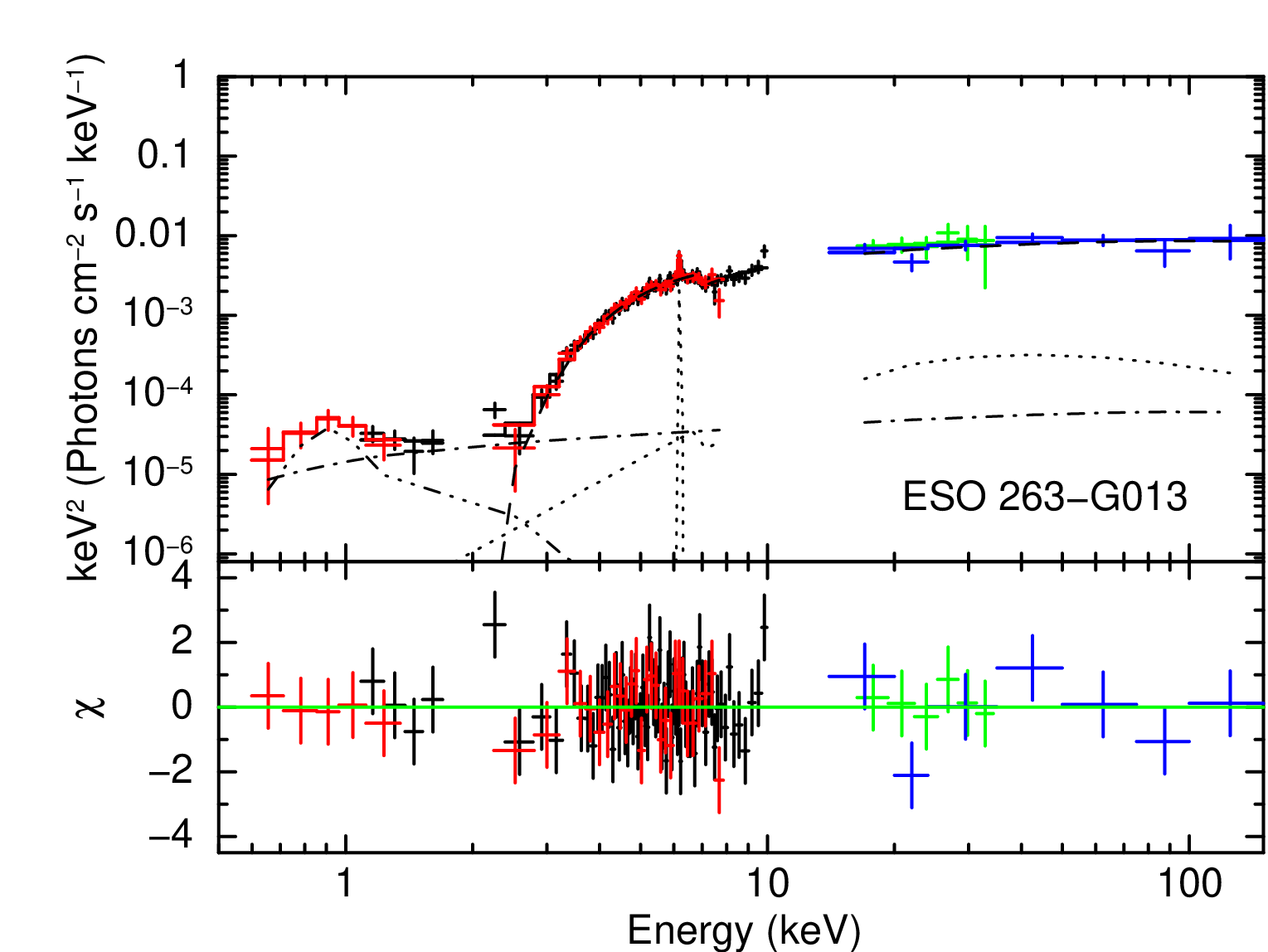}}
\resizebox{7.1cm}{!}{\includegraphics[scale=0.8,angle=-90]{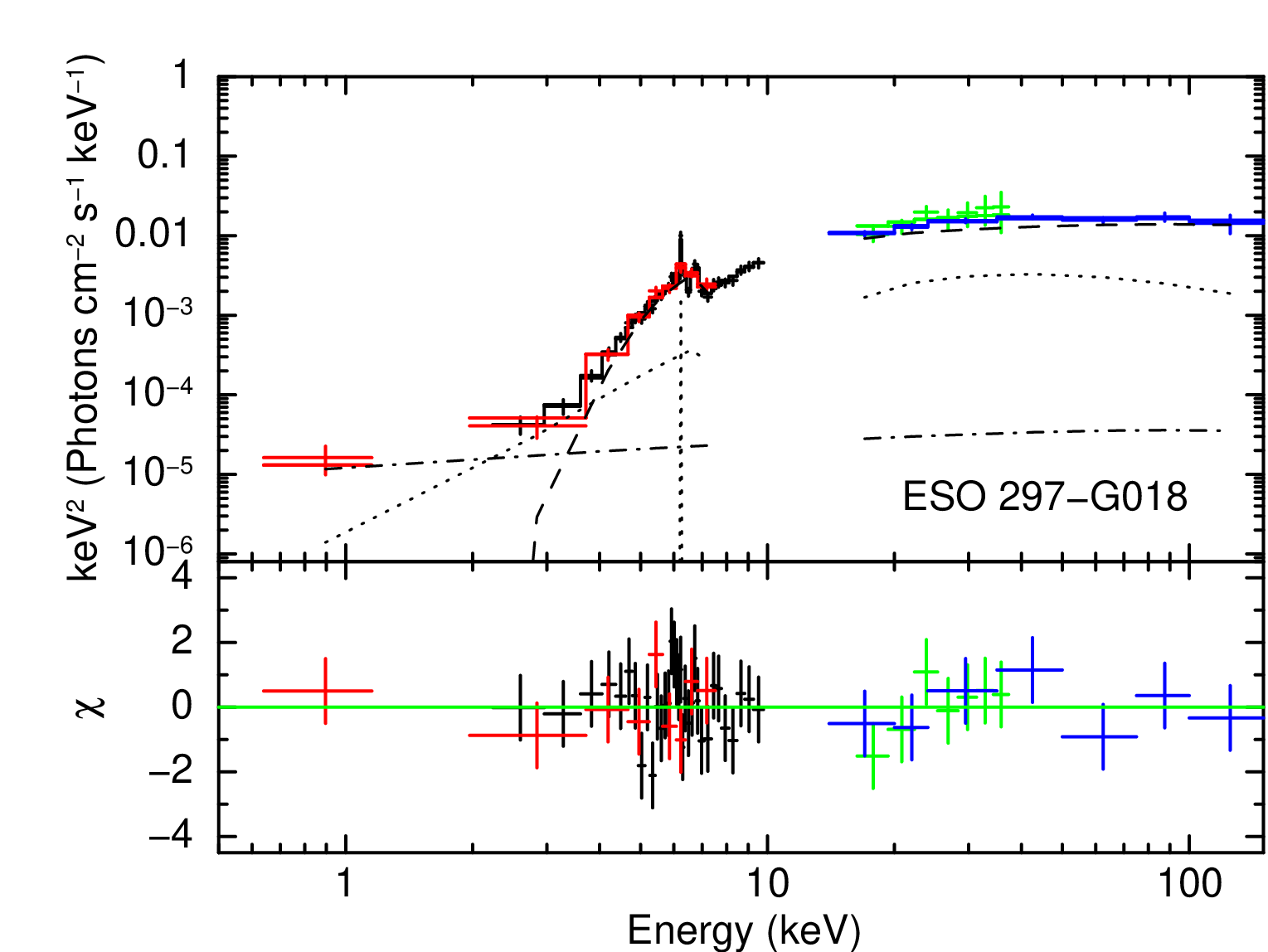}}
}\vspace{0.0cm}
\subfigure{
\resizebox{7.1cm}{!}{\includegraphics[scale=0.8,angle=-90]{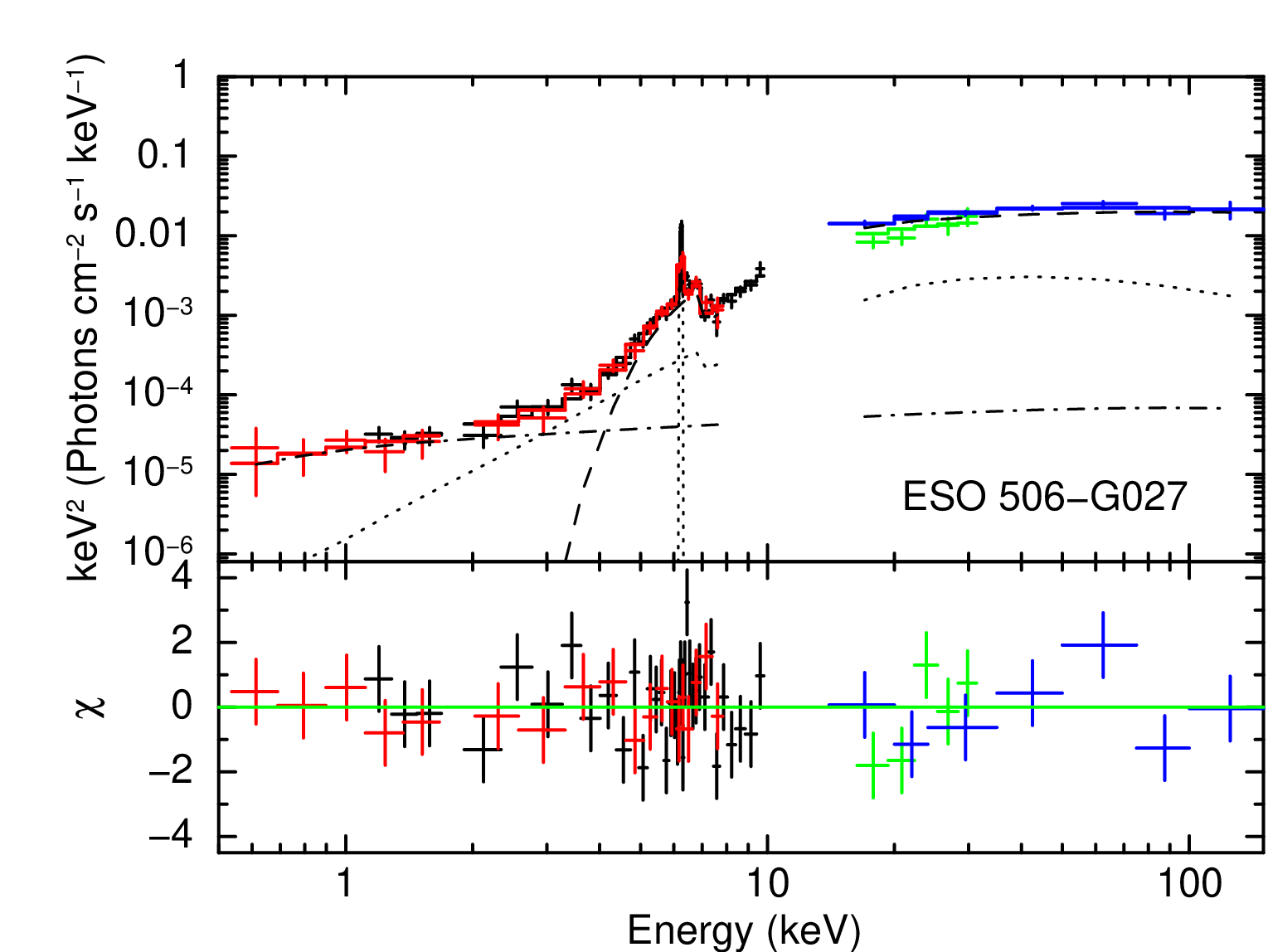}}
\resizebox{7.1cm}{!}{\includegraphics[scale=0.8,angle=-90]{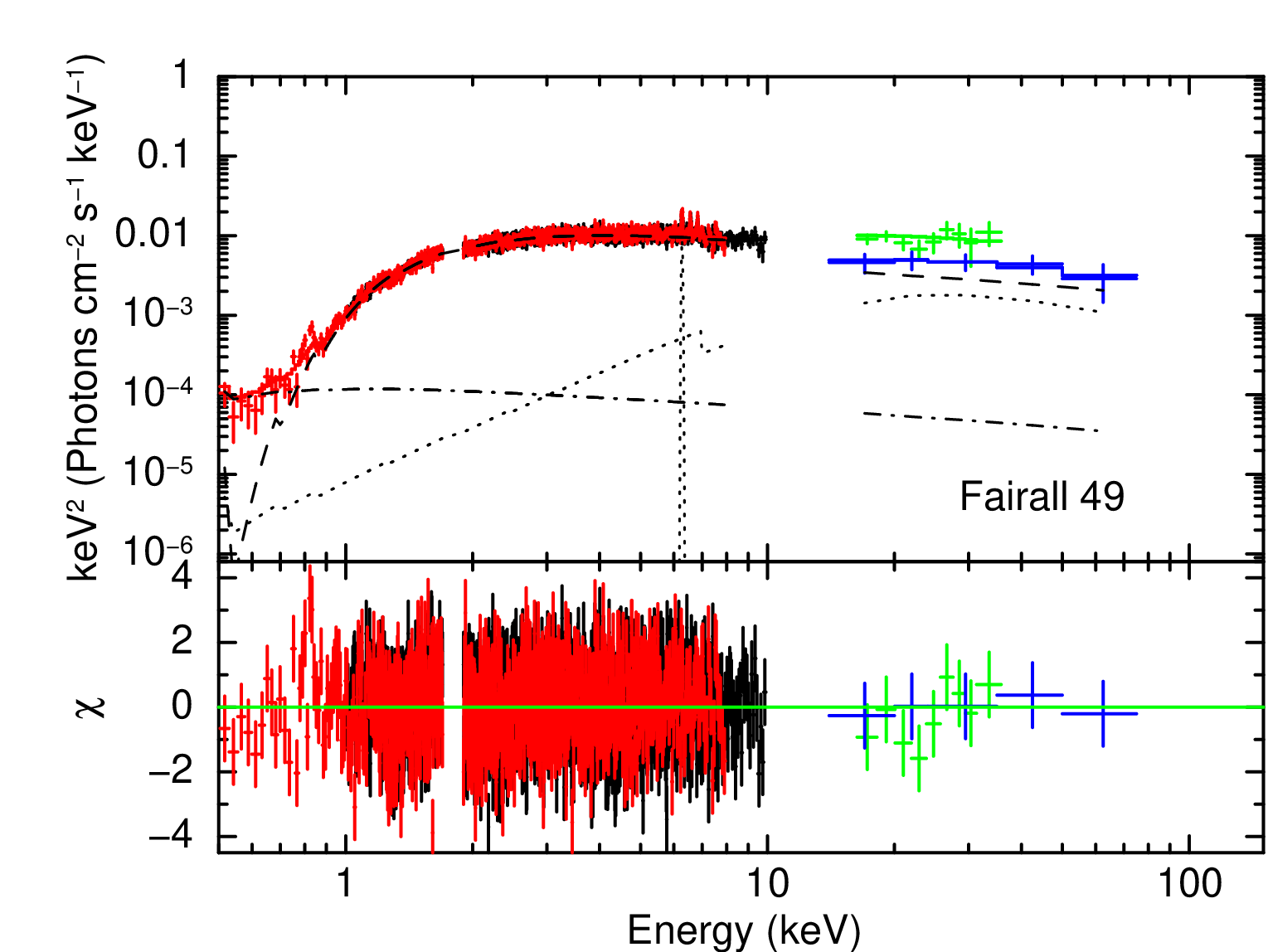}}
}\vspace{0.0cm}
\subfigure{
\resizebox{7.1cm}{!}{\includegraphics[scale=0.8,angle=-90]{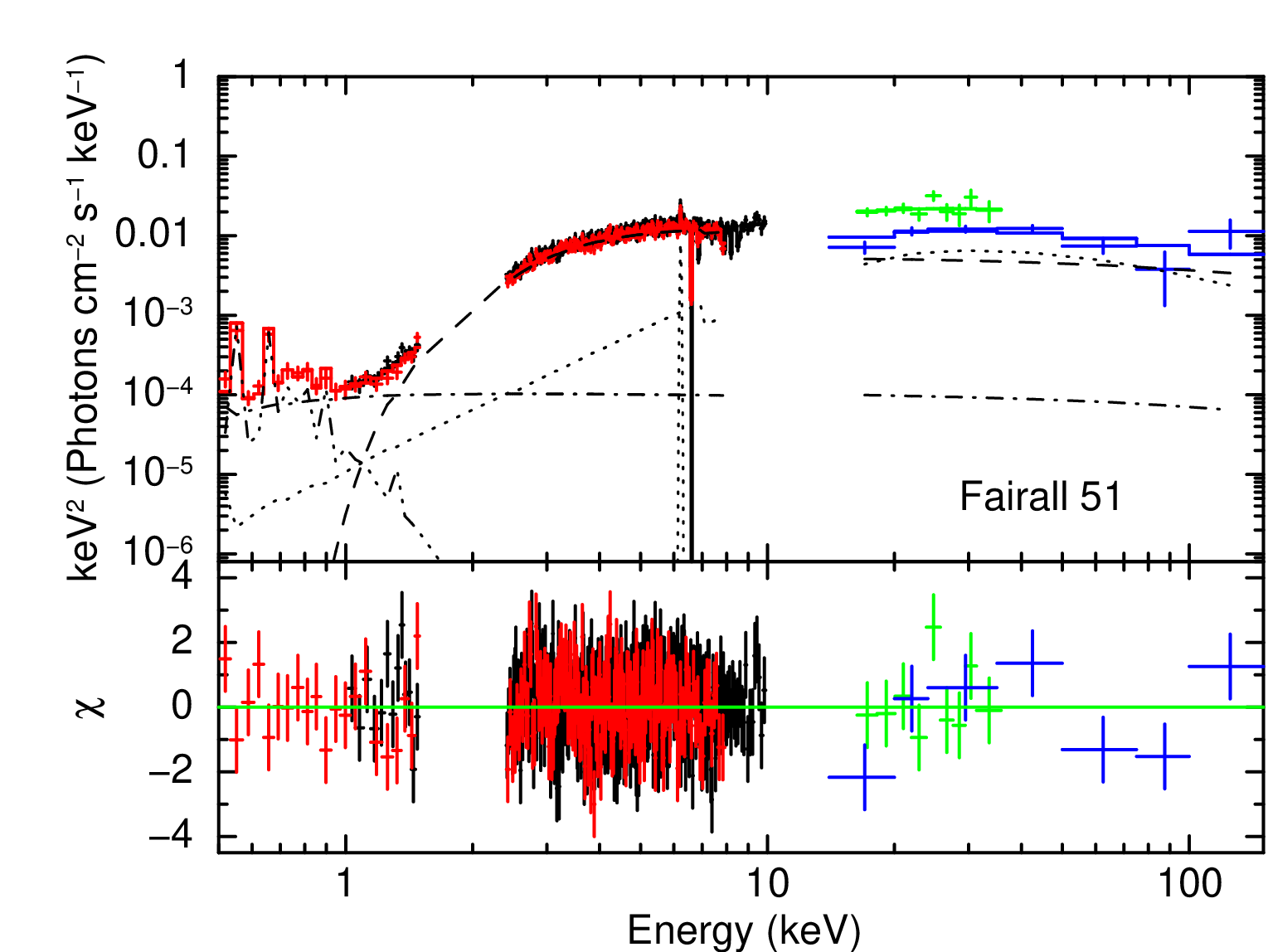}}
\resizebox{7.1cm}{!}{\includegraphics[scale=0.8,angle=-90]{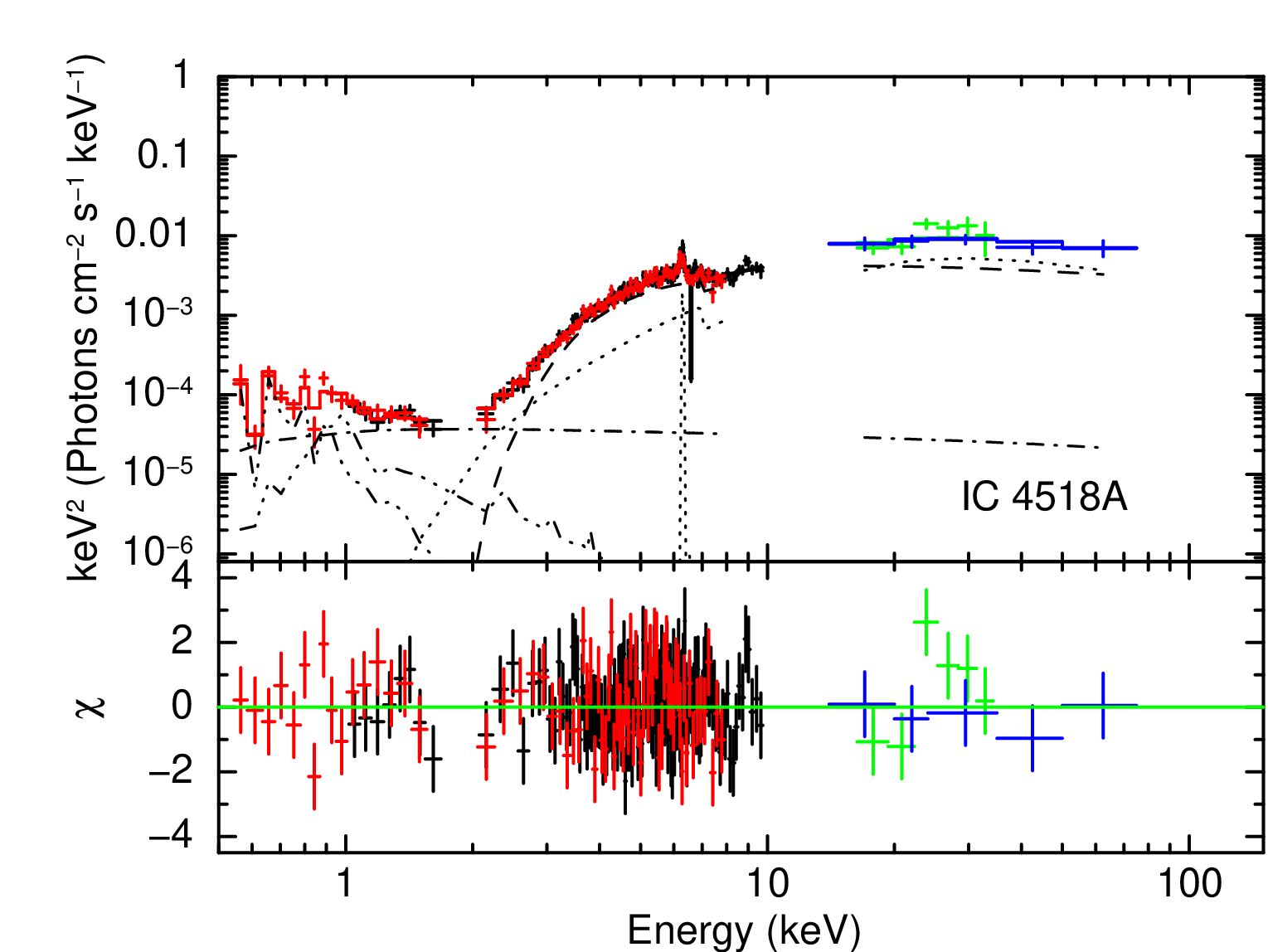}}
}\vspace{0.0cm}
\caption{
Continued.
}
\label{fig:spec}
\end{figure*}
\addtocounter{figure}{-1}
\begin{figure*}
\centering
\subfigure{
\resizebox{7.1cm}{!}{\includegraphics[scale=0.8,angle=-90]{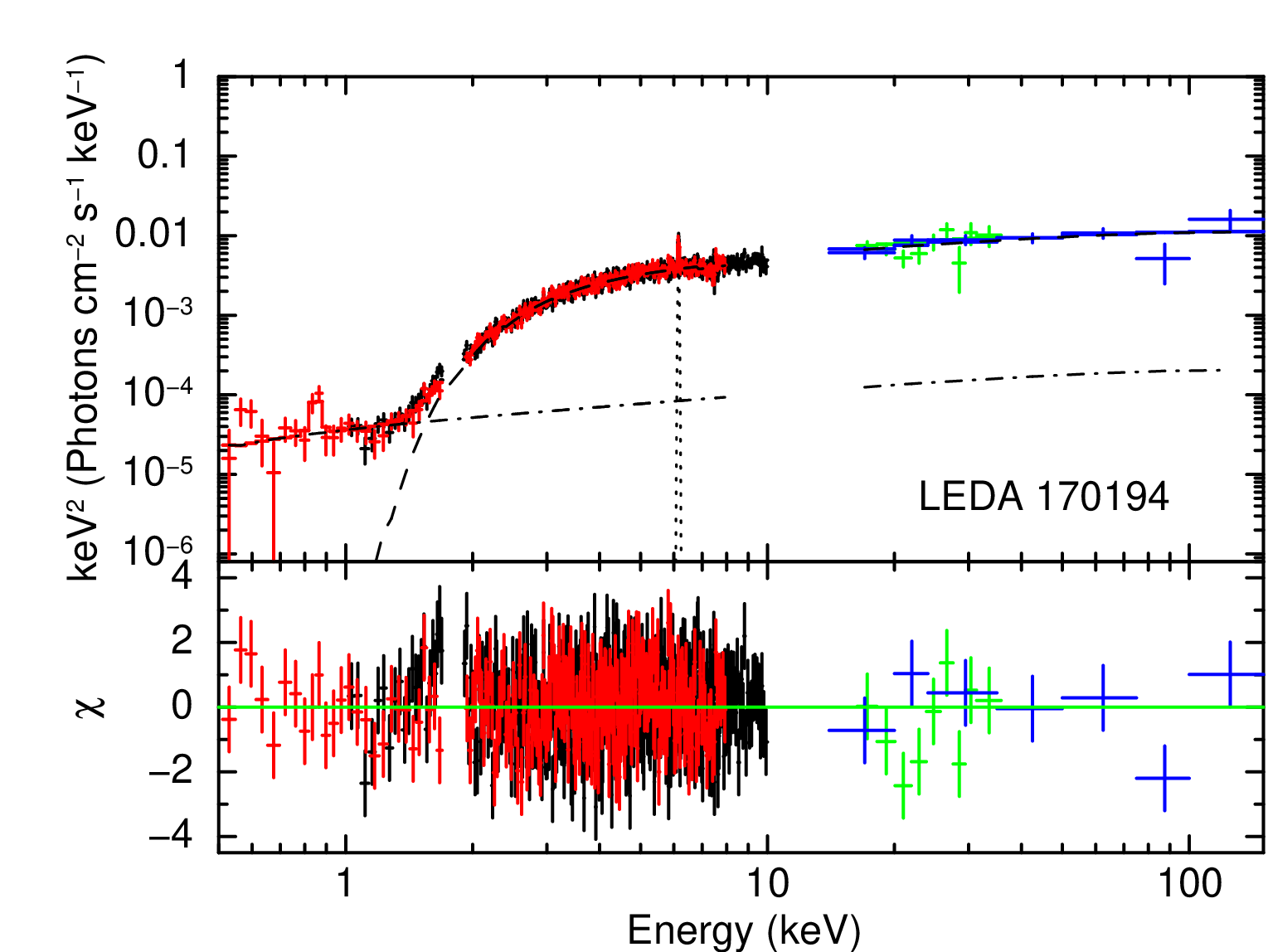}}
\resizebox{7.1cm}{!}{\includegraphics[scale=0.8,angle=-90]{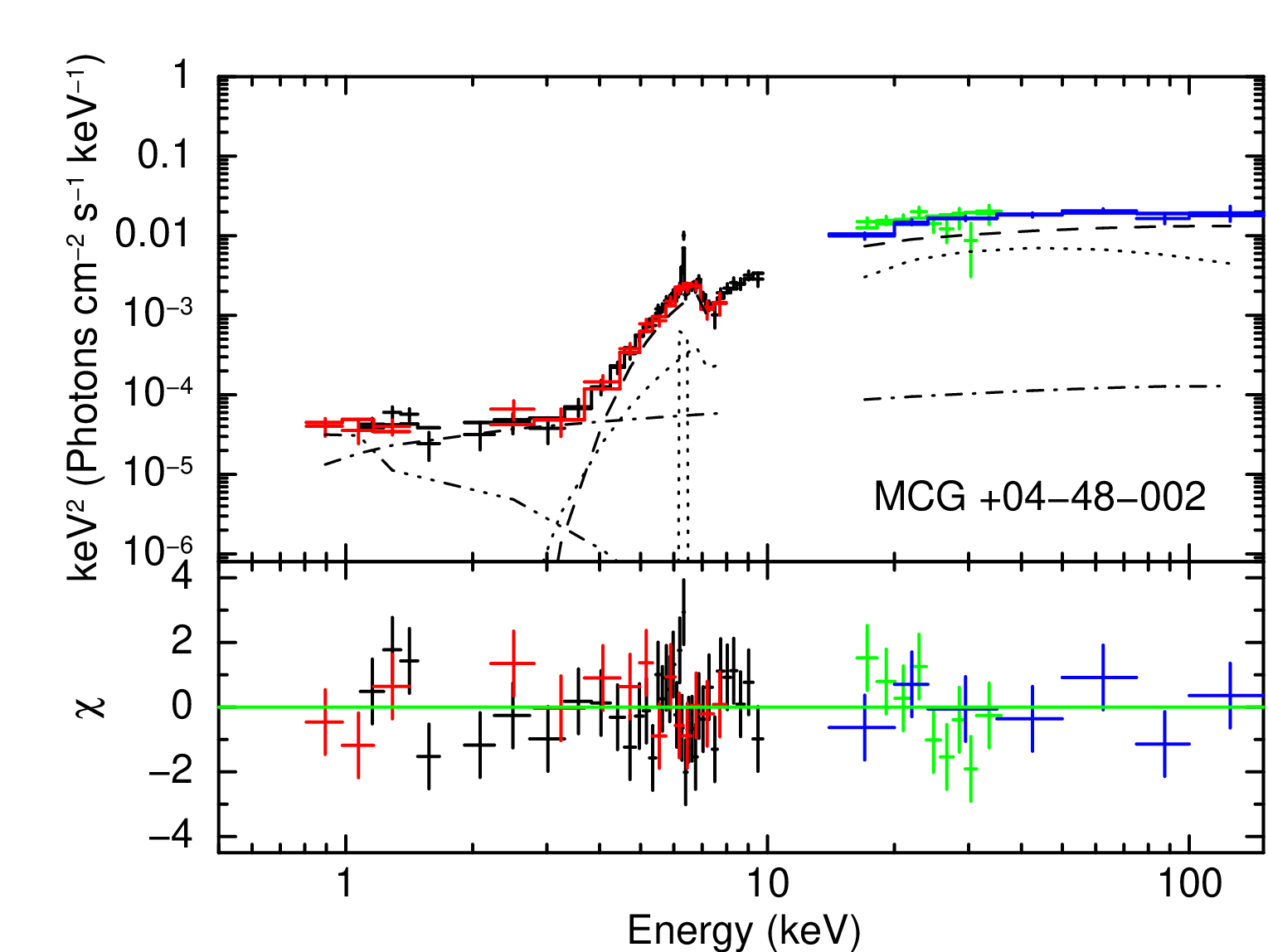}}
}\vspace{0.0cm}
\subfigure{
\resizebox{7.1cm}{!}{\includegraphics[scale=0.8,angle=-90]{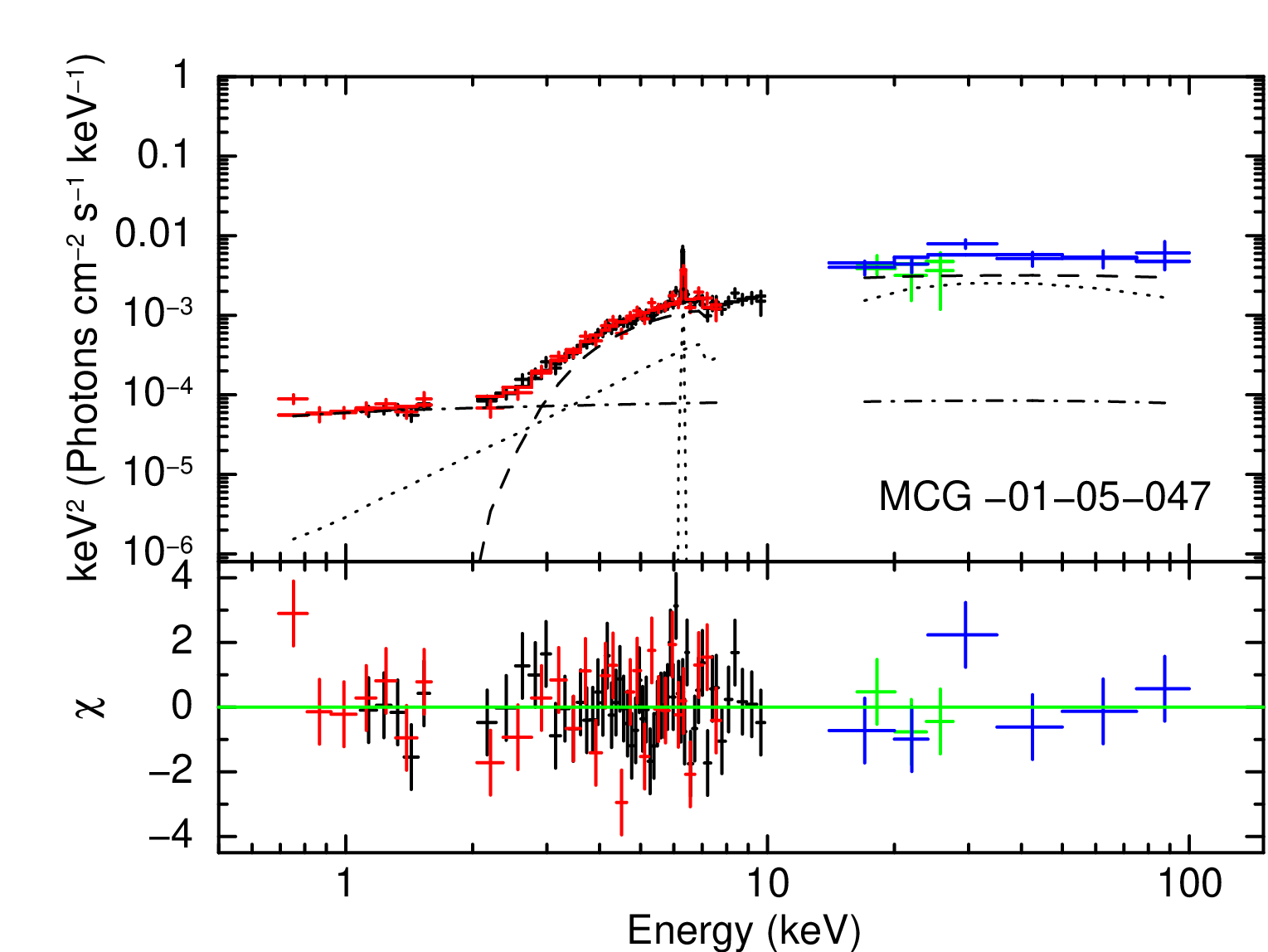}}
\resizebox{7.1cm}{!}{\includegraphics[scale=0.8,angle=-90]{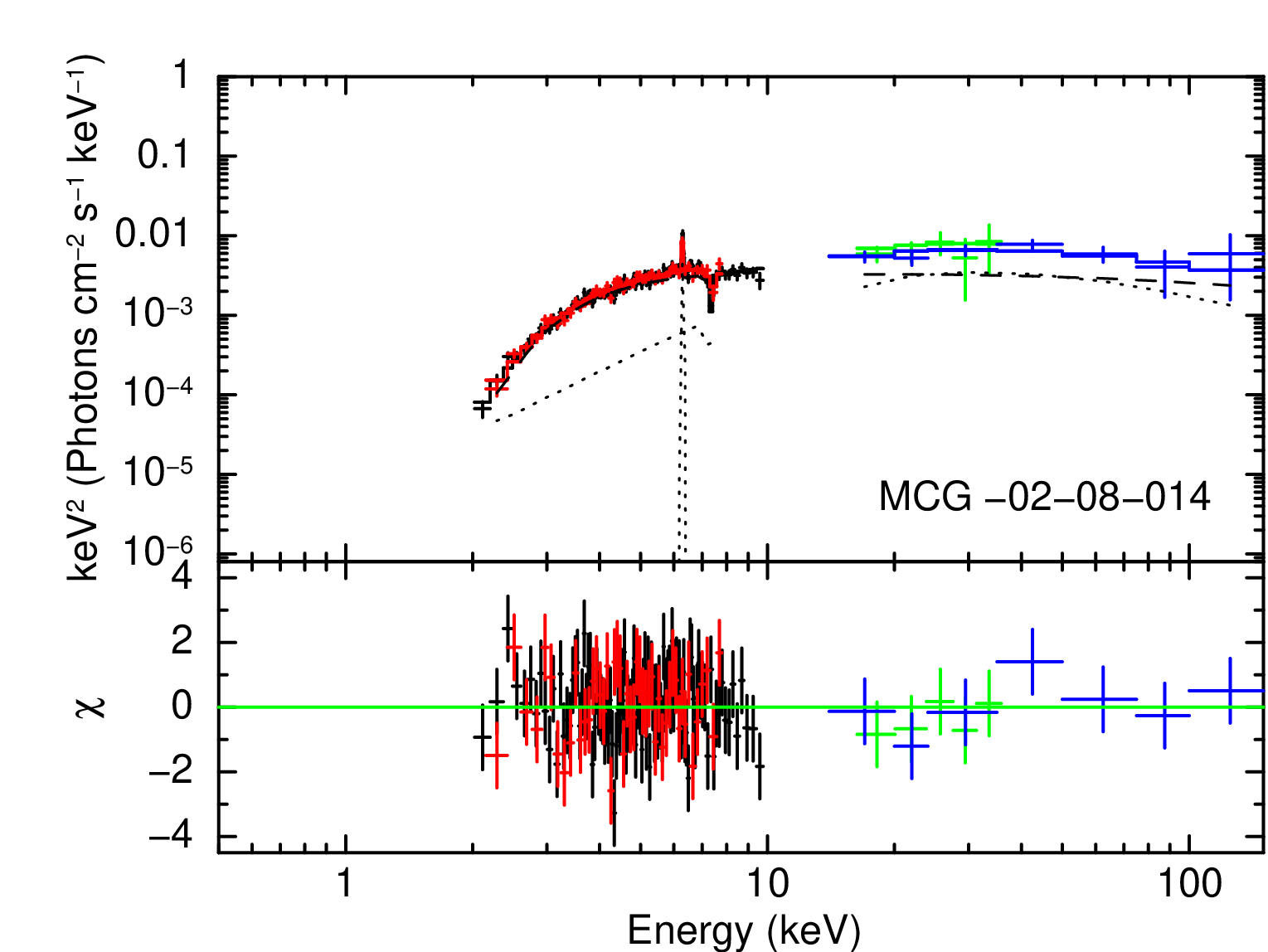}}
}\vspace{0.0cm}
\subfigure{
\resizebox{7.1cm}{!}{\includegraphics[scale=0.8,angle=-90]{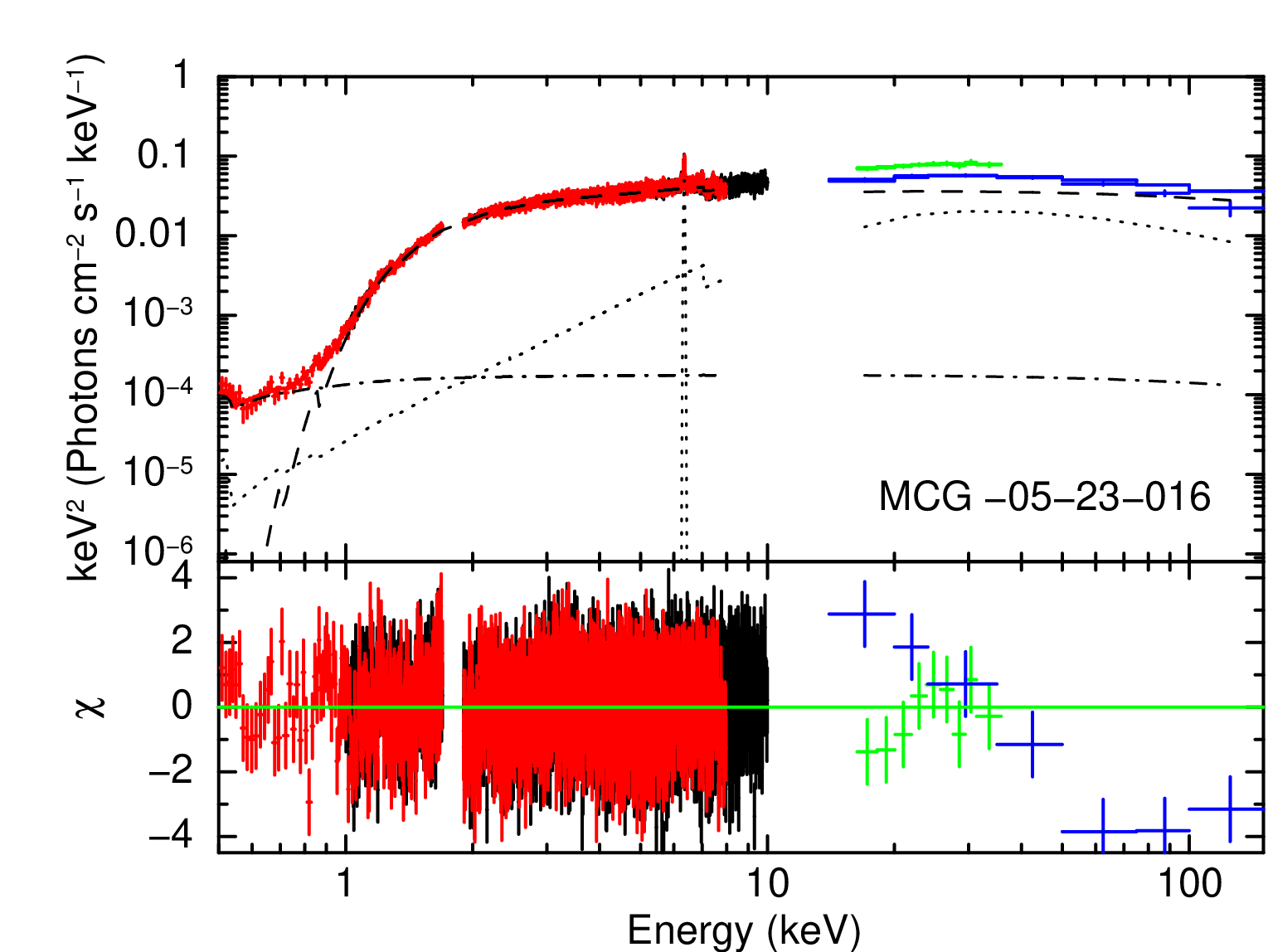}}
\resizebox{7.1cm}{!}{\includegraphics[scale=0.8,angle=-90]{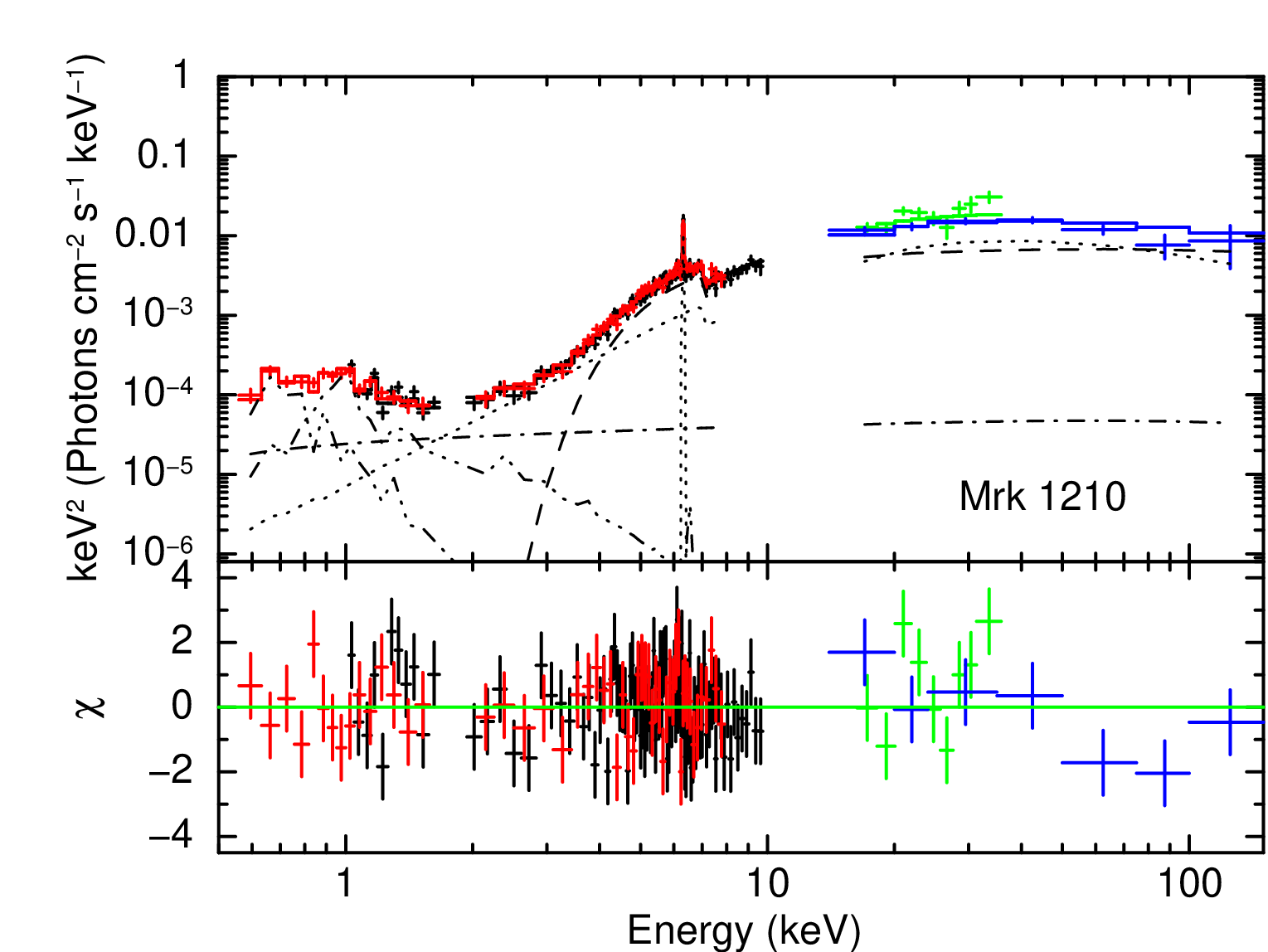}}
}\vspace{0.0cm}
\subfigure{
\resizebox{7.1cm}{!}{\includegraphics[scale=0.8,angle=-90]{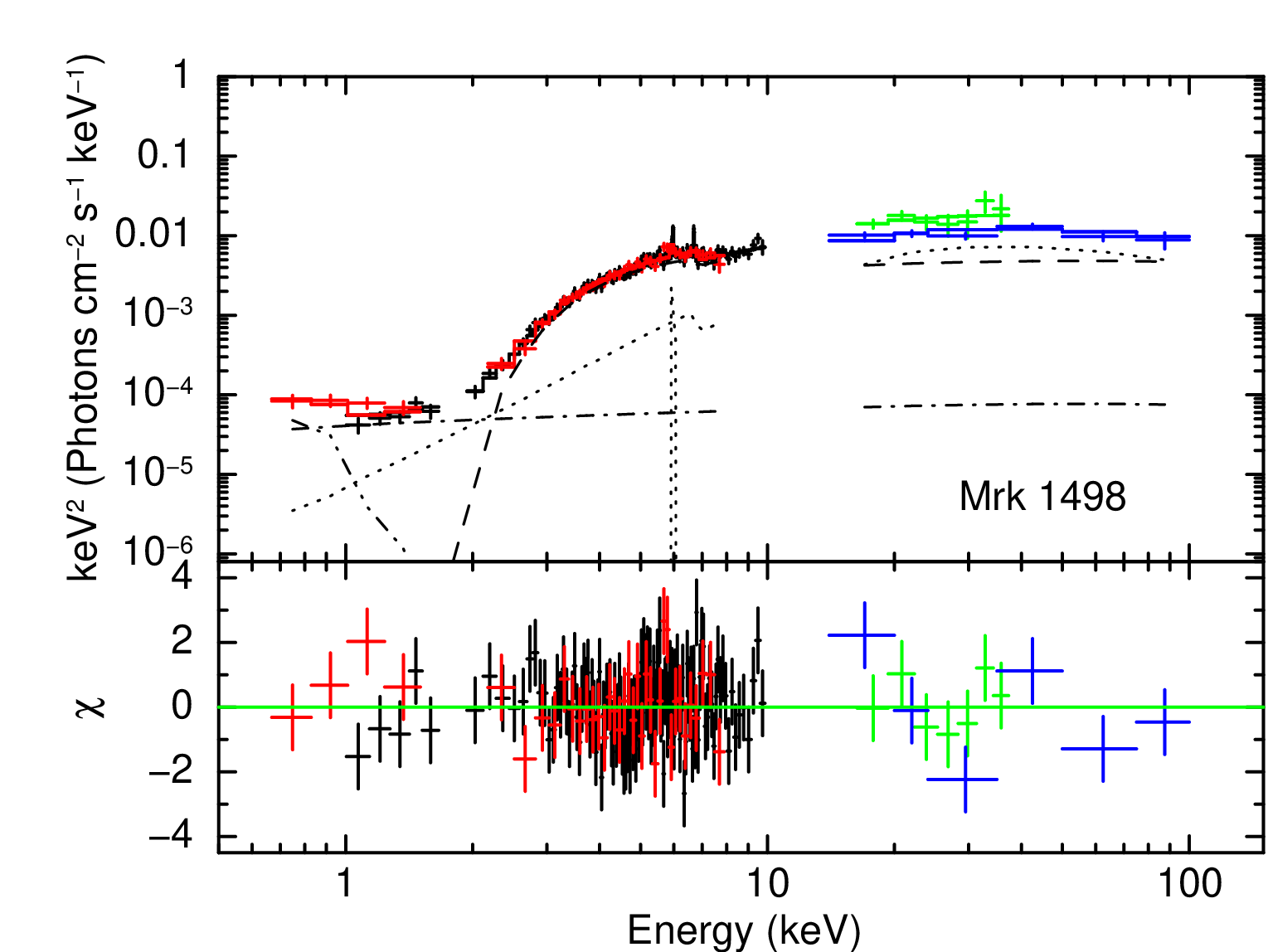}}
\resizebox{7.1cm}{!}{\includegraphics[scale=0.8,angle=-90]{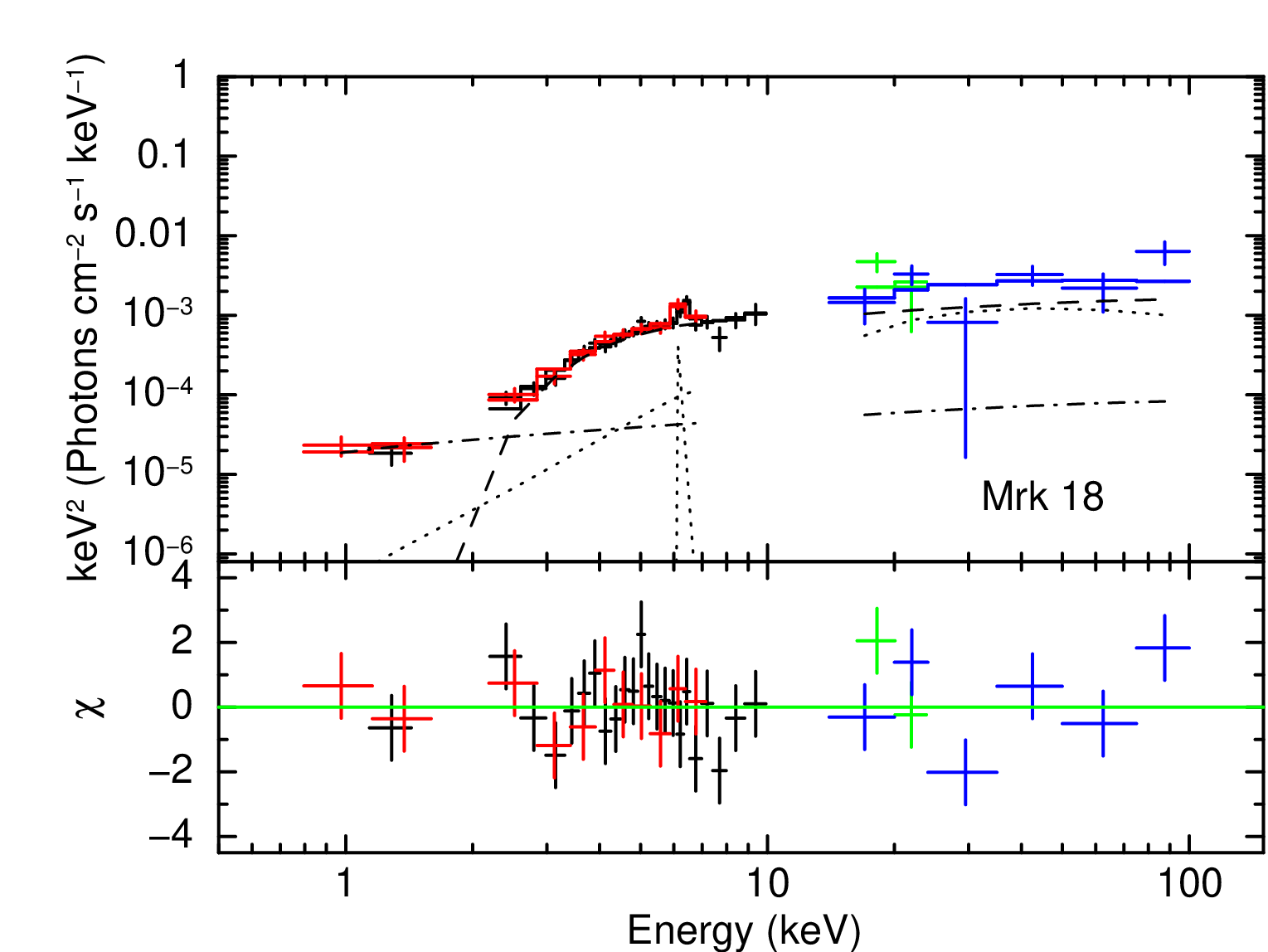}}
}\vspace{0.0cm}
\caption{
Continued.
}
\label{fig:spec}
\end{figure*}
\addtocounter{figure}{-1}
\begin{figure*}
\centering
\subfigure{
\resizebox{7.1cm}{!}{\includegraphics[scale=0.8,angle=-90]{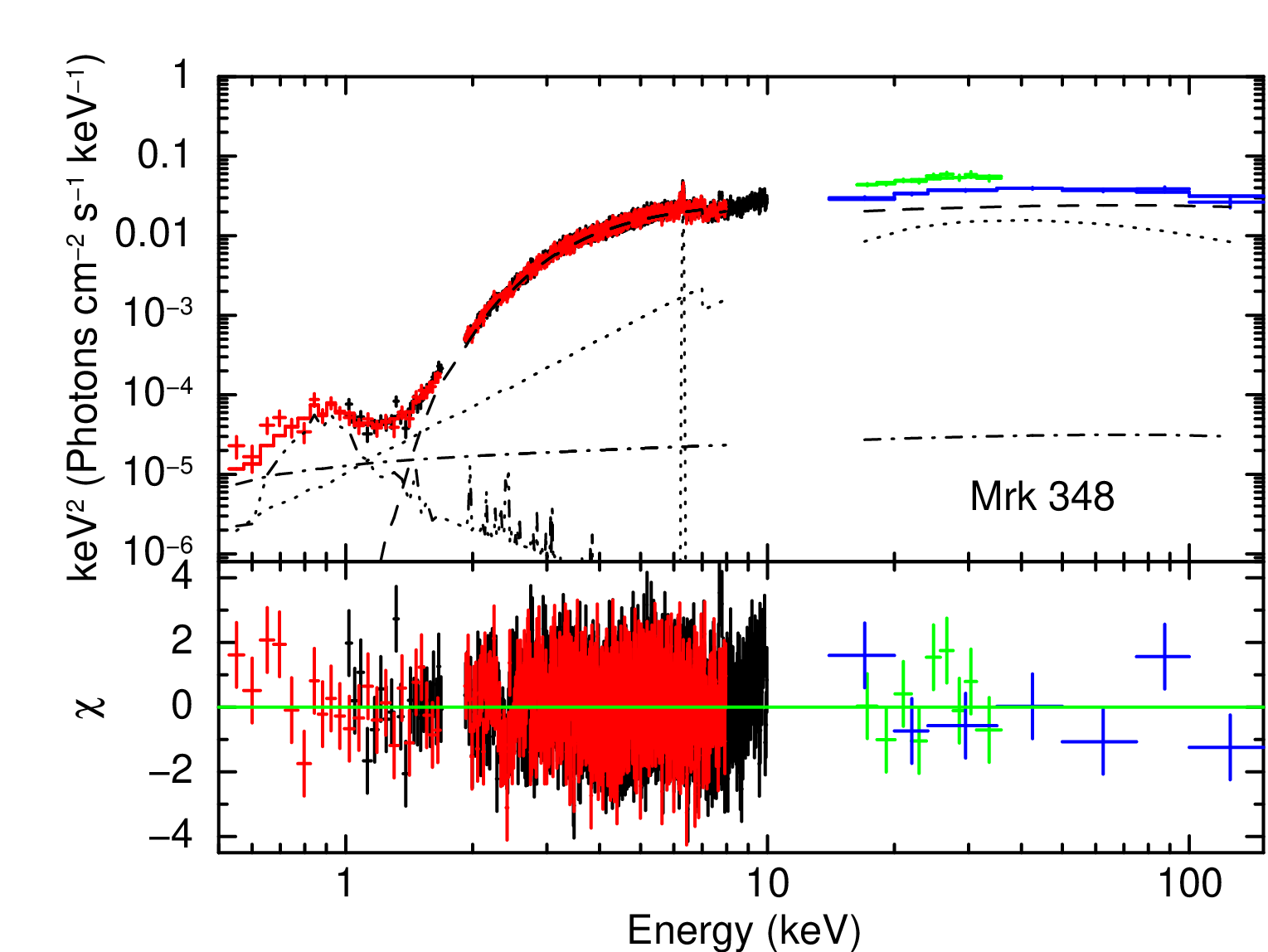}}
\resizebox{7.1cm}{!}{\includegraphics[scale=0.8,angle=-90]{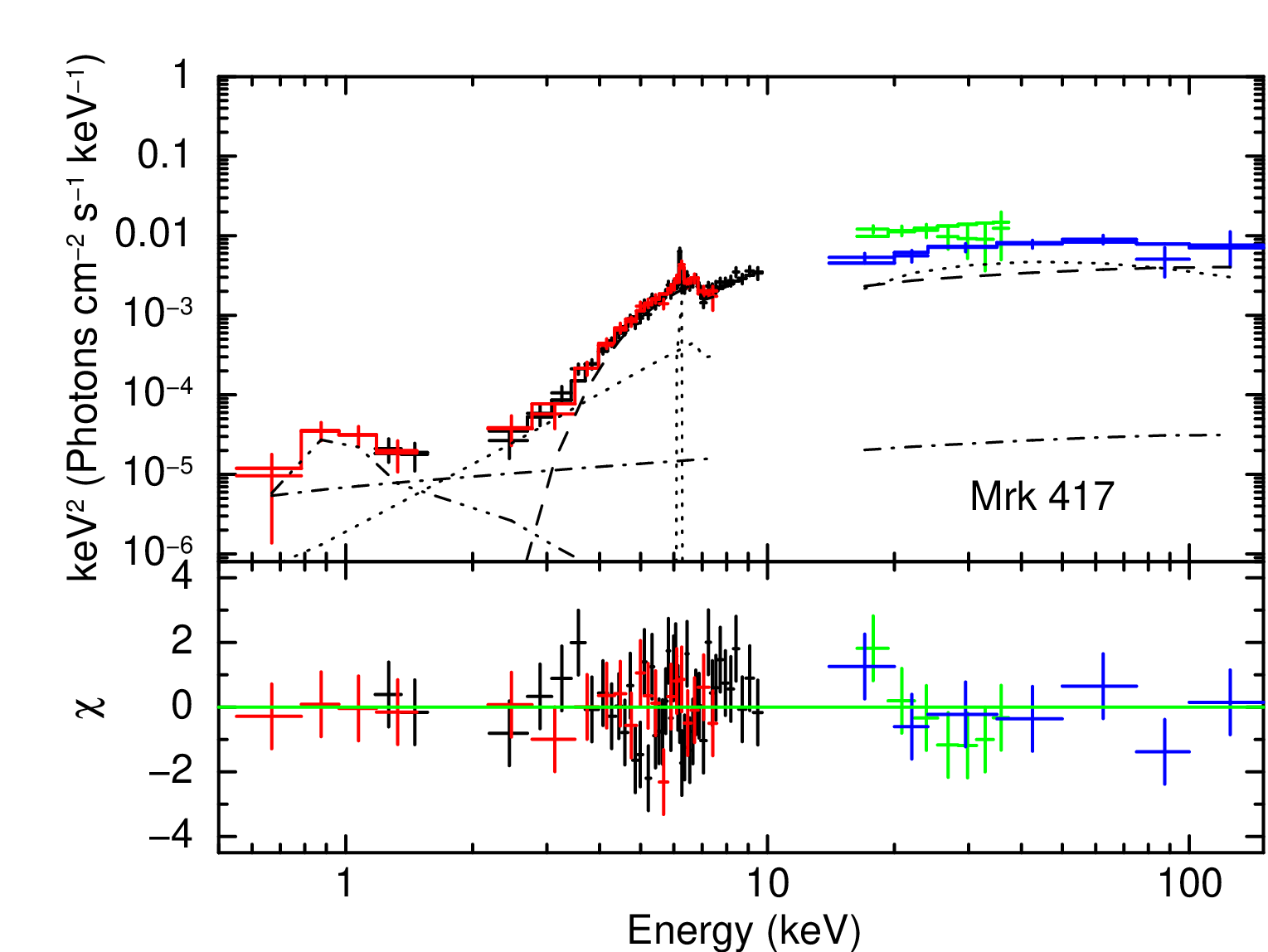}}
}\vspace{0.0cm}
\subfigure{
\resizebox{7.1cm}{!}{\includegraphics[scale=0.8,angle=-90]{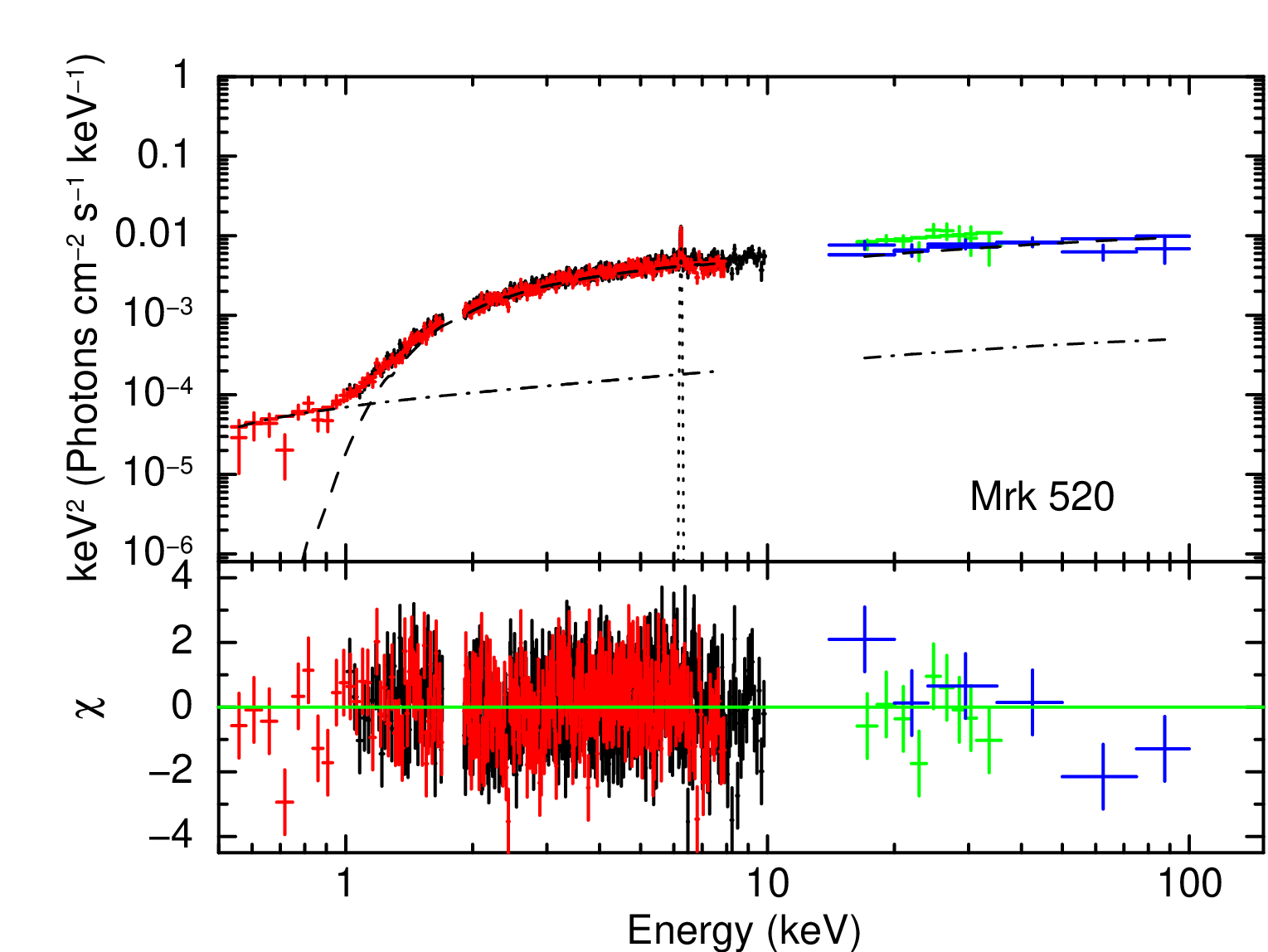}}
\resizebox{7.1cm}{!}{\includegraphics[scale=0.8,angle=-90]{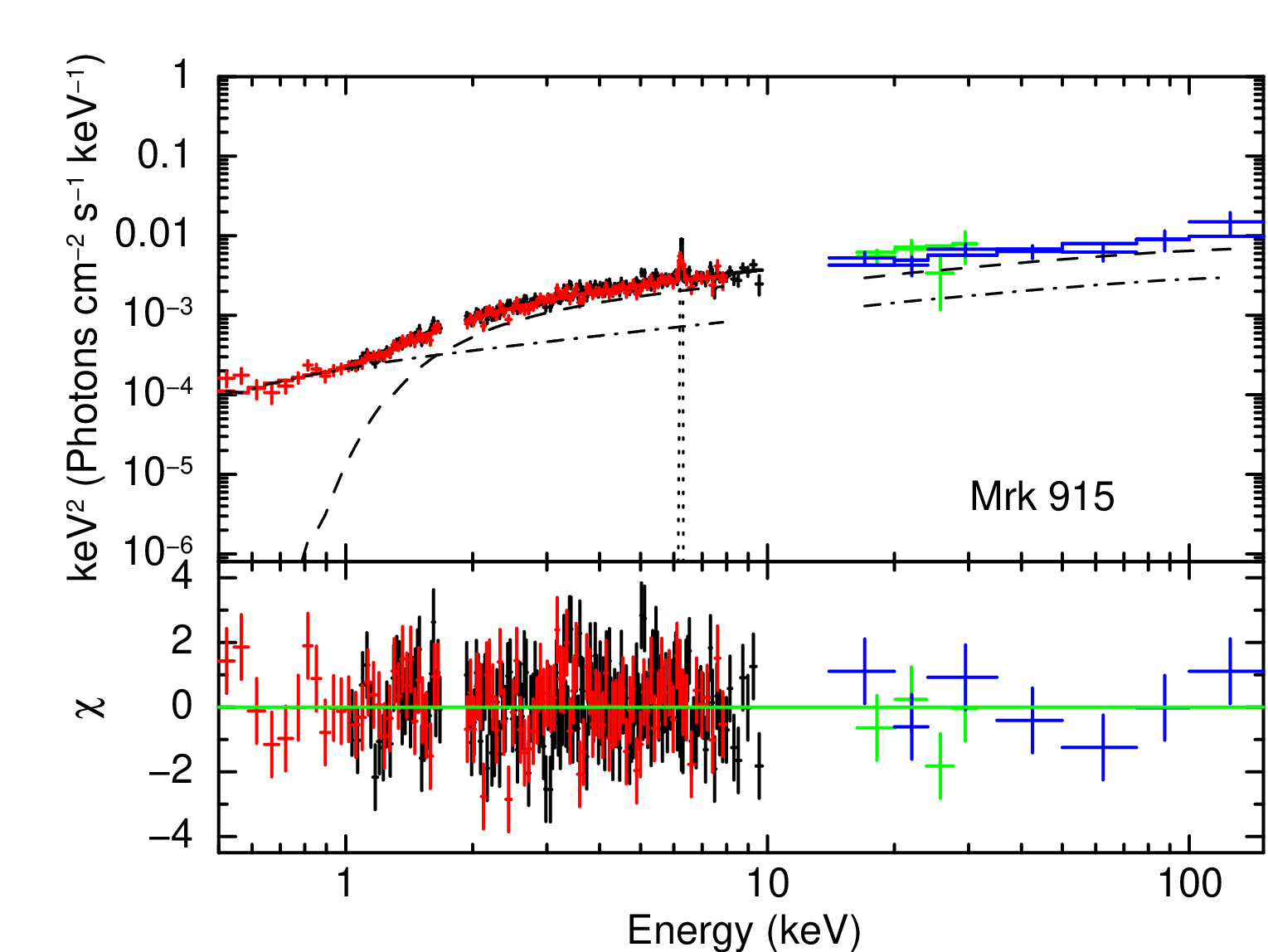}}
}\vspace{0.0cm}
\subfigure{
\resizebox{7.1cm}{!}{\includegraphics[scale=0.8,angle=-90]{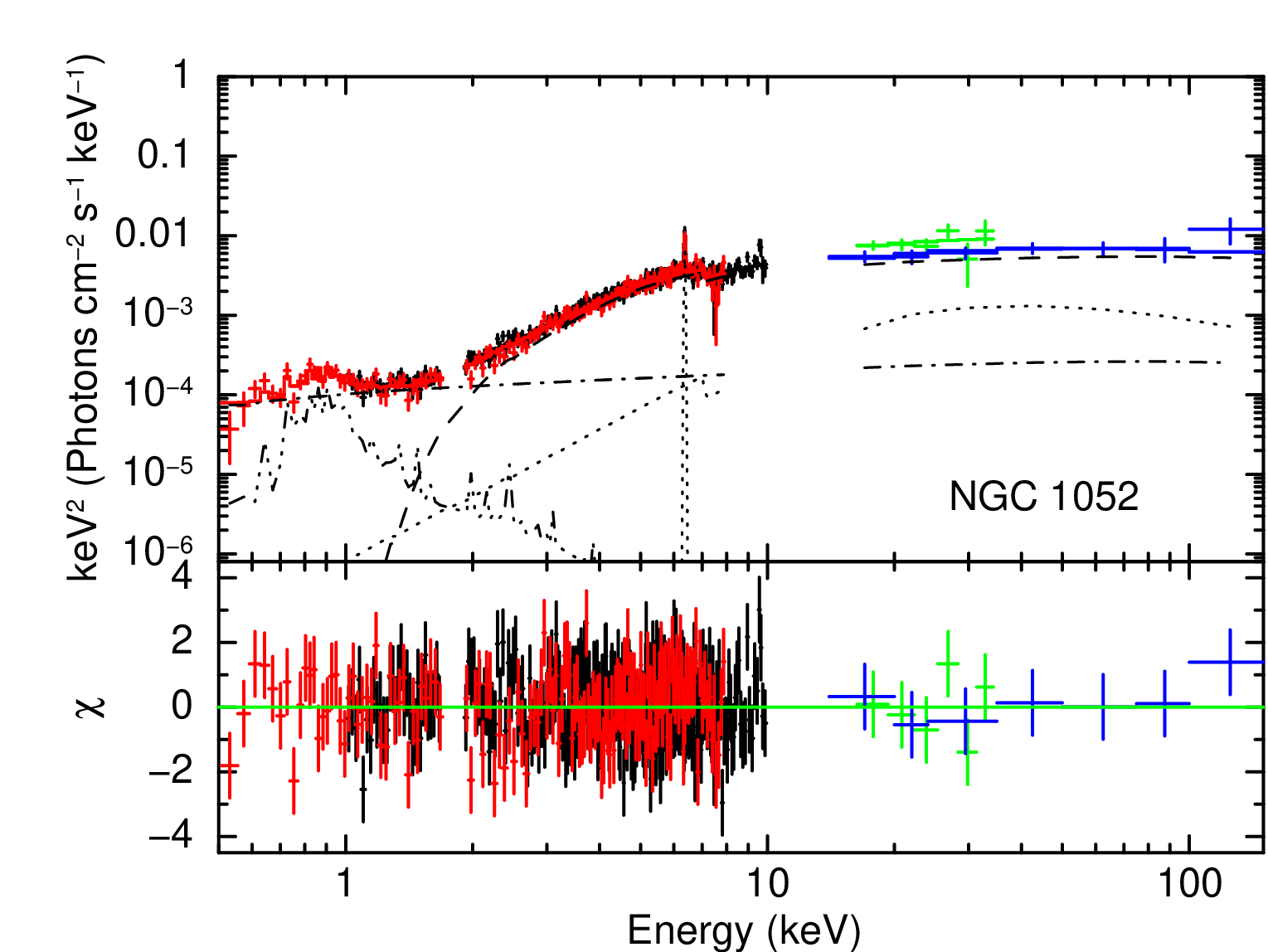}}
\resizebox{7.1cm}{!}{\includegraphics[scale=0.8,angle=-90]{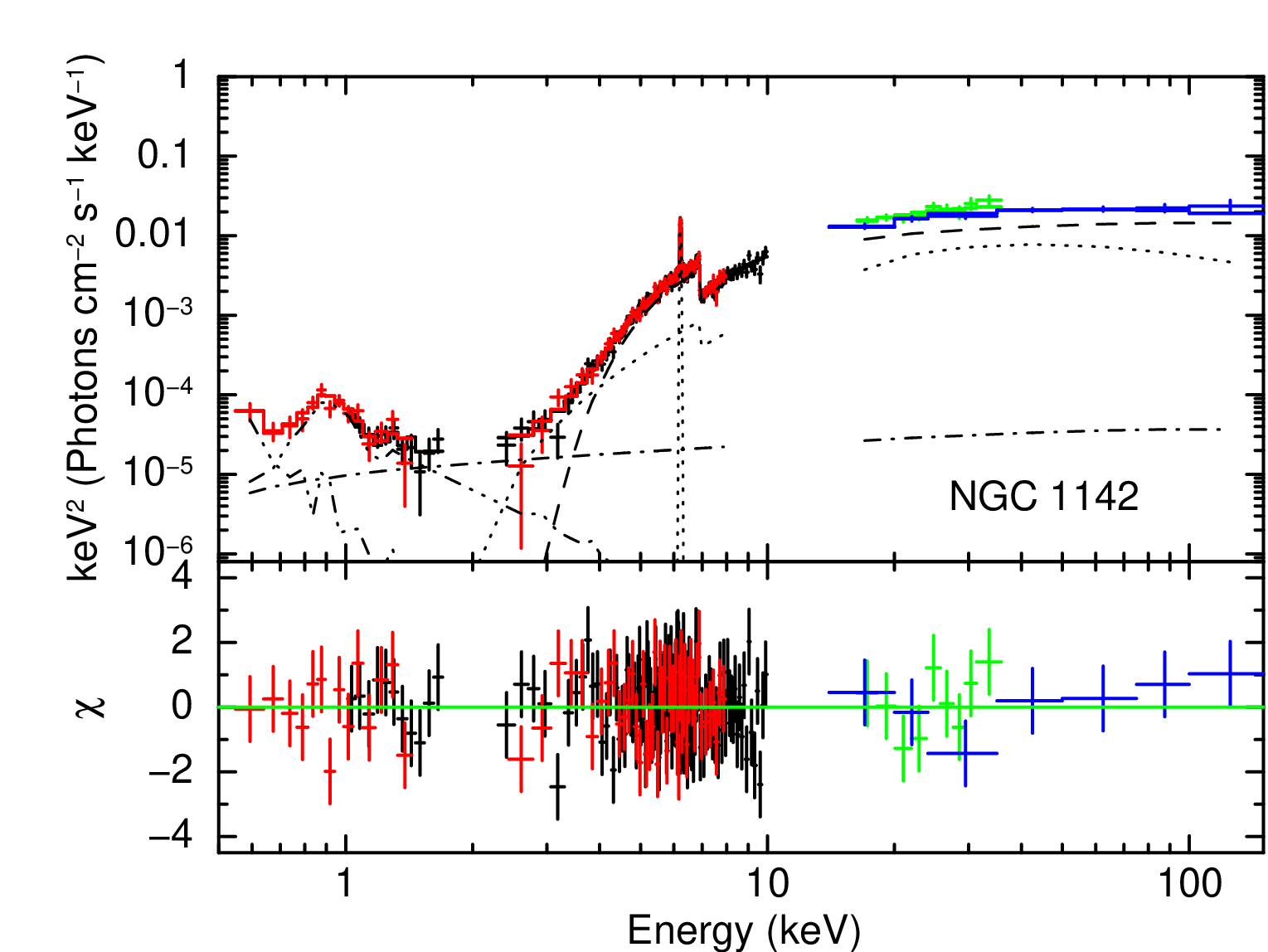}}
}\vspace{0.0cm}
\subfigure{
\resizebox{7.1cm}{!}{\includegraphics[scale=0.8,angle=-90]{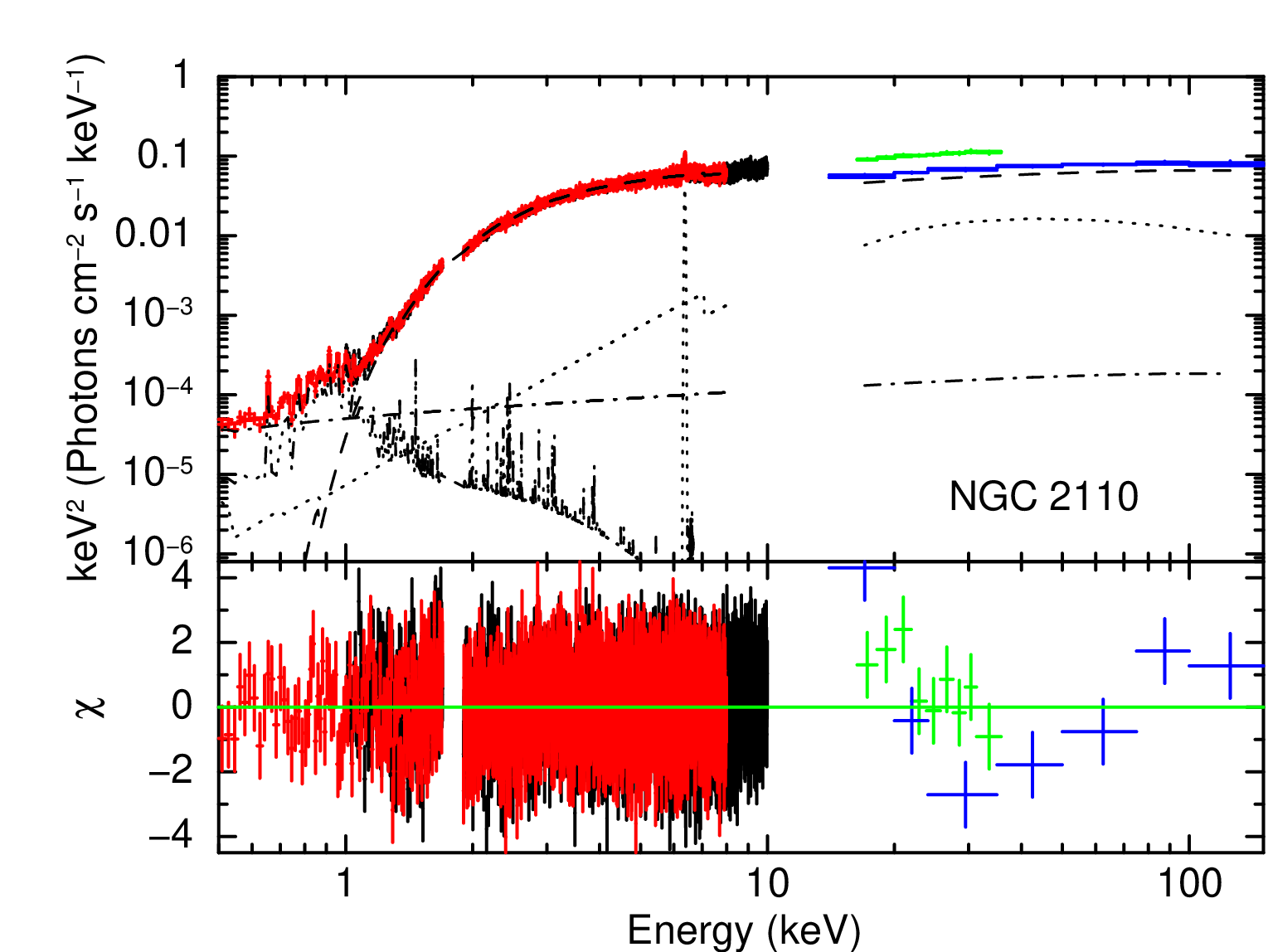}}
\resizebox{7.1cm}{!}{\includegraphics[scale=0.8,angle=-90]{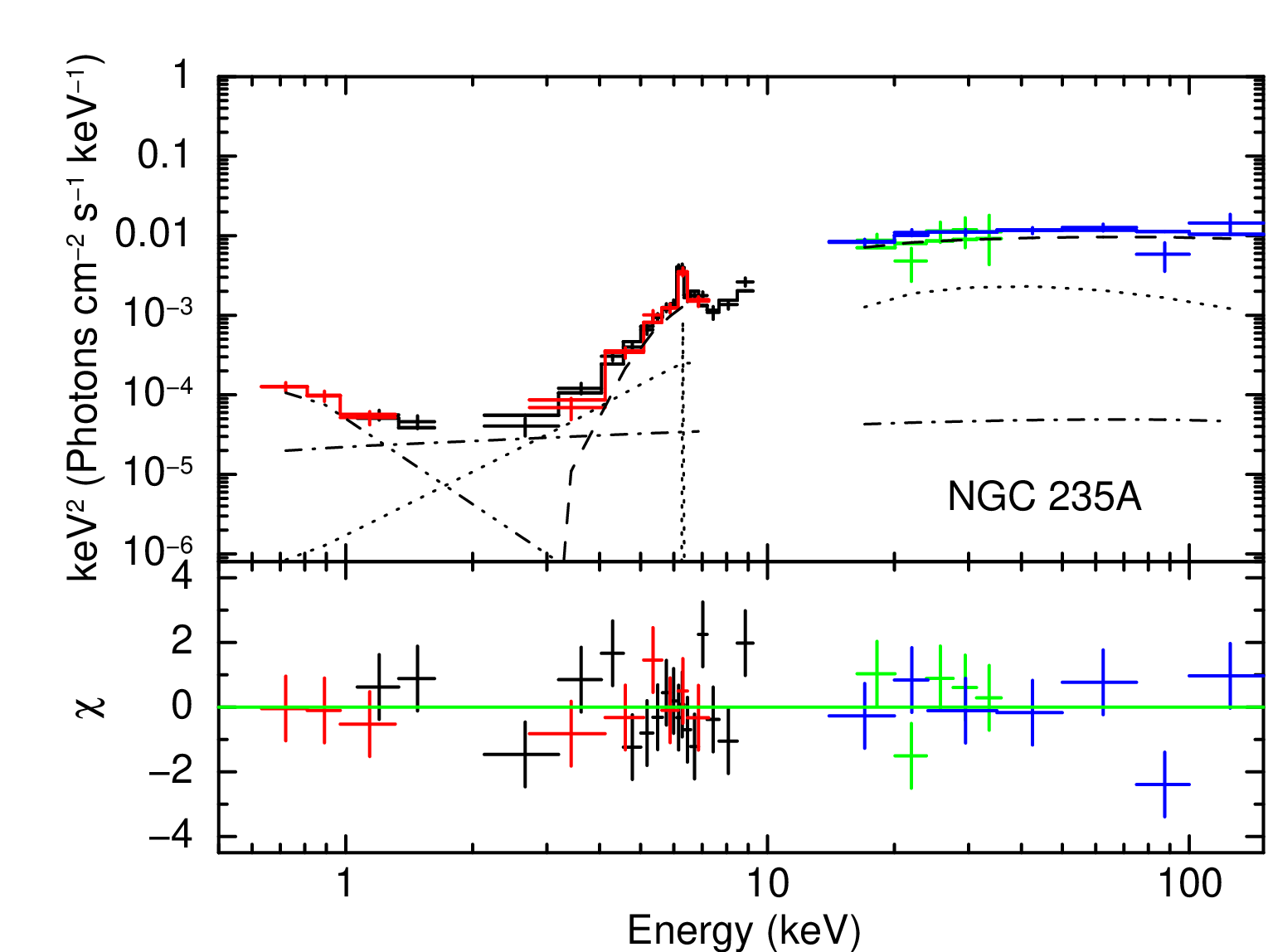}}
}\vspace{0.0cm}
\caption{
Continued.
}
\label{fig:spec}
\end{figure*}
\addtocounter{figure}{-1}
\begin{figure*}
\centering
\subfigure{
\resizebox{7.1cm}{!}{\includegraphics[scale=0.8,angle=-90]{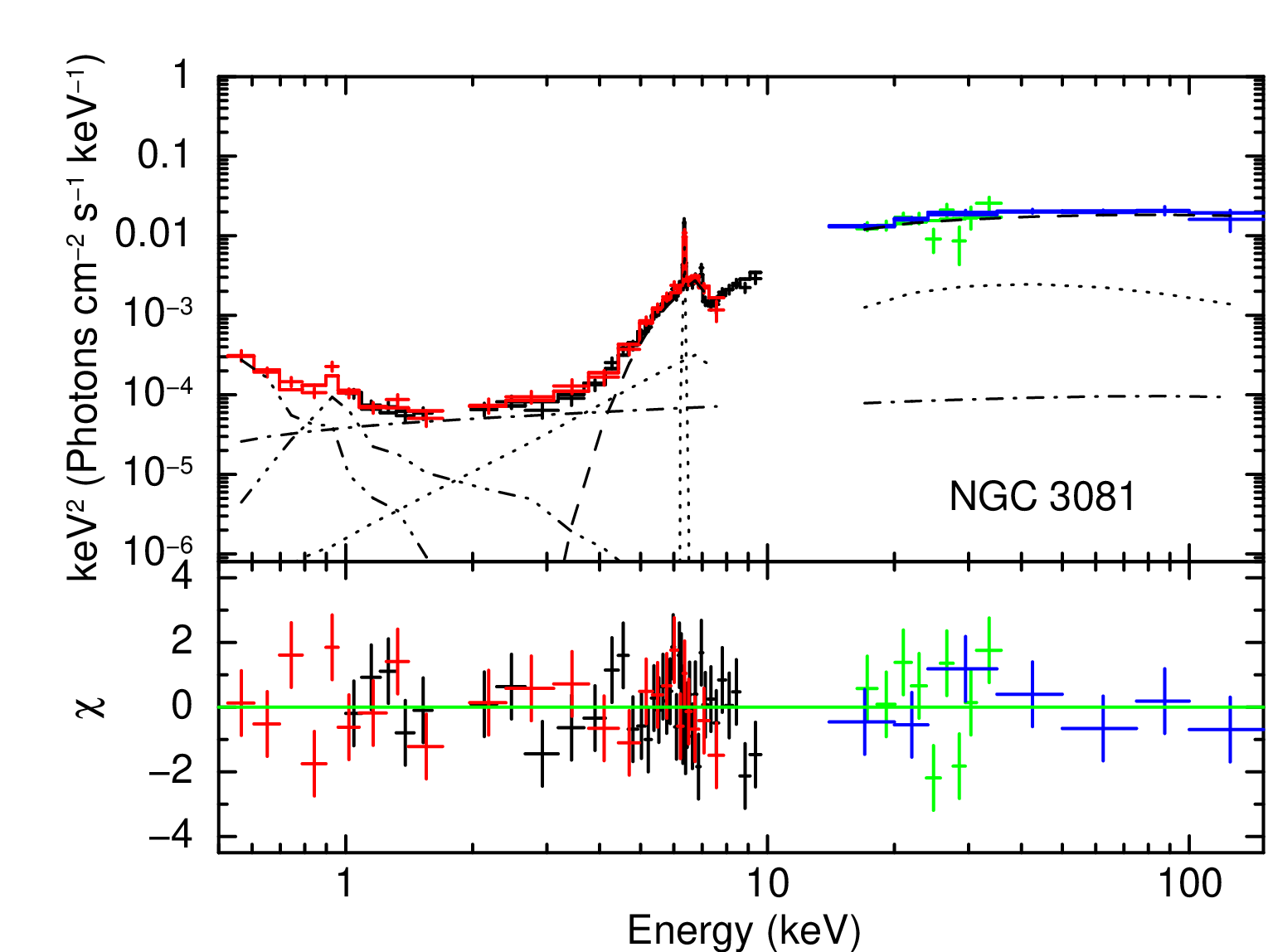}}
\resizebox{7.1cm}{!}{\includegraphics[scale=0.8,angle=-90]{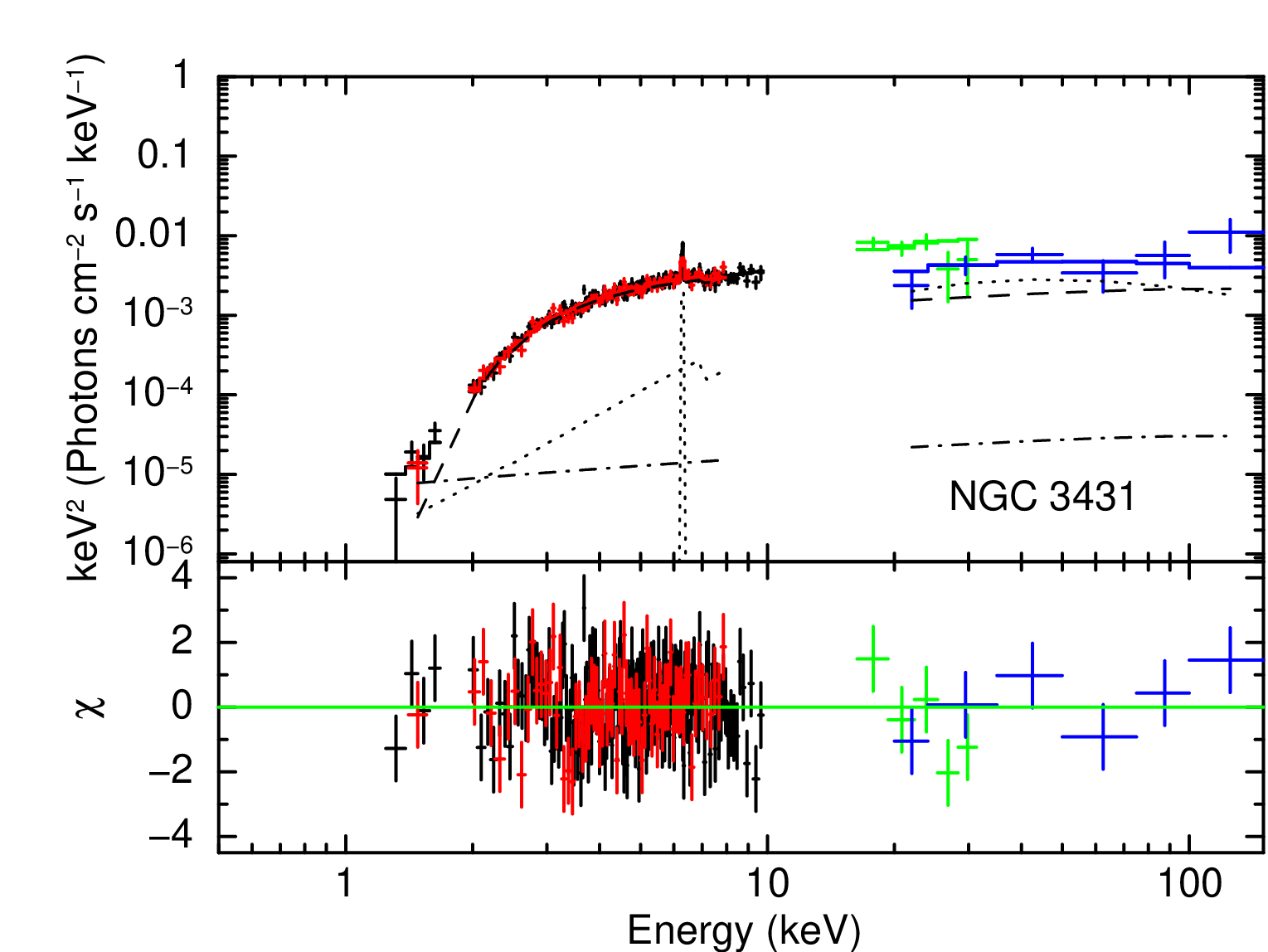}}
}\vspace{0.0cm}
\subfigure{
\resizebox{7.1cm}{!}{\includegraphics[scale=0.8,angle=-90]{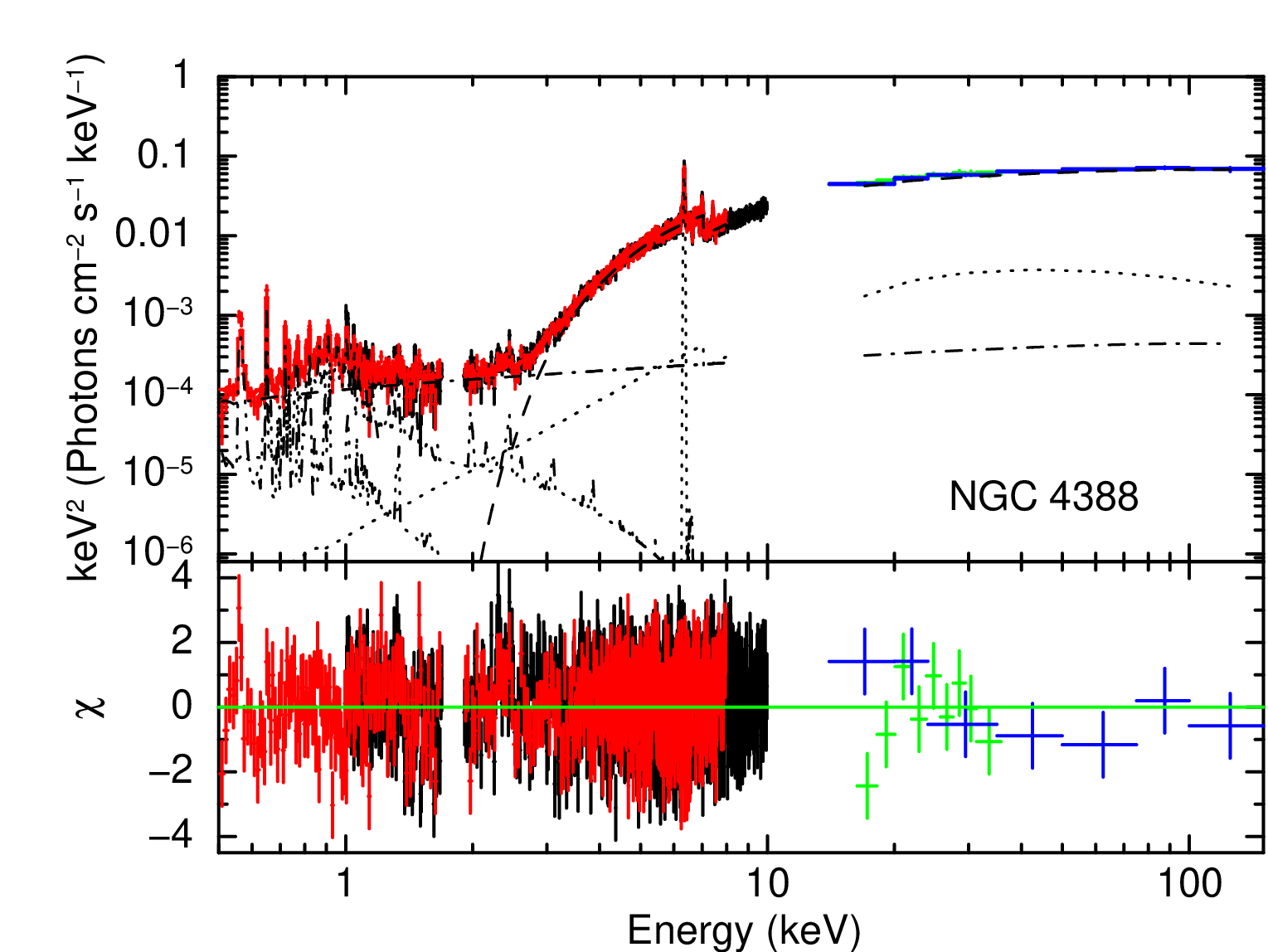}}
\resizebox{7.1cm}{!}{\includegraphics[scale=0.8,angle=-90]{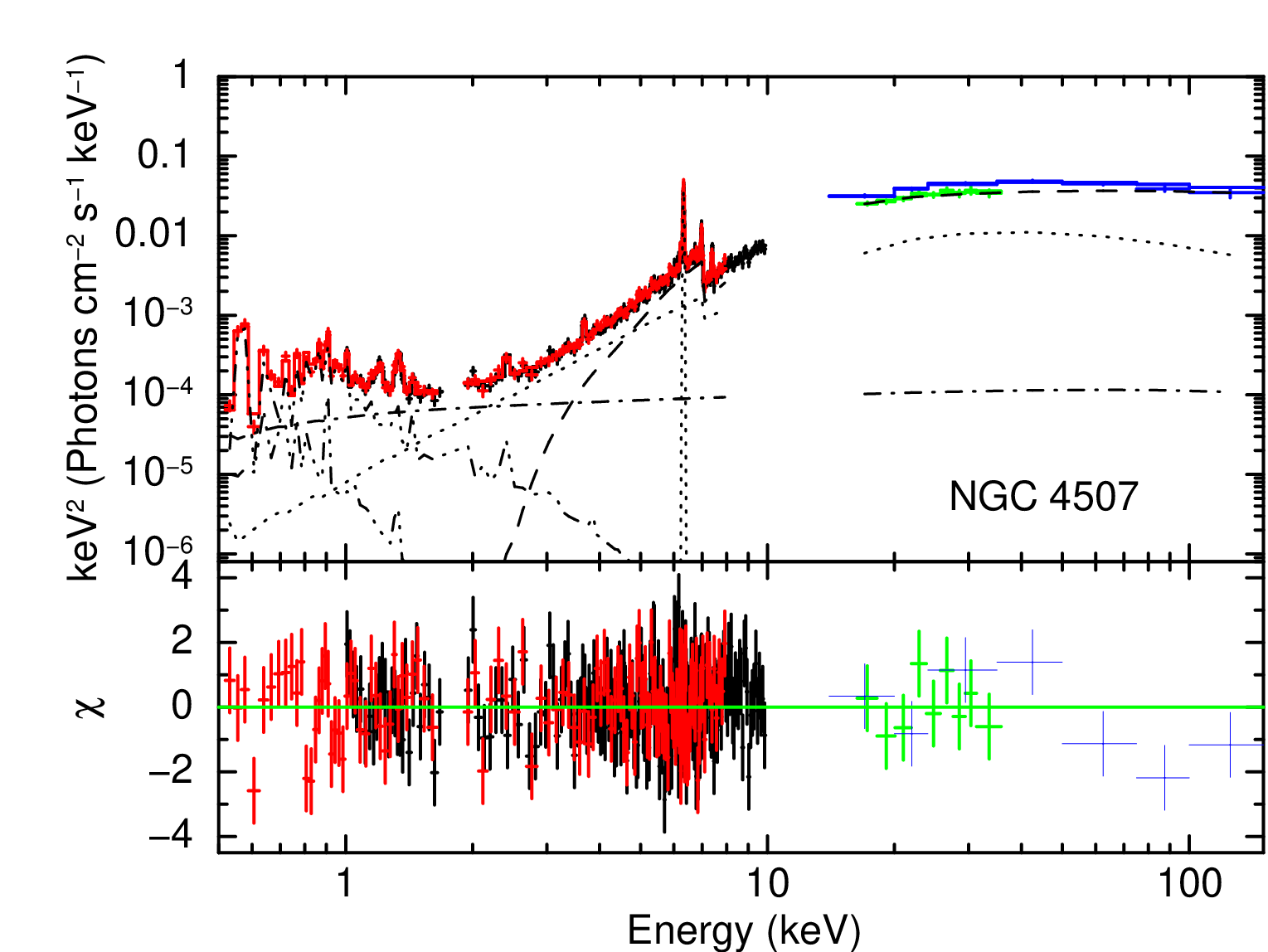}}
}\vspace{0.0cm}
\subfigure{
\resizebox{7.1cm}{!}{\includegraphics[scale=0.8,angle=-90]{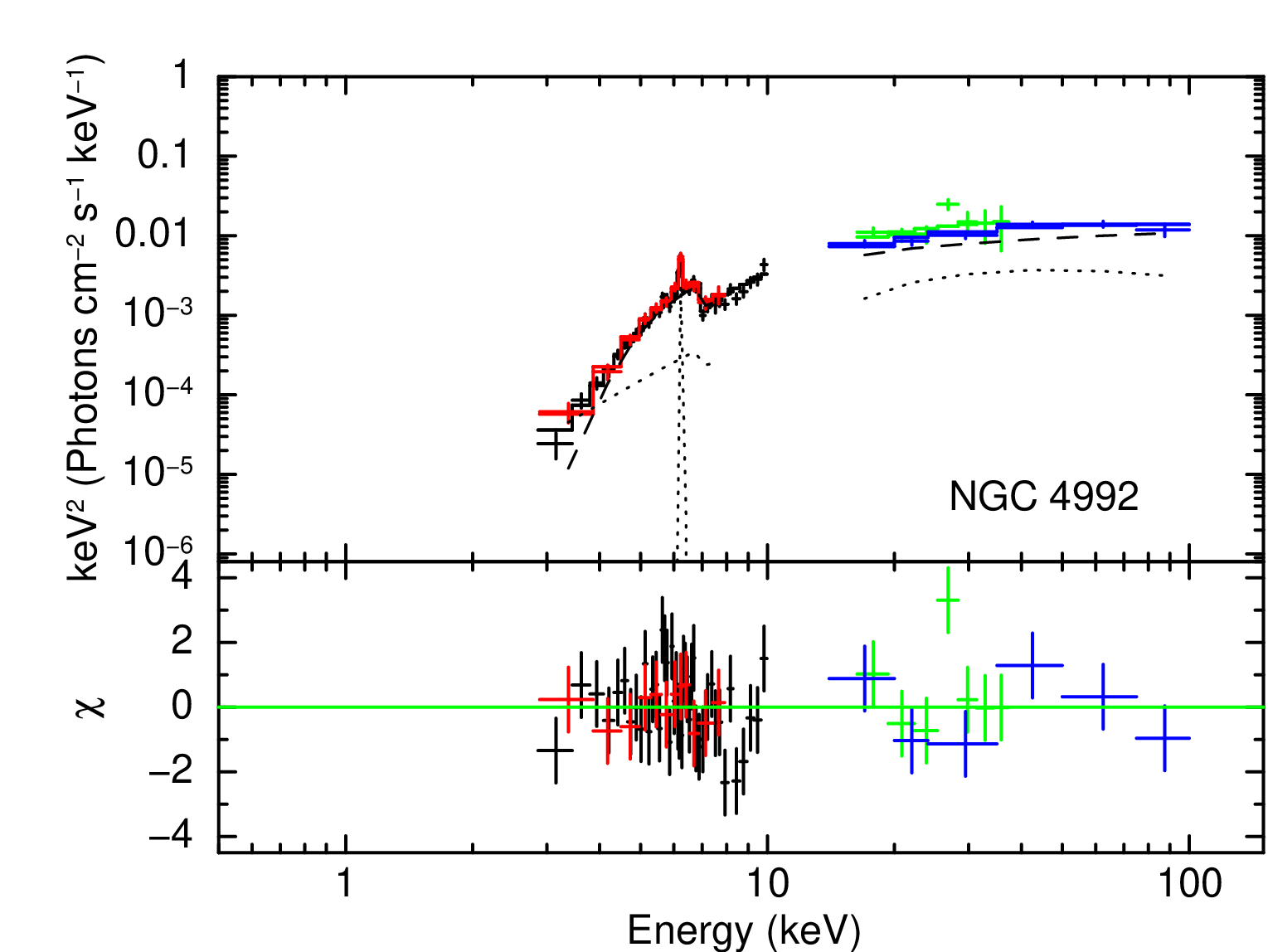}}
\resizebox{7.1cm}{!}{\includegraphics[scale=0.8,angle=-90]{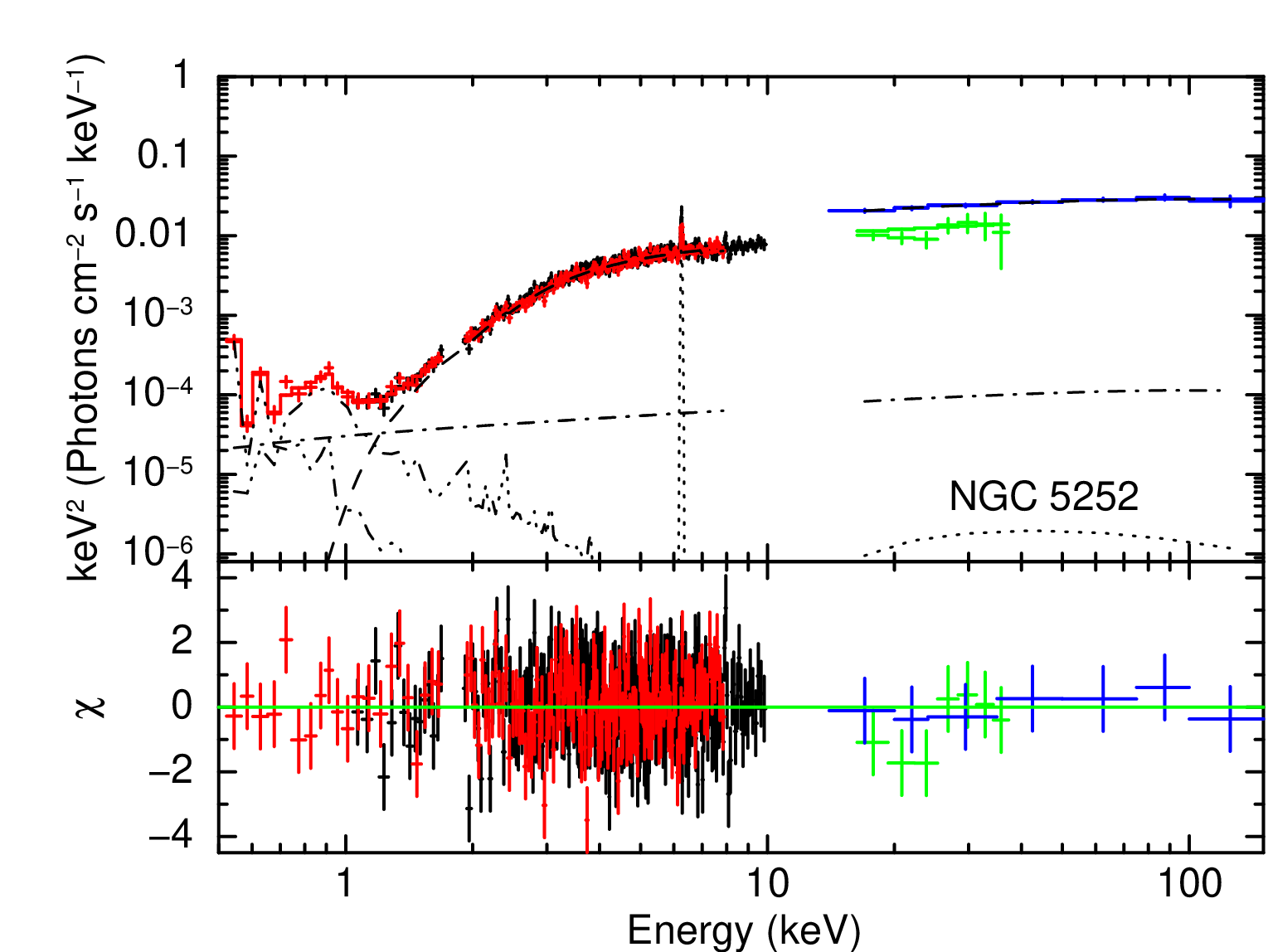}}
}\vspace{0.0cm}
\subfigure{
\resizebox{7.1cm}{!}{\includegraphics[scale=0.8,angle=-90]{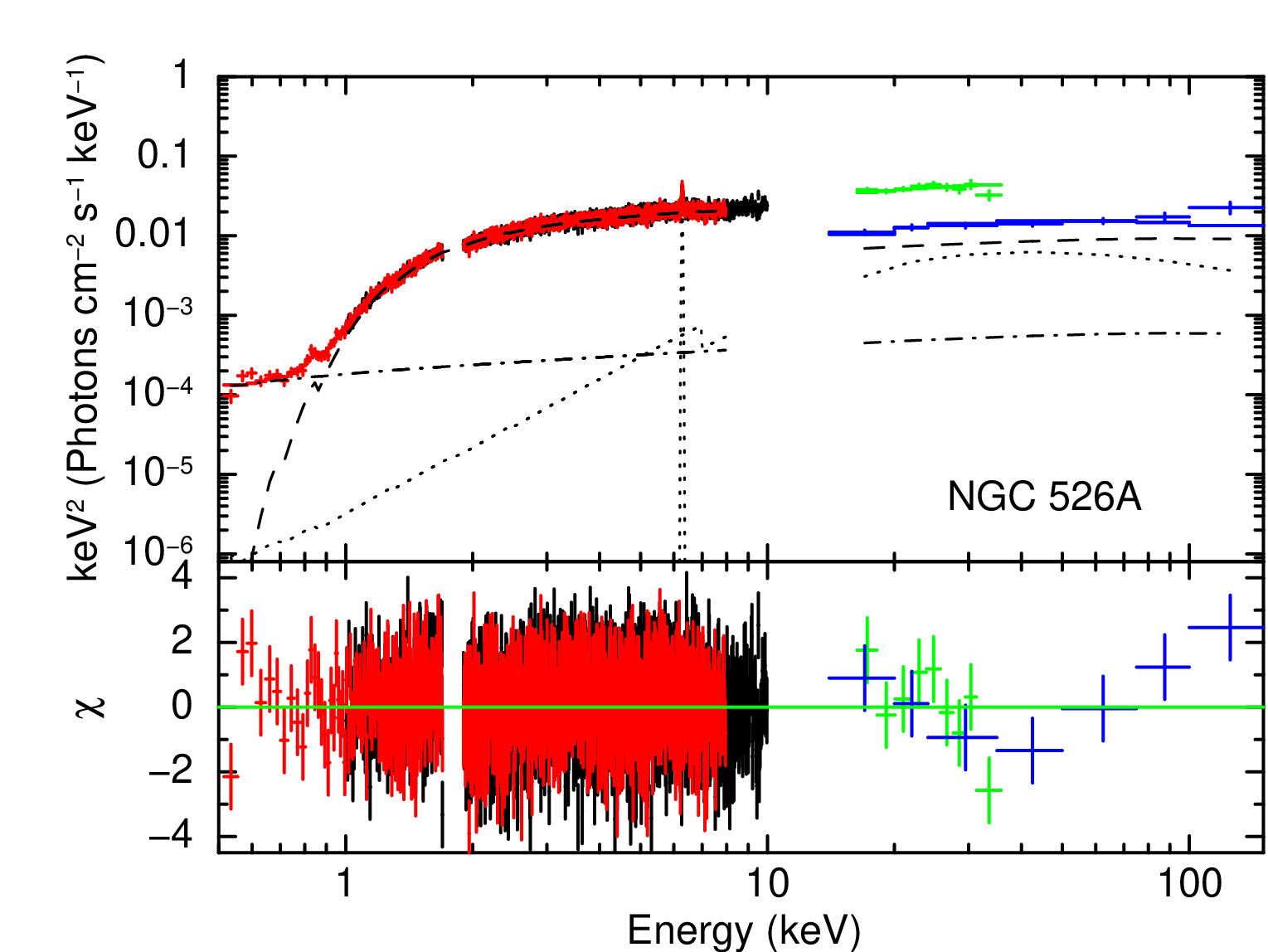}}
\resizebox{7.1cm}{!}{\includegraphics[scale=0.8,angle=-90]{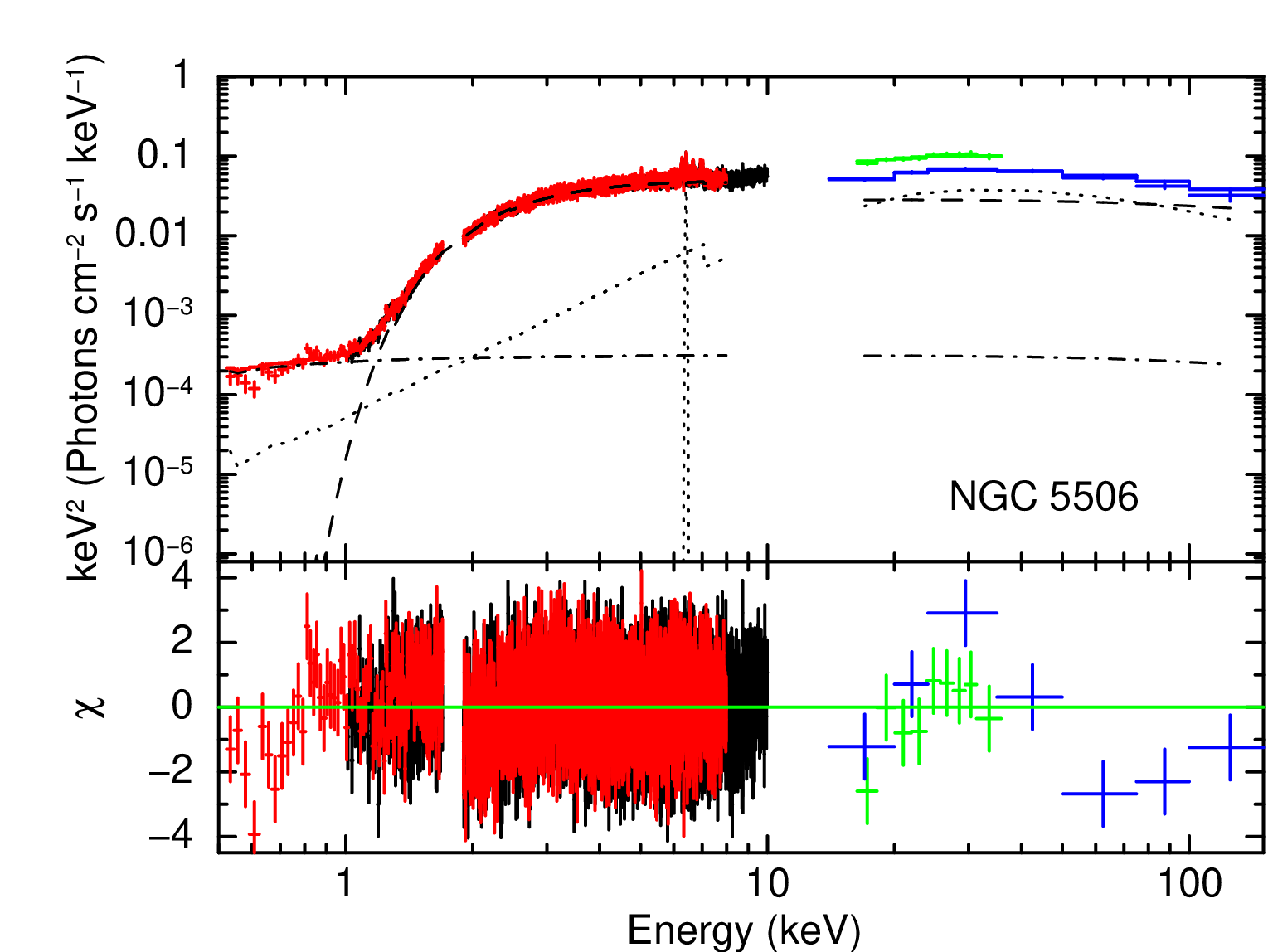}}
}\vspace{0.0cm}
\caption{
Continued.
}
\label{fig:spec}
\end{figure*}
\addtocounter{figure}{-1}
\begin{figure*}
\centering
\subfigure{
\resizebox{7.1cm}{!}{\includegraphics[scale=0.8,angle=-90]{fig_a41.ps}}
\resizebox{7.1cm}{!}{\includegraphics[scale=0.8,angle=-90]{fig_a42.ps}}
}\vspace{0.0cm}
\subfigure{
\resizebox{7.1cm}{!}{\includegraphics[scale=0.8,angle=-90]{fig_a43.ps}}
\resizebox{7.1cm}{!}{\includegraphics[scale=0.8,angle=-90]{fig_a44.ps}}
}\vspace{0.0cm}
\subfigure{
\resizebox{7.1cm}{!}{\includegraphics[scale=0.8,angle=-90]{fig_a45.ps}}
%\resizebox{7.1cm}{!}{\includegraphics[scale=0.8,angle=-90]{fig_a46.ps}}
}\vspace{0.0cm}
\caption{
Continued.
}
\label{fig:spec}
\end{figure*}

 % IRU 
\begin{figure*}
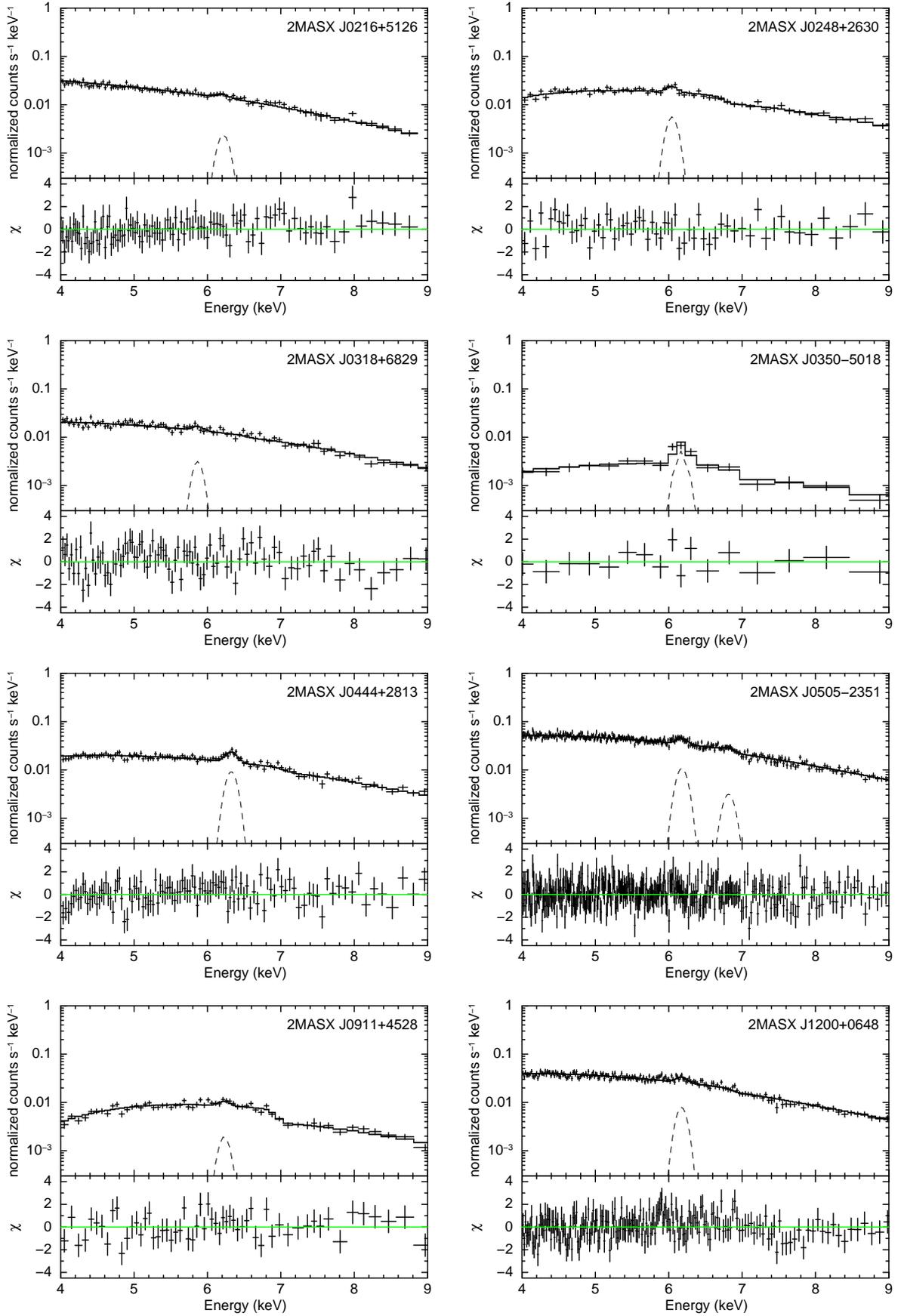

\centering
\subfigure{
\resizebox{8cm}{!}{\includegraphics[scale=0.8,angle=-90]{fig_b01.ps}}
\resizebox{8cm}{!}{\includegraphics[scale=0.8,angle=-90]{fig_b02.ps}}
}\vspace{0.0cm}
\subfigure{
\resizebox{8cm}{!}{\includegraphics[scale=0.8,angle=-90]{fig_b03.ps}}
\resizebox{8cm}{!}{\includegraphics[scale=0.8,angle=-90]{fig_b04.ps}}
}\vspace{0.0cm}
\subfigure{
\resizebox{8cm}{!}{\includegraphics[scale=0.8,angle=-90]{fig_b05.ps}}
\resizebox{8cm}{!}{\includegraphics[scale=0.8,angle=-90]{fig_b06.ps}}
}\vspace{0.0cm}
\subfigure{
\resizebox{8cm}{!}{\includegraphics[scale=0.8,angle=-90]{fig_b07.ps}}
\resizebox{8cm}{!}{\includegraphics[scale=0.8,angle=-90]{fig_b08.ps}}
}\vspace{0.0cm}
\caption{
Narrow band spectra and best-fitting models obtained in Section~\ref{sec:base_model}.
The FI-XISs spectrum is represented with the black crosses in the upper panel, whereas the fit residuals in 
the lower panel. 
The solid and dashed lines correspond to the total and detected eission lines, respectively.
}
\label{fig:narrow_spec}
\end{figure*}
\addtocounter{figure}{-1}
\begin{figure*}
\centering
\subfigure{
\resizebox{8cm}{!}{\includegraphics[scale=0.8,angle=-90]{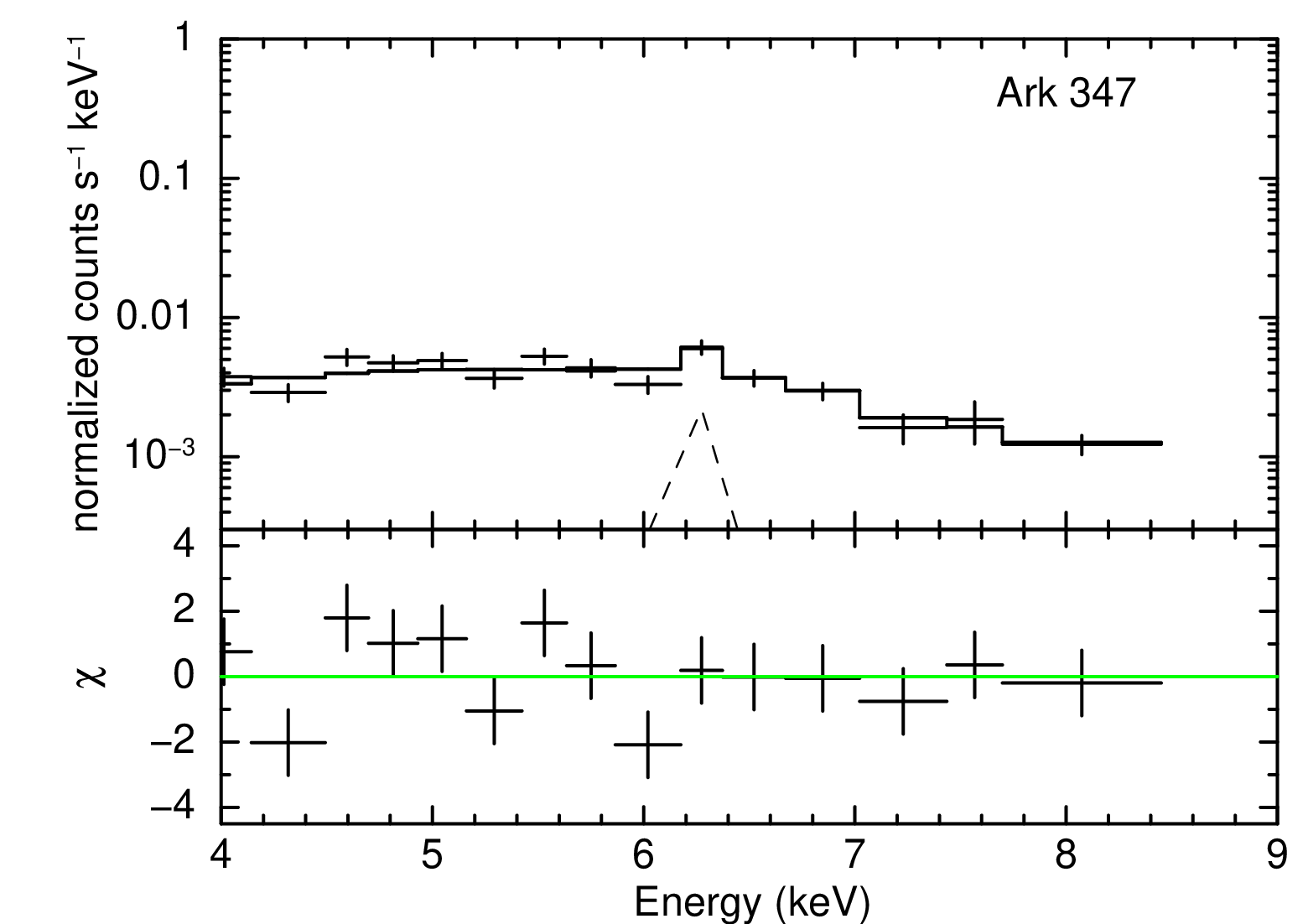}}
\resizebox{8cm}{!}{\includegraphics[scale=0.8,angle=-90]{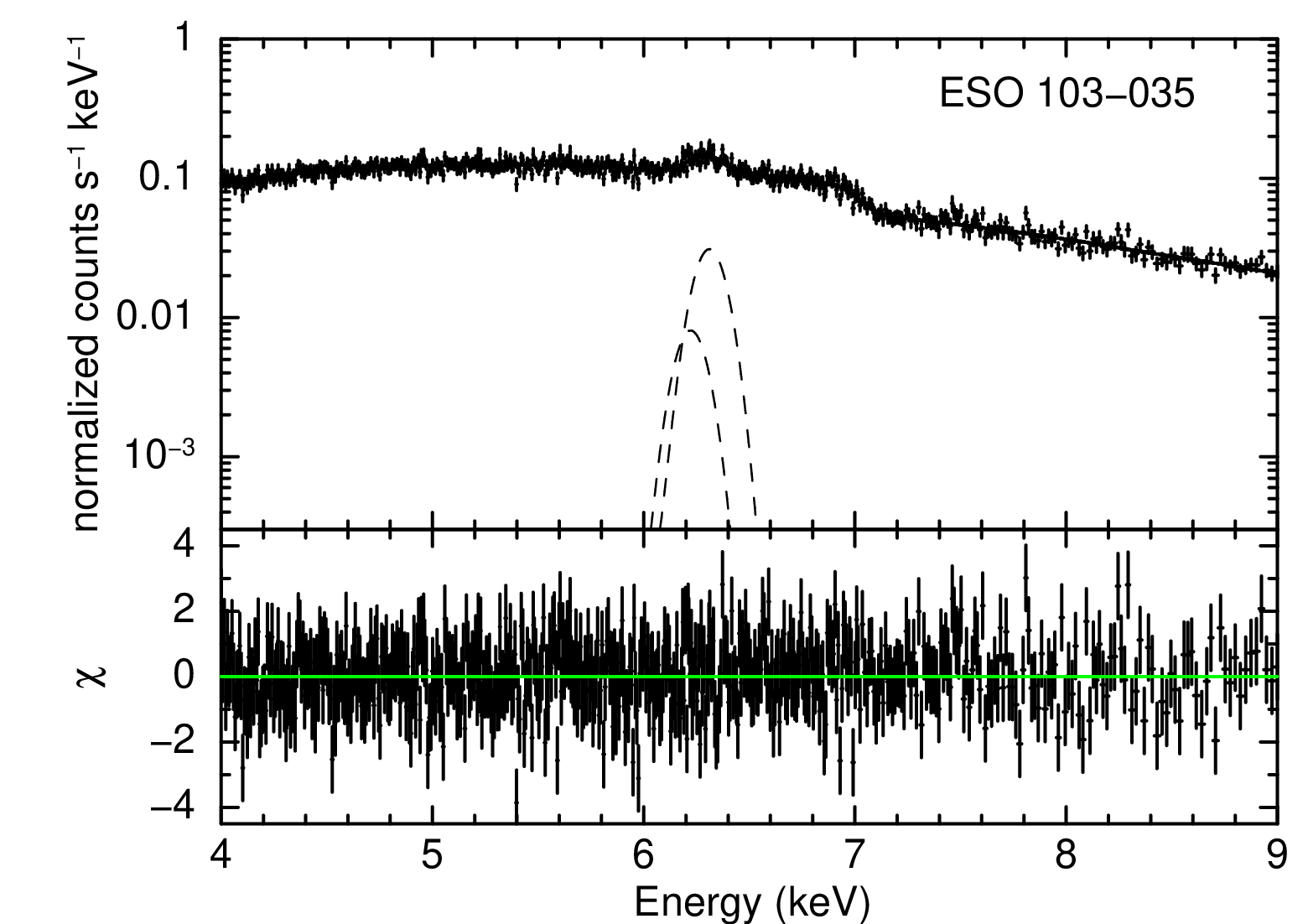}}
}\vspace{0.0cm}
\subfigure{
\resizebox{8cm}{!}{\includegraphics[scale=0.8,angle=-90]{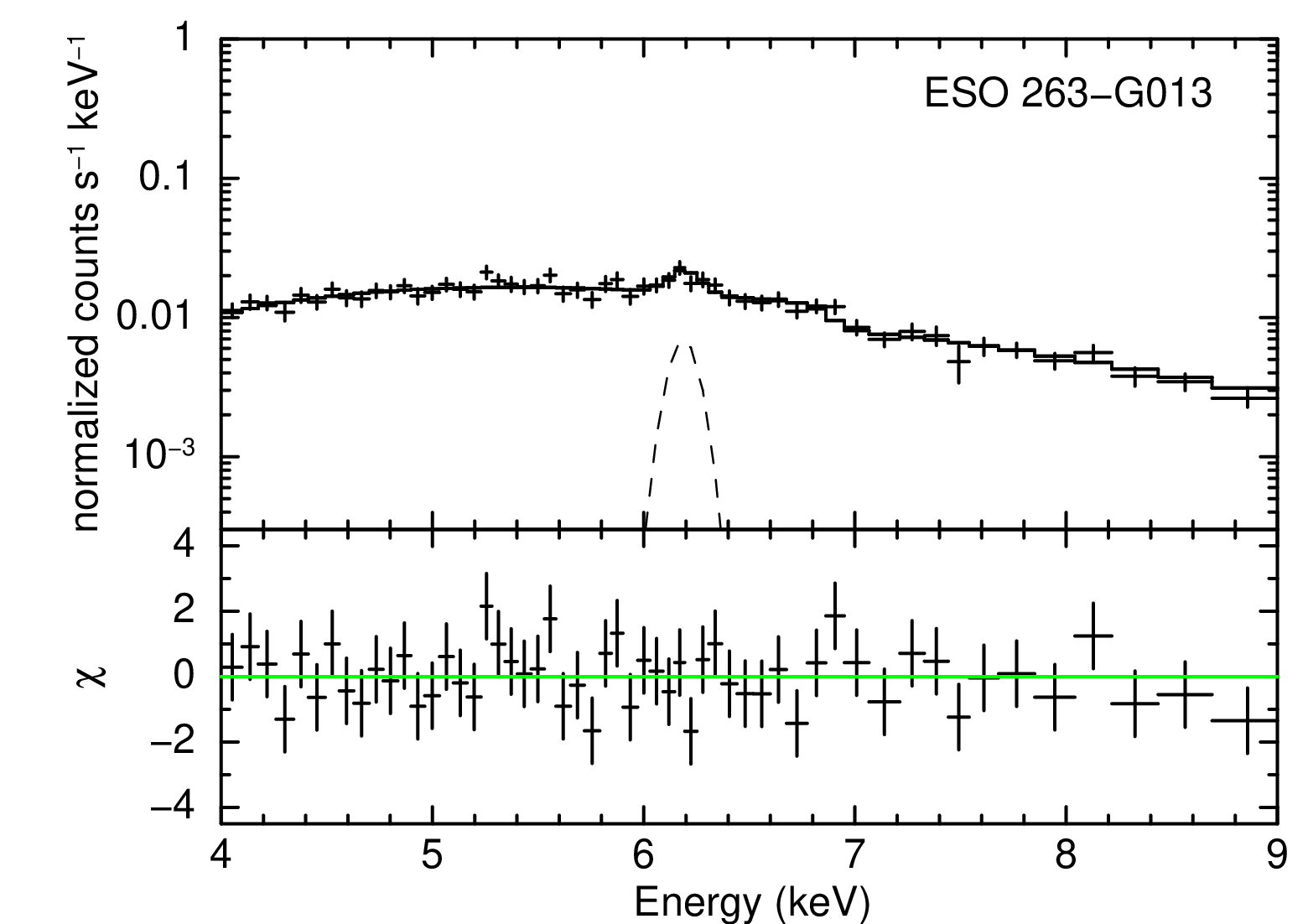}}
\resizebox{8cm}{!}{\includegraphics[scale=0.8,angle=-90]{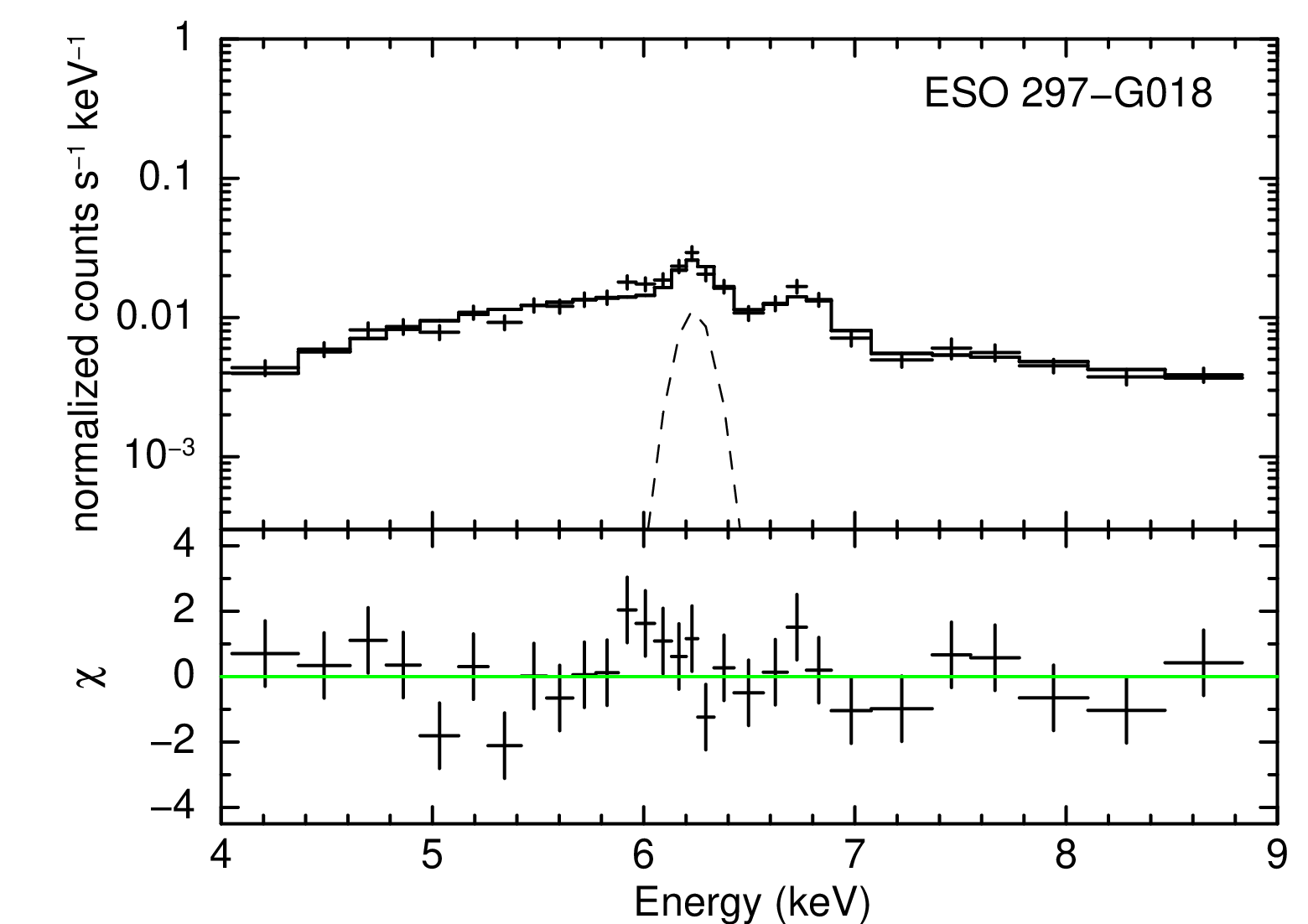}}
}\vspace{0.0cm}
\subfigure{
\resizebox{8cm}{!}{\includegraphics[scale=0.8,angle=-90]{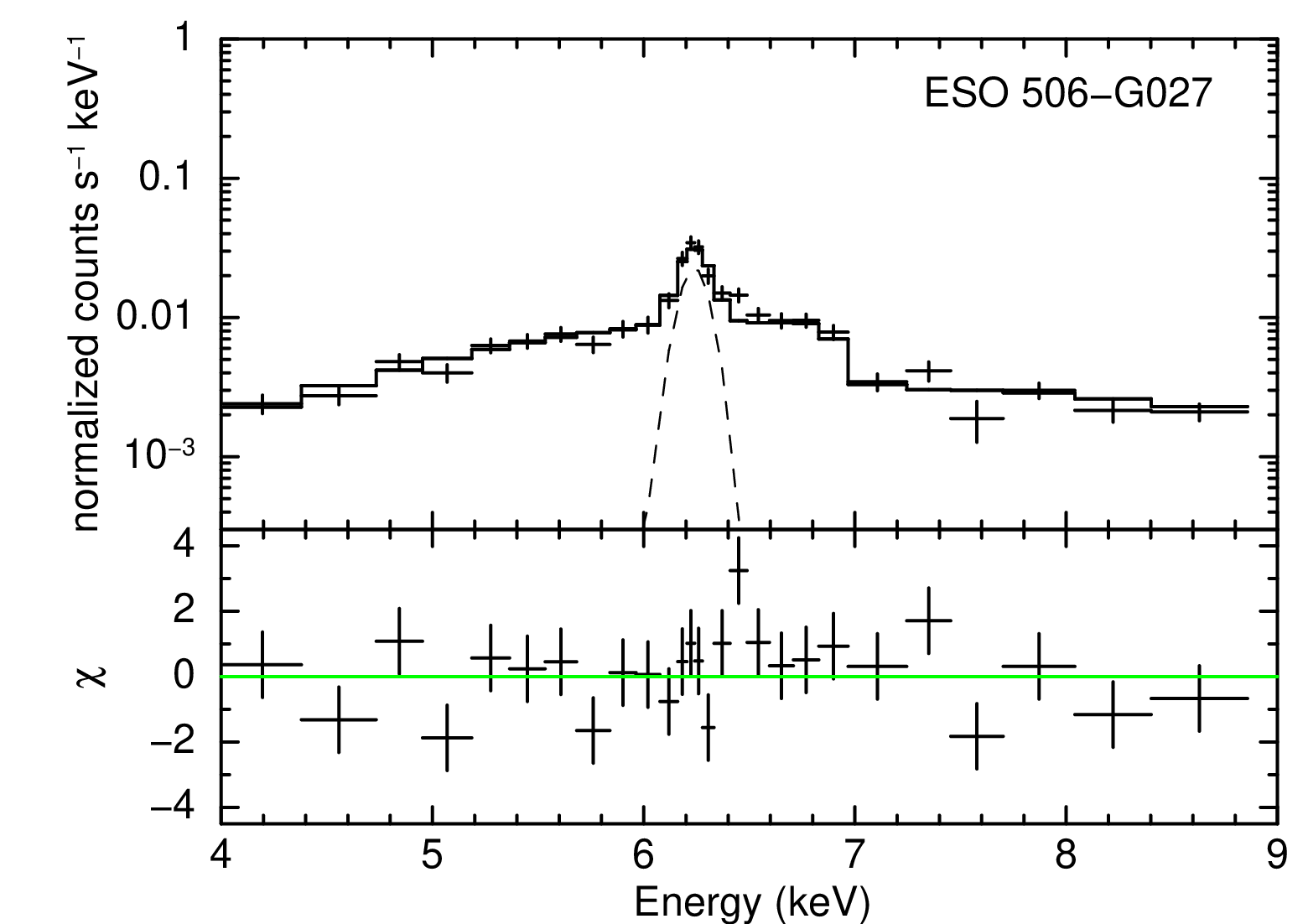}}
\resizebox{8cm}{!}{\includegraphics[scale=0.8,angle=-90]{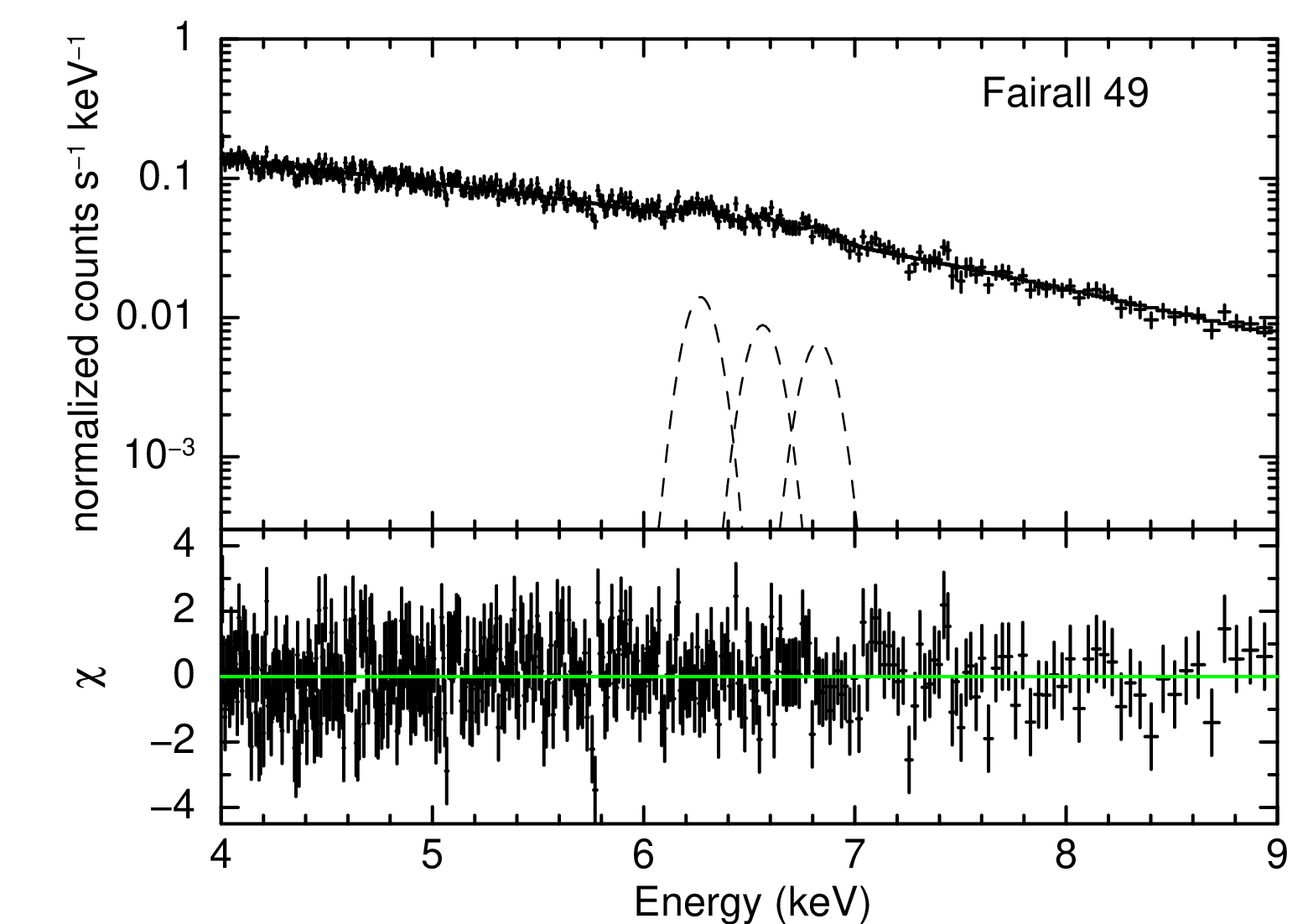}}
}\vspace{0.0cm}
\subfigure{
\resizebox{8cm}{!}{\includegraphics[scale=0.8,angle=-90]{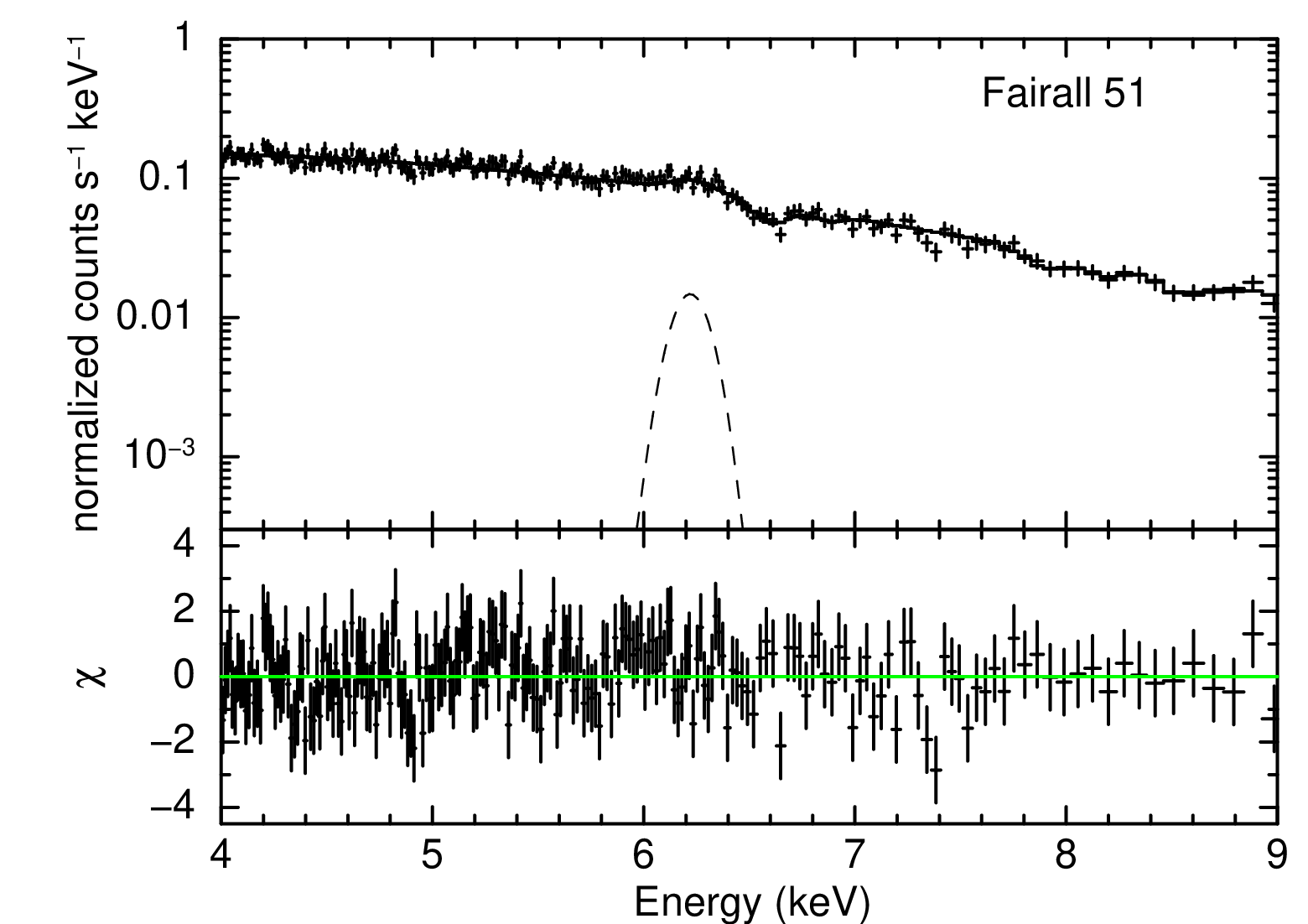}}
\resizebox{8cm}{!}{\includegraphics[scale=0.8,angle=-90]{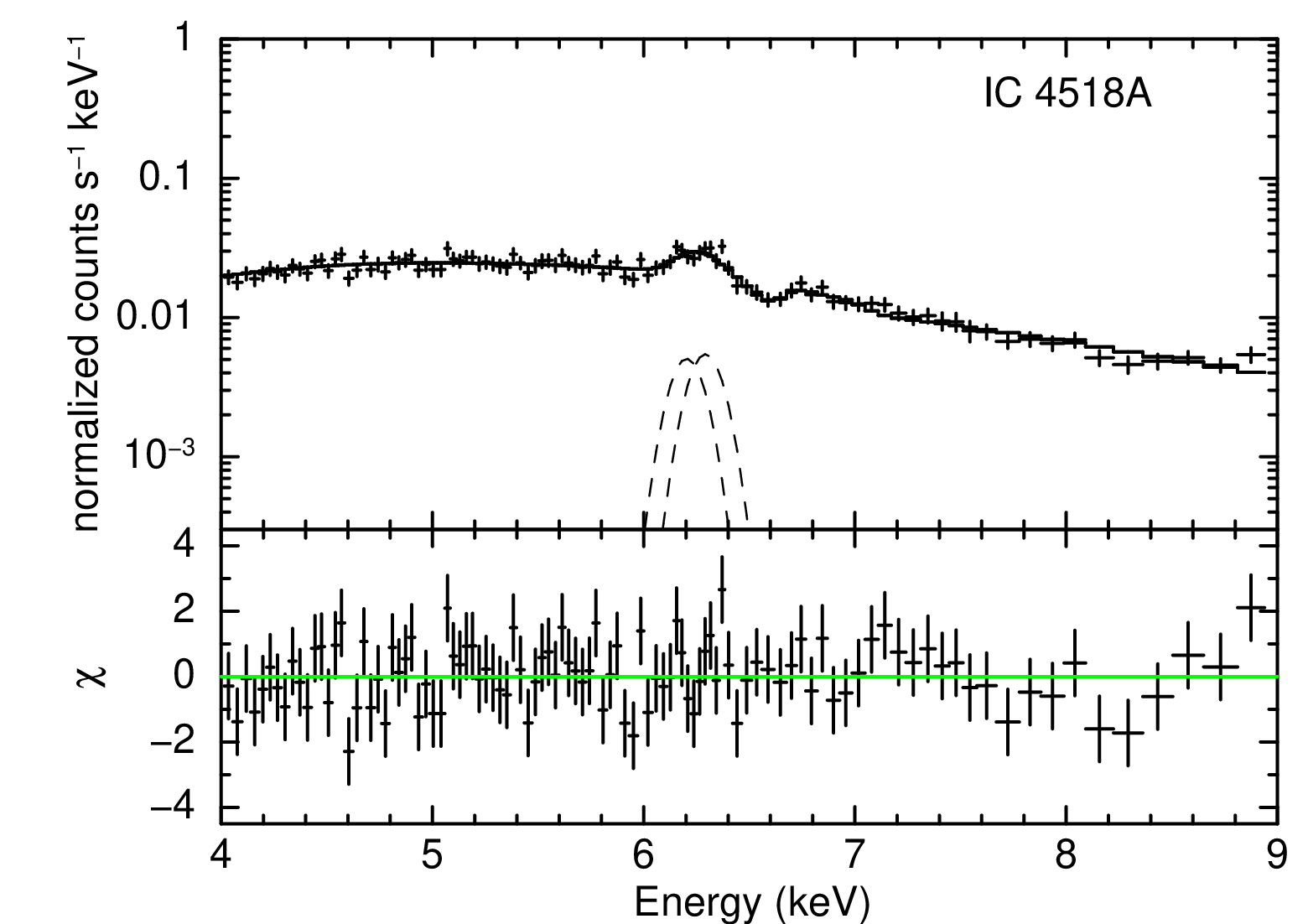}}
}\vspace{0.0cm}
\caption{
Continued.
}
\label{fig:narrow_spec}
\end{figure*}
\addtocounter{figure}{-1}
\begin{figure*}
\centering
\subfigure{
\resizebox{8cm}{!}{\includegraphics[scale=0.8,angle=-90]{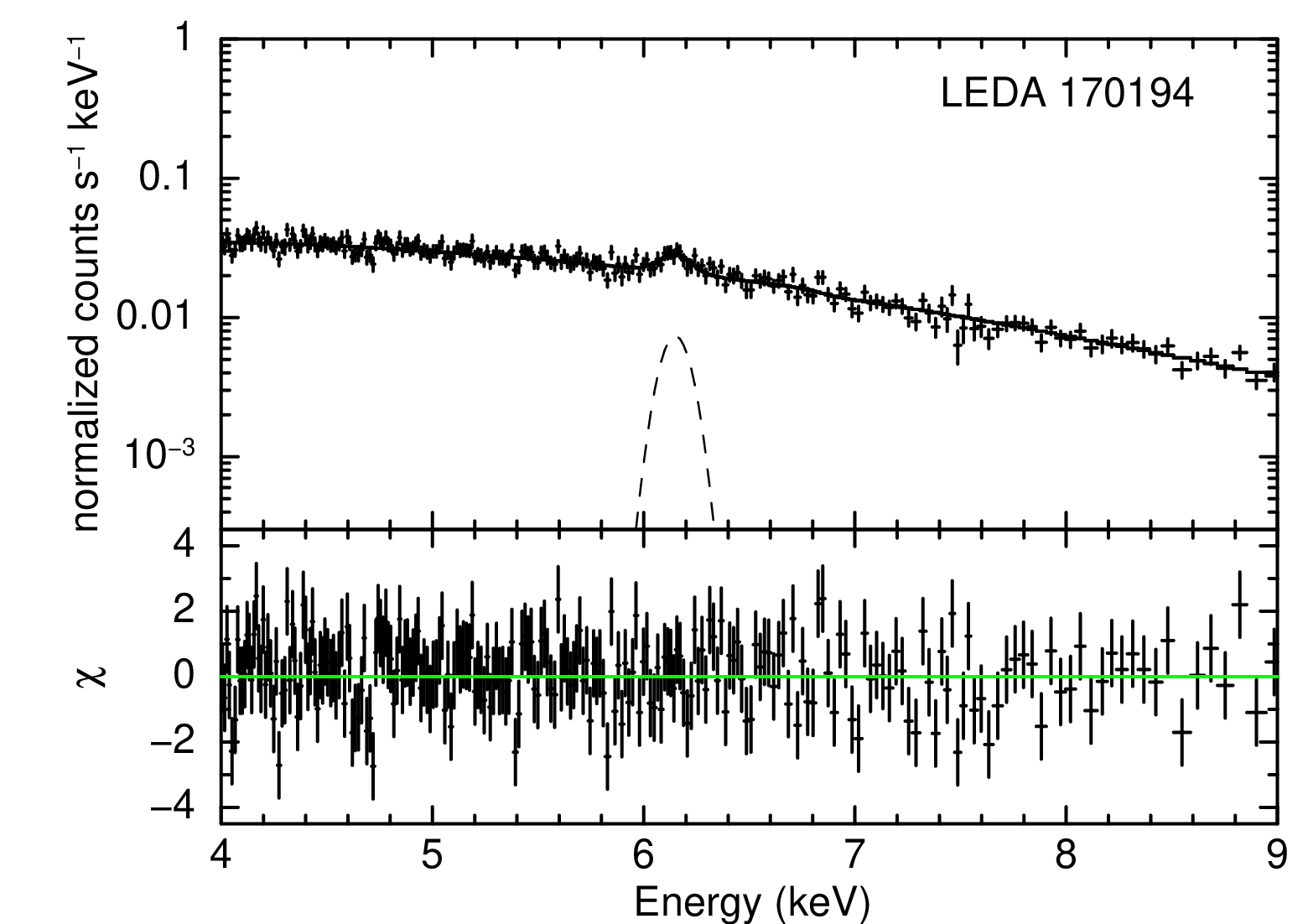}}
\resizebox{8cm}{!}{\includegraphics[scale=0.8,angle=-90]{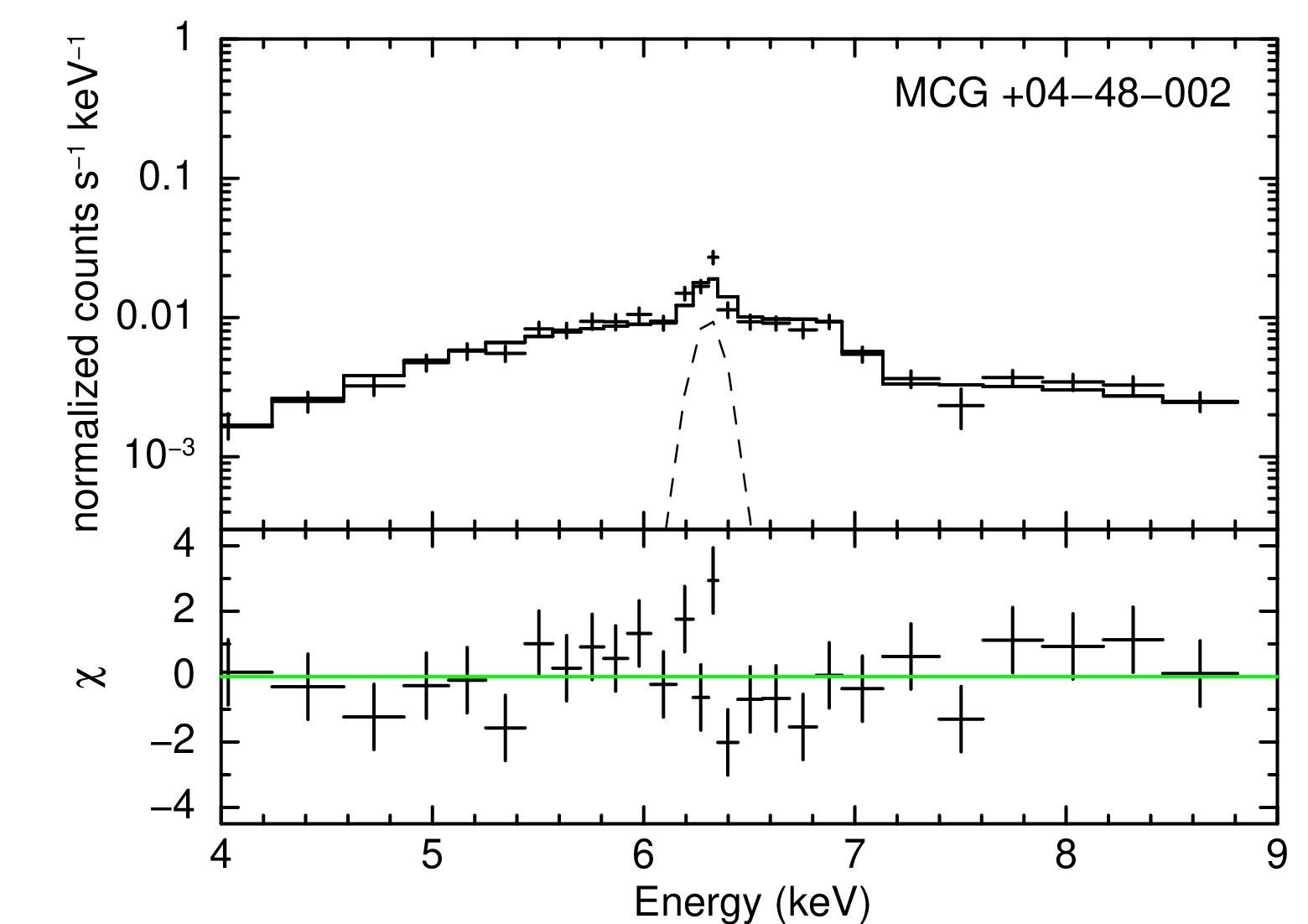}}
}\vspace{0.0cm}
\subfigure{
\resizebox{8cm}{!}{\includegraphics[scale=0.8,angle=-90]{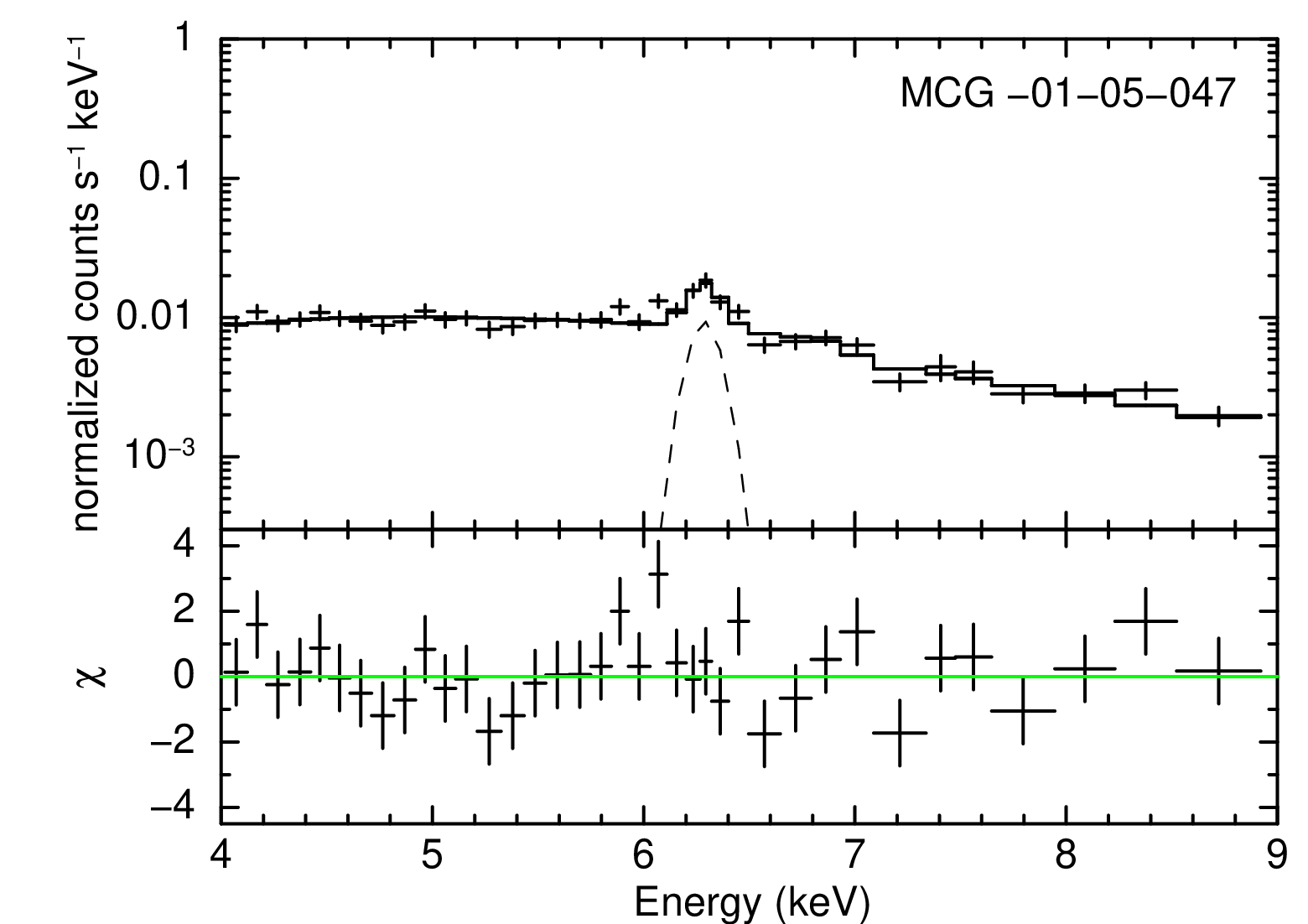}}
\resizebox{8cm}{!}{\includegraphics[scale=0.8,angle=-90]{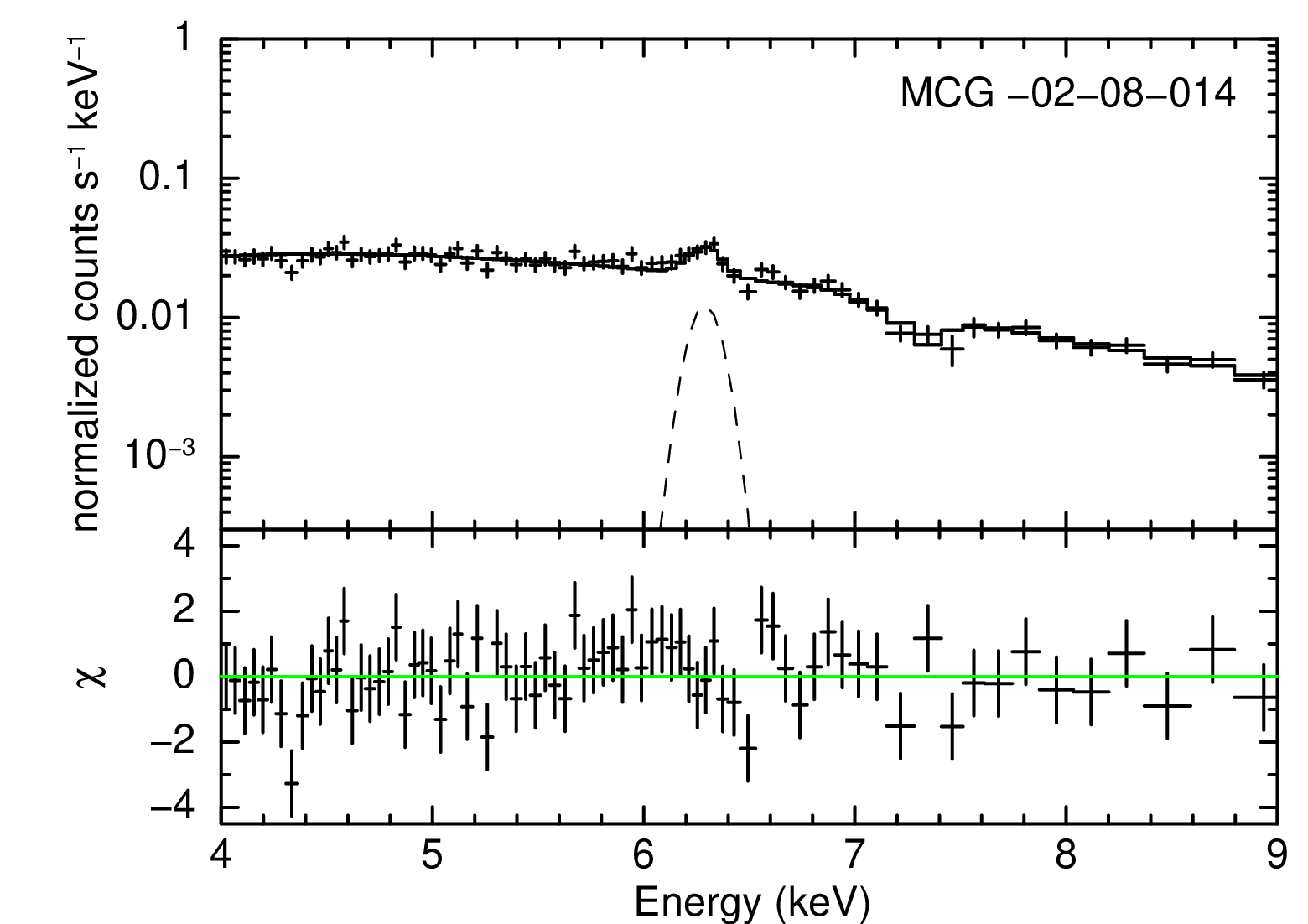}}
}\vspace{0.0cm}
\subfigure{
\resizebox{8cm}{!}{\includegraphics[scale=0.8,angle=-90]{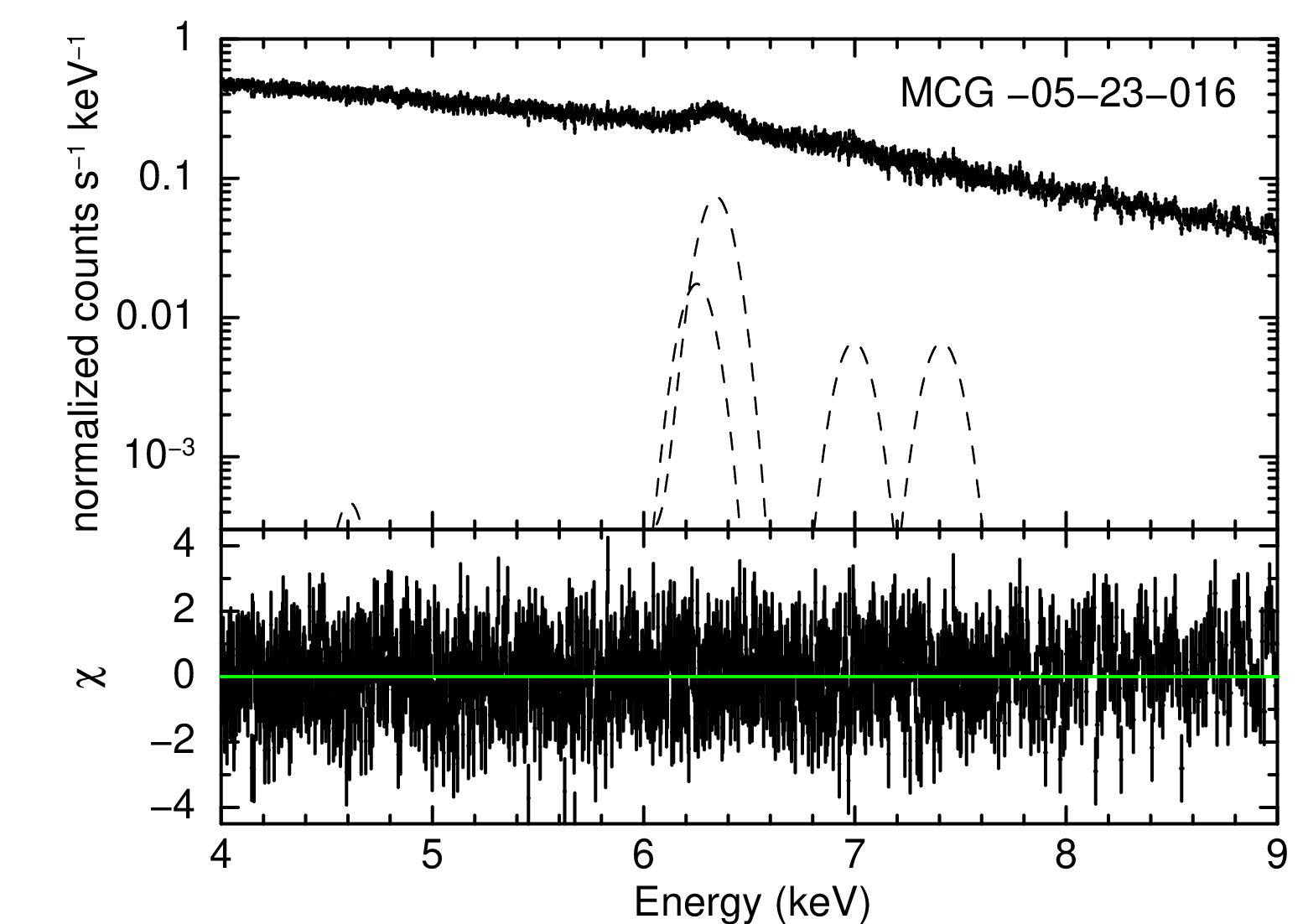}}
\resizebox{8cm}{!}{\includegraphics[scale=0.8,angle=-90]{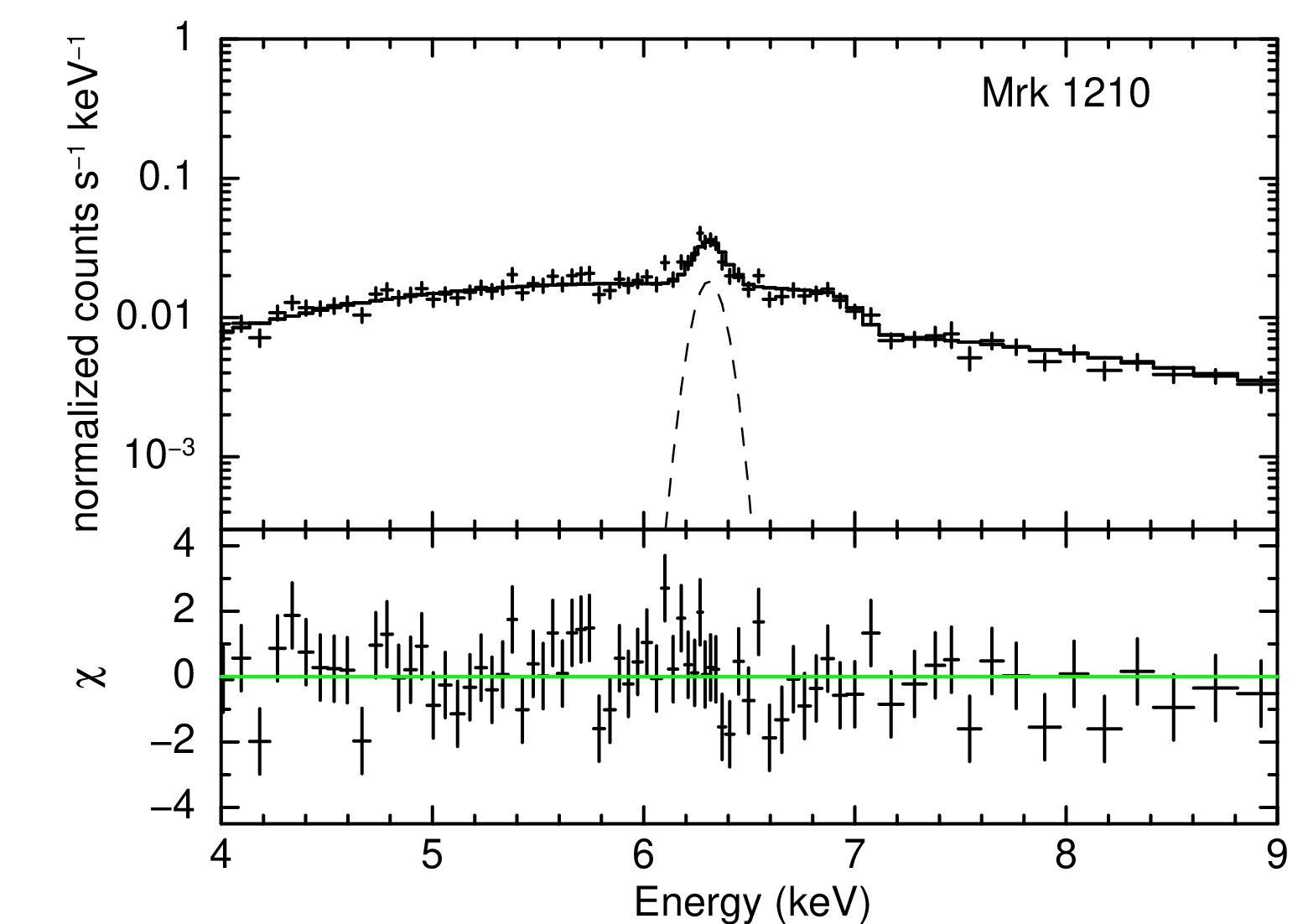}}
}\vspace{0.0cm}
\subfigure{
\resizebox{8cm}{!}{\includegraphics[scale=0.8,angle=-90]{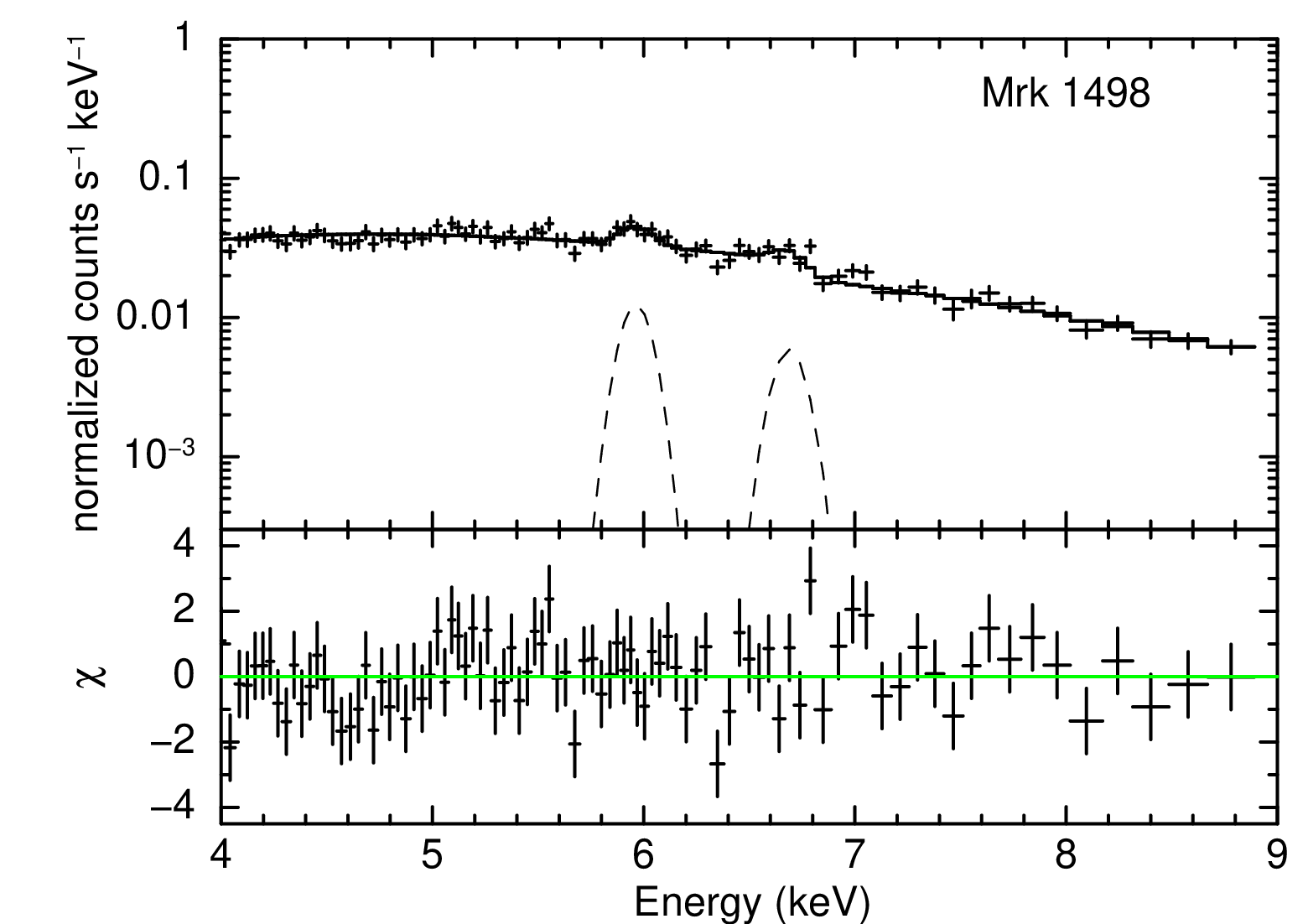}}
\resizebox{8cm}{!}{\includegraphics[scale=0.8,angle=-90]{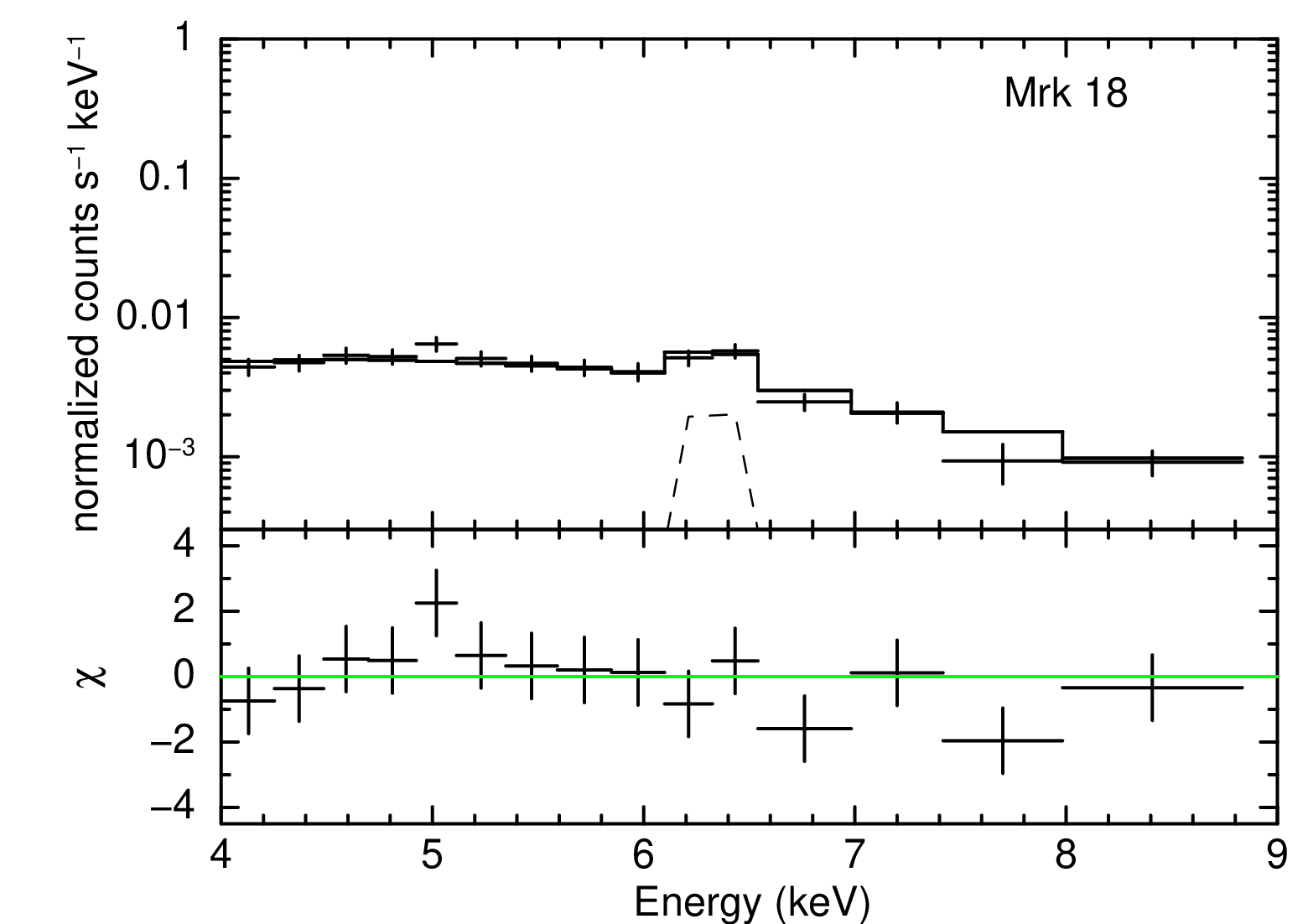}}
}\vspace{0.0cm}
\caption{
Continued.
}
\label{fig:narrow_spec}
\end{figure*}
\addtocounter{figure}{-1}
\begin{figure*}
\centering
\subfigure{
\resizebox{8cm}{!}{\includegraphics[scale=0.8,angle=-90]{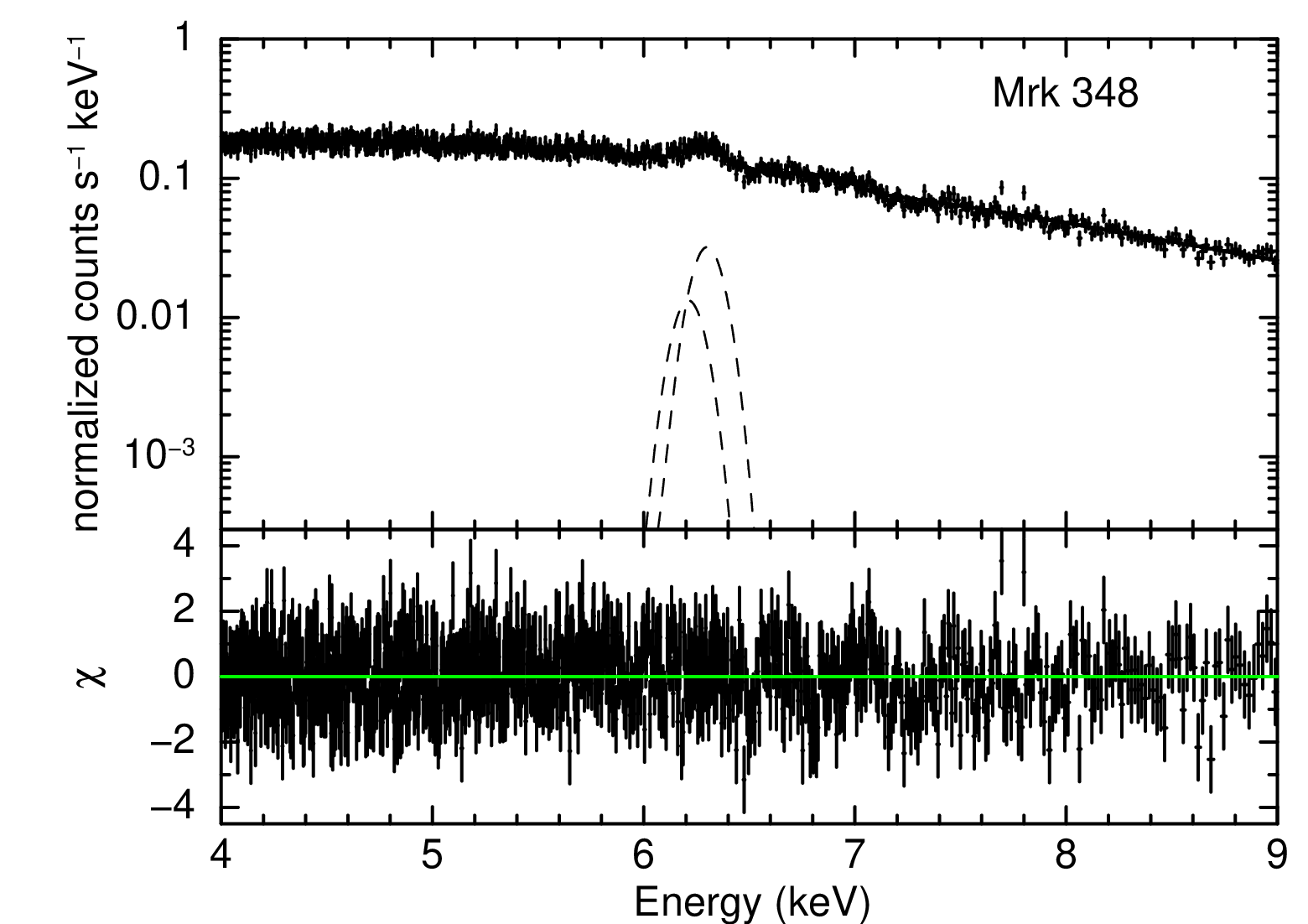}}
\resizebox{8cm}{!}{\includegraphics[scale=0.8,angle=-90]{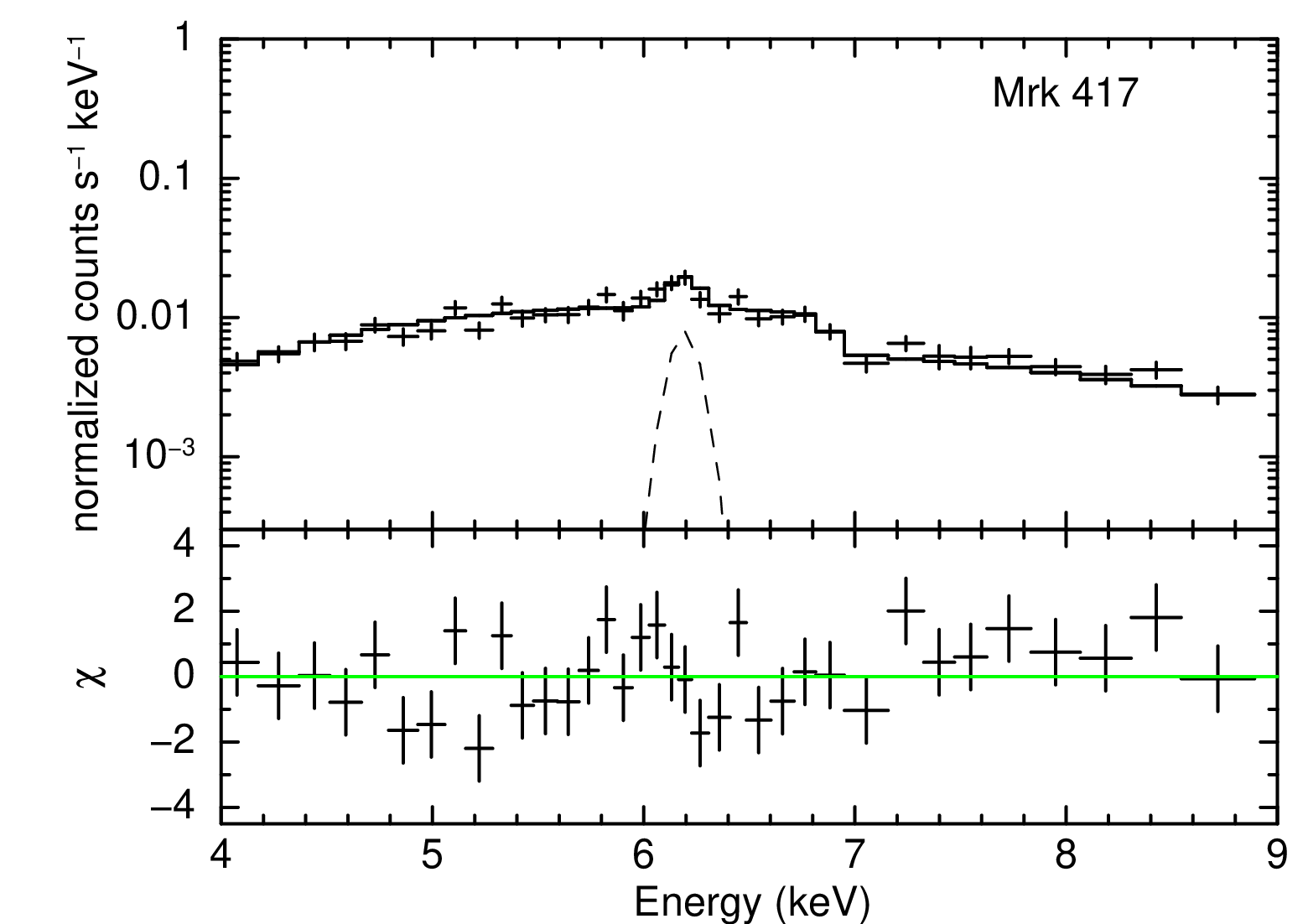}}
}\vspace{0.0cm}
\subfigure{
\resizebox{8cm}{!}{\includegraphics[scale=0.8,angle=-90]{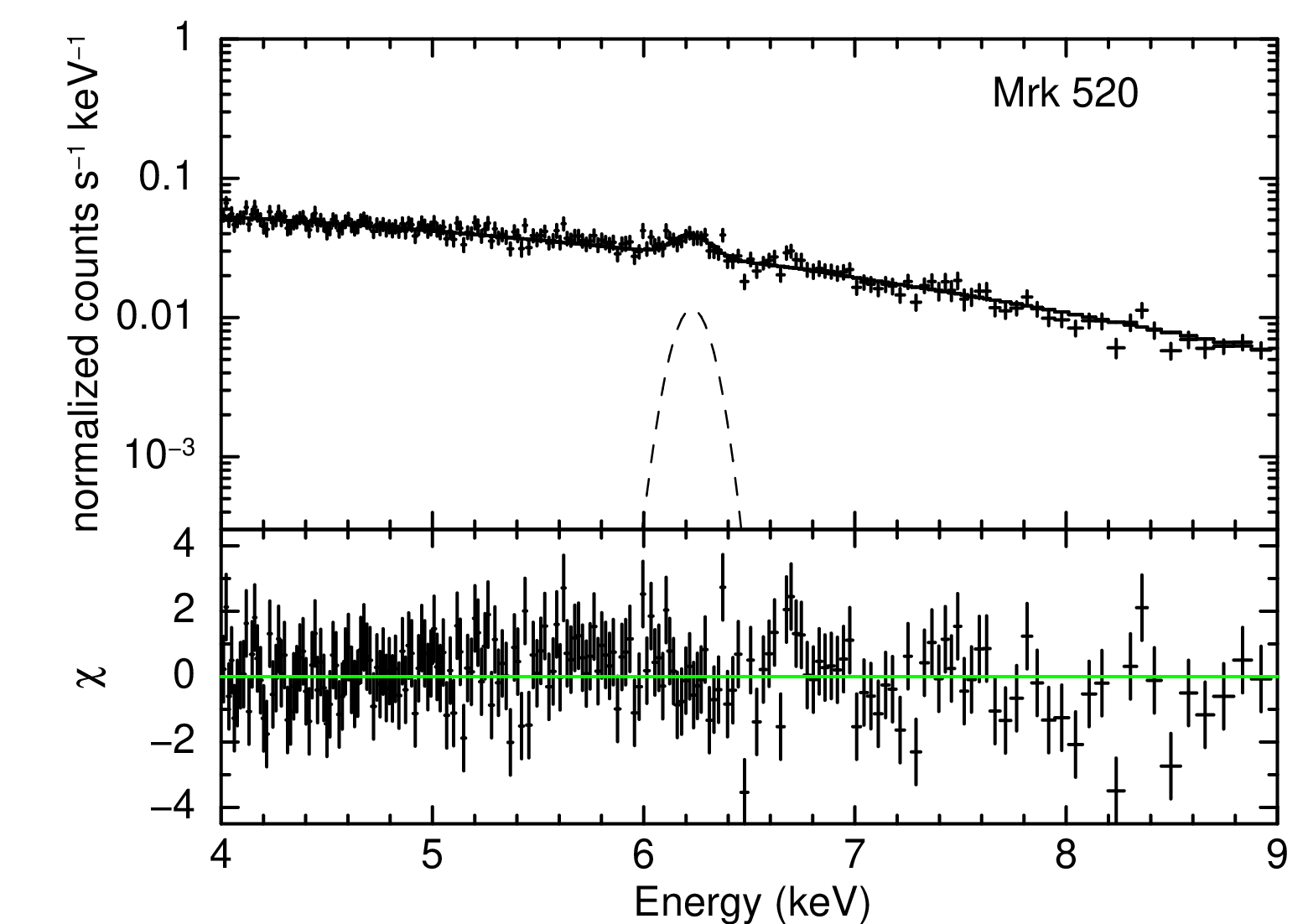}}
\resizebox{8cm}{!}{\includegraphics[scale=0.8,angle=-90]{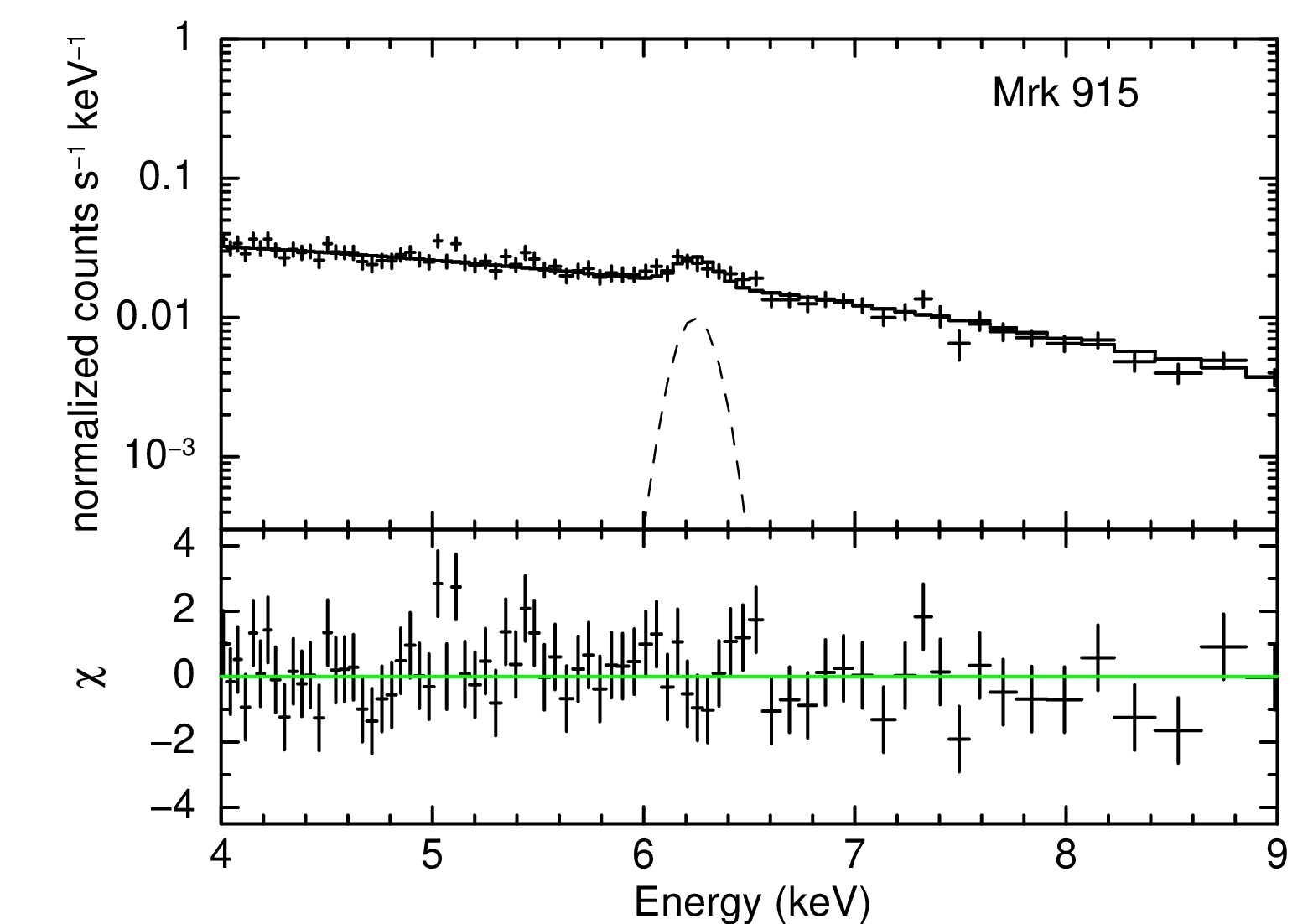}}
}\vspace{0.0cm}
\subfigure{
\resizebox{8cm}{!}{\includegraphics[scale=0.8,angle=-90]{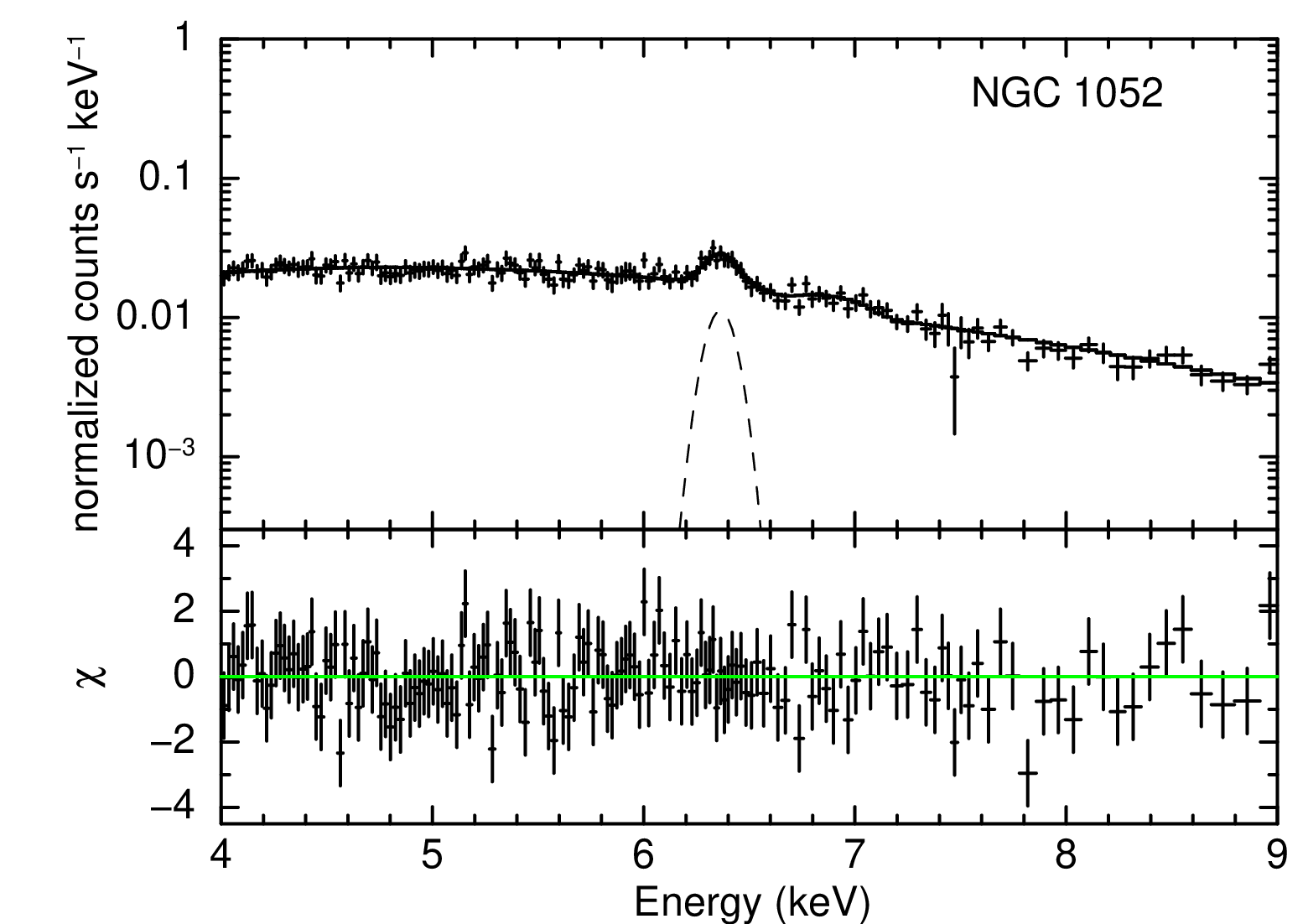}}
\resizebox{8cm}{!}{\includegraphics[scale=0.8,angle=-90]{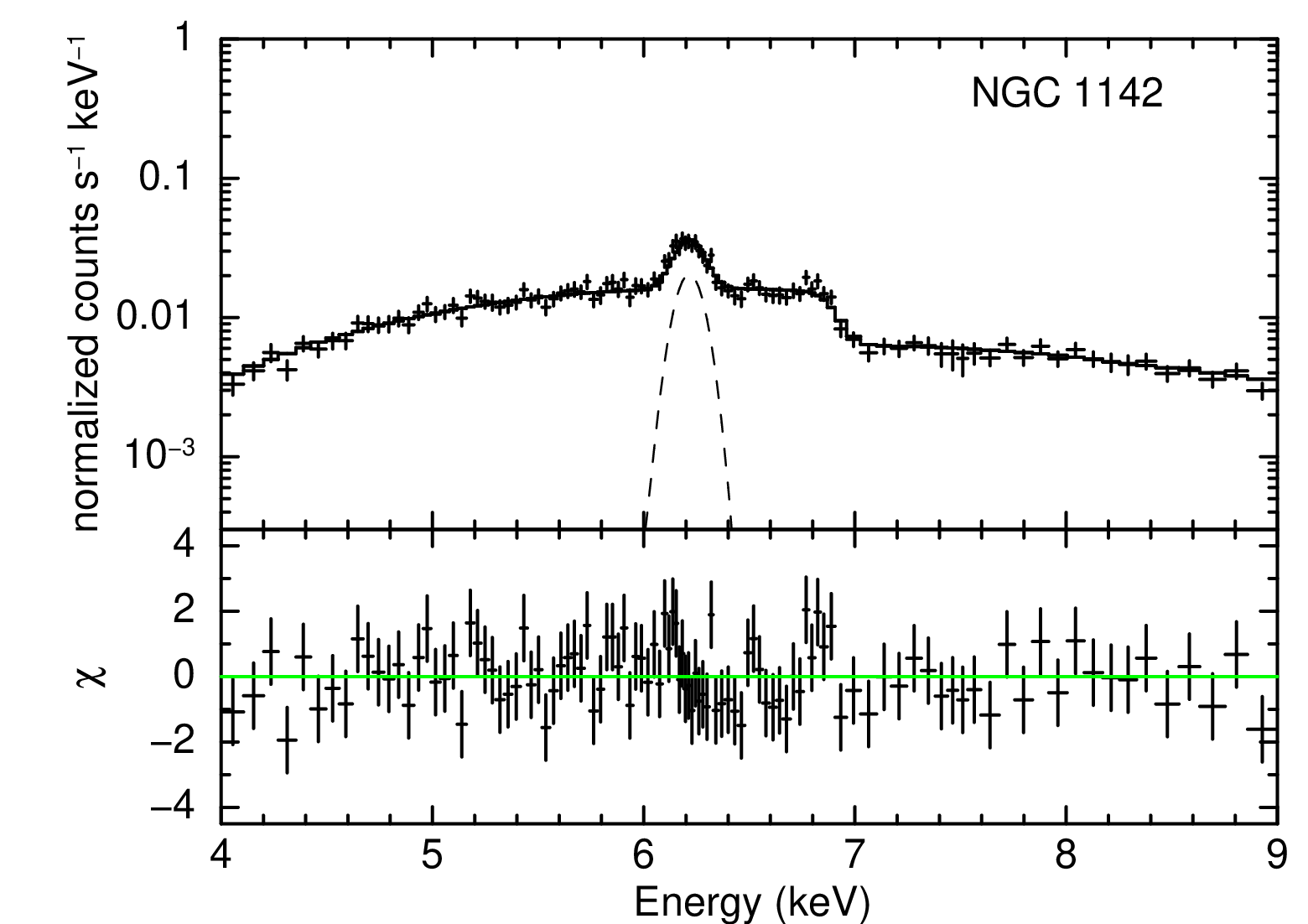}}
}\vspace{0.0cm}
\subfigure{
\resizebox{8cm}{!}{\includegraphics[scale=0.8,angle=-90]{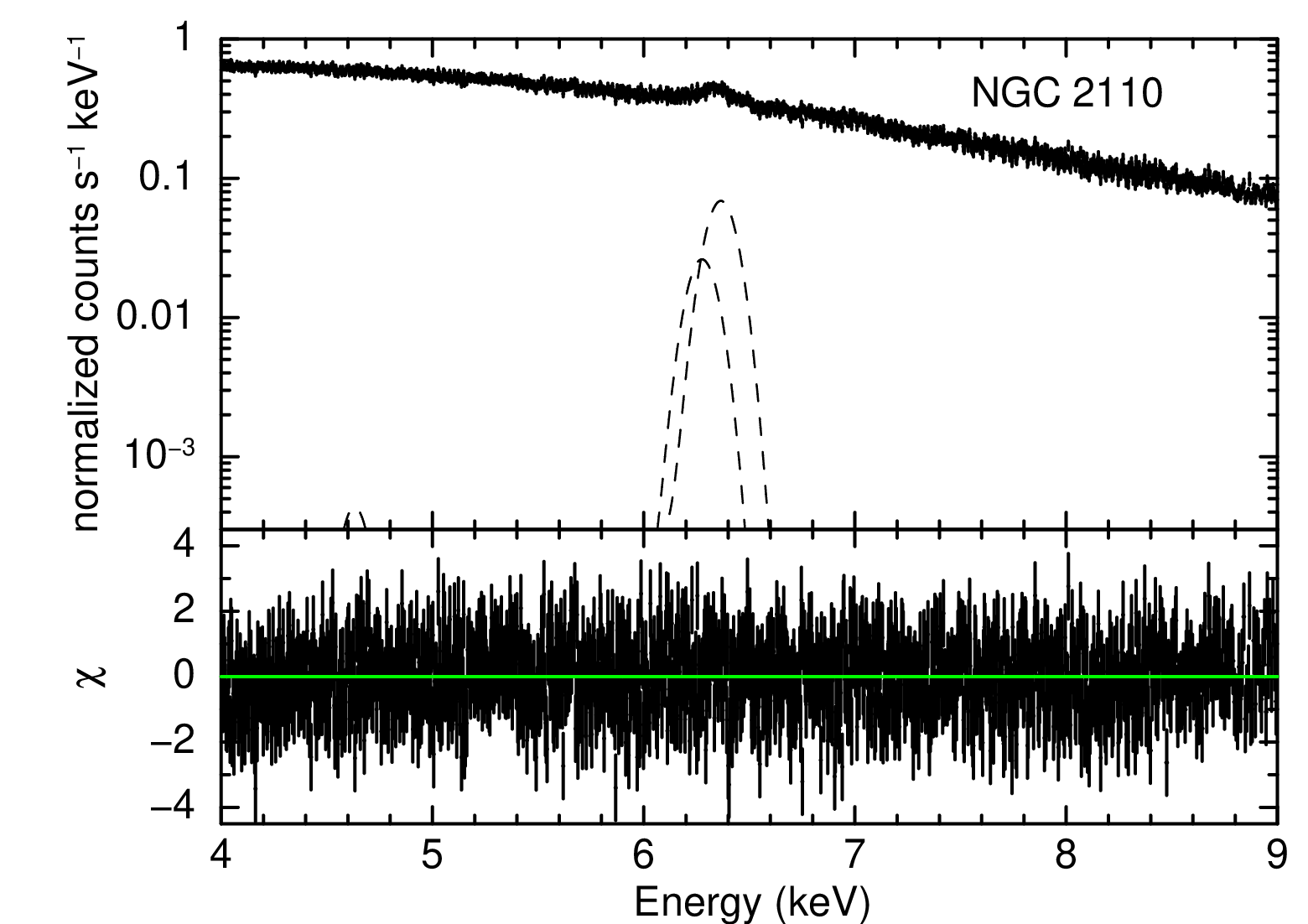}}
\resizebox{8cm}{!}{\includegraphics[scale=0.8,angle=-90]{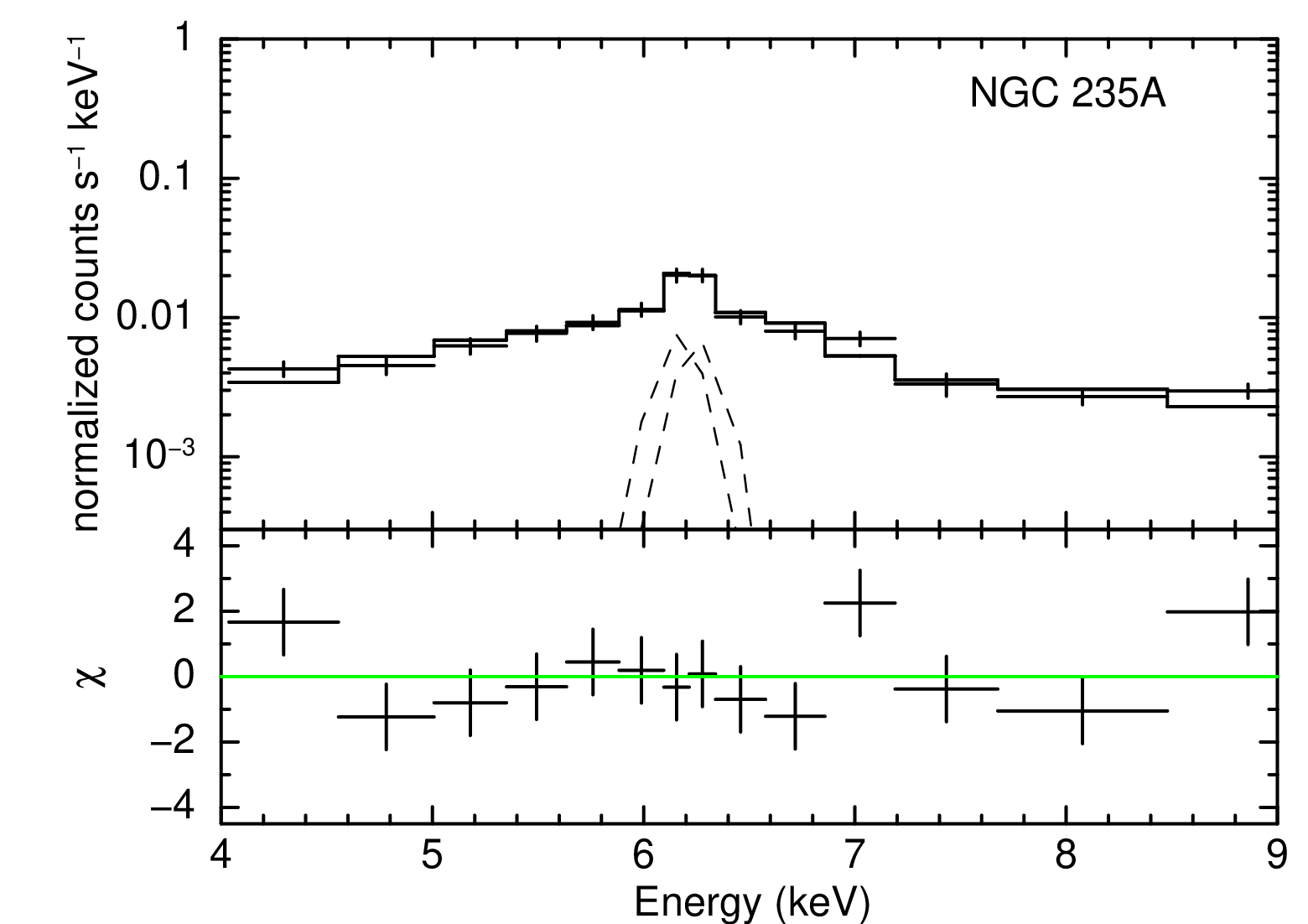}}
}\vspace{0.0cm}
\caption{
Continued.
}
\label{fig:narrow_spec}
\end{figure*}
\addtocounter{figure}{-1}
\begin{figure*}
\centering
\subfigure{
\resizebox{8cm}{!}{\includegraphics[scale=0.8,angle=-90]{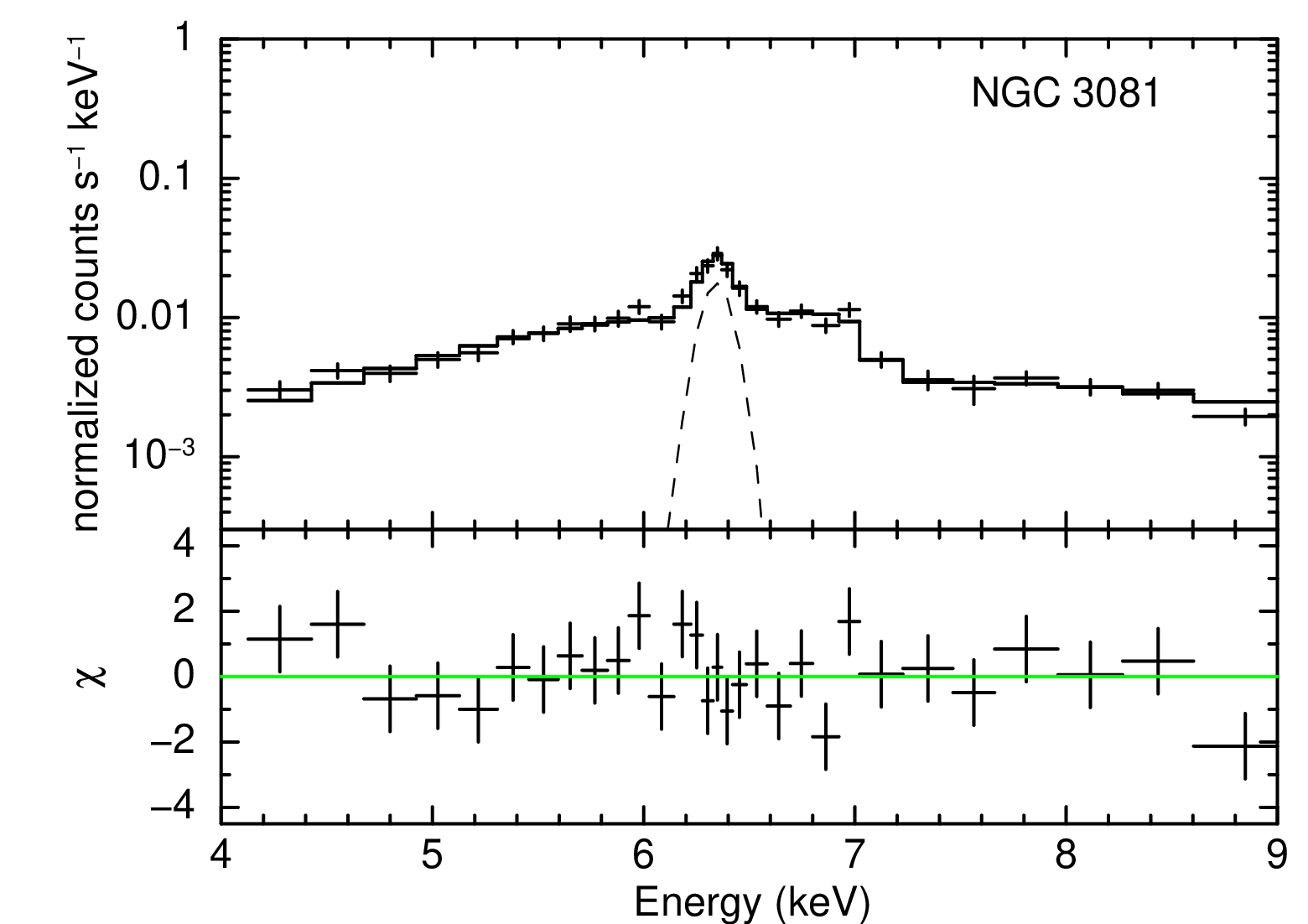}}
\resizebox{8cm}{!}{\includegraphics[scale=0.8,angle=-90]{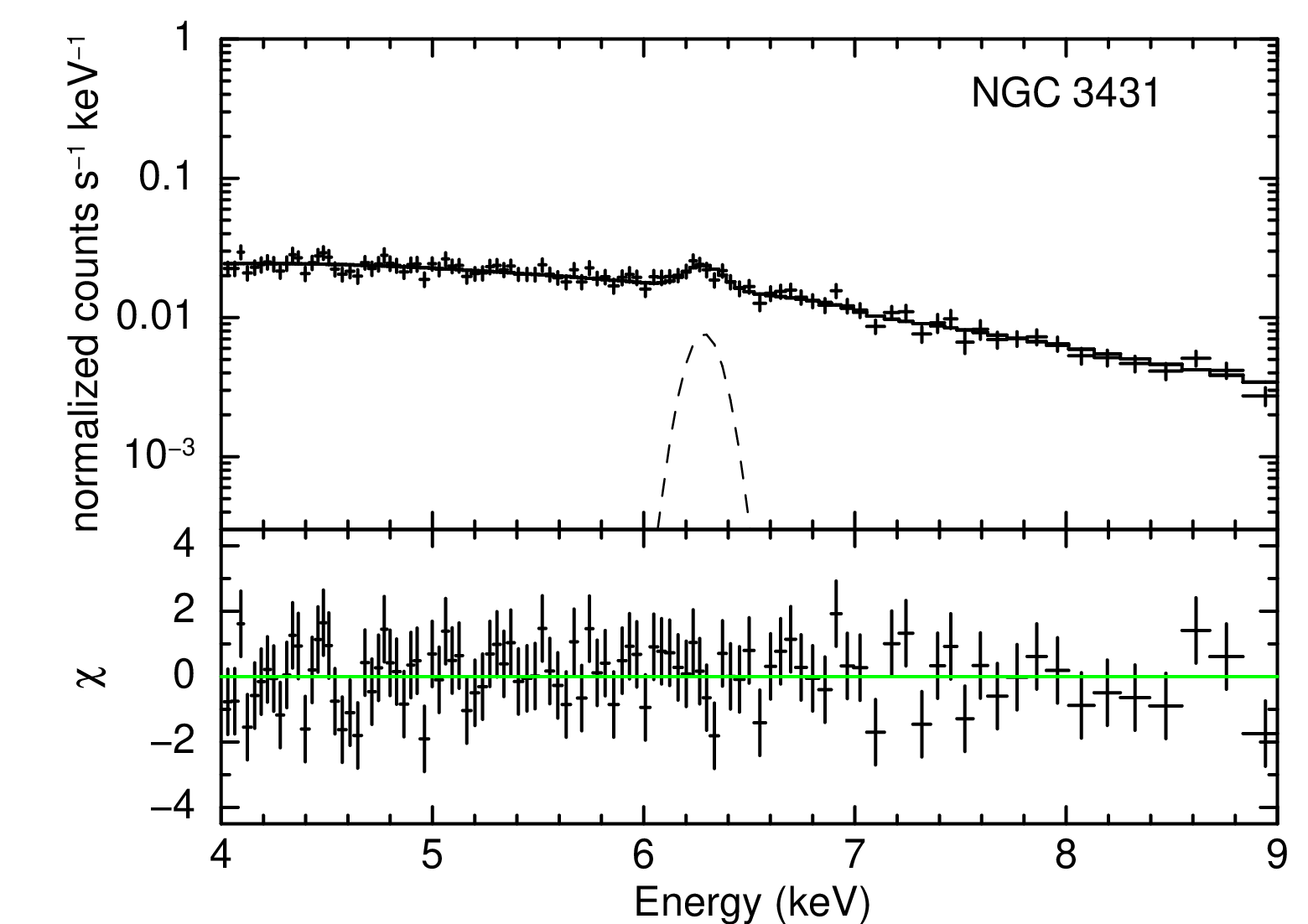}}
}\vspace{0.0cm}
\subfigure{
\resizebox{8cm}{!}{\includegraphics[scale=0.8,angle=-90]{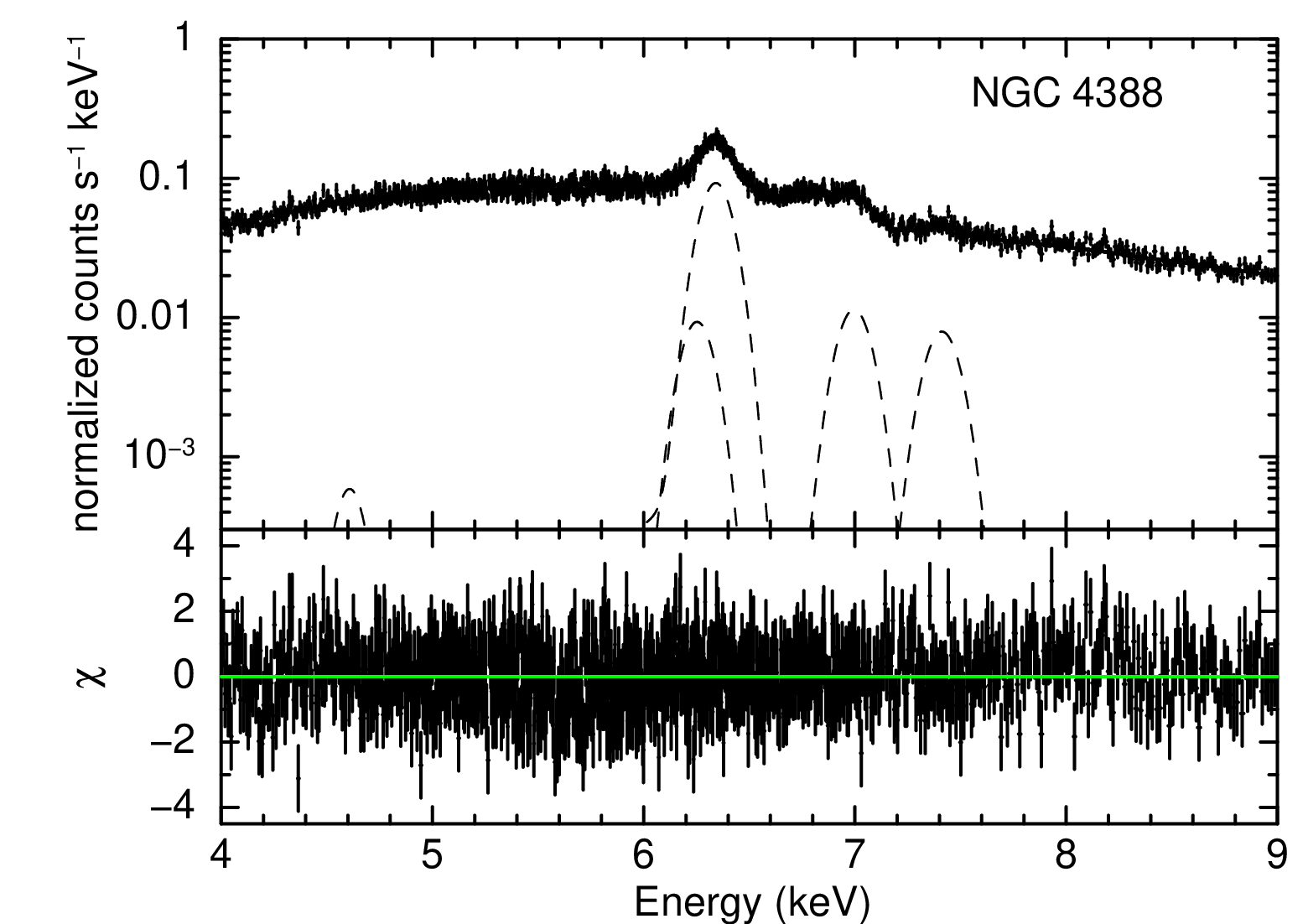}}
\resizebox{8cm}{!}{\includegraphics[scale=0.8,angle=-90]{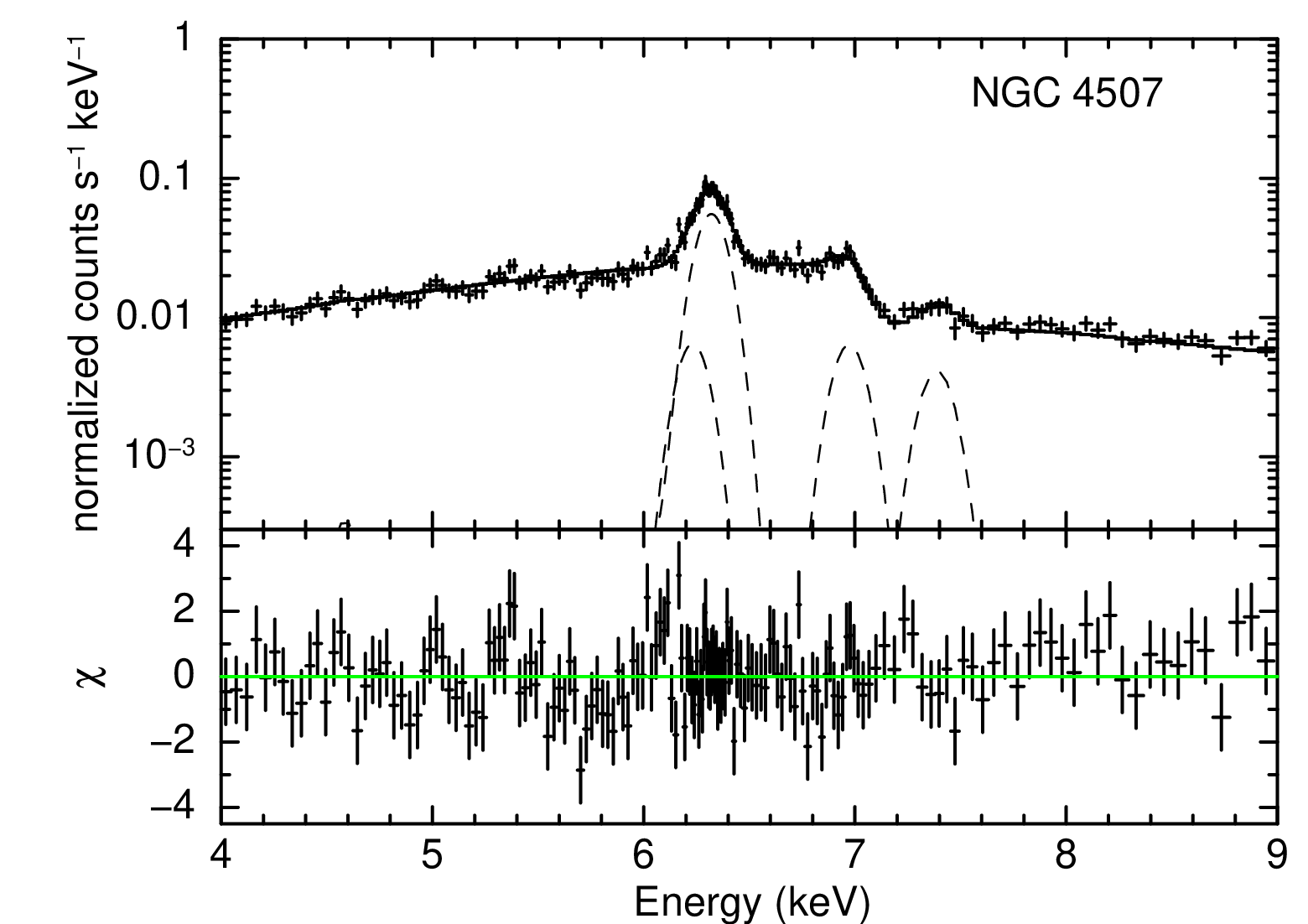}}
}\vspace{0.0cm}
\subfigure{
\resizebox{8cm}{!}{\includegraphics[scale=0.8,angle=-90]{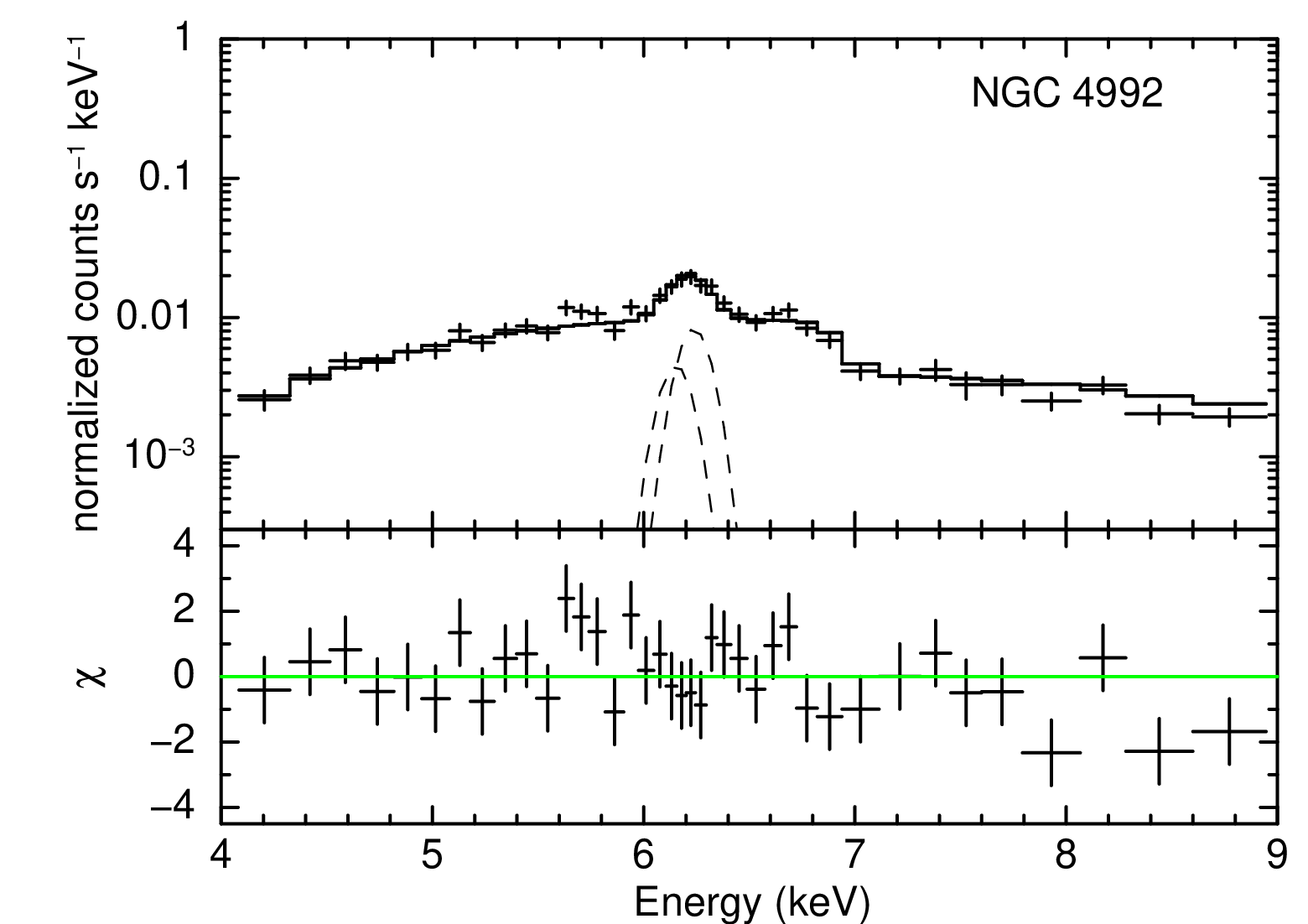}}
\resizebox{8cm}{!}{\includegraphics[scale=0.8,angle=-90]{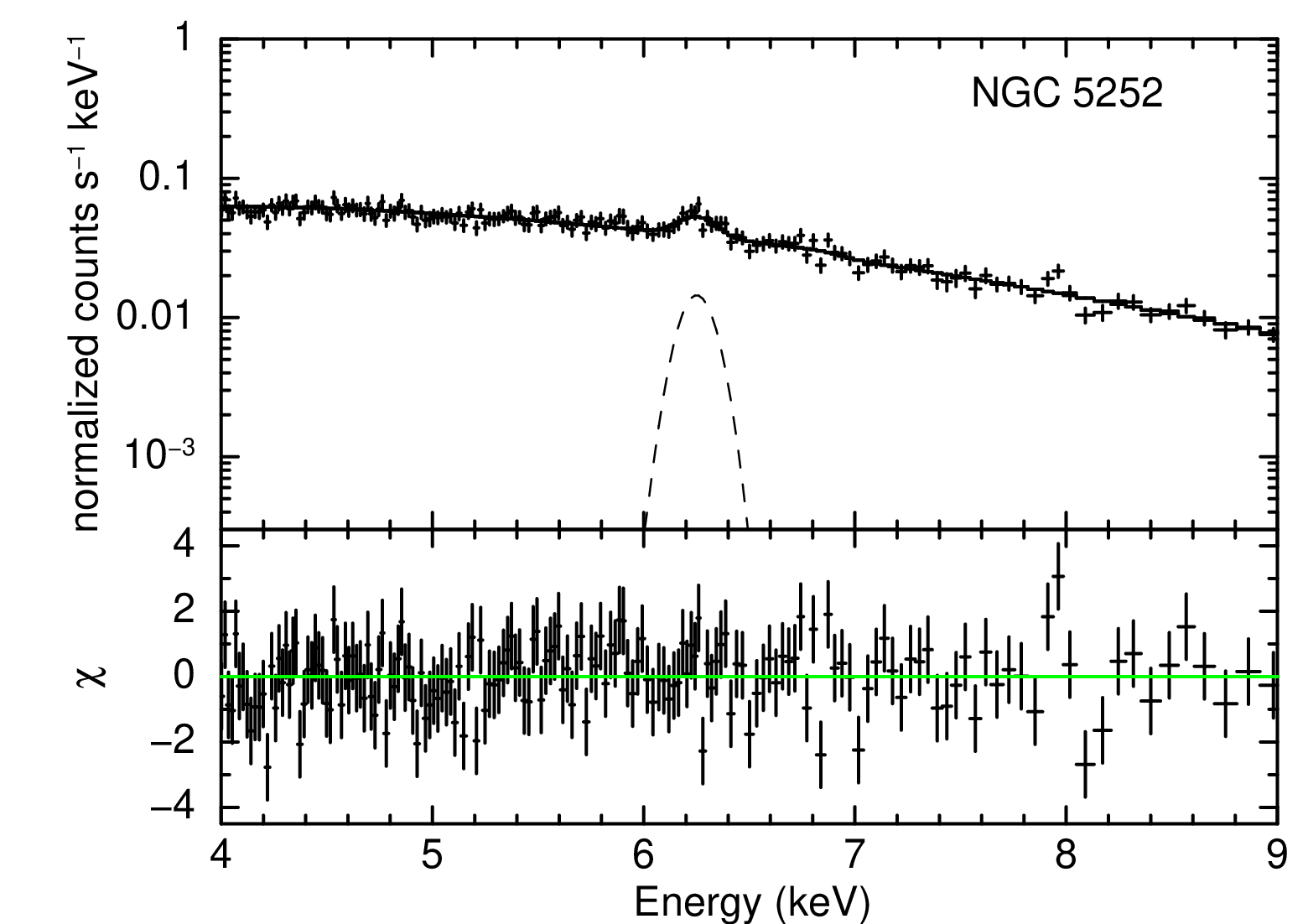}}
}\vspace{0.0cm}
\subfigure{
\resizebox{8cm}{!}{\includegraphics[scale=0.8,angle=-90]{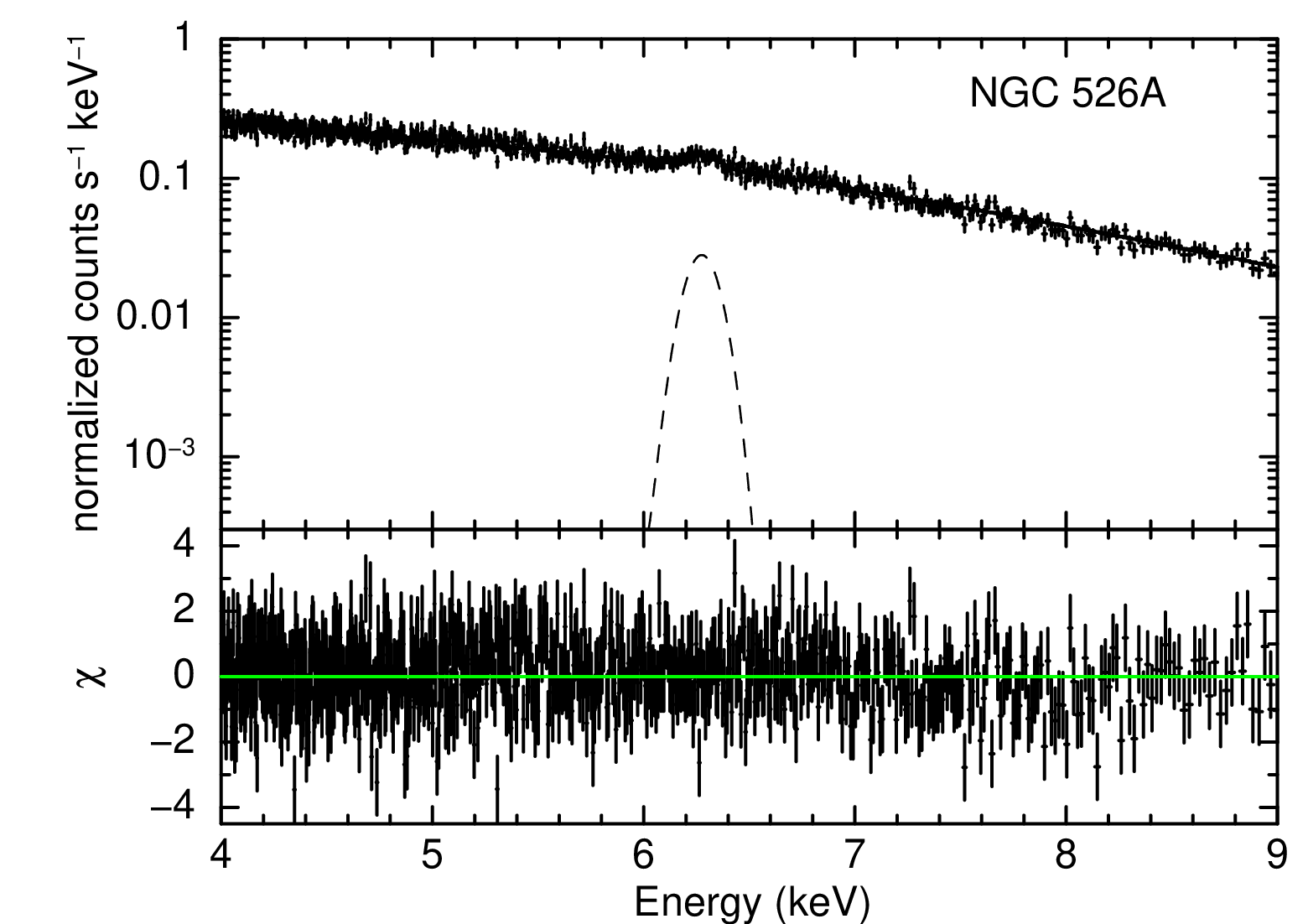}}
\resizebox{8cm}{!}{\includegraphics[scale=0.8,angle=-90]{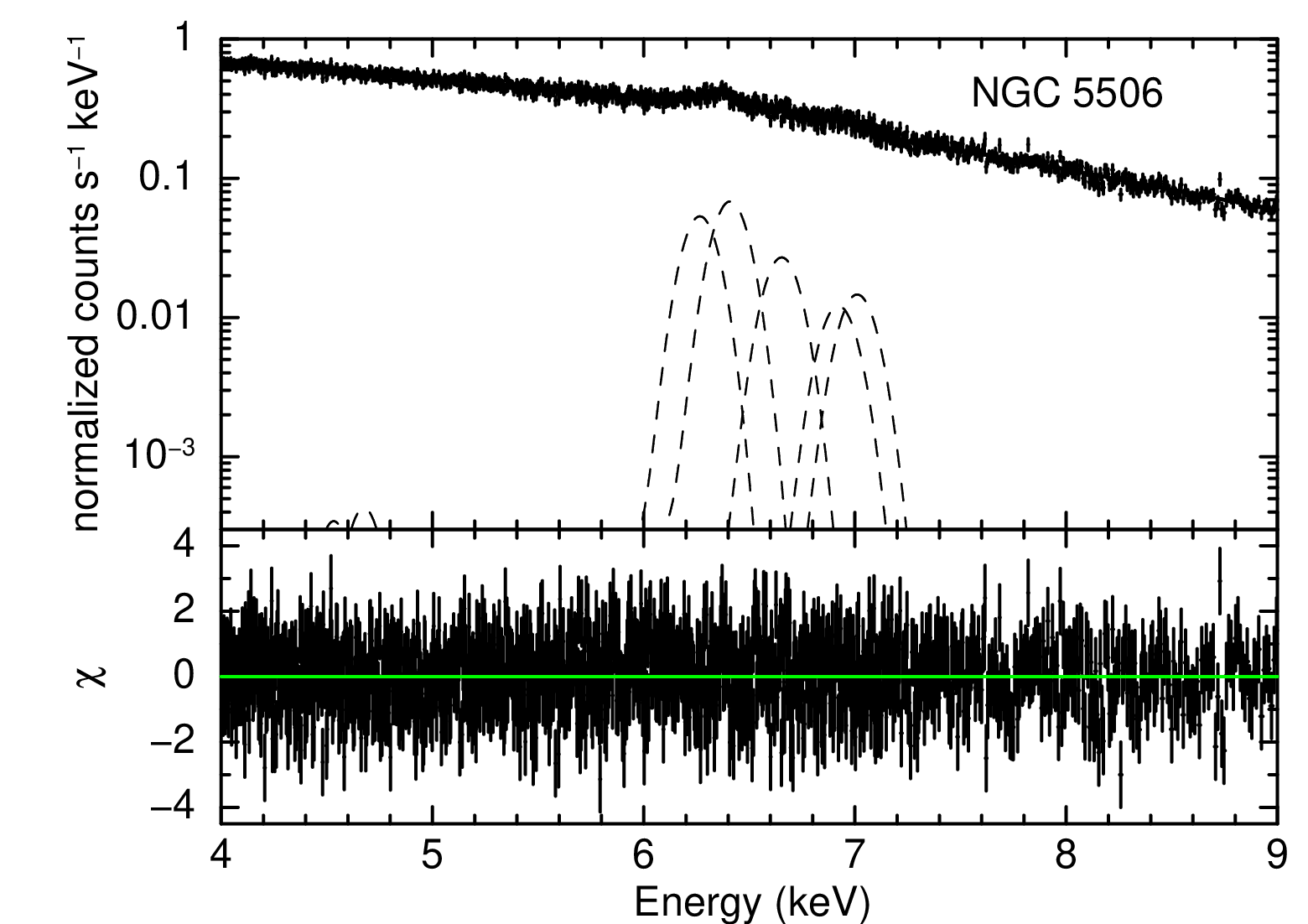}}
}\vspace{0.0cm}
\caption{
Continued.
}
\label{fig:narrow_spec}
\end{figure*}
\addtocounter{figure}{-1}
\begin{figure*}
\centering
\subfigure{
\resizebox{8cm}{!}{\includegraphics[scale=0.8,angle=-90]{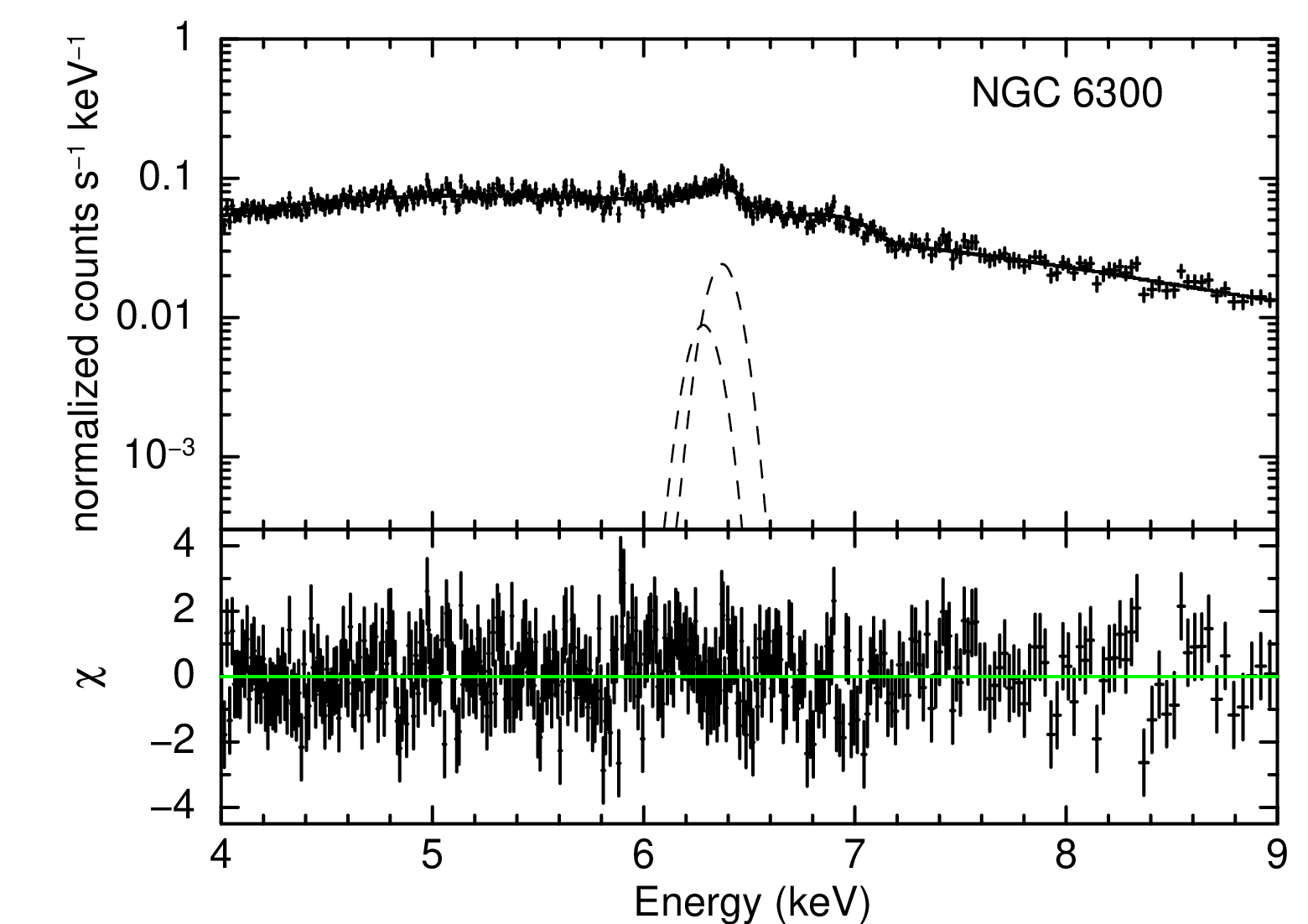}}
\resizebox{8cm}{!}{\includegraphics[scale=0.8,angle=-90]{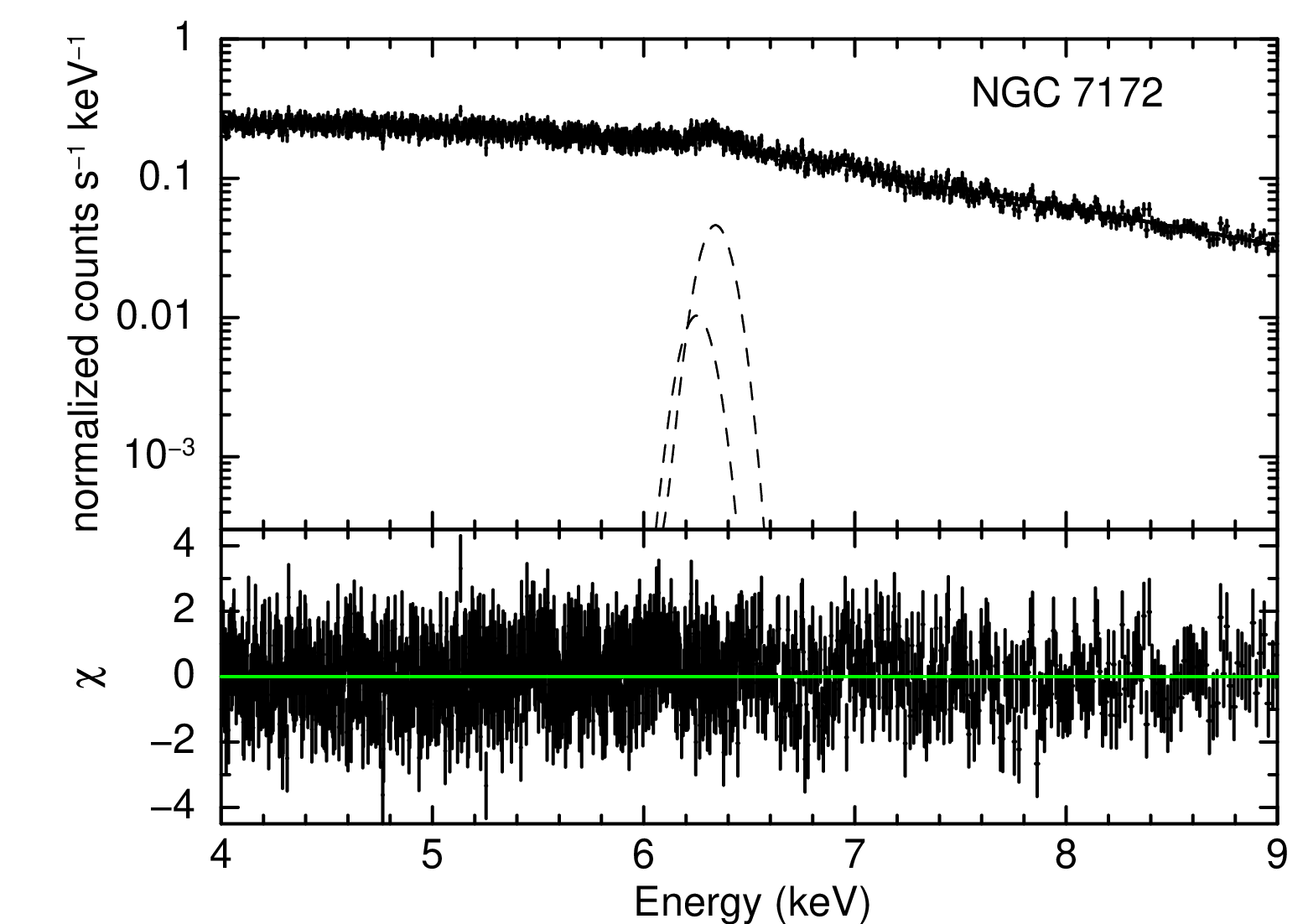}}
}\vspace{0.0cm}
\subfigure{
\resizebox{8cm}{!}{\includegraphics[scale=0.8,angle=-90]{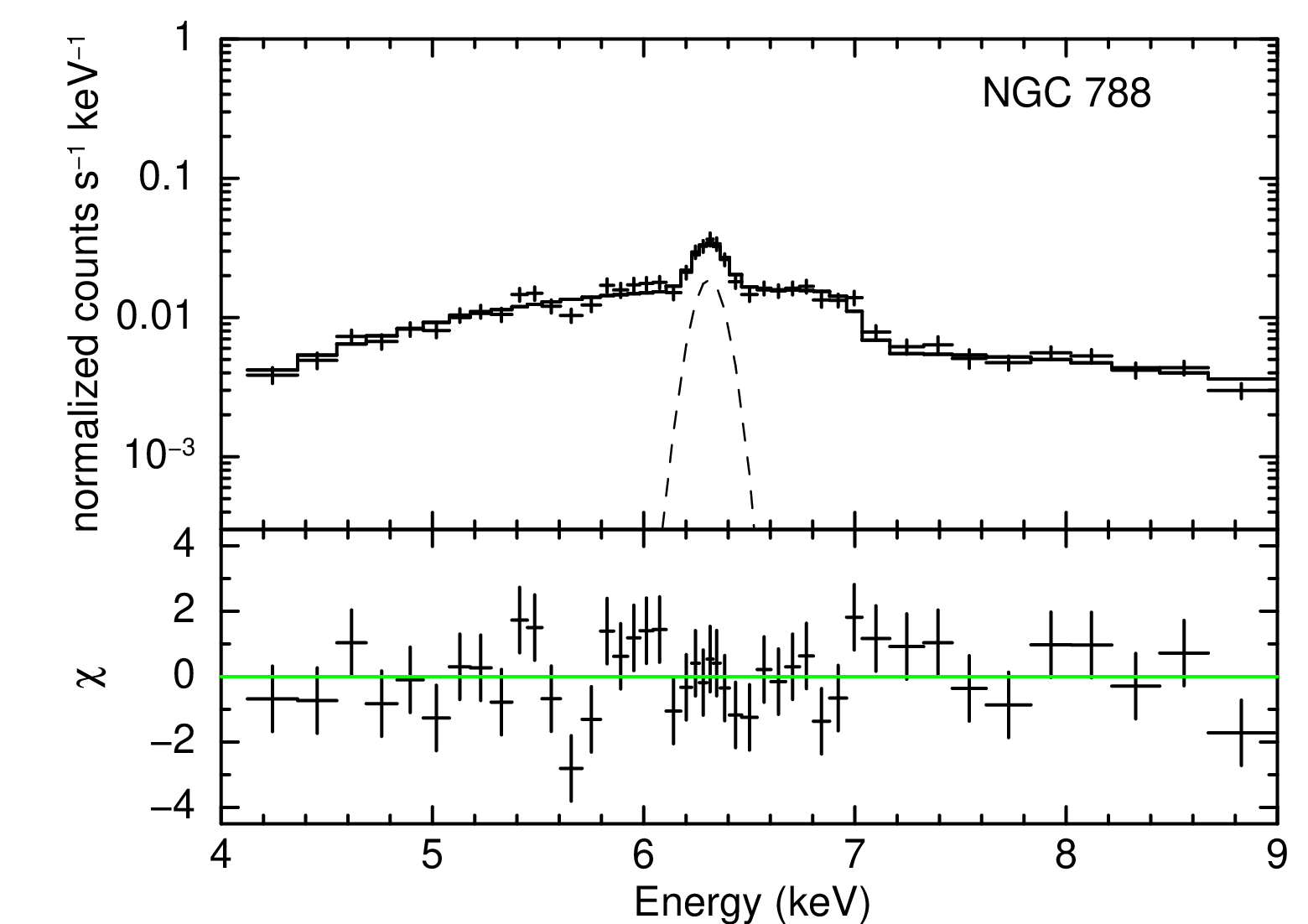}}
\resizebox{8cm}{!}{\includegraphics[scale=0.8,angle=-90]{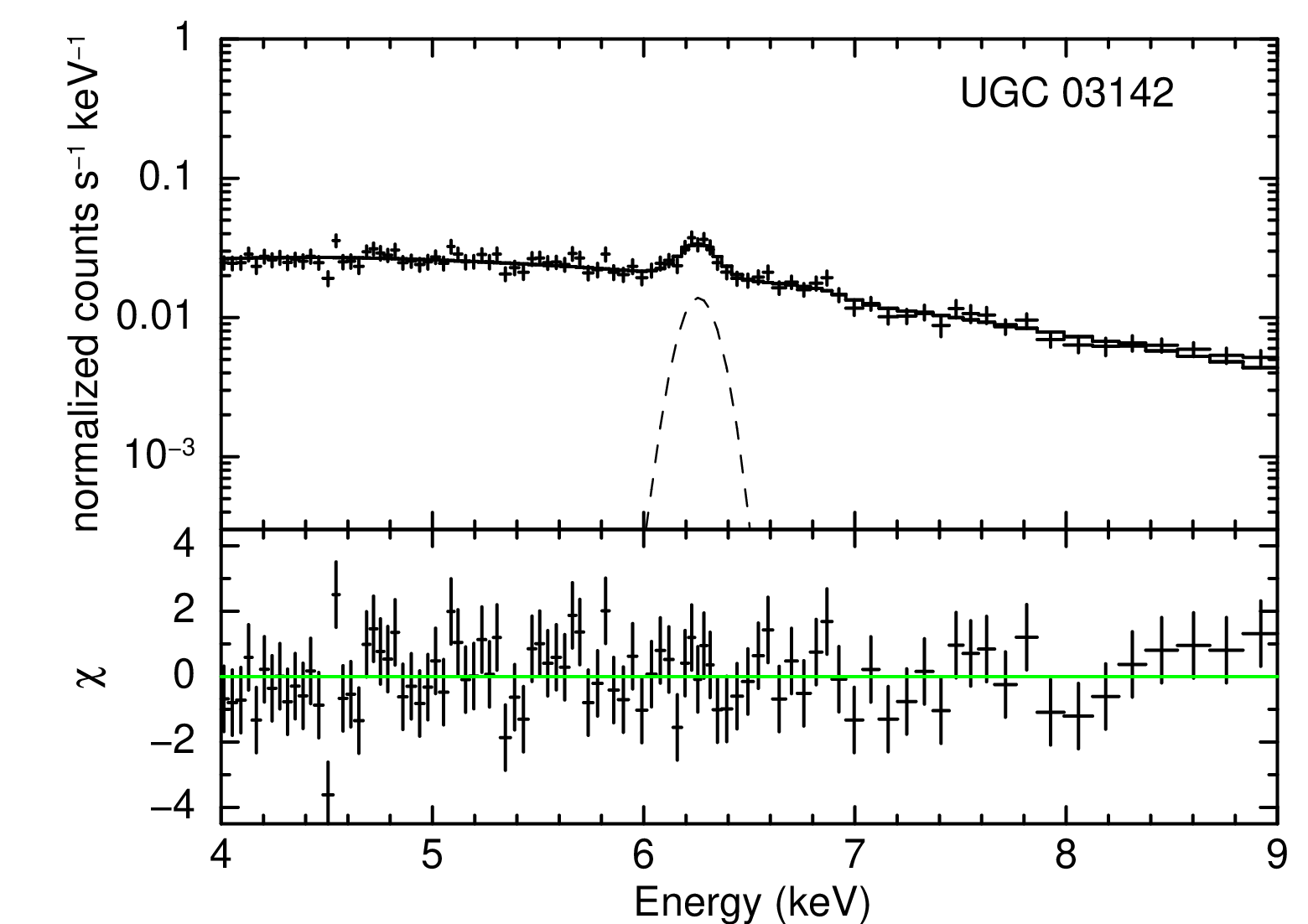}}
}\vspace{0.0cm}
\subfigure{
\resizebox{8cm}{!}{\includegraphics[scale=0.8,angle=-90]{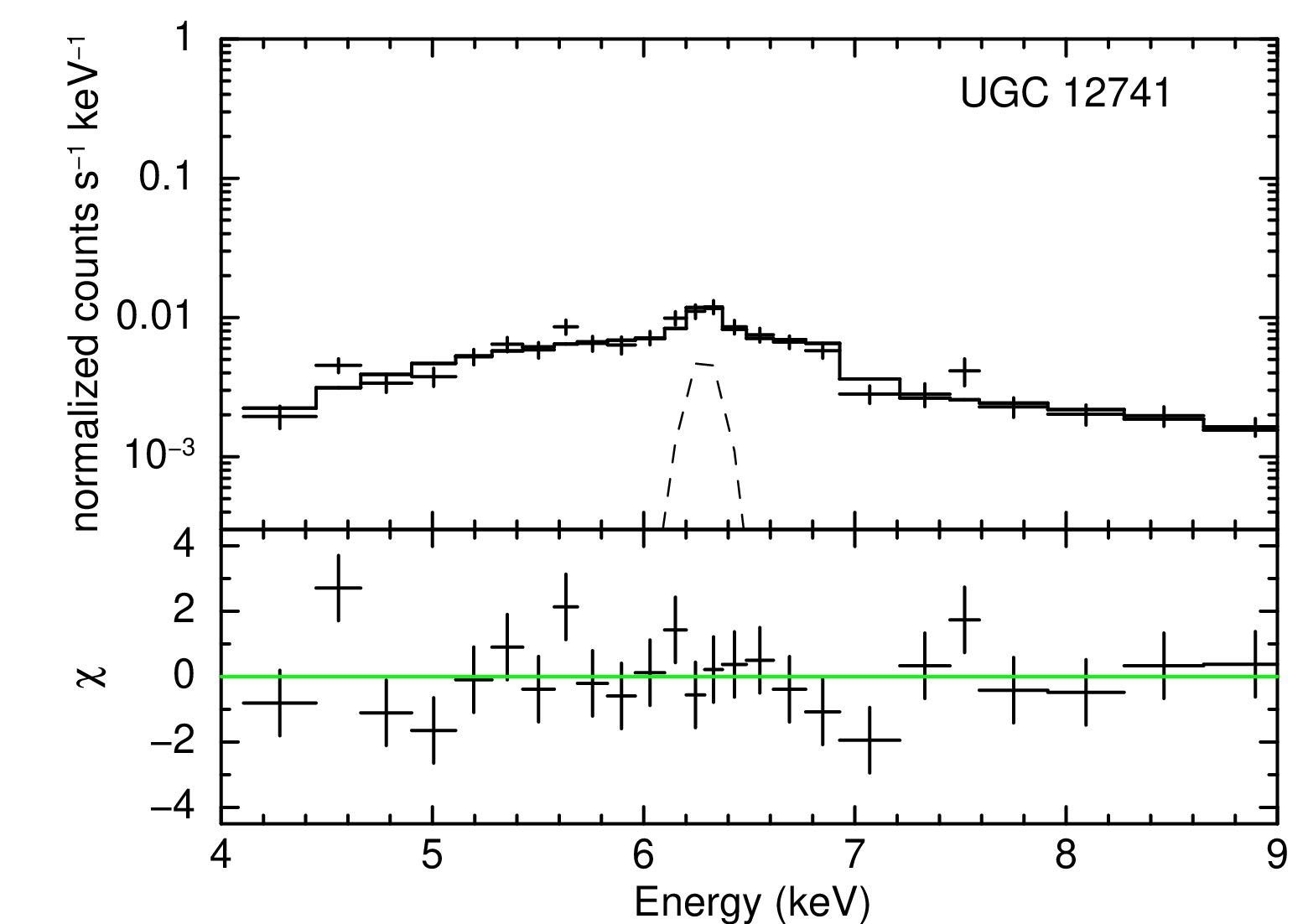}}
%\resizebox{8cm}{!}{\includegraphics[scale=0.8,angle=-90]{fig_b46.ps}}
}\vspace{0.0cm}
\caption{
Continued.
}
\label{fig:narrow_spec}
\end{figure*}

 % IRU 

\newpage 
\clearpage 
\clearpage 
\section{Spectra fitted with the relativistic reflection component} 
\begin{figure*}[h]
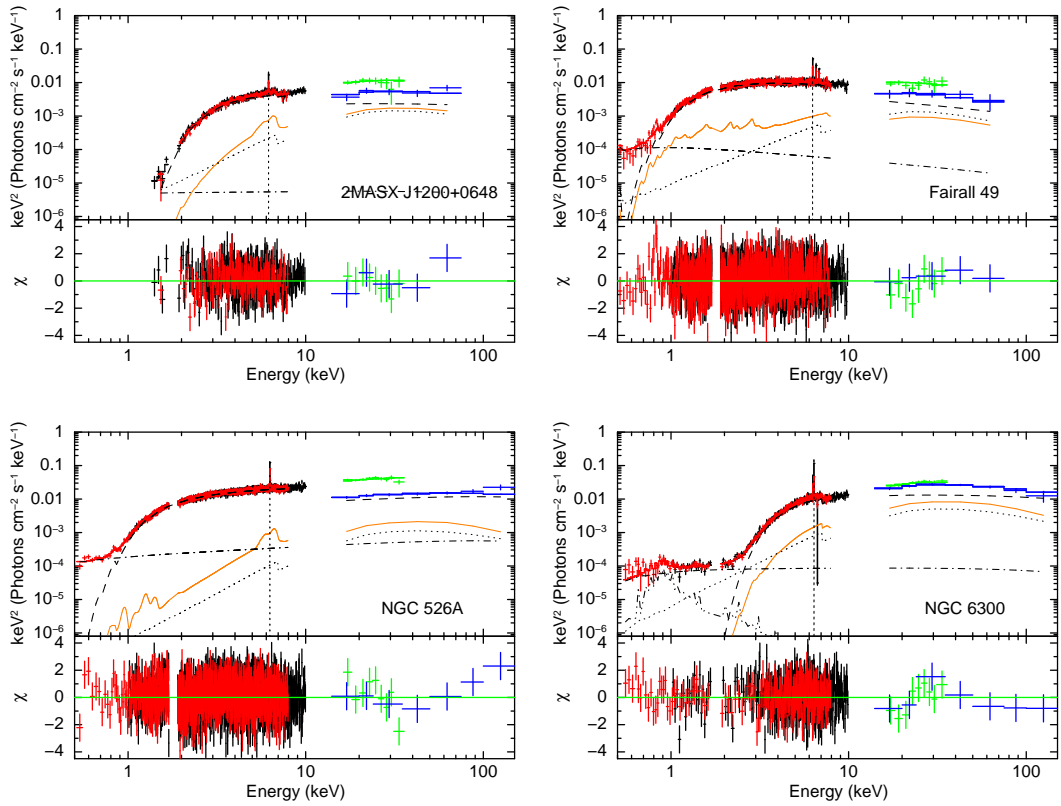

\centering
\subfigure{
\resizebox{7.1cm}{!}{\includegraphics[scale=0.8,angle=-90]{fig_c01.ps}}
\resizebox{7.1cm}{!}{\includegraphics[scale=0.8,angle=-90]{fig_c02.ps}}
}\vspace{0.0cm}
\subfigure{
% \resizebox{7.1cm}{!}{\includegraphics[scale=0.8,angle=-90]{fig_c03.ps}}
\resizebox{7.1cm}{!}{\includegraphics[scale=0.8,angle=-90]{fig_c05.ps}}
%\resizebox{7.1cm}{!}{\includegraphics[scale=0.8,angle=-90]{fig_c04.ps}}
\resizebox{7.1cm}{!}{\includegraphics[scale=0.8,angle=-90]{fig_c07.ps}}
}\vspace{0.0cm}
\caption{
Unfolded spectra and best-fitting models obtained in Section~\ref{sec:disk}.
The symbols and colors are the same as in Figure~\ref{fig:spec} but for the disk component represented
with the orange solid line.
}
\label{fig:disk}
\end{figure*}

  % IRU 
\clearpage 
\setlength{\tabcolsep}{0.20in}
\tabletypesize{\normalsize}
\begin{deluxetable*}{cccc}[H] 
\tablecaption{Disk parameters\label{tab:info_disk_para}}
\tablehead{
\colhead{Target Name} & \colhead{$r_{\rm in}$}  
& \colhead{$R_{\rm Disk}$} 
& \colhead{$-\Delta \chi^2$} \\
\colhead{(1)}  & \colhead{(2)}  & \colhead{(3)} & \colhead{(4)} 
} 
\thispagestyle{empty}
\startdata
 2MASX J0216+5126 & $ 100^{}_{-99}/100^{}_{-99}  $ & $ 0.00^{+0.15}_{}/0.00^{+0.09}_{}  $ & $ 0.0/0.0  $ \\
 2MASX J0248+2630 & $ 1^{+99}_{}/1^{+99}_{}  $ & $ 0.40^{+0.57}_{-0.40}/0.04^{+0.27}_{-0.04}  $ & $ 0.1/0.0  $ \\
 2MASX J0318+6829 & $ 10^{+22}_{-5}/10^{+23}_{-6}  $ & $ 0.62^{+0.32}_{-0.46}/0.55^{+0.16}_{-0.42}  $ & $ 1.7/1.6  $ \\
 2MASX J0350-5018 & $ 100^{}_{-99}/100^{}_{-99}  $ & $ 0.0^{+1.0}_{}/0.0^{+1.0}_{}  $ & $ 0.0/0.0  $ \\
 2MASX J0444+2813 & $ 25^{+19}_{-8}/25^{+19}_{-9}  $ & $ 1.00^{}_{-0.27}/1.00^{}_{-0.26}  $ & $ 5.3/5.7  $ \\
 2MASX J0505-2351 & $ 33^{+67}_{-32}/35^{+65}_{-34}  $ & $ 0.12^{+0.05}_{-0.12}/0.00^{+0.20}_{}  $ & $ 0.6/0.0  $ \\
 2MASX J0911+4528 & $ 100^{}_{-61}/100^{}_{-99}  $ & $ 0.22^{+0.19}_{-0.17}/0.22^{+0.35}_{-0.22}  $ & $ 1.6/0.7  $ \\
{\bf 2MASX J1200+0648} & ${\bf 29^{+24}_{-10}/20^{+23}_{-5}}  $ & ${\bf 0.62^{+0.18}_{-0.19}/0.51^{+0.13}_{-0.10} } $ & ${\bf 19.1/13.0 } $ \\
 Ark 347 & $ 1^{+99}_{}/1^{+99}_{}  $ & $ 0.00^{+0.54}_{}/0.0^{+1.0}_{}  $ & $ 0.0/0.0  $ \\
 ESO 103-035 & $ 4^{+3}_{-2}/4^{+3}_{-2}  $ & $ 0.17^{+0.08}_{-0.03}/0.25^{+0.09}_{-0.07}  $ & $ 5.8/5.7  $ \\
 ESO 263-G013 & $ 3^{+97}_{-2}/3^{+97}_{-2}  $ & $ 0.13^{+0.50}_{-0.13}/0.12^{+0.52}_{-0.12}  $ & $ 0.4/0.4  $ \\
 ESO 297-G018 & $ 63^{+37}_{-35}/64^{+36}_{-42}  $ & $ 0.58^{+0.42}_{-0.40}/0.48^{+0.52}_{-0.37}  $ & $ 2.4/1.8  $ \\
 ESO 506-G027 & $ 100^{}_{-99}/100^{}_{-99}  $ & $ 0.00^{+0.32}_{}/0.00^{+0.27}_{}  $ & $ 0.0/0.0  $ \\
{\bf Fairall 49} & $ 30^{+70}_{-17}/{\bf 4\pm2} $ & $ 0.11^{+0.09}_{-0.10}/{\bf 0.30^{+0.08}_{-0.06}}  $ & $ 1.0/{\bf 23.2}  $ \\
 Fairall 51 & $ 100^{}_{-56}/20^{+14}_{-7}  $ & $ 0.25^{+0.12}_{-0.09}/0.34^{+0.12}_{-0.14}  $ & $ 7.5/5.3  $ \\
 IC 4518A & $ 16^{+4}_{-5}/16^{+17}_{-5}  $ & $ 0.89^{+0.11}_{-0.36}/0.99^{+0.01}_{-0.43}  $ & $ 6.5/5.1  $ \\
 LEDA 170194 & $ 26^{+74}_{-25}/92^{+8}_{-91}  $ & $ 0.00^{+0.09}_{}/0.00^{+0.15}_{}  $ & $ 0.0/0.0  $ \\
 MCG +04-48-002 & $ 1^{+99}_{}/1^{+99}_{}  $ & $ 0.0^{+1.0}_{}/0.27^{+0.73}_{-0.27}  $ & $ 0.0/0.0  $ \\
 MCG -01-05-047 & $ 25^{+36}_{-9}/25^{+75}_{-23}  $ & $ 1.00^{}_{-0.58}/1.00^{}_{-0.47}  $ & $ 3.0/3.0  $ \\
 MCG -02-08-014 & $ 20^{+80}_{-9}/25^{+75}_{-10}  $ & $ 0.53^{+0.34}_{-0.33}/0.89^{+0.11}_{-0.41}  $ & $ 3.1/6.5  $ \\
 MCG -05-23-016 & $ 100^{}_{-36}/4^{+2}_{-3}  $ & $ 0.07\pm0.04/0.03\pm0.01 $ & $ 3.7/0.9  $ \\
 Mrk 1210 & $ 10^{+15}_{-5}/16^{+16}_{-9}  $ & $ 1.00^{}_{-0.39}/1.00^{}_{-0.21}  $ & $ 7.1/9.0  $ \\
 Mrk 1498 & $ 2\pm1/2^{+12}_{-1} $ & $ 0.96^{+0.04}_{-0.38}/1.00^{}_{-0.44}  $ & $ 6.5/6.3  $ \\
 Mrk 18 & $ 100^{}_{-99}/100^{}_{-99}  $ & $ 0.00^{+0.47}_{}/0.00^{+0.58}_{}  $ & $ 0.0/0.0  $ \\
 Mrk 348 & $ 77^{+23}_{-76}/18^{+82}_{-17}  $ & $ 0.00^{+0.06}_{}/0.00^{+0.06}_{}  $ & $ 0.1/0.1  $ \\
 Mrk 417 & $ 1^{+99}_{}/3^{+97}_{-2}  $ & $ 0.00^{+0.61}_{}/0.00^{+0.45}_{}  $ & $ 0.0/0.0  $ \\
 Mrk 520 & $ 46^{+54}_{-45}/30^{+70}_{-29}  $ & $ 0.02^{+0.27}_{-0.02}/0.00^{+0.14}_{}  $ & $ 0.1/0.0  $ \\
 Mrk 915 & $ 100^{}_{-99}/100^{}_{-99}  $ & $ 0.03^{+0.53}_{-0.03}/0.04^{+0.56}_{-0.04}  $ & $ 0.0/0.0  $ \\
 NGC 1052 & $ 100^{}_{-37}/100^{+-100}_{-100}  $ & $ 0.39^{+0.20}_{-0.17}/0.32^{+-0.32}_{-0.32}  $ & $ 4.3/2.7  $ \\
 NGC 1142 & $ 70^{+30}_{-69}/28^{+72}_{-27}  $ & $ 0.00^{+0.35}_{}/0.00^{+0.16}_{}  $ & $ 0.0/0.0  $ \\
 NGC 2110 & $ 25^{+75}_{-15}/1^{+6}_{}  $ & $ 0.05^{+0.03}_{-0.02}/0.08\pm0.04 $ & $ 1.1/1.3  $ \\
 NGC 235A & $ 11^{+31}_{-10}/1^{+99}_{}  $ & $ 0.86^{+0.14}_{-0.72}/0.73^{+0.27}_{-0.65}  $ & $ 1.5/1.3  $ \\
 NGC 3081 & $ 100^{}_{-87}/100^{}_{-88}  $ & $ 0.57^{+0.43}_{-0.30}/0.63^{+0.37}_{-0.32}  $ & $ 4.1/4.2  $ \\
 NGC 3431 & $ 25^{+19}_{-13}/7^{+5}_{-6}  $ & $ 0.66^{+0.32}_{-0.35}/1.00^{}_{-0.28}  $ & $ 2.5/5.6  $ \\
 NGC 4388 & $ 100^{}_{-99}/100^{}_{-99}  $ & $ 0.00^{+0.03}_{}/0.00^{+0.03}_{}  $ & $ 0.0/0.0  $ \\
 NGC 4507 & $ 1^{+6}_{}/100^{}_{-32}  $ & $ 0.59^{+0.28}_{-0.23}/0.28^{+0.18}_{-0.15}  $ & $ 8.8/3.9  $ \\
 NGC 4992 & $ 65^{+35}_{-21}/65^{+35}_{-21}  $ & $ 1.00^{}_{-0.20}/1.00^{}_{-0.20}  $ & $ 8.4/8.2  $ \\
 NGC 5252 & $ 5^{+95}_{-4}/40^{+60}_{-39}  $ & $ 0.00^{+0.08}_{}/0.00^{+0.06}_{}  $ & $ 0.0/0.0  $ \\
{\bf NGC 526A} & ${\bf 44^{+33}_{-11}}/33^{+67}_{-32}  $ & ${\bf 0.26^{+0.07}_{-0.03}}/0.00^{+0.05}_{}  $ & ${\bf 9.3}/0.0  $ \\
 NGC 5506 & $ 100^{}_{-36}/100^{}_{-99}  $ & $ 0.08^{+0.06}_{-0.04}/0.00^{+0.01}_{}  $ & $ 5.1/0.0  $ \\
{\bf NGC 6300} & ${\bf 2^{}_{-1}/2^{}_{-1}}  $ & ${\bf 0.39\pm0.11/0.56\pm0.16} $ & $ {\bf 9.7/13.7}  $ \\
 NGC 7172 & $ 100^{}_{-36}/100^{}_{-83}  $ & $ 0.10^{+0.03}_{-0.02}/0.07\pm0.06 $ & $ 3.5/1.1  $ \\
 NGC 788 & $ 3^{+97}_{-2}/3^{+97}_{-2}  $ & $ 0.21^{+0.55}_{-0.21}/0.19^{+0.57}_{-0.19}  $ & $ 0.2/0.1  $ \\
 UGC 03142 & $ 44^{+56}_{-43}/32^{+68}_{-31}  $ & $ 0.54^{+0.46}_{-0.54}/0.55^{+0.45}_{-0.55}  $ & $ 0.9/0.7  $ \\
 UGC 12741 & $ 100^{}_{-49}/100^{}_{-99}  $ & $ 0.49^{+0.51}_{-0.42}/0.46^{+0.54}_{-0.46}  $ & $ 1.4/1.0  $
\enddata
\tablecomments{(1) Galaxy name. 
(2) Inner radius in units of $r_{\rm g}$ for the assumed ionization parameters, $\xi = 10$ and $100$. 
(3) Equivalent reflection strength for the same ionization parameters as (2). 
(4) Difference of the chi-squred value before and after adding the disk components 
for the same ionization parameters as (2). 
AGNs for which fitting results are significantly improved by inclusion of the relativistic reflection 
components from a disk are represented in boldface. 
}
\end{deluxetable*}
 % IRU 
\section{Information of hydrogen column density}
%\begin{landscape}
%\LongTables
\renewcommand{\arraystretch}{1.13}
\setlength{\topmargin}{-2.5cm}
\tabletypesize{\scriptsize}
\setlength{\tabcolsep}{0.25in}
\begin{deluxetable*}{cccccccccccc}
\tablecaption{Information of hydrogen column density\label{tab:info_nh}}
\tablewidth{0pt}
\tablehead{\colhead{Target Name}
& \colhead{$N_{\rm H}$} 
& \colhead{Obs. date} 
& \colhead{Observatory} 
& \colhead{Ref. $N_{\rm H}$} 
\\
\colhead{(1)} & \colhead{(2)} & \colhead{(3)} & \colhead{(4)} &
\colhead{(5)} 
}
\startdata 
 2MASX J0216+5126 & $ 1.74^{+0.06}_{-0.07} $ & 2006-01-24 & {\it XMM-Newton} & 1 \\ % Winter+08 \\
 2MASX J0318+6829 & $ < 14 $ & 2006-01-29 & {\it XMM-Newton} & 1 \\ % Winter+08 \\
 2MASX J0505-2351 & $ 9.90\pm0.30 $ & 2009-08-06 & {\it XMM-Newton} & 2 \\ % Vasudevan+13 \\
 2MASX J0911+4528 & $ 48^{+28}_{-25} $ & 2006-04-10 & {\it XMM-Newton} & 1 \\ % Winter+08 \\
 2MASX J1200+0648 & $ 10.60^{+0.80}_{-1.01} $ & 2006-06-26 & {\it XMM-Newton} & 1 \\ % Winter+08 \\
 Ark 347 & $ 19.2^{+4.4}_{-3.5} $ & 2003-01-02 & {\it XMM-Newton} & 3 \\ %  Noguchi+09 \\
 ESO 103-035 & $ 18.9^{+0.6}_{-1.1} $ & 2002-03-15 & {\it XMM-Newton} & 4 \\ % De-Rosa+12 \\
 ESO 263-G013 & $ 25.7^{+1.5}_{-1.4} $ & 2007-06-14 & {\it XMM-Newton} & 4 \\ % De-Rosa+12 \\
 ESO 506-G027 & $ 66.0^{+5.0}_{-4.8} $ & 2006-01-24 & {\it XMM-Newton} & 3 \\ % Noguchi+09 \\
 Fairall 49 
  & $ 1.08\pm0.02 $ & 2001-03-05 & {\it XMM-Newton} & 5 \\ % Tripathi+13 \\
  & $ 1.06\pm0.02 $ & 2001-03-06 & {\it XMM-Newton} & 5 \\ % Tripathi+13 \\
  & $ 1.46\pm0.01 $ & 2013-09-04 & {\it XMM-Newton} & 6 \\ % Lobban+14 \\
  & $ 1.29^{+0.01}_{-0.02} $ & 2013-10-15 & {\it XMM-Newton} & 6 \\ % Lobban+14 \\
 Fairall 51
  & $ 3.43^{+0.29}_{-0.31} $ & 2013-09-05 & {\it Suzaku}  & this work \\ %  \\
  & $ 4.49^{+0.46}_{-0.49} $ & 2013-09-07 & {\it Suzaku}  & this work \\ %  \\
  & $ 2.67^{+1.09}_{-0.99} $ & 2013-09-13  & {\it Suzaku} & this work \\ %  \\
 IC 4518A & $ 14.0^{+3.0}_{-1.0} $ & 2006-08-07 & {\it XMM-Newton} & 4 \\ % De-Rosa+12 \\
 LEDA 170194 & $ 2.9^{+1.3}_{-0.3} $ & 2005-07-25 & {\it Chandra} & 4 \\ % De-Rosa+12 \\
 MCG +04-48-002 & $ 57.4^{+9.6}_{-6.6} $ & 2006-04-23 & {\it XMM-Newton} & 4 \\ % De-Rosa+12 \\
 MCG -01-05-047 & $ 26.3\pm1.0 $ & 2009-07-24 & {\it XMM-Newton} & 7 \\ % Trippe+11 \\
 MCG -05-23-016 
 & $ 1.25^{+0.29}_{-0.18} $ & 2000-11-14 & {\it Chandra} & 8 \\ % Balestra+04 \\
 & $ 1.94^{+0.38}_{-0.40} $ & 2001-05-13 & {\it XMM-Newton} & 8 \\ % Balestra+04 \\
 & $ 1.80\pm0.23 $        & 2001-12-01 & {\it XMM-Newton} & 8 \\ % Balestra+04 \\
 & $ 1.49\pm0.01 $        & 2013-06-01 & {\it Suzaku} & this work \\ % \\
 & $ 1.50\pm0.01 $        & 2013-06-05 & {\it Suzaku} & this work \\ %  \\
 Mrk 1210 
 & $ 17.8^{+7.8}_{-7.9} $ & 2001-05-05 & {\it XMM-Newton} & 9 \\ % Awaki+06 \\
 & $ 29.6^{+1.8}_{-1.7} $ & 2008-02-15 & {\it Chandra} & 10 \\ % Risaliti+10 \\
 & $ 25.5^{+3.3}_{-2.9} $ & 2008-02-17 & {\it Chandra} & 10 \\ % Risaliti+10 \\
 & $ 37.6^{+4.4}_{-4.6} $ & 2008-03-06 & {\it Chandra} & 10 \\ % Risaliti+10 \\
 Mrk 18 & $ 18.3^{+7.2}_{-5.7} $ & 2006-03-23 & {\it XMM-Newton} & 1 \\ % Winter+08 \\
 Mrk 348 
 & $ 13.4^{+0.20}_{-0.46} $ & 2002-07-18 & {\it XMM-Newton} & 11 \\ % Hernandez-Garcia+15 \\
 & $ 12.9\pm1.2 $ & 2013-01-04 & {\it XMM-Newton} & 11 \\ % Hernandez-Garcia+15 \\
 Mrk 417 & $ 54^{+25}_{-11} $ & 2006-06-15 & {\it XMM-Newton} & 1 \\ % Winter+08 \\
 NGC 1052 & $ 13.8^{+2.0}_{-1.8} $ & 2001-08-15 & {\it XMM-Newton} & 12 \\ % Hernandez-Garcia+14 \\
 & $ 5.3\pm1.5 $ & 2005-09-18 & {\it Chandra} & 12 \\ % Hernandez-Garcia+14 \\
 & $ 9.30^{+0.52}_{-0.51} $ & 2006-01-12 & {\it XMM-Newton} & 12 \\ % Hernandez-Garcia+14 \\
 & $ 8.96^{+0.43}_{-0.42} $ & 2009-01-14 & {\it XMM-Newton} & 12 \\ % Hernandez-Garcia+14 \\
 & $ 9.47^{+0.39}_{-0.38} $ & 2009-08-12 & {\it XMM-Newton} & 12 \\ % Hernandez-Garcia+14 \\
 NGC 1142 
& $ 47.0^{+3.5}_{-3.2} $ & 2006-01-28 & {\it XMM-Newton} & 13 \\ % LaMassa+12 \\
& $ 73.9^{+7.9}_{-7.0} $ & 2007-07-21  & {\it Suzaku} & this work \\ % \\ 
 NGC 2110 
 & $ 4.0\pm1.8 $ & 2001-12-19 & {\it Chandra} & 14 \\ % Marinucci+15 \\
 & $ < 4.5 $             & 2003-03-05 & {\it Chandra} & 14  \\ % Marinucci+15 \\
 & $ 3.90\pm0.4 $ & 2003-03-05 & {\it XMM-Newton} & 14 \\ % Marinucci+15 \\
 & $ 2.53\pm0.13 $ & 2012-08-31 & {\it Suzaku} & this work \\  
 & $ 4.0\pm0.4 $ & 2012-10-05 & {\it NuSTAR} & 14 \\ % Marinucci+15 \\
 & $ 4.0\pm0.7 $ & 2013-02-14 & {\it NuSTAR} & 14 \\ % Marinucci+15 \\
 NGC 4388 
 & $ 25.6^{+3.1}_{-2.9} $ & 2001-06-08 & {\it Chandra} & 13 \\ % LaMassa+12 \\
 & $ 24.3^{+1.1}_{-1.0} $ & 2002-12-12 & {\it XMM-Newton} & 13 \\ % LaMassa+12 \\
 NGC 4507 
 & $ 42.8^{+0.9}_{-0.7} $   & 2001-01-04 & {\it XMM-Newton} & 3 \\ % Noguchi+09 \\
 & $ 90\pm10 $ & 2005-07-25 & {\it Chandra} & 15 \\ % Sazonov+05 \\
 & $ 68.5^{+14.9}_{-9.6} $ & 2006-06-27 & {\it XMM-Newton} & 1 \\ % Winter+08 \\
 & $ 87^{+7}_{-8} $   & 2010-06-24 & {\it XMM-Newton} & 16 \\ % Marinucci+13 \\
 & $ 97\pm9 $ & 2010-07-03 & {\it XMM-Newton} & 16 \\ % Marinucci+13 \\
 & $ 76^{+10}_{-13} $ & 2010-07-13 & {\it XMM-Newton} & 16 \\ % Marinucci+13 \\
 & $ 94\pm11 $ & 2010-07-23 & {\it XMM-Newton} & 16 \\ % Marinucci+13 \\
 & $ 80^{+8}_{-6} $   & 2010-08-03 & {\it XMM-Newton} & 16 \\ % Marinucci+13 \\
 & $ 65\pm7 $   & 2010-12-02 & {\it Chandra} & 16 \\ % Marinucci+13 \\
 NGC 5252 & $ 2.32^{+0.13}_{-0.15} $ & 2003-08-11 & {\it Chandra} & 17 \\ % Dadina+10 \\
 NGC 526A & $ 1.14\pm0.26 $ & 2003-06-21 & {\it XMM-Newton} & 18 \\ % Brightman+11 \\
 NGC 5506 
 & $ 2.69^{+0.02}_{-0.03} $ & 2004-07-11 & {\it XMM-Newton} & 13 \\ % LaMassa+12 \\
 & $ 2.80^{+0.01}_{-0.02} $ & 2004-08-07 & {\it XMM-Newton} & 13 \\ % LaMassa+12 \\
 & $ 3.09\pm0.03 $ & 2006-08-08 & {\it Suzaku} & this work \\ %  \\
 & $ 3.16\pm0.03 $ & 2007-01-31 & {\it Suzaku} & this work \\ %  \\
 NGC 6300 
 & $ 25.4^{+4.3}_{-3.7} $ & 2001-03-02 & {\it XMM-Newton} & 11 \\ % Hernandez-Garcia+15 \\
 & $ 14.1^{+1.3}_{-2.0} $ & 2009-06-10 & {\it Chandra} & 11 \\ % Hernandez-Garcia+15 \\
 & $ 19.8^{+1.4}_{-2.7} $ & 2009-06-14 & {\it Chandra} & 11 \\ % Hernandez-Garcia+15 \\
 NGC 7172 
 & $ 8.45^{+0.36}_{-0.33} $ & 2002-11-18 & {\it XMM-Newton} & 11 \\ % Hernandez-Garcia+15 \\
 & $ 8.75\pm0.27 $ & 2004-11-11 & {\it XMM-Newton} & 11 \\ % Hernandez-Garcia+15 \\
 & $ 8.34^{+0.16}_{-0.15} $ & 2007-04-24 & {\it XMM-Newton} & 11 \\ % Hernandez-Garcia+15 \\
 NGC 788 
 & $ 44.4^{+8.7}_{-7.8} $ & 2009-09-06 & {\it Chandra} & 11 \\ % Hernandez-Garcia+15 \\
 & $ 50.3^{+6.1}_{-5.7} $ & 2010-01-15 & {\it XMM-Newton} & 11  % Hernandez-Garcia+15 
\enddata
\tablecomments{
(1) Galaxy name.
(2) Hydrogen column density of the neutral full-covering absorption model. Errors correspond to the 
90\% confidence interval. 
The confidence level of the errors compiled from \cite{Her14}, \cite{Her15}, \cite{LaM12}, 
and \cite{Saz05}, is not clear because it is not described.
(3) Observation date. 
(4) Observatory. 
(5) References for $N_{\rm H}$.\\
{\bf References.} 
(1) \cite{Win08}. 
(2) \cite{Vas13}. 
(3) \cite{Nog09}. 
(4) \cite{deR12}. 
(5) \cite{Tri13}. 
(6) \cite{Lob14}. 
(7) \cite{Tri11}. 
(8) \cite{Bal04}. 
(9) \cite{Awa06}. 
(10) \cite{Ris10}. 
(11) \cite{Her15}. 
(12) \cite{Her14}. 
(13) \cite{LaM12}. 
(14) \cite{Mar15}. 
(15) \cite{Saz05}. 
(16) \cite{Mar13}. 
(17) \cite{Dad10}. 
(18) \cite{Bri11a}. 
}
\end{deluxetable*}
%\clearpage
%\end{landscape}

%\fi %%%%%%%%%%%%%%%%%%%%%%%%%%%%%%%%%%%%%%

% \input{info_para.tex} % IRU 

% \input{info_nh.tex}
% \input{disk_para.tex} % IRU 

\end{document}